\documentclass[twoside,openright, titlepage, numbers=noenddot, headinclude, letterpaper, footinclude=true,cleardoublepage=empty, american,fontsize=11pt]{scrreprt} 

\newcommand{\mbf}{\boldsymbol}
\newcommand{\IntSum}{\ensuremath{\sum\!\!\!\!\!\!\! \int }} 

\makeatletter
\DeclareOldFontCommand{\rm}{\normalfont\rmfamily}{\textrm}
\DeclareOldFontCommand{\sf}{\normalfont\sffamily}{\textsf}
\DeclareOldFontCommand{\tt}{\normalfont\ttfamily}{\texttt}
\DeclareOldFontCommand{\bf}{\normalfont\bfseries}{\textbf}
\DeclareOldFontCommand{\it}{\normalfont\itshape}{\textit}
\DeclareOldFontCommand{\sl}{\normalfont\slshape}{\@nomath\sl}
\DeclareOldFontCommand{\sc}{\normalfont\scshape}{\@nomath\sc}
\makeatother

\newcommand{\tj}[6]{ \begin{pmatrix}
  #1 & #2 & #3 \\
  #4 & #5 & #6 
\end{pmatrix}}

\newcommand{\Gj}[6]{ \begin{Bmatrix}
  #1 & #2 & #3 \\
  #4 & #5 & #6 
\end{Bmatrix}}

\PassOptionsToPackage{eulerchapternumbers,listings, dottedtoc,%
				 pdfspacing,
				 subfig,parts}{classicthesis}										

\usepackage{ifthen}
\newboolean{enable-backrefs} 
\setboolean{enable-backrefs}{true} 
\usepackage{lmodern}
\usepackage{braket}
\usepackage{chngcntr}
\usepackage{epstopdf}
\usepackage{graphics}

\newcommand{\myTitle}{Information, Complexity, and Quantum Entanglement on Doubly Excited States of Helium Atom\xspace}
\newcommand{\mySubtitle}{A topological quantum correlation analysis of two-electron atoms.\xspace}
\newcommand{\myDegree}{Master of Science (M.Sc.) Physics Thesis\xspace}
\newcommand{\myName}{Juan Pablo Restrepo Cuartas\xspace}

\newcommand{\mySupervisor}{Advisor: Prof. Dr. Jos\'e Luis Sanz-Vicario\xspace}
\newcommand{\myFaculty}{Facultad de Ciencias Exactas y Naturales\xspace}
\newcommand{\myDepartment}{Instituto de F\'isica\xspace}
\newcommand{\myUni}{Universidad de Antioquia\xspace}

\newcommand{\myTime}{November -- 2013\xspace}
\newcommand{\myVersion}{}

\newcounter{dummy} 
\providecommand{\mLyX}{L\kern-.1667em\lower.25em\hbox{Y}\kern-.125emX\@}

\PassOptionsToPackage{latin9}{inputenc}	
 \usepackage{inputenc}				

 \usepackage{babel}					

\PassOptionsToPackage{square}{natbib}
 \usepackage{natbib}				

\PassOptionsToPackage{fleqn}{amsmath}		
 \usepackage{amsmath}

\PassOptionsToPackage{T1}{fontenc} 
	\usepackage{fontenc}                 
\usepackage{xspace} 
\usepackage{mparhack} 
\usepackage{fixltx2e} 
\PassOptionsToPackage{printonlyused,smaller}{acronym}
	\usepackage{acronym} 

\usepackage{tabularx} 
\usepackage{caption}
\usepackage{subfig}  

\usepackage{listings} 
\PassOptionsToPackage{hyperfootnotes=false,pdfpagelabels}{hyperref}
	\usepackage{hyperref}  
\PassOptionsToPackage{pdftex}{graphicx}
	\usepackage{graphicx} 

\newcommand{\backrefnotcitedstring}{\relax}
\newcommand{\backrefcitedsinglestring}[1]{(Cited on page~#1.)}
\newcommand{\backrefcitedmultistring}[1]{(Cited on pages~#1.)}
\ifthenelse{\boolean{enable-backrefs}}%
{%
		\PassOptionsToPackage{hyperpageref}{backref}
		\usepackage{backref} 
		   \renewcommand*{\backref}[1]{}  
		   \renewcommand*{\backrefalt}[4]{
		      \ifcase #1 %
		         \backrefnotcitedstring%
		      \or%
		         \backrefcitedsinglestring{#2}%
		      \else%
		         \backrefcitedmultistring{#2}%
		      \fi}%
}{\relax}    

\hypersetup{%
    colorlinks=true, linktocpage=true, pdfstartpage=3, pdfstartview=FitV,%
    breaklinks=true, pdfpagemode=UseNone, pageanchor=true, pdfpagemode=UseOutlines,%
    plainpages=false, bookmarksnumbered, bookmarksopen=true, bookmarksopenlevel=1,%
    hypertexnames=true, pdfhighlight=/O,
    urlcolor=webbrown, linkcolor=RoyalBlue, citecolor=webgreen, 
    pdftitle={\myTitle},%
    pdfauthor={\textcopyright\ \myName, \myUni, \myFaculty},%
    pdfsubject={},%
    pdfkeywords={},%
    pdfcreator={pdfLaTeX},%
    pdfproducer={LaTeX with hyperref and classicthesis}%
}   

\makeatletter
\@ifpackageloaded{babel}%
    {%
       \addto\extrasamerican{%
				}%
       \addto\extrasngerman{%
				}%
    }{\relax}
\makeatother

\usepackage{classicthesis} 
\numberwithin{equation}{chapter}

\usepackage[left=1.5in,right=1.5in,top=1in,bottom=1in]{geometry}

\usepackage{multirow}

\usepackage{tikz}
\usepackage{mathtools}
\usetikzlibrary{calc,through,backgrounds,shapes,snakes,decorations.shapes}

\begin{document}
\counterwithin{figure}{chapter}
\counterwithin{table}{chapter}
\frenchspacing
\raggedbottom
\selectlanguage{american} 
\pagenumbering{roman}
\pagestyle{plain}
\thispagestyle{empty}
\begin{center}
    \spacedlowsmallcaps{\myName} \\ \medskip                        

    \begingroup
        \color{Maroon}\spacedallcaps{\myTitle}
    \endgroup
\end{center}        

\begin{titlepage}
	\begin{addmargin}[-1cm]{-3cm}
    \begin{center}
        \large  

        \hfill

        \vfill

        \begingroup
            \spacedallcaps{\myTitle} \\ \bigskip
        \endgroup

        \spacedlowsmallcaps{\myName}

        \vfill

        \includegraphics[width=15cm]{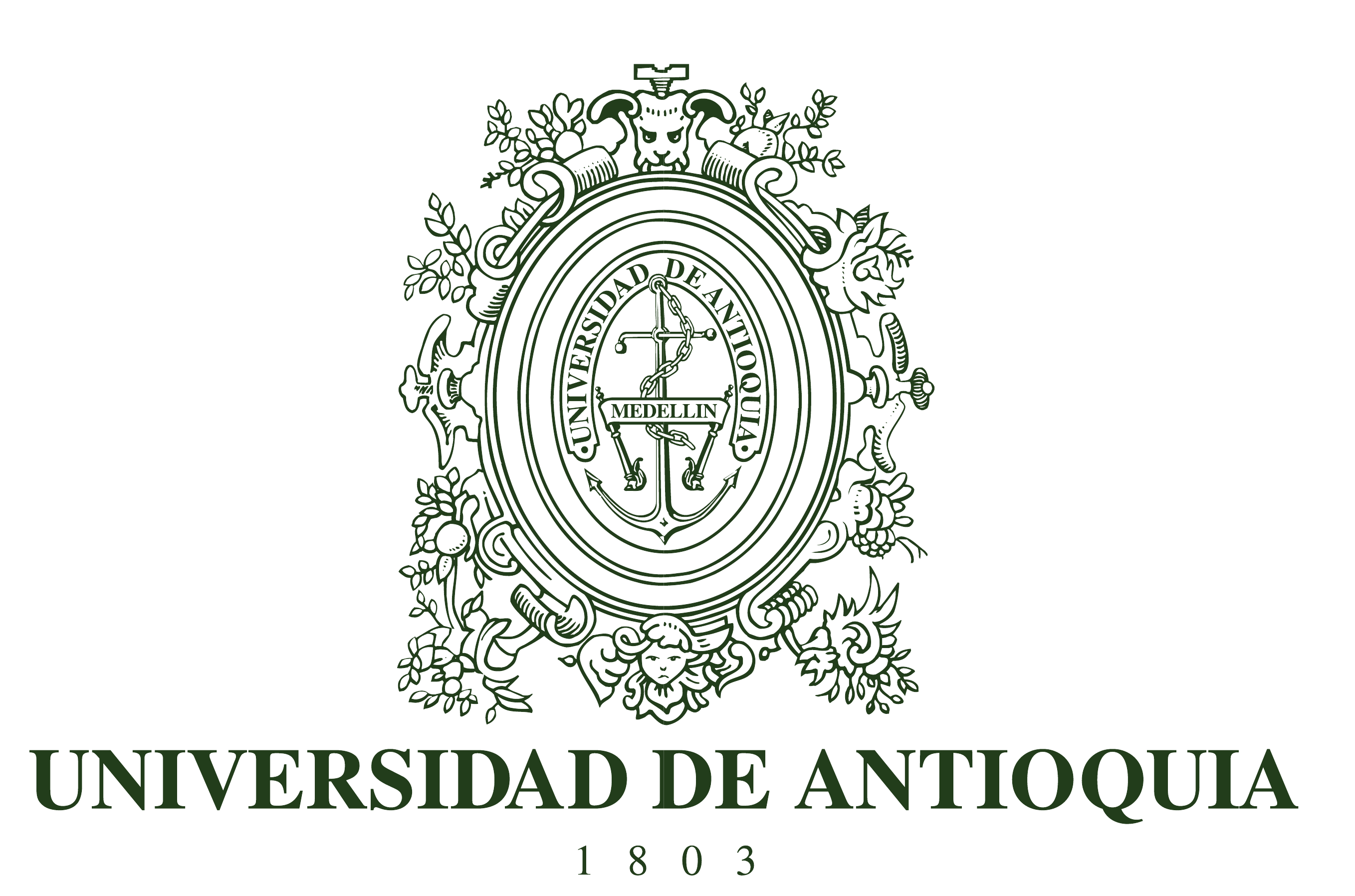} \\ \medskip
  \vfill
  \vfill
        \mySubtitle \\ \medskip   
         \vfill
         \vfill
        \myDegree \\
        \myDepartment \\                            
        \myFaculty \\
        \myUni \\ \bigskip

        \myTime\  \myVersion\\
          \vfill
          \vfill    
        \mySupervisor\\
        \vfill                      

    \end{center}  
  \end{addmargin}       
\end{titlepage}   
\thispagestyle{empty}

\hfill

\vfill

\noindent\myName: \textit{\myTitle,} \mySubtitle, 
\textcopyright\ \myTime

%
%
%
%
%

\cleardoublepage
\pdfbookmark[1]{Abstract}{Abstract}
\begingroup
\let\clearpage\relax
\let\cleardoublepage\relax
\let\cleardoublepage\relax

\chapter*{Abstract}
The electronic density $\rho({r})$ in atoms, molecules and solids is, in general, a distribution that can be observed experimentally, containing
spatial information projected from the total wave function. These density distributions can be though as probability distributions subject to the scrutiny of the analytical methods of information theory, namely, entropy measures, quantifiers for the complexity, or entanglement measures. Resonant
states in atoms have special properties in their wave functions, since although they pertain to the scattering continuum spectrum, they show
a strong localization of the density in regions close to the nuclei. Although the classification of resonant doubly excited states of He-like atoms in terms of labels of approximate quantum numbers have not been exempt from controversies, a well known proposal follows after the works by~\citep{Herrick1975,Lin1983}, with a labeling based on $K$, $T$, and $A$ numbers in the form $_{n_1}(K, T)^A_{n_2}$ for the Rydberg series of increasing $n_2$ and for a given ionization threshold He$^+$ ($N=n_1$).

In this work we intend to justify this kind of classification from the topological analysis of the one-particle $\rho({r})$ and two-particle $\rho(r_1, r_2)$ density distributions of the localized part of the resonances (computed with a Feshbach projection formalism and configuration interaction wave functions in terms of B-splines bases), using global quantifiers (Shannon) as well as local ones (Fisher information)~\citep{Lopez-Rosa2005,Lopez-Rosa2009,Lopez-Rosa2010}. For instance, the Shannon entropy is obtained after global integration of the density and the Fisher information contains local information on the gradient of the distribution. In addition, we also studied measures for the entanglement using the von Neumann and linear entropies~\citep{Manzano2010,Dehesa2012a,Dehesa2012b}, computed from the reduced one-particle density
matrix within our correlated configuration interaction approach.

We find in this study that global measures like the Shannon entropy hardly distinguishes among resonances in the whole Rydberg series. On the contrary, measures like the Fisher information, von Neumann and linear entropies are able to qualitatively discriminate the resonances, grouping them according to their $(K, T)^A$ labels.

\endgroup			

\vfill
\cleardoublepage

\begin{flushright}{\slshape    
We want to stand upon our own feet and look fair and square at the world -- its good facts, its bad facts, its beauties, and its ugliness; see the world as it is and be not afraid of it. Conquer the world by intelligence and not merely by being slavishly subdued by the terror that comes from it. The whole conception of God is a conception derived from the ancient Oriental despotisms. It is a conception quite unworthy of free men. When you hear people in church debasing themselves and saying that they are miserable sinners, and all the rest of it, it seems contemptible and not worthy of self-respecting human beings. We ought to stand up and look the world frankly in the face. We ought to make the best we can of the world, and if it is not so good as we wish, after all it will still be better than what these others have made of it in all these ages. A good world needs knowledge, kindliness, and courage; it does not need a regretful hankering after the past or a fettering of the free intelligence by the words uttered long ago by ignorant men. It needs a fearless outlook and a free intelligence. It needs hope for the future, not looking back all the time toward a past that is dead, which we trust will be far surpassed by the future that our intelligence can create.} \medskip
    --- \defcitealias{Russel1957}{Bertrand Russel}\citetalias{Russel1957} \citep{Russel1957}
\end{flushright}

\pdfbookmark[1]{Acknowledgments}{acknowledgments}

\bigskip

\begingroup
\let\clearpage\relax
\let\cleardoublepage\relax
\let\cleardoublepage\relax
\chapter*{Acknowledgments}
Foremost, I would like to express my sincere gratitude to my advisor Prof. Ph.D. Jos\'e Luis Sanz Vicario for the continuous support of my master study and research, for his patience, motivation, and immense knowledge and humanity. His guidance helped me in all the time of research and writing of this thesis.  In addition to my advisor, I would like to thank the rest of my thesis committee: Prof. Ph.D. Julio C\'esar Arce Clavijo and  Prof. Ph.D.  Karen Milena Fonseca Romero, for their insightful comments, and hard questions.

Thanks a lot to my friends in the UdeA Atomic and Molecular Physics Group: Fabiola G\'omez, Guillermo Guirales, Boris Rodr\'iguez, Alvaro Valdez, Leonardo Pachon,  Andr\'es Estrada, Carlos Florez, Melisa Dom\'inguez, Juliana Restrepo, Sebasti\'an Duque, Juan David Botero, Johan Tirana, Jairo David Garcia, for the stimulating discussions and for all the fun we have had in the last years.  Many thanks to Herbert Vinck and Ligia Salazar for their selfless friendship. 

My sincere thanks also goes to Ph.D. Juan Carlos Angulo Ib\'a\~nes of the University of Granada (Spain), for the internship opportunity in his group and leading me working on information theory measures.

I would like to express my sincere gratitude to the sponsors that make possible the Spain internship: The Cooperativa de Ahorro y Cr\'edito G\'omez Plata, the AUIP Asociaci\'on Iberoamericana de Posgrados, and the GFAM Group.

My deepest gratitude to Sergio Palacio. I greatly value his close friendship and I deeply appreciate his belief in me. I am also grateful with Natalia Bedoya, his girlfriend. Their constant support and care helped me overcome setbacks and stay focused on the very relevant things of life.

Finally, I would like to thank all my family: Fernando, David, M\'onica, Astrid, Rogelio and Daniela, and, Maria Fernanda; especially to my parents Pedro Restrepo  and Marina Cuartas, for giving birth to me at the first place and supporting me material and spiritually throughout my life.

\endgroup

\pagestyle{scrheadings}
\cleardoublepage
\refstepcounter{dummy}
\pdfbookmark[1]{\contentsname}{tableofcontents}
\setcounter{tocdepth}{2} 
\setcounter{secnumdepth}{3} 
\manualmark
\markboth{\spacedlowsmallcaps{\contentsname}}{\spacedlowsmallcaps{\contentsname}}
\tableofcontents 
\automark[section]{chapter}
\renewcommand{\chaptermark}[1]{\markboth{\spacedlowsmallcaps{#1}}{\spacedlowsmallcaps{#1}}}
\renewcommand{\sectionmark}[1]{\markright{\thesection\enspace\spacedlowsmallcaps{#1}}}
\clearpage

\begingroup 
    \let\clearpage\relax
    \let\cleardoublepage\relax
    \let\cleardoublepage\relax
    \refstepcounter{dummy}
    \pdfbookmark[1]{\listfigurename}{lof}
    \listoffigures

    \vspace*{8ex}

    \refstepcounter{dummy}
    \pdfbookmark[1]{\listtablename}{lot}
    \listoftables
        
    

    \vspace*{8ex}
       
    \refstepcounter{dummy}
    \pdfbookmark[1]{Acronyms}{acronyms}
    \markboth{\spacedlowsmallcaps{Acronyms}}{\spacedlowsmallcaps{Acronyms}}
    \chapter*{Acronyms}
    \begin{acronym}[HE]
        \acro{DES}{\textit{Doubly Excited States}}    
        \acro{CI}{\textit{Con\-fi\-gu\-ra\-tion Interaction}}
        \acro{SH}{\textit{Spherical Harmonics}}
        \acro{UML}{\textit{Unified Modeling Language}}
        \acro{bp}{\textit{Breakpoint}}
        \acro{pp}{\textit{Piecewise Polynomial}}
        \acro{a.u.}{\textit{Atomic Units}}
        \acro{WF}{\textit{Wave Function}}
        \acro{FM}{\textit{Feshbach Method}}
        \acro{HSL}{\textit{Herrick-Sinano\u{g}lu-Lin}}
        \acro{CGC}{\textit{Clebsch-Gordan Coefficients}}
        \acro{TISE}{\textit{Time Independent Schr\"odinger Equation}}
        \acro{HF}{\textit{Hartree-Fock}}
        \acro{DFT}{\textit{Density Functional Theory}}
        \acro{EPR}{\textit{Einstein-Podolsky-Rosen}}
    \end{acronym}                     
\endgroup

\cleardoublepage
\pagenumbering{arabic}
\cleardoublepage
P
\chapter{Introduction}\label{ch:introduction}
\section{\label{sec:pre}Preliminaries}

\begin{figure}[h]
\includegraphics[width=\textwidth]{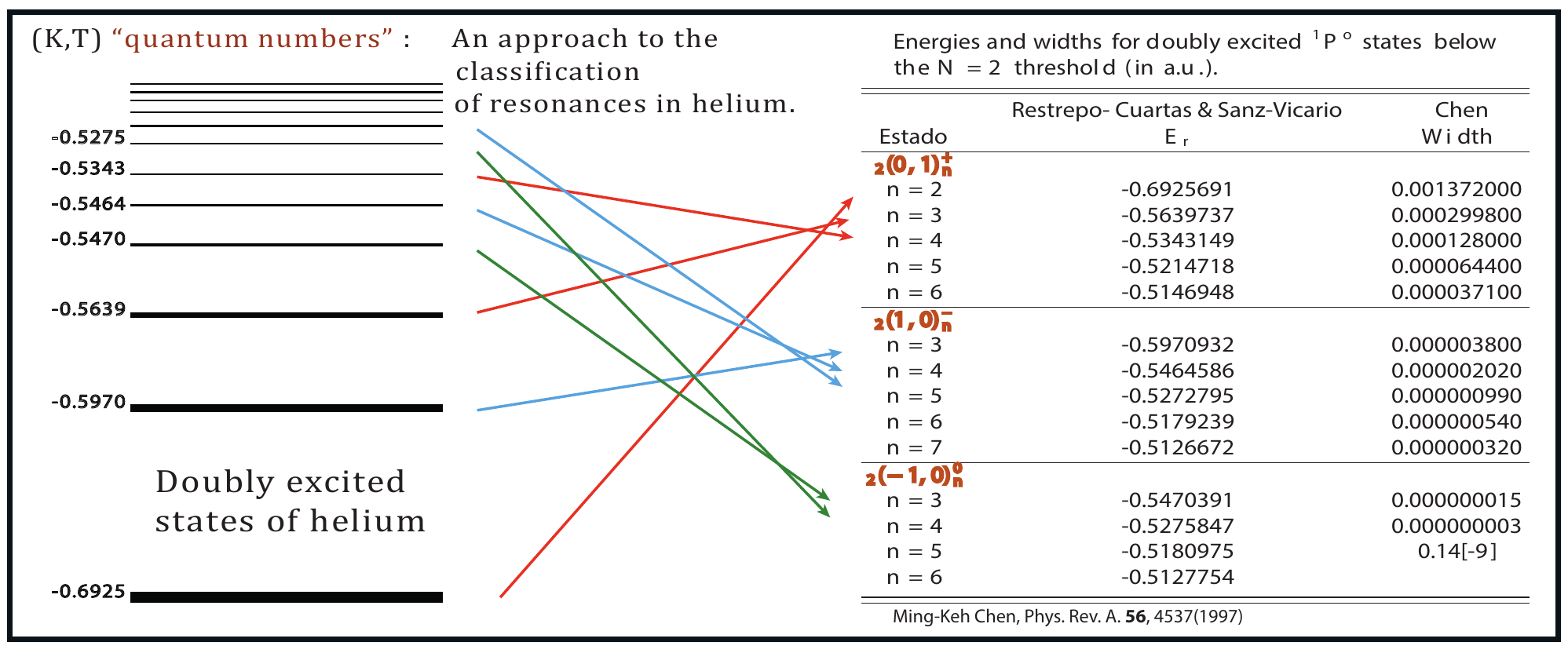}
\caption[Approximate  $(K,T)$ ``quantum numbers'' for doubly excited states. ]{\label{fig:clasification}The alternative and approximate $(K,T)$  ``quantum numbers'' that was proposed, by~\citep{Herrick1975,Lin1983}, as an attempt to classify  the \ac{DES} in helium atom.}
\end{figure}

The central physical phenomenon which lies behind all this work is called {\textit autoionization}. The very first sighting of this phenomenon dates back to 1935 by~\citeauthor{Beutler1935} in the context of the photoabsorption of rare gases atoms, and later again by ~\citep{Madden1963}  in the photoabsorption of atomic helium.  The modern understanding of autoionization considers  a sort of discrete states which are embedded in the continuum, these states are called resonances and they are genuine eigenstates of the interelectronic repulsion $1/r_{12}$. Even though each resonance may be characterized and described by three parameters: the energy position $E_r$, the shape parameter $q$ and the resonance width, a completely satisfactory classification scheme of resonances is still unavailable. The subject of classification of  \ac{DES} or resonances has been characterized by controversy and general disagreement. This work pretends to study the so called $(K,T)$ classification scheme of \ac{DES}~\citep{Herrick1975,Lin1983} by analysing  the topological features of the electronic density by means of measures of information theory, see figure~\ref{fig:clasification}. 

\subsection{\label{sec:resonanceselem}Resonances in elementary quantum mechanics: Resonance scattering from a double $\delta$-function potential}

In order to obtain a deep understanding on the concept of resonant state we may introduce a straightforward example  commonly encountered in almost all introductory course of quantum mechanics. Under certain conditions the scattering of a particle by a one-dimensional square potential barrier exhibits resonant behavior. It is also well known that the transmission coefficient  a one-dimensional double $\delta$-function potential also exhibits resonances at a series of energy values~\citep{Lapidus1982}. The problem involves the \ac{TISE}

\begin{equation}\label{eq:schroeq}
-\frac{\hbar^2}{2m}\frac{\partial^2 \psi(x)}{\partial x^2}+V(x)\psi(x)=E\psi(x),
\end{equation}
where $V(x)=V_0(\delta(x+a)+\delta(x-a))$ with $V_0$ an appropriate constant and $2a$ the separation between the barriers. The solution of the \ac{TISE} with this potential can be written as 
\begin{subequations}
\begin{align}\label{eq:ressol}
\psi_-&=e^{ikx}+re^{-ikx}\hspace{24pt}x<-a \\ 
\psi_0&=Ae^{ikx}+Be^{-ikx}\hspace{11pt}|x|\le a \\ 
\psi_+&=te^{ikx}\hspace{60pt}x> a
\end{align}
\end{subequations}
where the constants $A$, $B$,  $r$, and $t$ are determined by the boundary conditions
\begin{subequations}\label{eq:boundarycond}
\begin{align}
\psi_0(a)&=\psi_+(a), \\ 
\psi_-(-a)&=\psi_0(-a), 
\end{align}
\end{subequations}
and
\begin{subequations}\label{eq:derboundary}
\begin{align}
\psi_+^{\prime}(a)-\psi_0^{\prime}(a)=&(-\frac{2}{a_0})\psi_+(a), \\ 
\psi_0^{\prime}(-a)-\psi_-^{\prime}(-a)=&(-\frac{2}{a_0})\psi_-(-a),
\end{align}
\end{subequations}
where $a_0=\hbar^2/mV_0$.  By using  the equations~\eqref{eq:boundarycond} and~\eqref{eq:derboundary} we can eliminate the constants $A$ and $B$ and then solve for the amplitudes for transmission and reflection $r$ and $t$. The transmission and reflection coefficients are $T=|t|^2$ and $R=|r|^2$, where $R+T=1$. Finally, the transmission amplitude can be written as 

\begin{equation}\label{eq:transmission}
t=\frac{a_0^2k^2}{a_0^2k^2-2ia_0k+e^{4iak}-1}.
\end{equation}

\begin{figure}[h]
\includegraphics[width=\textwidth]{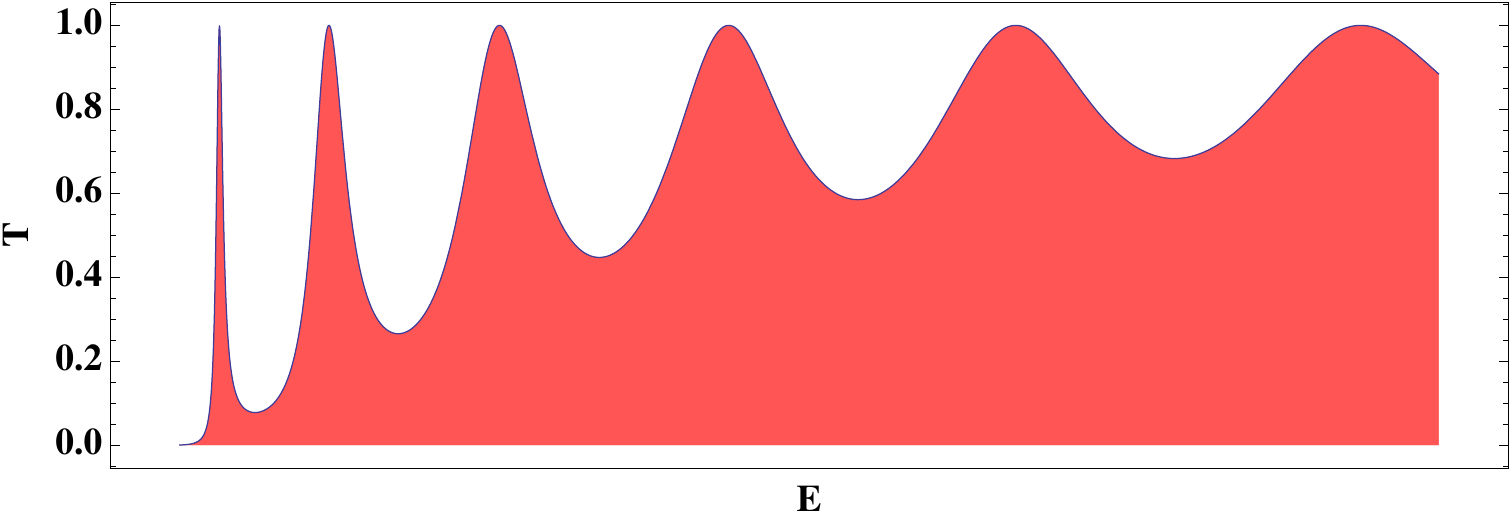}
\caption[Resonances in  a double $\delta$-function potential.]{\label{fig:elemres}An elementary example of resonances in the scattering form a double $\delta$-function potential. The transmission coefficient $T$ exhibits resonant behavior  for some values of energy.}
\end{figure}

The figure~\ref{fig:elemres} plots the transmission coefficient $T$ as a function of the energy $E=\hbar^2k^2/2m$. This illustrative and short discussion provides an intuitive picture of what is the meaning of a resonance state in quantum mechanical contexts. 

\section{\label{sec:outlook}Outlook}

This work is built up with three parts, five chapters and three appendices. The first part, composed by one chapter, is dedicated to the stationary calculation of the electronic structure, i.e., the calculation of energies and the two-electron \ac{CI}-\ac{WF} of  two-electron atoms using the \ac{TISE}. In chapter~\ref{ch:doublyhelium}  we describe the two-electron atoms and the methods that we have implemented to calculate their electronic structure, based on the construction of the two-electron \ac{CI}-\ac{WF} with the help of one-particle \ac{WF} which are obtained in the appendix~\ref{ch:hybsplines}; and particulary, we review the application of  \ac{FM} to obtain de \ac{DES} states of helium atom.

The second part, which is composed of two chapters  is dedicated to the topological description of the electronic density by means of measures of information theory.  In chapter~\ref{ch:tim} we calculate the two-dimensional electronic density for two-electron atoms and by integration over one radial dimension  we obtain the one-particle density. Additionally we introduce the Shannon entropy and the Fisher information as two measures of information theory in order to explore their topological implications over the densities of \ac{DES}. The chapter~\ref{ch:entanglement} is dedicated to the study of quantum entanglement in two-electron systems. There we will introduce its definition and the quantities which measure the amount of entanglement in helium. Finally we discuss the possible implications of the amount of entanglement in \ac{DES} and their classification.

The third part, in chapter~\ref{ch:conclusions} summarizes our conclusions and perspectives for future work. 

Finally, the appendices contain a detailed description of the numerical basis implemented in this work: B-splines, as well as some details of the analytical and numerical calculation of electronic structure of one-electron atoms, focussing only at bound states. Some supplementary figures are shown in appendix~\ref{ch:supgraphics}. 


\part{Two-electron systems: Stationary Approach}

\chapter{Doubly Excited States of Helium}\label{ch:doublyhelium}

\begin{figure}[h]
\centering
\includegraphics[width=0.75\textwidth]{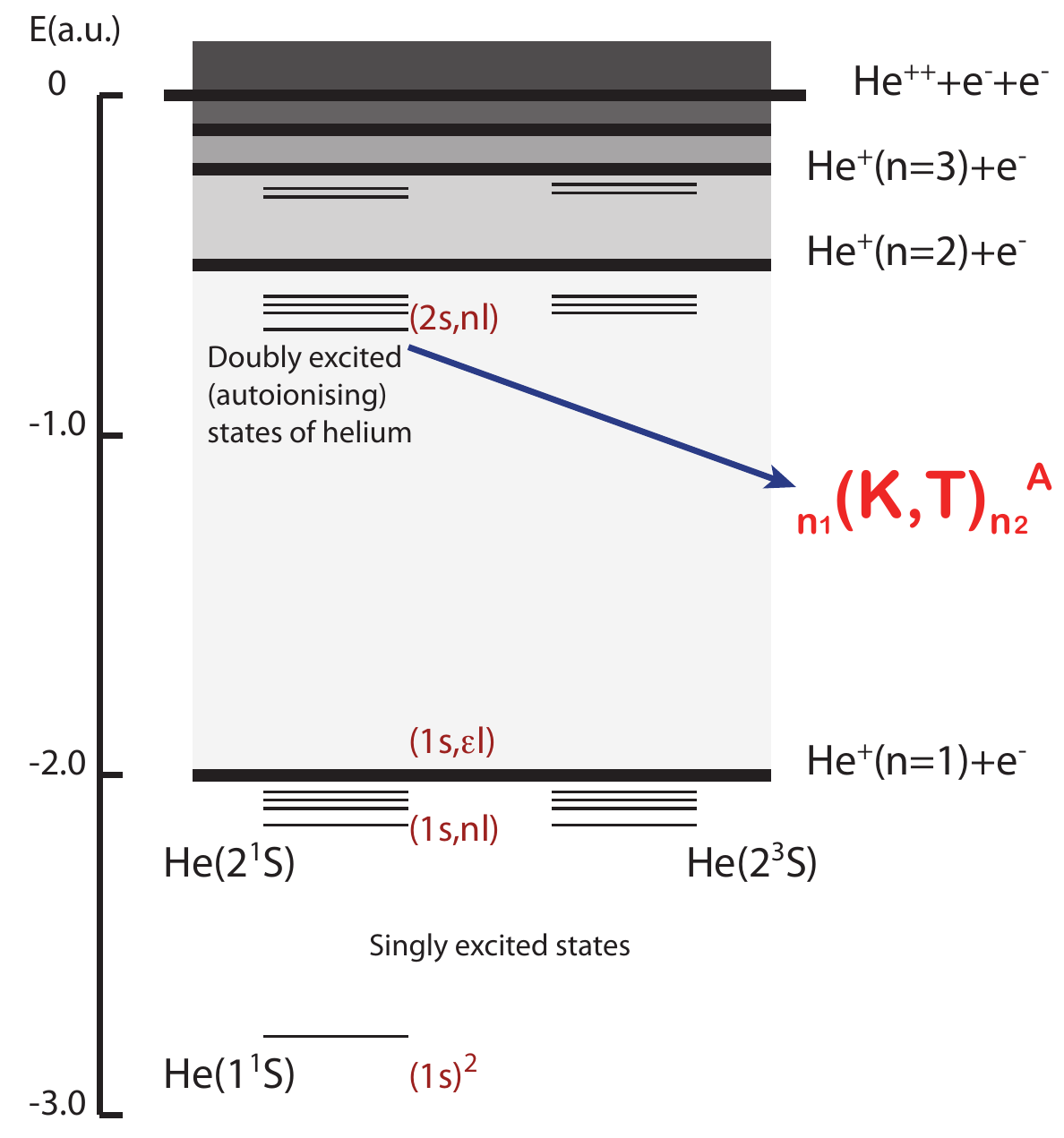}
\caption[The energy level spectrum of helium.]{\label{fig:hestates}The energy level spectrum of helium. The ground state $(1s^2)$ and the Rydberg series of singly excited states $(1s,nl)$ lie below the first ionization threshold (He$^+ (n=1) + e^-$)  with energy $E= -2.0$ a.u. Above this threshold and below the second ionization threshold (He$^+ (n=2) + e^-$) with energy $E= -0.5$ a.u. a Rydberg series of doubly excited states (naively denoted as $(2 l_1, n_2 l_2)$ and more properly with the~\citep{Lin1983, Lin1984} set of labels $_{n_1}(K,T)_{n_2}^A$. These \ac{DES} or resonances are immersed into the electronic continuum $(1s, \epsilon l)$ and they are degenerated to it, which ultimately causes its irreversible decay into the continuum once they become populated. A full Rydberg series of resonances appear just below each ionization threshold of the parent ion He$^+ (n),\hspace{5pt}2<n<\infty$.}
\end{figure}

The description of the electronic eigenspectrum of a two-electron atom involves, in addition to the bound and continuum states, another kind of quasibound eigenstates of the Hamiltonian. Similar to the infinite series of single excited states located below the first ionization threshold, Rydberg series of discrete \ac{DES} appear below the upper continuum thresholds associated to excited target configurations (for instance, He$^+$ $(n=2,3,4,\dots)+ e^-$ in the case of \ac{DES} in helium (see figure~\ref{fig:hestates}), i.e., these states are genuinely immersed in the continuum but share properties similar to the bound states (discrete energies and spatial quasi-localization). Since these quasibound states are coupled to the underlying continuum, they are indeed metastable states characterized with a finite lifetime \citep{Friedrich2005}. In principle, there is no external perturbation responsible for the decay of these metastable states into the degenerated continua (for instance, in helium, the decay from a \ac{DES} produces a final ionizing state, i.e.,  He$^{**} \to$ He$^+ + e^-$, a process which is called autoionization), but it is intrinsic to the two-electron Hamiltonian, in particular the electron correlation term $1/r_{12}$. 
Thus the good account of the properties of doubly excited states in two-electron atoms, also called resonances or autoionizing states, thoroughly depend on the proper description of the electron correlation. Consequently, these states and their properties cannot be described using simple models, like the independent particle model.  The actual existence of \ac{DES} was put forward experimentally by~\citep{Madden1963} and the first comprehensive theory of resonance phenomena was proposed by  \citep{Fano1961,Fano1983}. From then,  a vast amount of effort has been expended in the understanding of the atomic resonance phenomena and, in particular, the characterization of the resonance states (energy positions, lifetimes, time-dependent decay and its contribution to photoionization cross section,  electron distributions, classification using quantum labels, etc.)

Determining resonance energies and widths (or lifetimes) in atoms has been the subject of intense study and research for decades. Nevertheless, the classification of resonances within the same Rydberg series in terms of labels corresponding to approximate quantum numbers has been a matter of great controversy~\citep{Cooper1963, Nikitin1976, Lin1983,Lin1984,Lin1984a}. Indeed, at variance with the bound state, mostly labeled $(1s,1s)$ and singly excited states, easily labelled $(1s,nl)$, any attempt to describe the nature of  \ac{DES} using simple configurations  $(n_1l_1, n_2l_2)$ where  $n_1, n_2 > 1$ and $l_i=1,\dots,n_i-1$, has been ill-fated due to the general strong mixing of two-electron configurations in the description of each \ac{DES}.

Nowadays, the most successful proposal for the taxonomy of the different types of resonances is adopted from the work of~\citep{Lin1983, Lin1984}, after the pioneering work of~\citep{Herrick1975}. These approaches define a new set of approximate quantum numbers known as (K,T) numbers. As a result, these quasi-quantum numbers can describe, to a rather good approximation, the characteristics of the series of doubly excited states. In the following sections we shall discuss the \ac{CI} method to calculate the stationary eigenspectrum of helium-like ions. Additionally, we shall introduce the \ac{FM} which is one of the the most sophisticated theoretical tools to adequately deal with autoionizing states or resonances. Finally, we shall conclude with a review of the \ac{HSL} classification scheme. 
For the sake of clarity, figure~\ref{fig:hestates} shows a semi-quantitative energy spectrum of the He atom, indicating specifically the location of resonance states as Rydberg series below any ionization threshold.

\section{\label{sec:qmd}Stationary quantum-mechanical description of Helium-like atoms or ions}

Helium-like atoms cannot be solved analytically. The Hamiltonian of the system involves an inter-electronic interaction term which depends only upon the spatial separation between the electrons and it does not allow us to obtain a solution in terms of any known analytical function. Naturally, since the foundation of wave mechanics, a lot of diverse approaches have been intended to approximate the \ac{WF} for the ground state,  singly excited states, and resonances. So, we can list, among the most emblematic methods, the following ones: Hartree-Fock  and Multiconfigurational Hartree-Fock methods~\citep{FroeseFischer1973,FroeseFischer1977,FroeseFischer1978}, and the general \ac{CI} methods~\citep{Shavitt1977,Friedrich2005} which is our method to obtain the eigenspectrum of helium. 

\subsection{Hamiltonian of helium-like atoms}

Two-electron atoms consisting of a nucleus of mass $M$ and charge $Ze$ and two electrons, with mass $m$ and charge $e$, can be described in terms of the Coulomb interactions between the three charged particles. As in the case of one-electron atoms (see section~\eqref{sec:fullhy}), we can separate the motion of the centre of mass. Actually, for helium-like ions, this is a slightly more complicated procedure, that can be followed in~\citep{Bransden2003}. Therefore, denoting by $\mbf{r}_1$ and $\mbf{r}_2$ the relative coordinates of the two electrons with respect to the nucleus we can write the following two-particle Hamiltonian

\begin{align}\label{eq:Hham1}
H &=-\frac{\hbar^2}{2\mu}\nabla^2_{\mbf{r}_1}-\frac{\hbar^2}{2\mu}\nabla^2_{\mbf{r}_2}-\frac{\hbar^2}{M}\nabla^2_{\mbf{r}_1}\cdot\nabla^2_{\mbf{r}_2}-\frac{Ze^2}{(4\pi\epsilon_0)r_1}-\frac{Ze^2}{(4\pi\epsilon_0)r_2}\\ \nonumber
&+\frac{e^2}{(4\pi\epsilon_0)r_{12}},
\end{align}
where $\mu=mM/(m+M))$ is the reduced mass of an electron with respect to the nucleus and $r_{12}=|\mbf{r}_1-\mbf{r}_2|$.

We shall consider in our calculation an {\textit infinitely heavy} nucleus since in this work we are not in the pursuit of high precision calculations, that must include all corrections due to the finite mass. As a result, $\mu=m$ and the {\textit mass polarisation} term $(\hbar^2/M)\nabla^2_{\mbf{r}_1}\cdot\nabla^2_{\mbf{r}_2}$ can be neglected. Consequently, we can rewrite the expression~\eqref{eq:Hham1}, in \ac{a.u.}, as

\begin{equation}\label{eq:Hham2}
H = h(1) + h(2) + \frac{1}{r_{12}},
\end{equation}
where $h(i)= -\frac{\nabla^2_i}{2} -\frac{Z}{r_i}$. The Schr\"odinger equation reads

\begin{equation}\label{eq:schrotwo}
\left [h(1) + h(2) + \frac{1}{r_{12}}\right] \psi(\mbf{r}_1,\mbf{r}_2)=E\psi(\mbf{r}_1,\mbf{r}_2),
\end{equation}
and it  cannot be separated due to the presence of the $1/r_{12}$ term. For this reason, we cannot write the total two-particle \ac{WF} as a direct product of one-particle \ac{WF}s, i.e., the system is not separable or, in a more common and modern speaking terminology, it is  {\textit entangled}. 

\subsection{Two-fermion antisymmetric wave function }\label{sec:symmetrizationpos}

{\textit Pauli exclusion principle} asserts that in a system of identical fermions no more than one particle can have exactly the same single particle quantum numbers, this statement requires that the \ac{WF} of a two-electron system must be antisymmetric as a whole, i.e, the \ac{WF} must change its sign by a single permutation of the global electron coordinates (spatial plus spin), that is

\begin{equation}\label{eq;antisym}
\ket{\psi(2,1)} =-\ket{\psi(1,2)}.
\end{equation}
This is the so called {\textit Symmetrization Postulate} which states that: (a)~particles whose spin is an {\textit integer} multiple of $\hbar$ have only symmetric states (these particles are called {\textit bosons}); (b)~particles whose spin is a {\textit half odd-integer} multiple of $\hbar$ have only antisymmetric states (these particles are called {\textit fermions}); ({c})~partially symmetric states do no exist~\citep{Messiah1966, Ballentine1998}. Besides, this postulate is another form of the {\textit principle of indistinguishability of identical particles}.  Following to~\citep{Messiah1964} the principle states: {\textit "Dynamical states that differ only by a permutation of identical particles cannot be distinguished by any observation whatsoever"}.

Formally, in an {\textit independent particle model}, i.e., neglecting the term $1/r_{12}$ in the Hamiltonian~\eqref{eq:Hham2}, we can build up the two-electron \ac{WF} from the one-particle orbitals by means of the antisymmetrizing operator $\mathcal{\hat{A}}$ which may be defined as

\begin{equation}\label{eq;antisymket}
\ket{\psi(1,2)} =\mathcal{\hat{A}}[\ket{\phi(1)}\ket{\phi(2)}].
\end{equation}
For the case of two fermions, in which the {\textit Pauli exclusion principle} has a central role, the total \ac{WF} can be factorized into the spatial part (symmetric or antisymmetric) and the spin part (singlet or triplet), respectively. 

\begin{equation}\label{eq:wf1}
\ket{\psi(1,2)} = \ket{\phi(1, 2)}_{symmetric, antisymmetric} \otimes\ket{ \chi (1,2)}_{singlet,triplet}
\end{equation}

Hydrogen-like functions or one-particle functions, which are introduced in Appendix~\eqref{sec:fullhy}, may be written using the so called {\textit Dirac notation}, namely 

\begin{align}\label{eq:hylike}
\ket{\psi_{n,l,m_l,s, m_s}}&=\ket{\phi_{nlm_l}}\otimes\ket{\chi_{sm_s}}, \nonumber \\
&=\ket{n,l,m_l,s, m_s}
\end{align}
where, $\{n,l,m_l,s, m_s\}$ represent good quantum numbers corresponding to the commuting  operators $(\hat{H},\hat{L}^2, \hat{L}_z, \hat{S}^2, \hat{S}_z$) which completely describe the one-particle state. Since we deal with electrons, the spin quantum number has a definite constant {\textit half-integer (fermion)} number $s=1/2$, which is assumed hereafter in the formulas. Now, $\ket{\chi_{sm_s}}$ is the spin vector and  $\ket{\phi_{nlm_l}}$ is the orbital eigenstate which may be projected onto the space representation to obtain the corresponding \ac{WF} as

\begin{equation}\label{eq:orbitalfunc}
\braket{\mbf{r}|\phi_{n,l,m_l}}=\phi_{n,l,m_l}(r,\theta,\phi)=\frac{\mathcal{U}_{n,l}({r})}{r}\mathcal{Y}^l_{m_l}(\theta,\phi),
\end{equation}
where $\mathcal{Y}^l_{m_l}(\theta,\phi)$ is a spherical harmonic corresponding to the orbital angular momentum quantum number $l$ and to the magnetic quantum number $m_l$. The functions $\mathcal{U}_{n,l}({r})=rR_{n,l}({r})$ satisfy the reduced radial equation~\eqref{eq:radschr} quoted in Appendix~\ref{ch:hybsplines}.  So, the separation of equation~\eqref{eq:relativeeqn1} enables us to write the state vector as the full state vector in a partially mixed Dirac notation as

\begin{equation}\label{eq:hyket}
\braket{r|n,l,m_l,m_s}=\braket{r|\phi_{n,l,m_l},\chi_{m_s}}=\frac{\mathcal{U}_{n,l}({r})}{r}\ket{l,m_l,m_s}.
\end{equation}
Even though we make an abuse in the standard Dirac notation in equation~\eqref{eq:hyket}, it can be useful in the following. In addition, it is important to say that any {\textit inner product} involves an integration over the radial coordinate $r$, for $0\leq r<\infty$. Note that we drop the label $s$ from the spin ket $| \chi \rangle$ since $s=1/2$ always for the electrons. The spin projection $m_s$ can take the values $+\frac{1}{2}$ and $-\frac{1}{2}$  which corresponds to the two state vectors $\ket{\chi_\frac{1}{2}}\equiv\ket{\alpha}$ and $\ket{\chi_{-\frac{1}{2}}}\equiv\ket{\beta}$.

As stated above, the whole \ac{WF} of a two-electron atom must be antisymmetric against the permutation $1 \leftrightarrow 2$. Then, using the expression~\eqref{eq:wf1}, we may build up the following two antisymmetrized \ac{WF} by means of  the eigenstates of the global spin operators  $\hat{S}^2$ and $\hat{S}_z$  with $\hat{S}=\hat{s}_1+\hat{s}_2$ and $\hat{S}_z=(\hat{s}_1)_z+(\hat{s}_2)_z$ (spin {\textit singlet} function $S=0$ with $M_S=0$ and spin {\textit triplet} function $S=1$ with $M_S=1,0,-1$ ).

\begin{align}
\ket{\psi(1,2)}^{para}& =\ket{\Phi(1,2)}_{symmetric}\otimes\frac{1}{\sqrt{2}}[\ket{\alpha\beta}-\ket{\beta\alpha}], \label{eq:funpara}\\
\ket{\psi(1,2)}^{orto}& =\ket{\Phi(1,2)}_{antisymmetric}\otimes\begin{cases}
\ket{\alpha\alpha}, \\
\frac{1}{\sqrt{2}}[\ket{\alpha\beta}+\ket{\beta\alpha}], \\
\ket{\beta\beta}. 
\end{cases} \label{eq:funorto}
\end{align}
In the expression~\eqref{eq:funpara} we use a spatially symmetric vector state (commonly called {\textit para}) together with the singlet spin state vector, but in the equation~\eqref{eq:funorto} we use an antisymmetric vector state (commonly called {\textit orto}) together with the spin triplet symmetric state. 

Actually, in this work we only consider two-electron atoms with $LS$ coupling; this very special angular momentum coupling is adequate to describe atoms with small nuclear charge $Ze$. Particularly, the goal is to get an antisymmetric \ac{WF} with total angular momentum $L$ with projection $M_L$ and  total spin $S$ with projection $M_S$.  In order to accomplish our goal, we shall follow the graphical method described by~\citep{Lindgren1986}, although the same result may be readily obtained algebraically using Clebsch-Gordan angular momentum coupling coefficients, e.g.,~\citep{Edmonds1957}. First we couple the angular momenta of the two separated electrons, to build up the states of total orbital and spin angular momentum, as follows
\begin{align}
\ket{(l_al_b)LM_L}&=\sum_{m_l^am_l^b}\ket{l_am^a_l,l_bm^b_l}\braket{l_am^a_l,l_bm^b_l|LM_L},\label{eq:orbcoup} \\
\ket{SM_S}&=\sum_{m_s^am_s^b}\ket{m^a_s,m^b_s}\braket{s_am^a_s,s_bm^b_s|SM_S},\label{eq:spncoup}
\end{align}
where $\braket{l_am^a_l,l_bm^b_l|LM_L}$ and $\braket{s_am^a_s,s_bm^b_s|SM_S}$ are the {\textit vector-coupling coefficients} or \ac{CGC}~\citep{Ballentine1998,Lindgren1986}. Indeed, we may use, instead of \ac{CGC}, a more symmetrical quantity called the Wigner $3\text{-}j\text{-}$symbol which is defined as follows

\begin{equation}\label{eq:3jwigner}
\tj{j_1}{j_2}{j_3}{m_1}{m_2}{m_3}=(-1)^{j_1-j_2-m_3}(2j_3+1)^{-\frac{1}{2}}\braket{j_1m_1,j_2m_2|j_3-m_3}.
\end{equation}
It is easy to show that the $3j$-symbol vanishes, unless $m_1+m_2+m_3=0$. In addition, a non-vanishing $3j$-symbol must satisfy the triangular condition $|j_1-j_2|\leq j_3\leq j_1+j_2$.

Finally, using equations~\eqref{eq:orbcoup},~\eqref{eq:spncoup}, together with~\eqref{eq:hylike}, and~\eqref{eq;antisymket}, we can write  the antisymmetric state of the two-electron atom in $LS$ coupling as

\begin{align}\label{eq:ansymm1}
&\ket{\{n_al_a\hspace{4pt}n_bl_b\}LM_LSM_S}= \\ \nonumber
&\hspace{25pt}F\left[\ket{(n_al_a)_1(n_bl_b)_2LM_LSM_S}-\ket{(n_al_a)_2(n_bl_b)_1LM_LSM_S}\right]
\end{align}
where $F$ is a normalization factor to be calculated. The subscripts $1$ and $2$ refer to the individual electrons. The {\textit curly brackets} denote the antisymmetric combination, i.e.,  it implies the antisymmetric action of the projection operator

\begin{equation}\label{eq:antiop}
\hat{\mathcal{A}}=\frac{1}{N_t!}\sum_P(-1)^PP,
\end{equation}
where $N_t$ is the total number of particles and $P$ denotes one of the $N_t!$ permutations of the $N_t$ indexes for the particles. By means of the properties of symmetry of the \ac{CGC} or the $3j$-symbols, we may permute any two columns; an even permutation leaves the $3j$-symbol value invariant, but an odd permutation introduces the additional phase $(-1)^{j_1+j_2+j_3}$. For this reason, we get two phase factors in our case: $(-1)^{l_a+l_b+L}$ for the orbital part and $(-1)^{S+1}$ for the spin part. Consequently,  equation~\eqref{eq:ansymm1} can be recast in the form:

\begin{align}\label{eq:ansymm2}
&\ket{\{n_al_a\hspace{4pt}n_bl_b\}LM_LSM_S}= \\ \nonumber
&F\left[\ket{(n_al_a)_1(n_bl_b)_2LM_LSM_S}+(-1)^{l_a+l_b+L+S}\ket{(n_bl_b)_1(n_al_a)_2LM_LSM_S}\right].
\end{align}
If, $n_a=n_b$ and $l_a=l_b$, i.e., the electrons are said to be {\textit equivalent}, the normalization factor $F$ is equal to $1/2$ and the normalized antisymmetric function becomes  
$\ket{\{(nl)^2\}LM_LSM_S}=\ket{(nl)_1(nl)_2LM_LSM_S}$ for $S+L$ even. On the other hand, if, $n_a\neq n_b$ or $l_a\neq l_b$, i.e., the electrons are said to be {\textit non-equivalent}, and the normalization factor $F$ is equal to $1/\sqrt{2}$.

\subsection{Configuration interaction (CI) method}\label{sec:cimethod}

Our chosen method to get successful and accurate solutions, both for eigenstates and eigenenergies, to the \ac{TISE}, equation~\eqref{eq:schrotwo}, is the so called {\textit configuration interaction}~\ac{CI} method~\citep{Shavitt1977,Szabo1989}. As mentioned above, approximate solutions to the $N$-electron problem may be achieved using different methods, for instance, the \ac{HF} method. The \ac{HF} method retains the simplicity of solving the total \ac{WF} in terms of a single Slater determinant in which each orbital is optimized by solving the one-particle Fock operator, which averages the interaction with the other electrons. Nevertheless,
the \ac{HF} method is not able to describe the full electron correlation~\citep{Szabo1989,Friedrich2005}. The \ac{CI} is in essence a variational many-electron method which built up the \ac{WF} as a huge linear combination of antisymmetrized configurations constructed with \ac{HF} or hydrogenic orbitals (appropriately coupled to yield the correct total angular momenta $L$ and $S$). 

A general calculation using the \ac{CI} scheme can be understood as an optimization of the trial \ac{CI}-\ac{WF} constructed with a large combination of $N$ different configurations, i.e., it is a {\textit linear combination or superposition} of a large number of antisymmetrized two-electron functions based on products of spin-orbitals, then configurations in the form of~\eqref{eq:ansymm2} are built up~\citep{Sherrill1999,Cramer2004,Friedrich2005}. Therefore, the general  \ac{CI}-\ac{WF}  can be written as

\begin{equation}\label{eq:ciwfunc}
\ket{\Psi^{CI}}=\sum_{i=1}^N C_i\ket{\psi_i},
\end{equation}
where $C_i$ are the variational expansion coefficients, for the $i^{th}$ configuration, subject to optimization. As previously mentioned, each member of the expansion is defined by

\begin{equation}\label{eq:ansymm3}
\ket{\psi_i}=\ket{\{n_a^il_a^i\hspace{4pt}n_b^il_b^i\}LM_LSM_S}.
\end{equation}

Afterwards, the variational method asserts that the eigenvalues and eigenfunctions of the Hamiltonian~\eqref{eq:Hham2} can be approximated by seeking the conditions under which the following functional $E[\Psi^{CI}]$ will be stationary, this is

\begin{equation}\label{eq:varstat}
\delta E[\Psi^{CI}]=0,
\end{equation}
with
\begin{align}\label{eq:varstat1}
E[\Psi^{CI}]&=\frac{\braket{\Psi^{CI}|H|\Psi^{CI}}}{\braket{\Psi^{CI}|\Psi^{CI}}},\\ \nonumber
&=\frac{\sum_{ij}^NC_iC^{*}_j\braket{\psi_i|H|\psi_j}}{\sum_i^NC_iC^{*}_i}.
\end{align}

The \ac{CI} method is conceptually the most straightforward method to solve the \ac{TISE}.  It is said that \ac{CI} constitutes an "exact theory" in the limit of an infinite basis of configurations.  In practice, however, the matrix equations are not exact because the expansion in equation~\eqref{eq:ciwfunc} must be truncated to a finite number $N$ of terms. Therefore, if we include a large enough number of configurations, the diagonalization of the Hamiltonian in the truncated subspace can give a very good approximation to the exact eigenenergies and eigenstates since the electron correlation due to the Coulomb term $1/r_{ij}$ is better described using this kind of many particle methods. However, our \ac{CI}-\ac{WF} does not contain terms including the inter-electronic coordinate,  $r_{12}$, i.e., the trial \ac{WF}, we say, is not explicitly correlated. Instead, the proper description of the correlation term $1/r_{12}$ is achieved by the angular mixing of configurations in the \ac{CI}-\ac{WF}. Other \ac{CI} schemes might include a trial \ac{WF} which are explicitly correlated, i.e.,containing functions of the coordinate $r_{12}$. These explicitly correlated schemes have in general faster radial and angular convergence with the number of configurations. However, its computational implementation is much more involved, and lacks the simplicity of configurations based on direct products of orbitals. 
The classical (explicitly uncorrelated) \ac{CI} method has been implemented in a vast range of applications or calculations both in atomic and molecular physics. For instance, in the $H^-$ ion by~\citep{Chang1991}, in He atom by~\citep{Bachau1984,Castro-Granados2012, vanderHart1992a}, in $Li$ and $Li$-like atoms by~\citep{Cardona2010}, in $Be$ atom by~\citep{Chang1989}, in $N^{3+}$ and $N^{5+}$ by~\citep{vanderHart1992a,vanderHart1992b}, in $Mg$ atom by~\citep{Tang1990}, in $Mg^-$ and $Ca^-$ by~\citep{Sanz-Vicario2008}, to mention only a few.
 
To summarize the whole procedure, we first solve the one-electron problem in the parent ion in order to obtain the basis set of orbitals for different angular momenta $l=0(s),1({p}),2(d),3(f),\dots$; the orbitals themselves can be expanded in terms of a basis set. In our case, the latter basis consist of B-splines (see Appendix~\ref{ch:hybsplines}). Secondly, we construct the two-electron variational \ac{CI}-\ac{WF} with antisymmetrized configurations out of the set of orbitals, accordingly to the $LS$ coupling. Once the matrix elements of the total two-electron Hamiltonian are calculated, we solve the generalized eigenvalue problem to obtain the eigenvalues and the corresponding eigenvectors.  
  
\subsection{Hamiltonian matrix elements}\label{sec:rmatxelem}

As suggested above, the variational theorem requires the optimization of the average value of the Hamiltonian operator $\braket{\Psi^{CI}|H|\Psi^{CI}}$ with respect to the expansion coefficients $C_i$ in equation~\eqref{eq:ciwfunc}. This variational optimization is equivalent to solve the matrix eigenvalue problem in an algebraic subspace spanned by the basis of configurations, see \citep{Levine2008}. In order to solve this eigenvalue problem, the Hamiltonian matrix elements  $H_{ij}=\braket{\psi_{i}|H|\psi_{j}}$  with $i,j=1,\dots,N$ must be calculated. Using the equations~\eqref{eq:Hham2} and~\eqref{eq:ansymm3} they read

\begin{align}\label{eq:hammatrix1}
H_{ij}&=\braket{\psi_i|H|\psi_j}, \\ \nonumber
&=\bra{\{n_al_a\hspace{4pt}n_bl_b\}_iLM_LSM_S}\left(h(1) + h(2) + \frac{1}{r_{12}}\right)\ket{\{n_cl_c\hspace{4pt}n_dl_d\}_jLM_LSM_S}, \\ \nonumber
&=E_{i}\delta_{ij}+E_i\delta_{ij}+\left(\frac{1}{r_{12}}\right)_{ij}.
\end{align}
where $i\equiv\{n_al_a,n_bl_b\}$, $j\equiv\{n_cl_c,n_dl_d\}$, and $\left(1/r_{12}\right)_{ij}$ is the matrix element of the inter-electronic Coulomb operator. The Kronecker deltas in equation~\eqref{eq:hammatrix1} suggest that we are dealing with orthogonal orbitals. The Hamiltonian matrix $\mbf{H}$ is a dense matrix, i.e., it is not a sparse matrix, which we must diagonalize;  many optimised algorithms are available to do this task. We accomplish the diagonalization procedure with the help of the routine \texttt{DSYEV} included in the \texttt{LAPACK} library~\citep{Anderson1999}.

\subsubsection{Matrix elements for the interelectronic Coulomb repulsion.}

Since the trial \ac{CI}-\ac{WF} is built up as antisymmetric products of spin orbitals, once we have the one-particle energies, we only need to calculate the matrix elements for the electron-electron Coulomb interaction, that read

\begin{align}\label{eq:intermatrix}
\left(\frac{1}{r_{12}}\right)_{ij}&=\braket{\psi_i|\left(\frac{1}{r_{12}}\right)|\psi_j},\\ \nonumber
&=\bra{\{n_a^il_a^i\hspace{4pt}n_b^il_b^i\}LM_LSM_S}\left(\frac{1}{r_{12}}\right)\ket{\{n_c^jl_c^j\hspace{4pt}n_d^jl_d^j\}LM_LSM_S}.
\end{align}
Even though the indices $i$ and $j$ are too redundant, we have kept them in order to emphasize our \ac{CI} approach. In this context they denote two different specific configurations. Hereafter, these indices will be removed from the notation.

For a two-electron system with two non-equivalent electrons, following the equation~\eqref{eq:ansymm2}, the antisymmetric \ac{WF} with quantum numbers $n_al_a$ and $n_bl_b$ reads,

\begin{align}\label{eq:ansymm33}
&\ket{\{n_al_a\hspace{4pt}n_bl_b\}LM_LSM_S}= \\ \nonumber
&\frac{1}{\sqrt{2}}\left[\ket{(n_al_a)_1(n_bl_b)_2LM_LSM_S}+(-1)^{l_a+l_b+L+S}\ket{(n_bl_b)_1(n_al_a)_2LM_LSM_S}\right],
\end{align}
and the antisymmetric \ac{WF} for the case of two-equivalent electrons is

\begin{equation}\label{eq:ansymm34}
\ket{\{(nl)^2\}LM_LSM_S}=\ket{(nl)_1(nl)_2LM_LSM_S}\hspace{15pt}L+S\hspace{5pt}even.
\end{equation}

Accordingly, the electron-electron Coulomb interaction matrix elements between
antisymmetric \ac{WF} will be in terms of basic integrals in the form:

\begin{equation}\label{eq:matrixnonanty}
(r_{12}^{-1})^{\aleph}=\bra{(n_al_a)_1(n_bl_b)_2LM_LSM_S}r_{12}^{-1}\ket{(n_cl_c)_1(n_dl_d)_2LM_LSM_S},
\end{equation}
where the superscript $\aleph$ denotes that we are using non-antisymmetrized functions.
To begin with the calculations of the matrix elements~\eqref{eq:matrixnonanty}, which in principle involves integrations over $\mbf{r}_1$ and $\mbf{r}_2$, we must introduce a commonly used multipole expansion of the inter-electronic correlation term

\begin{equation}\label{eq:intermulti}
r_{12}^{-1}=\sum_l\frac{r^l_{<}}{r^{l+1}_{>}}P_l(cos\theta_{12}),
\end{equation}
where $P_l(cos\theta_{12})$ is the Legendre polynomial of order $l$ whose argument is the cosine of the inter-electronic angle~\citep{Abramowitz1965}; $r_{<}$ is the lesser between $r_1$ and $r_2$, $r_>$ is the greater. 

As previously introduced, we use a mixed Dirac notation to separate the radial part form the orbital and spin angular momentum part in the matrix elements as follows

\begin{align}\label{eq:matrixnonanty2}
(r_{12}^{-1})_{ab,cd}^{\aleph}&=\sum_l\int_0^\infty dr_1\int_0^\infty dr_2\mathcal{U}_{n_{a}l_{a}}(r_{1})\mathcal{U}_{n_{b}l_{b}}(r_{2})\frac{r^l_{<}}{r^{l+1}_{>}}\mathcal{U}_{n_{c}l_{c}}(r_{1})\mathcal{U}_{n_{d}l_{d}}(r_{2})\\ \nonumber
&\times\bra{(l_al_b)LM_LSM_S}P_l(cos\theta_{12})\ket{(l_cl_d)LM_LSM_S},\\ \nonumber
&=\sum_lR^{l}(ab,cd)\braket{(LS)_{ab}|P_l(cos\theta_{12})|(LS)_{cd}}.
\end{align}

At first, in order to calculate the orbital-spin part of the matrix element which are written now as  $\braket{(LS)_{ab}|P_l(cos\theta_{12})|(LS)_{cd}}$, we shall use the graphical method described in~\citep{Lindgren1986}; this method is based on the {\textit Spherical Tensor Operators Theory}. Now, using the addition theorem for \ac{SH}, we write the Legendre polynomial $P_{l}(cos\theta_{12})$ in terms of products of \ac{SH}

\begin{equation}\label{eq:addthe1}
P_{l}(cos\theta_{12})=\frac{2\pi}{2l+1}\sum_{m}\mathcal{Y}_{m}^{l}(\theta_{1},\phi_{1})\mathcal{Y}_{m}^{*l}(\theta_{2},\phi_{2}),
\end{equation}
In addition, the definition of  the ``$\mathbf{C}$ tensor'', having components 
\begin{subequations}\label{eq:ccomp144}
\begin{align}
C_{q}^{k}&=\sqrt{\frac{2\pi}{2k+1}}\mathcal{Y}_{q}^{k}(\theta,\phi), \label{eq:ccomp14} \\
C_{-q}^{k}&=\sqrt{\frac{2\pi}{2k+1}}\mathcal{Y}_{-q}^{k}(\theta,\phi), \label{eq:ccomp15}
\end{align}
\end{subequations}
the relation of parity symmetry for the \ac{SH}: $\mathcal{Y}_{q}^{*k}(\theta,\phi)=(-1)^{q}\mathcal{Y}_{-q}^{k}(\theta,\phi)$, and finally, the canonical form of \textit{tensor scalar product} of two tensors, that is defined by  $\mathbf{t}^{k}(1)\cdot\mathbf{u}^{k}(2)=\sum_{q}(-1)^{q}t_{q}^{k}(1)u_{-q}^{k}(2)$, allow us to rewrite (\eqref{eq:addthe1}) as

\begin{equation}\label{eq:addthe3}
P_{l}(cos\theta_{12})=\sum_{m}(-1)^{m}C_{m}^{l}(1)C_{-m}^{l}(2)=\mathbf{C}^{l}(1)\cdot\mathbf{C}^{l}(2).
\end{equation}

Consequently, we are able to write the inter-electronic interaction operator in a rather useful form

\begin{equation}\label{eq:intermulti2}
r_{12}^{-1}=\sum_l(-1)^l(2l+1)^{\frac{1}{2}}\frac{r^l_{<}}{r^{l+1}_{>}}\{\mathbf{C}^{l}(1)\mathbf{C}^{l}(2)\}^{0}_{0},
\end{equation}
where we have also used the expression 

\begin{equation}\label{eq:tensororank}
\{\mathbf{t}^{k}(1)\mathbf{u}^{k}(2)\}^{0}_{0}=(-1)^{k}(2k+1)^{-\frac{1}{2}}\mathbf{t}^{k}(1)\cdot\mathbf{u}^{k}(2),
\end{equation}
and here $\{\mathbf{t}^{k}(1)\mathbf{u}^{k}(2)\}^{0}_{0}$ is a scalar operator (or a tensor of rank zero) as expected for the Coulomb repulsion $1/r_{12}$. Finally, for a two-electron atom (helium isoelectronic series)  the matrix element of equation~\eqref{eq:matrixnonanty2} may be written as

\begin{equation}\label{eq:nonantmatrix}
(r_{12}^{-1})_{ab,cd}^{\aleph}=\sum_l(-1)^l(2l+1)^{\frac{1}{2}}R^{l}(ab,cd)\braket{(LS)_{ab}|\{\mathbf{C}^{l}(1)\mathbf{C}^{l}(2)\}^{0}_{0}|(LS)_{cd}}.
\end{equation}

\subsubsection{\label{sec:orbitalspinopelem}Two-particle orbital-spin angular momentum matrix element}

In the first place, we shall evaluate the general matrix element of the compound tensor of rank $K$, defined as $\hat{g}_{12}=\gamma(r_1,r_2)\{\mathbf{t}^{k_{1}}(1)\mathbf{u}^{k_{2}}(2)\}^{K}_{Q}$, which reads

\begin{equation}\label{eq:maelemten12}
\braket{ab|\gamma(r_1,r_2)\{\mathbf{t}^{k_{1}}(1)\mathbf{u}^{k_{2}}(2)\}^{K}_{Q}|cd},
\end{equation}

\noindent where $a, b, c, d$ denote the uncoupled one-electron states of equation~\eqref{eq:hyket}, this is
\[\ket{a}=\ket{n_al_am^{a}_{l}m^{a}_{s}}\] 
The arbitrary function  $\gamma(r_1,r_2)$ depends on the  radial coordinates of the two electrons.  The matrix element \eqref{eq:maelemten12} may be calculated using the {\textit Wigner-Eckart theorem}~\citep{Edmonds1957,Lindgren1986, Ballentine1998} to represent integrals over the angular coordinates in the following way:

 \tikzset{
paint/.style={draw=#1!50!black, fill=#1!60!black}
}

\begin{align}\label{eq:maelemten13}
\braket{ab|\hat{g}_{12}|cd}&=R(ab,cd)\braket{ab|\{\mathbf{t}^{k_{1}}(1)\mathbf{u}^{k_{2}}(2)\}^{K}_{Q}|cd},\\ \nonumber 
&=\int_{0}^{\infty}\int_{0}^{\infty} P_{a}(r_{1})P_{b}(r_{2})\gamma(r_{1},r_{2})P_{c}(r_{1})P_{d}(r_{2})dr_{1}dr_{2} \\ \nonumber 
&\times\braket{l_a||\mathbf{t}^{k_{1}}||l_c}\braket{l_b||\mathbf{u}^{k_{2}}||l_d}\\ \nonumber
&\begin{tikzpicture}[decoration={
shape width=.45cm, shape height=.25cm,
shape=isosceles triangle, shape sep=.55cm,
shape backgrounds}]
\draw (2,3) node[left] {-} -- (6,3) node[right] {+};
\draw (2,0) node[below] {$l_{c}m^{c}_{l}$} -- (2,5) node[above] {$l_{a}m^{a}_{l}$};
\draw (6,0) node[below] {$l_{d}m^{d}_{l}$} -- (6,5) node[above] {$l_{b}m^{b}_{l}$};
\draw [line width=3pt] (4,1.5) -- (4,3) node[above] {+};
\draw (4,0) node[below] {KQ} -- (4,1.5);
\draw [paint=purple,decorate] (3,3) node[above] {$k_{1}$} -- (3.5,3);
\draw [paint=purple,decorate] (5,3) node[above] {$k_{2}$} -- (5.5,3);
\draw [paint=purple,decorate] (2,4) -- (2,4.5); 
\draw [paint=purple,decorate] (6,4) -- (6,4.5);
\draw (7.5,0) node[below] {$s_{c}m^{c}_{s}$} -- (7.5,5) node[above] {$s_{a}m^{a}_{s}$};
\draw (9.5,0) node[below] {$s_{d}m^{d}_{s},$} -- (9.5,5) node[above] {$s_{b}m^{b}_{s}$};
\node (a) at (1,2.5) {$\times$};
\end{tikzpicture}
\end{align}
here $\braket{l_a||\mathbf{t}^{k}||l_c}$ is the reduced matrix element which is independent of $m_{k}$ and 
$R(ab,cd)=\int_{0}^{\infty}\int_{0}^{\infty} \mathcal{U}_{a}(r_{1})\mathcal{U}_{b}(r_{2})\gamma(r_{1},r_{2})\mathcal{U}_{c}(r_{1})\mathcal{U}_{d}(r_{2})dr_{1}dr_{2}$.
By the way, the operator~\eqref{eq:intermulti2} is a tensor of rank zero, i.e., we must set $K=0$ and $Q=0$. The corresponding $\gamma$ function for the electron-electron interaction is

\begin{equation}\label{eq:gamma13}
\gamma^k(r_1,r_2)=\frac{r^k_{<}}{r^{k+1}_{>}},
\end{equation}
where $r_<$ is the lesser between $r_1$ and $r_2$, and $r_>$  is the greater.  Using the graphical identity

\begin{equation}\label{eq:graph1}
\begin{tikzpicture}[decoration={
shape width=.45cm, shape height=.25cm,
shape=isosceles triangle, shape sep=.55cm,
shape backgrounds}]
\draw (0,2) node[left] {$l_{2}m_{2}$} -- (3,2) node[right] {$l_{1}m_{1}$};
\draw (1.5,0) node[below] {$00$} -- (1.5,2);
\draw [paint=purple,decorate] (0.75,2) -- (1.25,2);
\draw [paint=purple,decorate] (2.25,2) -- (2.75,2);
\node (a) at (4.5,1) {$= [l_{1}]^{-\frac{1}{2}}$};
\draw (6.5,1) node[left] {$l_{2}m_{2}$} -- (9.6,1) node[right] {$l_{1}m_{1},$};
\draw [paint=purple,decorate] (8,1) -- (7.5,1);
\end{tikzpicture}
\end{equation}
where $[l_1]^{-\frac{1}{2}}=(2l_1+1)^{-\frac{1}{2}}$, the uncoupled matrix element of the operator $r_{12}^{-1}$ may be written as

\begin{align}\label{eq:maelemten23}
\braket{ab|r_{12}^{-1}|cd} &= \sum_{k}(-1)^{k}\braket{ab|\gamma^k(r_1,r_2)\{\mathbf{C}^{k}(1)\mathbf{C}^{k}(2)\}^{0}_{0}|cd}\\ \nonumber
&= \sum_{k}(-1)^{k}R^k(ab,cd)\braket{l_a||\mathbf{C}^{k}||l_c}\braket{l_b||\mathbf{C}^{k}||l_d}\\ \nonumber
&\begin{tikzpicture}[decoration={
shape width=.45cm, shape height=.25cm,
shape=isosceles triangle, shape sep=.55cm,
shape backgrounds}]
\draw (1,3) node[left] {$l_{a}m^{a}_{l}$} -- (4,3) node[right] {$l_{c}m^{c}_{l}$};
\draw (1,0) node[left] {$l_{b}m^{b}_{l}$} -- (4,0) node[right] {$l_{d}m^{d}_{l}$};
\draw (2.5,0) node[below]{-} -- (2.5,3) node[above]{+} ;
\draw [paint=purple,decorate] (1.75,3) -- (1.25,3);
\draw [paint=purple,decorate] (1.75,0)  -- (1.25,0);
\draw [paint=purple,decorate] (2.5,1.5) -- (2.5,2);
\node (a) at (0,1.5) {$\times$};
\node (b) at (2.75,1.65) {$k$};
\end{tikzpicture} \\ \nonumber
&\begin{tikzpicture}
\draw (1,1.5) node[left] {$s_{a}m^{a}_{s}$} -- (4,1.5) node[right] {$s_{c}m^{c}_{s}$};
\draw (1,0) node[left] {$s_{b}m^{b}_{s}$} -- (4,0) node[right] {$s_{d}m^{d}_{s}.$};
\node (a) at (0,0.75) {$\times$};
\end{tikzpicture}
\end{align}
This is the matrix element of the inter-electronic  operator for uncoupled states  $a, b, c, d$. Now, we shall calculate the general coupled matrix element. The coupled states may be written

\begin{align}\label{eq:coupled13}
&\hspace{-1cm}\ket{(LS)_{cd}}=\ket{(l_cl_d)LM_LSM_S}= \ket{(l_cl_d)LM_L}\otimes\ket{(s_cs_d)SM_S}\\ \nonumber
&\hspace{-1cm}\begin{tikzpicture}[decoration={
shape width=.45cm, shape height=.25cm,
shape=isosceles triangle, shape sep=.55cm,
shape backgrounds}]
\node (a) at (0,1) {$=\sum_{m^{c}_{l}m^{d}_{l}}$};
\draw [line width=3pt] (3,1) node[above] {-} -- (3.5,1);
\draw (3.5,1)  -- (4,1) node[right] {$LM_{L}$} ;
\draw (3,1)  -- (2,2) node[left] {$l_{c}m^{c}_{l}$};
\draw (3,1)  -- (2,0) node[left] {$l_{d}m^{d}_{l}$};
\draw [paint=purple,decorate] (2.5,1.5) -- (2.75,1.25);
\draw [paint=purple,decorate] (2.5,0.5) -- (2.25,0.25);
\node (a) at (6,1) {$\sum_{m^{c}_{s}m^{d}_{s}}$};
\draw [line width=3pt] (9,1) node[above] {-} -- (9.5,1);
\draw (9.5,1)  -- (10,1) node[right] {$SM_{S}$} ;
\draw (9,1)  -- (8,2) node[left] {$s_{c}m^{c}_{s}$};
\draw (9,1)  -- (8,0) node[left] {$s_{d}m^{d}_{s}$};
\draw [paint=purple,decorate] (8.5,1.5) -- (8.75,1.25);
\draw [paint=purple,decorate] (8.5,0.5) -- (8.25,0.25);
\node ({c}) at (11.5,1) {$\ket{cd},$};
\end{tikzpicture} 
\end{align}

\noindent and

\begin{align}\label{eq:coupled23}
&\hspace{-1cm}\bra{(LS)_{ab}}=\bra{(l_al_b)LM_LSM_S}= \bra{(l_al_b)LM_L}\otimes\bra{(s_as_b)SM_S}\\ \nonumber
&\hspace{-1cm}\begin{tikzpicture}[decoration={
shape width=.45cm, shape height=.25cm,
shape=isosceles triangle, shape sep=.55cm,
shape backgrounds}]
\node (a) at (0,1) {$=\sum_{m^{a}_{l}m^{b}_{l}}$};
\draw (2,1) node[left] {$LM_{L}$} -- (2.5,1);
\draw [line width=3pt] (2.5,1) -- (3,1) node[above] {+};
\draw (3,1)  -- (4,2) node[right] {$l_{a}m^{a}_{l}$};
\draw (3,1)  -- (4,0) node[right] {$l_{b}m^{b}_{l}$};
\draw [paint=purple,decorate] (3.5,1.5) -- (3.25,1.25);
\draw [paint=purple,decorate] (3.5,0.5) -- (3.75,0.25);
\node (a) at (6,1) {$\sum_{m^{a}_{s}m^{b}_{s}}$};
\draw (8,1) node[left] {$SM_{S}$} -- (8.5,1);
\draw [line width=3pt] (8.5,1) -- (9,1) node[above] {+};
\draw (9,1)  -- (10,2) node[right] {$s_{a}m^{a}_{s}$};
\draw (9,1)  -- (10,0) node[right] {$s_{b}m^{b}_{s}$};
\draw [paint=purple,decorate] (9.5,1.5) -- (9.25,1.25);
\draw [paint=purple,decorate] (9.5,0.5) -- (9.75,0.25);
\node ({c}) at (11.25,1) {$\bra{ab},$};
\end{tikzpicture} 
\end{align}

\noindent At this point, we can combine expressions~\eqref{eq:maelemten23},~\eqref{eq:coupled13}, and~\eqref{eq:coupled23} to obtain the coupled matrix elements 

\begin{align}\label{eq:maelemten33}
(r_{12}^{-1})_{ab,cd}^{\aleph}&=\sum_k(-1)^kR^k(ab,cd)\bra{(LS)_{ab}}\{\mathbf{C}^{k}(1)\mathbf{C}^{k}(2)\}^{0}_{0}\ket{(LS)_{cd}}\\ \nonumber
&= \sum_{k}(-1)^{k}R^k(ab,cd)\braket{l_a||\mathbf{C}^{k}||l_c}\braket{l_b||\mathbf{C}^{k}||l_d}\\ \nonumber
&\begin{tikzpicture}[decoration={
shape width=.35cm, shape height=.20cm,
shape=isosceles triangle, shape sep=.55cm,
shape backgrounds}]
\draw (0,1) node[left] {$LM_{L}$} -- (0.5,1);
\draw [line width=3pt] (0.5,1) -- (1,1) node[above] {+};
\draw (1,1)  -- (2,2);
\draw (1,1)  -- (2,0);
\draw (2,0) node[below] {-}   -- (2,2) node[above] {+} ;
\draw [paint=purple,decorate] (1.75,1.75) -- (1.25,1.25);
\draw [paint=purple,decorate] (1.25,0.75) -- (1.5,0.5);
\draw [paint=purple,decorate] (1.75,0.25) -- (1.5,0.5);
\draw [line width=3pt] (3,1) node[above] {-} -- (3.5,1);
\draw (3.5,1)  -- (4,1) node[right] {$LM_{L}$};
\draw (3,1)  -- (2,2);
\draw (3,1)  -- (2,0);
\draw [paint=purple,decorate] (2.5,1.5) -- (2.75,1.25);
\draw [paint=purple,decorate] (2.5,0.5) -- (2.25,0.25);
\draw [paint=purple,decorate] (2,1) -- (2,1.5);
\node (aa) at (2.75,1.65) {$l_c$};
\node (bb) at (2.75,0.35) {$l_d$};
\node (cc) at (1.2,1.65) {$l_a$};
\node (dd) at (1.2,0.35) {$l_b$};
\node (dd) at (2.25,1) {$k$};
\node (dd) at (-1.5,1) {$\times$};
\end{tikzpicture} \\ \nonumber
&\begin{tikzpicture}[decoration={
shape width=.35cm, shape height=.20cm,
shape=isosceles triangle, shape sep=.55cm,
shape backgrounds}]
\draw (0,1) node[left] {$SM_{S}$} -- (0.5,1);
\draw [line width=3pt] (0.5,1) -- (1,1) node[above] {+};
\draw (1,1)  -- (2,2);
\draw (1,1)  -- (2,0);
\draw [paint=purple,decorate] (1.5,1.5) -- (1.25,1.25);
\draw [paint=purple,decorate] (1.5,0.5) -- (1.75,0.25);
\draw [line width=3pt] (3,1) node[above] {-} -- (3.5,1);
\draw (3.5,1)  -- (4,1) node[right] {$SM_{S}.$} ;
\draw (3,1)  -- (2,2);
\draw (3,1)  -- (2,0);
\draw [paint=purple,decorate] (2.5,1.5) -- (2.75,1.25);
\draw [paint=purple,decorate] (2.5,0.5) -- (2.25,0.25);
\node (aa) at (2.75,1.65) {$s_c$};
\node (bb) at (2.75,0.35) {$s_d$};
\node (cc) at (1.2,1.65) {$s_a$};
\node (dd) at (1.2,0.35) {$s_b$};
\node (dd) at (-1.5,1) {$\times$};
\end{tikzpicture} 
\end{align}
The spin and the orbital angular momentum of the electrons are coupled separately and their graphical diagrams obey the following identities:  

\begin{equation}\label{eq:coupled355}
\begin{tikzpicture}[decoration={
shape width=.35cm, shape height=.20cm,
shape=isosceles triangle, shape sep=.55cm,
shape backgrounds}]
\draw (0,1) node[left] {$SM_{S}$} -- (0.5,1);
\draw [line width=3pt] (0.5,1) -- (1,1) node[above] {+};
\draw (1,1)  -- (2,2);
\draw (1,1)  -- (2,0);
\draw [paint=purple,decorate] (1.5,1.5) -- (1.25,1.25);
\draw [paint=purple,decorate] (1.5,0.5) -- (1.75,0.25);
\draw [line width=3pt] (3,1) node[above] {-} -- (3.5,1);
\draw (3.5,1)  -- (4,1) node[right] {$SM_{S}$} ;
\draw (3,1)  -- (2,2);
\draw (3,1)  -- (2,0);
\draw [paint=purple,decorate] (2.5,1.5) -- (2.75,1.25);
\draw [paint=purple,decorate] (2.5,0.5) -- (2.25,0.25);
\node (aa) at (2.75,1.65) {$s_c$};
\node (bb) at (2.75,0.35) {$s_d$};
\node (cc) at (1.2,1.65) {$s_a$};
\node (dd) at (1.2,0.35) {$s_b$};
\node (dd) at (5.5,1) {$= 1,$};
\end{tikzpicture} 
\end{equation}

\begin{equation}\label{eq:coupled45}
\begin{tikzpicture}[decoration={
shape width=.35cm, shape height=.20cm,
shape=isosceles triangle, shape sep=.55cm,
shape backgrounds}]
\draw (0,1) node[left] {$LM_{L}$} -- (0.5,1);
\draw [line width=3pt] (0.5,1) -- (1,1) node[above] {+};
\draw (1,1)  -- (2,2);
\draw (1,1)  -- (2,0);
\draw (2,0) node[below] {-}   -- (2,2) node[above] {+} ;
\draw [paint=purple,decorate] (1.75,1.75) -- (1.25,1.25);
\draw [paint=purple,decorate] (1.25,0.75) -- (1.5,0.5);
\draw [paint=purple,decorate] (1.75,0.25) -- (1.5,0.5);
\draw [line width=3pt] (3,1) node[above] {-} -- (3.5,1);
\draw (3.5,1)  -- (4,1) node[right] {$LM_{L}$};
\draw (3,1)  -- (2,2);
\draw (3,1)  -- (2,0);
\draw [paint=purple,decorate] (2.5,1.5) -- (2.75,1.25);
\draw [paint=purple,decorate] (2.5,0.5) -- (2.25,0.25);
\draw [paint=purple,decorate] (2,1) -- (2,1.5);
\node (aa) at (2.75,1.65) {$l_c$};
\node (bb) at (2.75,0.35) {$l_d$};
\node (cc) at (1.2,1.65) {$l_a$};
\node (dd) at (1.2,0.35) {$l_b$};
\node (dd) at (2.25,1) {$k$};

\node (ee) at (5.5,1) {$=$};

\draw (6,0) node[left] {+} -- (9.464,0) node[right] {-};
\draw (6,0)  -- (7.732,1);
\draw (6,0)  -- (7.732,3);
\draw (7.732,3) -- (9.464,0);
\draw (7.732,1) -- (9.464,0);
\draw (7.732,1) node[below] {-} -- (7.732,3) node[above] {+};
\draw [paint=purple,decorate] (6.866,1.5) -- (7.066,1.846) ;
\draw [paint=purple,decorate] (8.598,1.5) -- (8.798,1.154) ;
\draw [paint=purple,decorate] (6.866,0.5) -- (6.966,0.557) ;
\draw [paint=purple,decorate] (8.598,0.5) -- (8.498,0.557) ;
\draw [paint=purple,decorate] (7.732,2) -- (7.732,2.5) ;
\draw [paint=purple,decorate] (7.732,0) -- (7.832,0) ;
\node (ff) at (8.85,1.65) {$l_c$};
\node (gg) at (8.25,0.35) {$l_d$};
\node (hh) at (6.6,1.65) {$l_a$};
\node (ii) at (7.2,0.35) {$l_b$};
\node (jj) at (7.95,2.1) {$k$};
\node (kk) at (7.732,-0.3)  {$L$};
\end{tikzpicture} 
\end{equation}

\begin{equation}\label{eq:coupled55}
\begin{tikzpicture}[decoration={
shape width=.35cm, shape height=.20cm,
shape=isosceles triangle, shape sep=.55cm,
shape backgrounds}]
\draw (0,0) node[left] {+} -- (3.464,0) node[right] {-};
\draw (0,0)  -- (1.732,1);
\draw (0,0)  -- (1.732,3);
\draw (1.732,3) -- (3.464,0);
\draw (1.732,1) -- (3.464,0);
\draw (1.732,1) node[below] {-} -- (1.732,3) node[above] {+};
\draw [paint=purple,decorate] (0.866,1.5) -- (1.066,1.846) ;
\draw [paint=purple,decorate] (2.598,1.5) -- (2.798,1.154) ;
\draw [paint=purple,decorate] (0.866,0.5) -- (0.966,0.557) ;
\draw [paint=purple,decorate] (2.598,0.5) -- (2.498,0.557) ;
\draw [paint=purple,decorate] (1.732,2) -- (1.732,2.5) ;
\draw [paint=purple,decorate] (1.732,0) -- (1.832,0) ;
\node (aa) at (2.85,1.65) {$l_c$};
\node (bb) at (2.25,0.35) {$l_d$};
\node (cc) at (0.6,1.65) {$l_a$};
\node (dd) at (1.2,0.35) {$l_b$};
\node (dd) at (1.95,2.1) {$k$};
\node (dd) at (1.732,-0.3)  {$L$};
\node (ff) at (6,1) {$=(-1)^{l_b+l_c+L+k}\Gj{l_a}{l_b}{L}{l_d}{l_c}{k}$,};
\end{tikzpicture} 
\end{equation}
where $\Gj{l_a}{l_b}{L}{l_d}{l_c}{k}$ is the $6j$-symbol~\citep{Lindgren1986,Edmonds1957,Landau1977}. Then the basic formula for the Coulomb matrix elements for unsymmetrized configurations in the bra and ket is

\begin{align}\label{eq:maelemten45}
(r_{12}^{-1})_{ab,cd}^{\aleph}&= \sum_{k}R^{k}(ab,cd)\braket{a||\mathbf{C}^{k}||c}\braket{b||\mathbf{C}^{k}||d}\\ \nonumber
&\times(-1)^{l_b+l_c+L}\Gj{l_a}{l_b}{L}{l_d}{l_c}{k}.
\end{align}

\noindent Now, using the following expression for the reduced matrix element

\begin{equation}
\braket{l||\mathbf{C}^{k}||l^{\prime}}= (-1)^{l}\left[(2l+1)(2l^{\prime}+1)\right]^{\frac{1}{2}}\tj{l}{k}{l^{\prime}}{0}{0}{0},
\end{equation}
we can rewrite the matrix element as

\begin{align}\label{eq:maelemten55}
(r_{12}^{-1})_{ab,cd}^{\aleph}&= \sum_{k}R^k(ab,cd)(-1)^{l_a+l_c+L}\\ \nonumber
&\times\left[(2l_a+1)(2l_b+1)(2l_c+1)(2l_d+1)\right]^{\frac{1}{2}}\\ \nonumber
&\times\tj{l_a}{k}{l_c}{0}{0}{0}\tj{l_b}{k}{l_d}{0}{0}{0}\Gj{l_a}{k}{l_c}{l_d}{L}{l_b},
\end{align}
by means of $\tj{j_2}{j_1}{j_3}{m_2}{m_1}{m_3}=(-1)^{j_1+j_2+j_3}\tj{j_1}{j_2}{j_3}{m_1}{m_2}{m_3}$, we have
 
 \begin{align}\label{eq:maelemten65}
(r_{12}^{-1})_{ab,cd}^{\aleph}&= \sum_{k}R^k(ab,cd)(-1)^{l_b+l_d+L}\\ \nonumber
&\times\left[(2l_a+1)(2l_b+1)(2l_c+1)(2l_d+1)\right]^{\frac{1}{2}}\\ \nonumber
&\times\tj{l_a}{l_c}{k}{0}{0}{0}\tj{l_b}{l_d}{k}{0}{0}{0}\Gj{l_a}{k}{l_c}{l_d}{L}{l_b}.
\end{align}
At this point, we can use $\tj{j_1}{j_2}{j_3}{-m_1}{-m_2}{-m_3}=(-1)^{j_1+j_2+j_3}\tj{j_1}{j_2}{j_3}{m_1}{m_2}{m_3}$ and $(-1)^k=(-1)^{-k}$, for k an integer, to obtain
 
 \begin{align}\label{eq:maelemten75}
(r_{12}^{-1})_{ab,cd}^{\aleph}&= \sum_{k}R^k(ab,cd)(-1)^{L-k}\\ \nonumber
&\times \left[(2l_a+1)(2l_b+1)(2l_c+1)(2l_d+1)\right]^{\frac{1}{2}}\\ \nonumber
&\times\tj{l_a}{l_c}{k}{0}{0}{0}\tj{l_b}{l_d}{k}{0}{0}{0}\Gj{l_a}{k}{l_c}{l_d}{L}{l_b}.
\end{align}

\subsubsection{Radial Matrix Elements}

The radial integral which we have denoted $R^k(ab,cd)$ as a factorized term in the inter-electronic Coulomb matrix element is given by
\begin{equation}\label{eq:radialelem}
R^k(ab,cd)=\int_{0}^{\infty}\int_{0}^{\infty} \mathcal{U}_{a}(r_{1})\mathcal{U}_{b}(r_{2})\frac{r^k_<}{r^{k+1}_>}\mathcal{U}_{c}(r_{1})\mathcal{U}_{d}(r_{2})dr_{1}dr_{2},
\end{equation}
where we have used the relation $\mathcal{U}_{i}({r})=rR_i({r})$. Now, in order to get the solution of equation~\eqref{eq:radialelem}, we may define the function $Y^k({r})$~\citep{Bachau2001,McCurdy2004,Castro-Granados2012}.

\begin{align}\label{eq:functionary}
Y_{bd}^k({r})&=r\int_{0}^{\infty} \mathcal{U}_{b}(r^{\prime})\frac{r^k_<}{r^{k+1}_>}\mathcal{U}_{d}(r^{\prime})dr^{\prime}, \\ \nonumber
&=\int_{0}^{r}  \mathcal{U}_{b}(r^{\prime})\left(\frac{r^{\prime}}{r}\right)^k\mathcal{U}_{d}(r^{\prime})dr^{\prime}+\int_{r}^{\infty}  \mathcal{U}_{b}(r^{\prime})\left(\frac{r}{r^{\prime}}\right)^{k+1}\mathcal{U}_{d}(r^{\prime})dr^{\prime},
\end{align}
with this definition we can rewrite the radial integral~\eqref{eq:radialelem} as
\begin{equation}\label{eq:radialelem2}
R^k(ab,cd)=\int_{0}^{\infty} \mathcal{U}_{a}({r})\frac{Y^{k}_{bd}({r})}{r}\mathcal{U}_{c}({r})dr,
\end{equation}

We need a two-step way to calculate the radial integral~\eqref{eq:radialelem}. Firstly we compute the function $Y^{k}_{bd}({r})$ and immediately we insert it into the equation~\eqref{eq:radialelem2}. Anyway, one may try to compute this function by different methods. An efficient method is to solve the associated Poisson's equation. Actually, we can rewrite the integral form of equation~\eqref{eq:functionary} as a differential equation for  $Y^{k}_{bd}({r})$ using the {\textit Leibniz integral rule}~\citep{Abramowitz1965}

\begin{align}\label{eq:leibnizrule}
\text{if}\hspace{20pt}F(x)&=\int_{a(x)}^{b(x)}f(x,t)dt,\\ \nonumber
\Longrightarrow\hspace{10pt}\frac{dF(x)}{dx}&=f(x,b(x))\frac{db(x)}{dx}-f(x,a(x))\frac{da(x)}{dx}+\int_{a(x)}^{b(x)}\frac{\partial}{\partial x}f(x,t)dt.
\end{align}
Firstly, we calculate the first and second derivative of $Y^{k}_{bd}({r})$

\begin{align}
\frac{d}{dr}Y^{k}_{bd}({r})&=-\frac{k}{r}\int_{0}^{r}  \mathcal{U}_{b}(r^{\prime})\left(\frac{r^{\prime}}{r}\right)^k\mathcal{U}_{d}(r^{\prime})dr^{\prime}\label{eq:firstfuny}\\ \nonumber
&+\frac{k+1}{r}\int_{r}^{\infty}  \mathcal{U}_{b}(r^{\prime})\left(\frac{r}{r^{\prime}}\right)^{k+1}\mathcal{U}_{d}(r^{\prime})dr^{\prime},\\ 
\frac{d^2}{dr^2}Y^{k}_{bd}({r})&=-\frac{2k+1}{r}\mathcal{U}_b({r})\mathcal{U}_d({r}) \label{eq:secfuny} \\ \nonumber
&+\frac{k(k+1)}{r^2}\int_{0}^{r}  \mathcal{U}_{b}(r^{\prime})\left(\frac{r^{\prime}}{r}\right)^k\mathcal{U}_{d}(r^{\prime})dr^{\prime}\\ \nonumber 
&+\frac{k(k+1)}{r^2}\int_{r}^{\infty}  \mathcal{U}_{b}(r^{\prime})\left(\frac{r}{r^{\prime}}\right)^{k+1}\mathcal{U}_{d}(r^{\prime})dr^{\prime}, 
\end{align}
after that, combining the equations~\eqref{eq:functionary} and~\eqref{eq:secfuny}, we obtain the ordinary differential equation

\begin{equation}\label{eq:poissonyy}
\frac{d^2}{dr^2}Y^{k}_{bd}({r})=\frac{k(k+1)}{r^2}Y^{k}_{bd}({r})-\frac{2k+1}{r}\mathcal{U}_b({r})\mathcal{U}_d({r}).
\end{equation}
This is a non-homogeneous one-dimensional Poisson's equation which must satisfy the following boundary conditions

\begin{subequations}\label{eq:condiniy}
\begin{align}
Y^{k}_{bd}(0)&=0,\label{eq:condiniy1}\\
Y^{k}_{bd}(L)&=\frac{1}{L^k}\int_0^L\mathcal{U}_b(r^\prime)r^{\prime k}\mathcal{U}_b(r^\prime)dr^\prime.\label{eq:condiniy2}
\end{align}
\end{subequations}
Given that we are solving the problem within a finite box, the upper limit in the integration is $L$ instead of infinity. Moreover, our numerical implementation to obtain the solution of this differential equation is to expand $Y^{k}_{bd}({r})$ in the same basis of B-splines used to solve the one-electron Schr\"odinger equation, see section~\eqref{sec:hybsplines}. Nevertheless, our basis of B-splines satisfies only the first of these boundary conditions. To solve the equation~\eqref{eq:poissonyy}, with both of the boundary conditions~\eqref{eq:condiniy}, we may use the Green's function~\citep{McCurdy2004,Jackson1998} for two-point boundary conditions, in the interval $(0,L)$

\begin{equation}\label{eq:greenfnps}
G(r,r^\prime)=\frac{r^k_<}{r^{k+1}_>}-\frac{r^kr^{\prime k}}{L^{2k+1}},
\end{equation}
which satisfies the the equation 

\begin{equation}\label{eq:greeneqps}
\left(\frac{d^2}{dr^2}-\frac{k(k+1)}{r^2}\right)G(r,r^\prime)=-\frac{2k+1}{r}\delta(r-r^\prime).
\end{equation}
In the first place, we must seek a solution to the function $Y^{k}_{bd}({r})^{(0)}$, that  satisfies the boundary condition~\eqref{eq:condiniy1}, but actually, at $r=L$ satisfies $Y^{k}_{bd}({L})^{(0)}=0$ instead of the condition~\eqref{eq:condiniy2}. We expand this function in the basis of B-splines, which are functions that satisfies the boundary conditions that they vanish at $0$ and $L$,

\begin{equation}\label{eq:ybsplines}
Y^{k}_{bd}({r})^{(0)}=\sum_{i}C^k_iB_i({r}).
\end{equation}
Replacing this expansion into the equation~\eqref{eq:poissonyy}, we obtain

\begin{equation}\label{eq:eqexpandbs}
\left(\frac{d^2}{dr^2}-\frac{k(k+1)}{r^2}\right)\sum_{i}C^k_jB_j({r})=-\frac{2k+1}{r}\mathcal{U}_b({r})\mathcal{U}_d({r}).
\end{equation}
Multiplying by one of the B-splines from the left and integrating over $r$ gives the following algebraic matrix equation for the coefficients $C^k_j$

\begin{align}\label{eq:poissonM}
\sum_jT^k_{ij}C^k_j=(2k+1)U_i^{bd},\\ \nonumber
\mbf{T}^k\mbf{C}^k=(2k+1)\mbf{U}^{bd},
\end{align}
the latter written in compact matrix form, where 
\begin{equation}\label{eq:poissonM1}
T^k_{ij}=-\int_0^LB_i({r})\left(\frac{d^2}{dr^2}-\frac{k(k+1)}{r^2}\right)B_j({r})dr
\end{equation}
and
\begin{equation}\label{eq:poissonM2}
U_i^{bd}=\int_0^LB_i({r})\frac{1}{r}\mathcal{U}_b({r})\mathcal{U}_d({r})dr.
\end{equation}

Equation~\eqref{eq:poissonM} has the solution (the second line in compact matrix form)

\begin{align}\label{eq:poissonM3}
C^k_i&=(2k+1)\sum_j(T^k)^{-1}_{ij}U_j^{bd},\\ \nonumber
\mbf{C}^k&=(2k+1)(\mbf{T}^k)^{-1}\mbf{U}^{bd},
\end{align}
in this way we have the solution to $Y^{k}_{bd}({r})^{(0)}$. In order to calculate the actual solution $Y^{k}_{bd}({r})$ of the Poisson's equation~\eqref{eq:poissonyy}, with the proper boundary conditions~\eqref{eq:condiniy}, we need to add a term which is a solution of the homogeneous equation. Using the Green's function~\eqref{eq:greenfnps}, we may add an exact expression which is analogous to its second term, this is

\begin{align}\label{eq:soltoty}
Y^{k}_{bd}({r})&=Y^{k}_{bd}({r})^{(0)}+\frac{r^{k+1}}{L^{2k+1}}\int_0^L \mathcal{U}_b({r^\prime})r^{\prime k}\mathcal{U}_d({r^\prime}),\\ \nonumber
&=(2k+1)\sum_{ij}B_i({r})(T^k)^{-1}_{ij}U_j^{bd}+\frac{r^{k+1}}{L^{2k+1}}Q^k_{bd},\\ \nonumber
&=(2k+1)\mbf{B}^T({r})(\mbf{T}^k)^{-1}\mbf{U}_{bd}+\frac{r^{k+1}}{L^{2k+1}}Q^k_{bd},
\end{align}
the latter written in compact matrix form, and where 

\begin{equation}\label{eq:qintegral}
Q^k_{bd}=\int_0^L \mathcal{U}_b({r^\prime})r^{\prime k}\mathcal{U}_d({r^\prime}).
\end{equation}
Therefore, we have now a solution to the function $Y^{k}_{bd}({r})$ satisfying the correct boundary conditions~\eqref{eq:condiniy}. Finally, we substitute it back into the original expression for the radial two-electron integral~\eqref{eq:radialelem2}, in order to obtain its solution 

\begin{align}\label{eq:radialelem3}
R^k(ab,cd)&=\int_{0}^{L} \mathcal{U}_{a}({r})\frac{Y^{k}_{bd}({r})}{r}\mathcal{U}_{c}({r})dr, \\ \nonumber
&=\int_{0}^{L} \mathcal{U}_{a}({r})\frac{1}{r}\left[(2k+1)\sum_{ij}B_i({r})(T^k)^{-1}_{ij}U_j^{bd}+\frac{r^{k+1}}{L^{2k+1}}Q^k_{bd}\right]  \mathcal{U}_{c}({r})dr,\\ \nonumber
&=(2k+1)\sum_{ij}\left[\int_{0}^{L}B_i({r})\frac{1}{r}\mathcal{U}_{a}({r})\mathcal{U}_{c}({r})dr\right](T^k)^{-1}_{ij}U_j^{bd}\\ \nonumber
&+\frac{1}{L^{2k+1}}Q^k_{bd}\left[\int_{0}^{L}\mathcal{U}_{a}({r})r^k\mathcal{U}_{c}({r})dr\right].
\end{align}
using the equations~\eqref{eq:poissonM2} and~\eqref{eq:qintegral}, we finally arrive at the expression

\begin{align}\label{eq:radialelem4}
R^k(ab,cd)&=(2k+1)\sum_{ij}U_i^{ac}(T^k)^{-1}_{ij}U_j^{bd}+\frac{1}{L^{2k+1}}Q^k_{ac}Q^k_{bd}, \\ \nonumber
&=(2k+1)\mbf{U}^T_{ac}(\mbf{T}^k)^{-1}\mbf{U}_{bd}+\frac{1}{L^{2k+1}}Q^k_{ac}Q^k_{bd}.
\end{align}

A comparative table showing  the computational validity of equation~\eqref{eq:radialelem4} can be found in references~\citep{Bachau2001,Castro-Granados2012}.

\subsubsection{Inter-electronic Coulomb matrix elements between antisymmetric configurations}

To summarize, we have calculated the matrix elements of $1/r_{12}$ between non-antisymmetric configurations. Now we must take into account the antisymmetrization in the configurations for the bra and ket states, along with the property of {\textit equivalent} or  {\textit non-equivalent} electrons in the configurations, that affects the form of the \ac{WF}.
 
\subsubsection*{Equivalent---Equivalent Electrons}

\begin{align}\label{eq:eqeqcoul}
(r^{-1}_{12})_{aa,cc}&=\braket{\{(n_al_a)^2\}LS|r^{-1}_{12}|\{(n_cl_c)^2\}LS},\\ \nonumber
&=(r_{12}^{-1})_{ab,cd}^{\aleph}=\braket{(n_al_a)_1(n_al_a)_2LS|r^{-1}_{12}|(n_cl_c)_1(n_cl_c)_2LS},\\ \nonumber
&=\sum_{k}R^k(aa,cc)(-1)^{L-k}(2l_a+1)(2l_c+1)\\ \nonumber
&\times\tj{l_a}{l_c}{k}{0}{0}{0}^2\Gj{l_a}{k}{l_c}{l_c}{L}{l_a}.
\end{align}
In a similar manner, all combinations of {\textit equivalent} and {\textit non-equivalent} two-electron \ac{WF}s, for the general antisymmetrized matrix element, may be written, as a result, in terms of the non-antisymmetrized matrix elements~\eqref{eq:maelemten75}.  

\subsubsection*{Equivalent---Non-equivalent Electrons}

\begin{align}\label{eq:eqeqcoul1}
(r^{-1}_{12})_{aa,cd}&=\braket{\{(n_al_a)^2\}LS|r^{-1}_{12}|\{n_cl_c\hspace{4pt}n_dl_d\}LS},\\ \nonumber
&=\bra{(n_al_a)_1(n_al_a)_2LS}r^{-1}_{12}\frac{1}{\sqrt{2}}\left[\ket{(n_cl_c)_1(n_dl_d)_2LS}\right. \\ \nonumber
&+\left.(-1)^{l_c+l_d+L+S}\ket{(n_dl_d)_1(n_cl_c)_2LS}\right],\\ \nonumber
&=\frac{1}{\sqrt{2}}(r_{12}^{-1})_{aa,cd}^{\aleph}+\frac{1}{\sqrt{2}}(-1)^{l_c+l_d+L+S}(r_{12}^{-1})_{aa,dc}^{\aleph},\\ \nonumber
&=\frac{1}{\sqrt{2}}(2l_a+1)[(2l_c+1)(2l_d+1)]^{\frac{1}{2}}\sum_{k}\left[R^k(aa,cd)(-1)^{L-k}\right.\\ \nonumber
&+(-1)^{l_c+l_d-k+S}\left.R^k(aa,dc)\right]\\ \nonumber
&\times\tj{l_a}{l_c}{k}{0}{0}{0}\tj{l_a}{l_d}{k}{0}{0}{0}\Gj{l_a}{k}{l_c}{l_d}{L}{l_a}.\\ \nonumber
\end{align}

\subsubsection*{Non-equivalent---Equivalent Electrons}

\begin{align}\label{eq:eqeqcoul2}
(r^{-1}_{12})_{ab,cc}&=\braket{\{n_al_a\hspace{4pt}n_bl_b\}LS|r^{-1}_{12}|\{(n_cl_c)^2\}LS},\\ \nonumber
&=\frac{1}{\sqrt{2}}\left[\bra{(n_al_a)_1(n_bl_b)_2LS}+(-1)^{l_a+l_b+L+S}\bra{(n_bl_b)_1(n_al_a)_2LS}\right] \\ \nonumber
&\times r^{-1}_{12}\ket{(n_cl_c)_1(n_cl_c)_2LS},\\ \nonumber
&=\frac{1}{\sqrt{2}}(r_{12}^{-1})_{ab,cc}^{\aleph}+\frac{1}{\sqrt{2}}(-1)^{l_a+l_b+L+S}(r_{12}^{-1})_{ba,cc}^{\aleph},\\ \nonumber
&=\frac{1}{\sqrt{2}}(2l_c+1)[(2l_a+1)(2l_b+1)]^{\frac{1}{2}}\sum_{k}\left[R^k(ab,cc)(-1)^{L-k}\right.\\ \nonumber
&+(-1)^{l_a+l_b-k+S}\left.R^k(ba,cc)\right]\\ \nonumber
&\times\tj{l_a}{l_c}{k}{0}{0}{0}\tj{l_b}{l_c}{k}{0}{0}{0}\Gj{l_a}{k}{l_c}{l_c}{L}{l_b}.\\ \nonumber
\end{align}

\subsubsection*{Non-equivalent---Non-equivalent Electrons}

\begin{align}\label{eq:eqeqcoul3}
(r^{-1}_{12})_{ab,cd}&=\braket{\{n_al_a\hspace{4pt}n_bl_b\}LS|r^{-1}_{12}|\{n_cl_c\hspace{4pt}n_dl_d\}LS},\\ \nonumber
&=\frac{1}{2}\left[\bra{(n_al_a)_1(n_bl_b)_2LS}+(-1)^{l_a+l_b+L+S}\bra{(n_bl_b)_1(n_al_a)_2LS}\right] \\ \nonumber
&\times r^{-1}_{12}\left[\ket{(n_cl_c)_1(n_dl_d)_2LS}+(-1)^{l_c+l_d+L+S}\ket{(n_dl_d)_1(n_cl_c)_2LS}\right] ,\\ \nonumber
&=\frac{1}{2}\left[(r_{12}^{-1})_{ab,cd}^{\aleph}+(-1)^{l_c+l_d+L+S}(r_{12}^{-1})_{ab,dc}^{\aleph}\right.\\ \nonumber
&+\left.(-1)^{l_a+l_b+L+S}(r_{12}^{-1})_{ba,cd}^{\aleph}+\frac{1}{\sqrt{2}}(-1)^{l_a+l_b+l_c+l_d}(r_{12}^{-1})_{ba,dc}^{\aleph}\right].\\ \nonumber
&=\frac{1}{2}[(2l_a+1)(2l_b+1)(2l_c+1)(2l_d+1)]^{\frac{1}{2}}\\ \nonumber
&\times\sum_{k}\left[\left[R^k(ab,cd)(-1)^{L-k}+(-1)^{l_a+l_b+l_c+l_d+L-k}R^k(ba,dc)\right]\right.\\ \nonumber
&\times\tj{l_a}{l_c}{k}{0}{0}{0}\tj{l_b}{l_d}{k}{0}{0}{0}\Gj{l_a}{k}{l_c}{l_d}{L}{l_b}\\ \nonumber
&+\left[R^k(ab,dc)(-1)^{l_c+l_d+S-k}+(-1)^{l_a+l_b+S-k}R^k(ba,cd)\right]\\ \nonumber
&\left.\times\tj{l_a}{l_d}{k}{0}{0}{0}\tj{l_b}{l_c}{k}{0}{0}{0}\Gj{l_a}{k}{l_d}{l_c}{L}{l_b}\right].
\end{align}

\newpage
\begin{table}[h]
\centering
\begin{tabular}{ccccccc}
\hline\hline
 &$^{1}S^{e}$ & $^{3}S^{e}$ & $^{1}P^{o}$ & $^{3}P^{o}$ & $^{1}D^{e}$ & $^{3}D^{e}$\\ \hline 
 $1$ & $-2.903509$ & $-2.175228$ & $-2.123797$ & $-2.133156$ & $-2.055619$ & $-2.055635$ \\
 $2$ & $-2.145961$ & $-2.068689$ & $-2.055131$ & $-2.058078$ & $-2.031278$ & $-2.031287$ \\
 $3$ & $-2.061268$ & $-2.036510$ & $-2.031061$ & $-2.032321$ & $-2.020011$ & $-2.020016$ \\
 $4$ & $-2.033582$ & $-2.022607$ & $-2.019896$ & $-2.020543$ & $-2.013884$ & $-2.013888$ \\
 $5$ & $-2.021165$ & $-2.015345$ & $-2.013818$ & $-2.014187$ & $-2.010177$ & $-2.010179$ \\
 $6$ & $-2.014535$ & $-2.011064$ & $-2.010137$ & $-2.010361$ & $-2.007701$ & $-2.007702$ \\
 $7$ & $-2.010570$ & $-2.008295$ & $-2.007659$ & $-2.007815$ & $-2.005394$ & $-2.005396$ \\
 $8$ & $-2.007951$ & $-2.005958$ & $-2.005263$ & $-2.005437$ & $-2.002107$ & $-2.002112$ \\
 $9$ & $-2.005547$ & $-2.002974$ & $-2.001502$ & $-2.001865$ & &  \\
 $10$ &  & & $-2.001948$ &  &  &  \\ \hline\hline
\end{tabular}
\caption[Energies of singly excited states of symmetries $^{1,3}P^{o}$, $^{1,3}S^{e}$, $^{1,3}D^{e}$ of helium atom]{\label{tab:singly1}Energies (in \ac{a.u.}) for the bound states in helium, below the first ionization threshold  He$^+(n=1)$, $E=-2.0$ \ac{a.u.}, for the symmetries $^{1,3}S^e$, $^{1,3}P^o$ and $^{1,3}D^e$, computed with a variational CI method using two-electron configurations in terms of hydrogenic orbitals computed with a basis of B-splines. The basis of configurations is described in tables~\ref{tab:config23} and~\ref{tab:config24}.}
\end{table}

\subsection{Calculations for bound states of helium atom}
In our work the \ac{CI} approach  is based on the expansion in terms of antisymmetrized products of atomic orbitals, the latter expanded in B-splines polynomials defined within a finite box of length $L$. B-splines have been widely used in the last years and for a fuller description the reader is referred to~\citep{Bachau2001}. An almost precise ground state energy for helium atom may be obtained using B-splines with an exponential \ac{bp} (knot) sequence, see section~\ref{sec:expseq}, and $25$ B-splines of order $k=7$, generating one-electron orbitals with $l\le4$, with a full \ac{CI}-\ac{WF} of 2500 configurations, which yields the energy $-2.903509$~\ac{a.u.} to be compared with results reported by~\citep{Pekeris1958}, $-2.903742$~\ac{a.u.}

In table~\eqref{tab:singly1} we include all the calculated eigenenergies for the ground state $^1S^e$ and the singly excited states of helium located below the first ionization threshold for
the spectroscopic symmetries  $^{1,3}S^e$, $^{1,3}P^o$ and 
$^{1,3}D^e$. The tables~\eqref{tab:config23} and~\eqref{tab:config24} show the configurations and its number used in all calculations for all symmetries of bound states of helium atom.

\begin{table}
\centering
\begin{tabular}{ccccc}
\hline\hline
Symmetry & Configurations & $n_1^{max}$ & $n_2^{max}$ & Number of Conf. \\ \hline

\multirow{5}{*}{ $^{1}S^{e}$} &ss&25&25&325\\
&pp&26&26&325\\
&dd&27&27&325\\
&ff&28&28&325\\ 
&gg&29&29&325\\  \hline
Total&&&&1625 \\ \hline
\multirow{4}{*}{ $^{1}P^{o}$} &sp&25&26&625\\
&pd&26&27&625\\
&df&27&28&625\\
&fg&28&29&625\\ \hline
Total&&&&2500 \\ \hline
\multirow{7}{*}{ $^{1}D^{e}$} &sd&25&27&625\\
&pp&26&26&325\\
&pf&26&28&625\\
&dd&27&27&325\\
&dg&27&289&625\\
&ff&28&28&325\\
&gg&29&29&325\\ \hline
Total&&&&3175 \\ \hline\hline
\end{tabular}
\caption[Number of configurations used in each calculation for the symmetries $^{1}S^{e}$, $^{1}P^{o}$, $^{1}D^{e}$.]{\label{tab:config23}Number of configurations of the type $(n_1 l_1, n_2 l_2)$ included in CI calculations to obtain variational energies quoted in table~\ref{tab:singly1} for spectroscopic states $^1S^e$, $^1P^o$ and $^1D^e$. The second column refers to angular configurations ($l_1,l_2$) (compatible with the total symmetry) and $n_i^{max}$ refers to the highest {\em effective} principal quantum number of the hydrogenic orbitals, i.e.,  for $s$ orbitals $n$=$1,...,n_s^{max}$; for $p$ orbitals, $n$=$2,...,n_p^{max}$ and so on.}
\end{table}\begin{table}
\centering
\begin{tabular}{ccccc}
\hline\hline
Symmetry & Configurations & $n_1^{max}$ & $n_2^{max}$ & Number of Conf. \\ \hline
\multirow{5}{*}{ $^{3}P^{o}$} &ss&24&25&300\\
&pp&25&26&300\\
&dd&26&27&300\\
&ff&27&28&300\\
&gg&28&29&300\\ \hline
Total&&&&1500 \\ \hline
\multirow{4}{*}{ $^{3}P^{o}$} &sp&25&26&625\\
&pd&26&27&625\\
&df&27&28&625\\
&fg&28&29&625\\ \hline
Total&&&&2500 \\ \hline
\multirow{7}{*}{ $^{3}D^{e}$} &sd&25&27&625\\
&pp&25&26&300\\
&pf&26&28&625\\
&dd&26&27&300\\
&dg&27&29&625\\
&ff&27&28&300\\
&gg&28&29&300\\ \hline
Total&&&&3075 \\ \hline\hline
\end{tabular}
\caption[Number of configurations used in each calculation for the symmetries $^{3}S^{e}$, $^{3}P^{o}$, $^{3}D^{e}$.]{\label{tab:config24}Number of configurations of the type $(n_1 l_1, n_2 l_2)$ included in CI calculations to obtain variational energies quoted in table~\ref{tab:singly1} for spectroscopic states $^3S^e$, $^3P^o$ and $^3D^e$. The second column refers to angular configurations ($l_1,l_2$) (compatible with the total symmetry) and $n_i^{max}$ refers to the highest {\em effective} principal quantum number of the hydrogenic orbitals, i.e.,  for $s$ orbitals $n$=$1,...,n_s^{max}$; for 
$p$ orbitals, $n$=$2,...,n_p^{max}$ and so on.   }
\end{table}

\section{\label{sec:feshbach}The projection operator formalism}

Autoionization is a dynamical process of decay that occurs in the continuum spectra of atoms and molecules. It belongs to a general class of phenomena known as {\textit Auger effect} where a quantum physical system "seemingly" spontaneously decays into a partition of its  constituent parts~\citep{Drake2006}. The Auger effect, in two-electron atoms, has three variations, inter alia, {\textit autoionization} where a neutral or positively charged composite system decays into an electron and a residual ion, {\textit autodetachment}  where the original system is a negative ion, and {\textit radiative stabilization} or {\textit radiative decay}, where the system decays to an autoinization state of lower energy, or a true bound state. It is worth noting that even though the autoionization process is rigorously a part of the scattering continuum, a formalism elaborated by~\citeauthor{Feshbach1962} can be introduced  whereby the resonant case can be transformed into a bound-like problem with the scattering elements built around it.  The Feshbach's projection operator formalism~\citep{Feshbach1962} has been a widely used method to describe resonance phenomena. It is possible to find a vast literature on its application to atomic and molecular electronic structure,~\citep{Temkin1985a} and references therein. Nevertheless, its practical implementation has been mostly reduced to atomic systems with two and three electrons. A detailed study of the application of the stationary Feshbach method in helium has been performed by~\citep{Sanchez1995}. Also, after the pioneering work of Temkin and Bathia on three-electron systems \citep{Temkin1985b}, the Feshbach formalism has been recently revisited and applied to the $Li$ and $He$ atoms in our group \citep{Cardona2010, Castro-Granados2013}, with a complete formal implementation of the method.

\subsection{Implementation of the Feshbach formalism}

\begin{figure}[h]
\includegraphics[width=\textwidth]{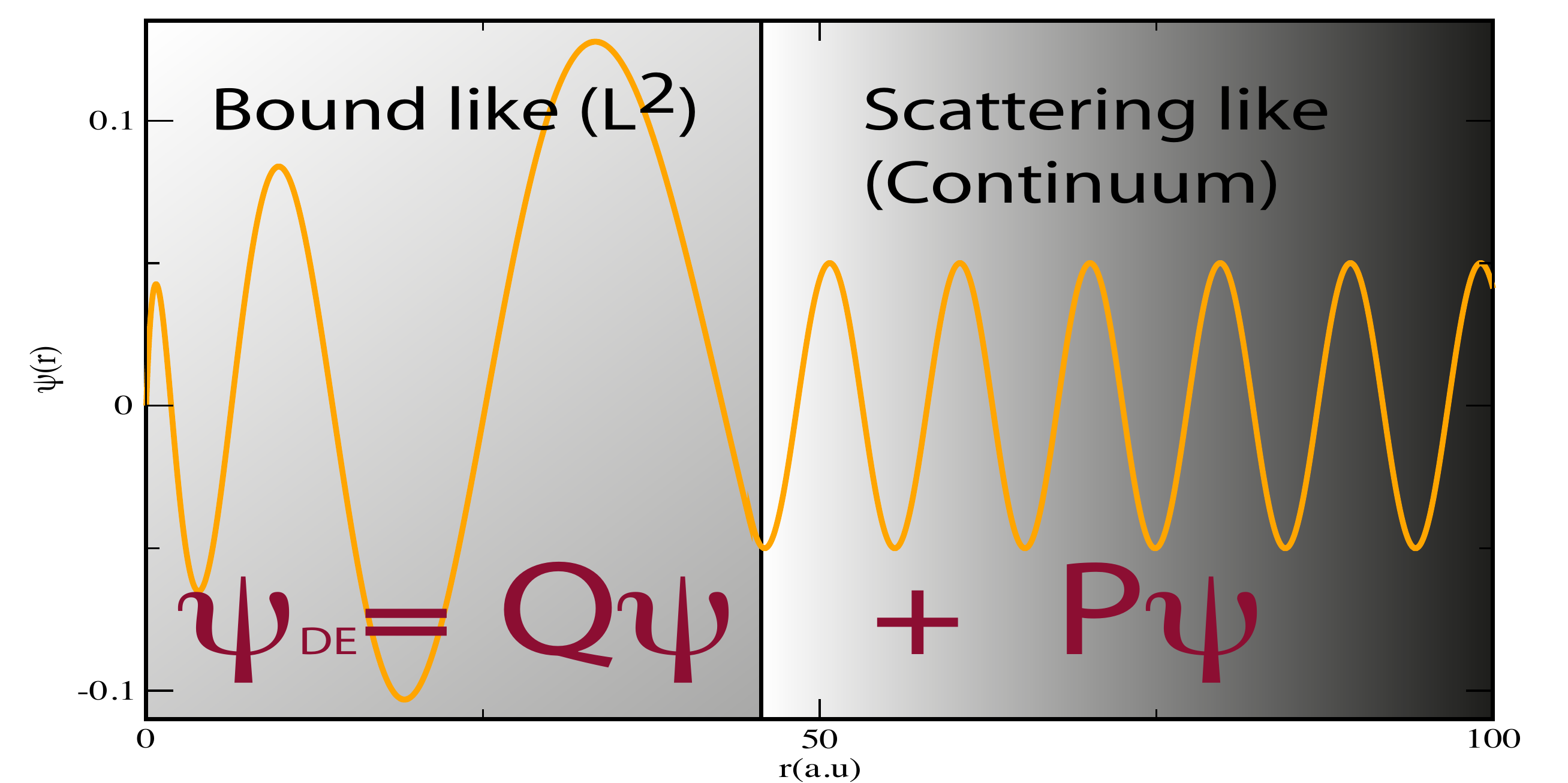}
\caption[Schematic wave function of a doubly excited state.]{\label{fig:functionFB}Schematic form of the wave function corresponding to an atomic Feshbach resonance. The total resonance wave function splits into an inner radial localized part $\mathcal{Q}\Psi$ and an outer scattering-like part $\mathcal{P}\Psi$, and the latter part does not vanish asymptotically for $r \to \infty$. The localized part carries most of the distinguishable topological information that allows us to discriminate properties among different resonances in helium. }
\end{figure}

The basic idea of the Feshbach projection operator formalism is based on the definition of projection operators $\mathcal{P}$ and $\mathcal{Q}$ which separate $\Psi$ into scattering-like $P\Psi$ and quadratically integrable or bound-like  $Q\Psi$ parts, yielding $\Psi=Q\Psi+P\Psi$; and satisfying the projection operator conditions

\begin{subequations}\label{eq:projprop}
\begin{align}
\mathcal{P}+\mathcal{Q}=1&\hspace{15pt}\text{completeness,}\label{eq:projpropa}\\ 
\mathcal{P}^2=\mathcal{P},\hspace{5pt}\mathcal{Q}^2=\mathcal{Q}&\hspace{15pt}\text{idempotency, and} \\ 
\mathcal{P}\mathcal{Q}=\mathcal{Q}\mathcal{P}=0&\hspace{15pt}\text{orthogonality.} 
\end{align}
\end{subequations}
Additionally, the projected wave functions must also satisfy the asymptotic boundary conditions 
 
\begin{subequations}\label{eq:projprop1}
\begin{align}
\lim_{r_i \to \infty}\mathcal{P}\Psi=\Psi \\
\lim_{r_i \to \infty}\mathcal{Q}\Psi=0,
\end{align}
\end{subequations}
where the latter expression indicates the confined nature of the localized part of the resonance. This Feshbach splitting of the continuum resonance wave function can be drawn in schematic form as in figure~\ref{fig:functionFB}.

By  replacing the splitting form  of the total wave function $\Psi$=$\mathcal{Q}\Psi+\mathcal{P}\Psi$ into the time independent Schr\"odinger equation $H \Psi$=$E\Psi$, it is straightforward to obtain the following equations for the bound-like and the {\em non-resonant} scattering-like parts~\citep{Cardona2010}

\begin{subequations}\label{eq:qpfunction}
\begin{eqnarray}
(\mathcal{Q}H\mathcal{Q}-\mathcal{E}_n ) \mathcal{Q}\Phi_n=0 \label{eq:qfunction}\\ 
(  \mathcal{P} H'  \mathcal{P}-E)  \mathcal{P} \Psi^{0} = 0 \label{eq:pfunction},
\end{eqnarray}
\end{subequations}
where $H'$ is the operator containing the atomic Hamiltonian plus an optical potential devoid of any resonant contribution  from the state $\mathcal{Q}\Phi_s$ with energy $\mathcal{E}_s$, i.e., $H'$=$H+ V^{n\ne s}_{opt}$ where 

\begin{equation}
V^{n\ne s}_{opt}=\IntSum_{n \ne s} \mathcal{P}H\mathcal{Q} \frac{|\Phi_n\rangle \langle \Phi_n|}{E-\mathcal{E}_n} \mathcal{Q}H\mathcal{P}.
\end{equation}

In a similar manner the Hamiltonian splits into four different terms by means of the projection operators (by using the completeness
identity~\eqref{eq:projpropa})
\begin{equation}\label{eq:splitH}
H=\mathcal{Q}H\mathcal{Q}+\mathcal{P}H\mathcal{P}+\mathcal{Q}H\mathcal{P}+\mathcal{P}H\mathcal{Q}, 
\end{equation}
where the last two terms are responsible for the coupling between both halfspaces which ultimately causes the resonant decay into the continuum. In practice one starts by solving equation \eqref{eq:qfunction} for the $\mathcal{Q}$ space with a \ac{CI} method to obtain a first approximation to the location of resonant states and it implies to use 

\begin{equation}
\mathcal{Q} = \mathcal{Q}_1 \mathcal{Q}_2 = (1 - \mathcal{P}_1) (1 -\mathcal{P}_2) = 1 - \mathcal{P}_1 - \mathcal{P}_2 + \mathcal{P}_1 \mathcal{P}_2=1-\mathcal{P}, \label{eq:opoone1}
\end{equation}
then
\begin{equation}
\mathcal{P}=\mathcal{P}_1+\mathcal{P}_2-\mathcal{P}_1\mathcal{P}_2.\label{eq:opoone}
\end{equation}
where $\mathcal{Q}_i$ and $\mathcal{P}_i$ are one-particle projection operators. In this work we are restricted to doubly excited states lying below the second ionization threshold of the He atom, so that 
\begin{equation}\label{eq:popertwo}
\mathcal{P}_i =| \phi_{1s} ({\textbf r}_i) \rangle \langle \phi_{1s} ({\textbf r}_i)|. 
\end{equation}
Therefore, the $\mathcal{Q}$ operator removes all those configurations containing the $1s$ orbital, then avoiding the variational collapse to the ground state ($1s^2$), to singly excited states ($1s n\ell$) and removing also the single ionization continuum ($1s\epsilon \ell$). As a result the lowest variational energies of the $\mathcal{Q}H\mathcal{Q}$ eigenvalue problem correspond to the doubly excited states or resonances, that were immersed in the single ionization continuum. 

\subsection{Resonant $\mathcal{Q}$-halfspace.}
We have performed \ac{CI} calculations for the $\mathcal{Q}H\mathcal{Q}$ doubly excited resonant space using the same configurational basis set that for bound states, but now, removing the $1s$ orbital as a direct effect of the projection operator $\mathcal{Q}$ using   the   equations~\eqref{eq:opoone1},~\eqref{eq:opoone} and~\eqref{eq:popertwo}. Therefore we are able to obtain 19 \ac{DES} of symmetry $^1S^e$,  26 states for $^1P^o$, and 25 states for $^1D^e$, using 1360, 1634 and 1909 configurations, respectively. On the other hand, we get 17 \ac{DES} of symmetry $^3S^e$,  27 states for $^3P^o$, and 24 states for $^3D^e$, using 1246, 1634 and 1840 configurations, respectively. In order to  illustrate the accuracy of our computation, we show  in tables~\ref{tab:se},~\ref{tab:po1}, and~\ref{tab:de1} a comparison of the our calculated energies of \ac{DES} below the $N_{th}=2$ with the previously calculated by~\citep{Chen1997} using the saddle-point complex rotation method. From these table we can finally conclude that our results are in close agreement with the reported ones by~\citeauthor{Chen1997}. Otherwise, we have also calculated the \ac{CI}-\ac{WF} of these \ac{DES}. These \ac{WF} will be used as the starting point to build up the two-dimensional two-particle density in the next chapter.  We will postpone the analysis of the accuracy of our calculations of the \ac{WF}, by comparing the density with others reported previously in the literature, until there.  

\begin{table}
\centering
\begin{tabular}{ccccccc}
\hline\hline
& &\multicolumn{2}{c}{$^{1}S^{e}$} && \multicolumn{2}{c}{$^{3}S^{e}$}\\
\cline{3-4}\cline{6-7}
& & This work &  \citep{Chen1997} & &  This work  & \citep{Chen1997}  \\ \hline 
     $_{2}(1,0)^{+.-}_{2}$  && $-0.7787708$  & $-0.777870$   && \\
     $_{2}(1,0)^{+.-}_{3}$  && $-0.5900951$  & $-0.589896$   &&    $-0.6026003$ & $-0.602577$\\
     $_{2}(1,0)^{+.-}_{4}$  && $-0.5449365$   & $-0.544882$  &&    $-0.5488467$ & $-0.548841$\\
     $_{2}(1,0)^{+.-}_{5}$  && $-0.5267032$ & $-0.526687$    &&    $-0.5284149$ & $-0.528414$\\
     $_{2}(1,0)^{+.-}_{6}$  && $-0.5176390$  & $-0.517641$   &&    $-0.5185441$ & $-0.518546$\\ \hline
     $_{2}(-1,0)^{+.-}_{2}$  && $-0.6223959$ & $-0.621810$    && \\
     $_{2}(-1,0)^{+.-}_{3}$  && $-0.5481972$  & $-0.548070$   &&    $-0.5597603$ & $-0.559745$ \\
     $_{2}(-1,0)^{+.-}_{4}$  && $-0.5277724$ & $-0.527707$    &&    $-0.5325090$ & $-0.532505$  \\
     $_{2}(-1,0)^{+.-}_{5}$  && $-0.5181340$ & $-0.518100$   &&    $-0.5205454$ &  $-0.520549$ \\
     $_{2}(-1,0)^{+.-}_{6}$  && $-0.5127799$  & $-0.512762$    &&     $-0.5141647$ & $-0.514180$ \\
     \hline\hline
\end{tabular}
\caption[Energy positions of resonant doubly excited states of helium symmetries $^{1,3}S^{e}$]{\label{tab:se}Energy positions (in \ac{a.u.}) of resonant doubly excited states of helium located below the second ionization threshold He$^+$ $(n_1=2)$ for the total symmetries $^{1,3}S^e$. Resonances are labelled according to the classification proposed by~\citep{Lin1983} using $_{n_1}(K,T)_{n_2}^A$;  $^{2s+1}L^\pi$. The notation must be understood as $A=1$ for the symmetry  $^{1}S^{e}$, and  $A=-1$ for symmetry $^{3}S^{e}$.}
\end{table}

\begin{table}
\centering
\begin{tabular}{ccccccc}
\hline\hline
& & \multicolumn{2}{c}{$^{1}P^{o}$} & & \multicolumn{2}{c}{$^{3}P^{o}$}\\
\cline{3-4}\cline{6-7}
& & This work & \citep{Chen1997} &  & This work & \citep{Chen1997}  \\ \hline 
& &\multicolumn{2}{c}{ $_{2}(0,1)^{+}_{n}$} &  &\multicolumn{2}{c}{$_{2}(1,0)^{+}_{n}$}  \\ \cline{3-4}\cline{6-7}
     $n=2$  && $-0.6927496$ &$-0.693069 $  &&   $-0.7614841$ &$-0.760489$\\
     $n=3$  && $-0.5640090$  &$-0.564074$  &&   $-0.5849286$ &$-0.584671$\\
     $n=4$  && $-0.5343290$ &$-0.534358$   &&   $-0.5429314$ &$-0.542837$ \\
     $n=5$  && $-0.5214789$  &$-0.521501$  &&   $-0.5257518$ &$-0.525711$ \\
     $n=6$  && $-0.5146989$   &$-0.514732$ &&   $-0.5171160$  &$-0.517107$\\ \hline
     && \multicolumn{2}{c}{$_{2}(1,0)^{-}_{n}$}  && \multicolumn{2}{c}{$_{2}(0,1)^{+}_{n}$}  \\ \cline{3-4}\cline{6-7}
     $n=3$  && $-0.5970953$  &$-0.597074$  &&    $-0.5790297$& $-0.579030$\\
     $n=4$  && $-0.5464858$  &$-0.546490$  &&    $-0.5395578$& $-0.539558$ \\
     $n=5$  && $-0.5272924$   &$-0.527295$ &&    $-0.5239415$& $-0.523946$\\
     $n=6$  && $-0.5179307$   &$-0.517939$ &&    $-0.5160618$ & $-0.516079$\\
     $n=7$  && $-0.5126711$   &$-0.512679$ &&    $-0.5115092$& $-0.511547$\\ \hline
      &&\multicolumn{2}{c}{ $_{2}(-1,0)^{0}_{n}$}  && \multicolumn{2}{c}{$_{2}(-1,0)^{+}_{n}$} \\ \cline{3-4}\cline{6-7}
     $n=3$  && $-0.5470914$  &$-0.547087$   &&    $-0.5488529$&$-0.548841$\\
     $n=4$  && $-0.5276130$   &$-0.527613$  &&    $-0.5286420$ &$-0.528637$\\
     $n=5$  && $-0.5181138$   &$-0.518115$  &&    $-0.5187098$ &$-0.518708$ \\
     $n=6$  && $-0.5127857$   &$-0.512789$  &&    $-0.5131517$ &$-0.513155$\\ \hline\hline
\end{tabular}
\caption[Energy positions of resonant doubly excited states of helium symmetries $^{1,3}P^{o}$]{\label{tab:po1}Energy positions (in \ac{a.u.}) of resonant doubly excited states of helium located below the second ionization threshold He$^+$ $(n_1=2)$ for the total symmetries $^{1,3}P^o$. Resonances are labelled according to the classification proposed by~\citep{Lin1983} using $_{n_1}(K,T)_{n_2}^A$;  $^{2s+1}L^\pi$.}
\end{table}

 \begin{table}
\centering
\begin{tabular}{ccccccc}
 \hline\hline
&&\multicolumn{2}{c}{$^{1}D^{e}$} && \multicolumn{2}{c}{$^{3}D^{e}$}\\ 
\cline{3-4}\cline{6-7}
 && This work & \citep{Chen1997} &&   This work & \citep{Chen1997}  \\ \hline 
     $_{2}(1,0)^{+.-}_{2}$  && $-0.7026974$    & $-0.701830$  && \\
     $_{2}(1,0)^{+.-}_{3}$  && $-0.5693682$    & $-0.569193$ &&    $-0.5838054$ & $-0.583784$\\
     $_{2}(1,0)^{+.-}_{4}$  && $-0.5367840$    & $-0.539715$ &&    $-0.5416857$ & $-0.541679$ \\
     $_{2}(1,0)^{+.-}_{5}$  && $-0.5227632$    & $-0.522737$ &&    $-0.5250186$ & $-0.525018$\\
     $_{2}(1,0)^{+.-}_{6}$  && $-0.5154448$    & $-0.515451$ &&    $-0.5166775$ & $-0.516687$\\ \hline
     $_{2}(0,1)^{0}_{3}$    && $-0.5564146$     & $-0.556417$&&    $-0.5606819$ & $-0.560684$\\
     $_{2}(0,1)^{0}_{4}$    && $-0.5315042$     & $-0.531506$&&    $-0.5334602$ & $-0.533462$\\
     $_{2}(0,1)^{0}_{5}$    && $-0.5201091$     & $-0.520114$&&    $-0.5211252$  & $-0.521130$\\
     $_{2}(0,1)^{0}_{6}$    && $-0.5139367$     & $-0.513950$&&    $-0.5145241$  & $-0.514540$\\ \hline
     $_{2}(-1,0)^{0}_{4}$   && $-0.5292883$     & $-0.529292$&&    $-0.5293086$  & $-0.529312$\\
     $_{2}(-1,0)^{0}_{5}$   && $-0.5189966$     & $-0.519000$&&    $-0.5190130$  & $-0.519016$\\
     $_{2}(-1,0)^{0}_{6}$   && $-0.5133034$     & $-0.513310$&&    $-0.5133149$  & $-0.513322$\\\hline\hline
\end{tabular}
\caption[Energy positions of resonant doubly excited states of helium symmetries $^{1,3}D^{e}$]{\label{tab:de1}Energy positions (in \ac{a.u.}) of resonant doubly excited states of helium located below the second ionization threshold He$^+$ $(n_1=2)$ for the total symmetries $^{1,3}D^e$. Resonances are labelled according to the classification proposed by~\citep{Lin1983} using $_{n_1}(K,T)_{n_2}^A$;  $^{2s+1}L^\pi$. The notation must be understood as $A=1$ for the symmetry  $^{1}D^{e}$, and  $A=-1$ for symmetry $^{3}D^{e}$.}
\end{table}


\cleardoublepage


\part{Measures of Information theory applied to the analysis of doubly excited states of  two-electron atoms or ions}

\chapter{Measures of Information theory based on the electron density}\label{ch:tim}

\section{\label{sec:twodensity}Two-electron distribution function}

The two-electron density function or distribution function $\rho(\mathbf{r}_{1},\mathbf{r}_{2})$ is defined as the probability to find an electron at point $\mathbf{r}_{1}$ and another at point $\mathbf{r}_{2}$.  The two-electron density carries almost all the information about quantum correlations of a compound system~\citep{Ezra1982,Ezra1983}. In the following sections we calculate this distribution function by means of a two-electron operator.

\subsection{\label{sec:densityop}Two-electron density function operator}

The two-electron distribution function  $\rho(\mathbf{r}_{1},\mathbf{r}_{2})$  is the expectation value of the operator  $\hat{G}(\mathbf{r}_{1},\mathbf{r}_{2})$ which has the following form in the position representation for an atom or ion with N electrons~\citep{Ellis1996}

\begin{equation}\label{eq:densityop}
\hat{G}(\mathbf{r}_{1},\mathbf{r}_{2})=\sum_{i=1}^{N}\sum_{j=1}^{i-1}\frac{1}{2}\left[\delta^{3}(\mathbf{u}_{i}-\mathbf{r}_{1})\delta^{3}(\mathbf{u}_{j}-\mathbf{r}_{2})+\delta^{3}(\mathbf{u}_{i}-\mathbf{r}_{2})\delta^{3}(\mathbf{u}_{j}-\mathbf{r}_{1})\right].
\end{equation}
The integral of the two-electron distribution function $\rho$ gives the number of electron pairs, i.e., $\int  \bra{ \Psi}  \hat{G}({\textbf r}_1,{\textbf r}_2) \ket{ \Psi}d{\textbf r}_1d{\textbf r}_2 = N(N-1)/2$. In our approach, if we consider the atom prepared in a \ac{CI} pure state $\ket{\Psi^{CI}}$, then $\rho(\mathbf{r}_{1},\mathbf{r}_{2})=\braket{\Psi^{CI}|\hat{G}(\mathbf{r}_{1},\mathbf{r}_{2})|\Psi^{CI}}$ is a rather complicated function of six coordinates, three for each electron coordinate.  On the other hand, the three coordinates that specify the orientation of the atom or ion in space are irrelevant for our present purpose. Hence, we can average or integrate over the three of the coordinates, i.e., the Euler angles. Consequently, we obtain a two-particle operator which only depends on three relevant variables $r_{1}$ ,$r_{2}$, and the interelectronic angle $\theta$

\begin{align}\label {eq:redop}
\hat{G}(r_{1},r_{2},\theta)&=\sum_{j<i}\frac{1}{2}\left[\delta(u_{i}-r_{1})\delta(u_{j}-r_{2})+\delta(u_{i}-r_{2})\delta(u_{j}-r_{1})\right] \\ \nonumber
&\times \delta(cos\theta_{ij}-cos\theta).
\end{align}
Then, the two-electron density in terms of the internal variables can be recast in the form of the expectation value of the operator~\eqref{eq:redop}
\begin{equation}\label{eq:twodens1}
\rho(r_{1},r_{2},\theta)= \braket{\Psi^{CI}|\hat{G}(r_{1},r_{2},\theta)|\Psi^{CI}}, 
\end{equation}
and the density in equation~\eqref{eq:twodens1} is normalized to the number of electron pairs

\begin{equation}\label{eq:rednorm}
\int_{0}^{\infty}dr_{1}\int_{0}^{\infty}dr_{2}\int_{-1}^{1}d(cos\theta)\rho(r_{1},r_{2},\theta) = N(N-1)/2.
\end{equation}
In this form, we have a rotational-invariant two-electron density with no dependence on the total azimuthal quantum numbers $M_L$ and $M_S$. This procedure to obtain the two-electron density is equivalent to that proposed by~\citep{Ezra1982,Ezra1983} where a formula for the density is derived following a different procedure. .

With the aim of evaluating $\rho$ using a general numerical \ac{CI}-\ac{WF}, it is convenient to express $\hat{G}$ in the language of {\textit spherical-tensor operator}. For this propose, the two-particle operator can be write as 

\begin{align}\label{eq:redop2}
\hat{G}(r_{1},r_{2},\theta)&=\sum_{j<i}\sum_{k=0}^{\infty}\frac{1}{4}(2k+1)P_{k}(cos\theta)P_{k}(cos\theta_{ij}) \\ \nonumber
&\times \left[\delta(u_{i}-r_{1})\delta(u_{j}-r_{2})+\delta(u_{i}-r_{2})\delta(u_{j}-r_{1})\right],
\end{align}
where we have used the completeness relation for Legendre polynomials \citep{Sepulveda2009}

\begin{equation}\label{eq:comleg}
\delta(cos\theta_{ij}-cos\theta)=\sum_{k=0}^{\infty}\frac{1}{2}(2k+1)P_{k}(cos\theta)P_{k}(cos\theta_{ij}). 
\end{equation}

Now, the addition theorem for \ac{SH}, see equation~\eqref{eq:addthe1}, enables us to write $P_{k}(cos\theta_{ij})$ in terms of products of \ac{SH}

\begin{equation}\label{eq:addthe}
P_{k}(cos\theta_{ij})=\frac{2\pi}{2k+1}\sum_{q}\mathcal{Y}_{q}^{k}(\theta_{i},\phi_{i})\mathcal{Y}_{q}^{*k}(\theta_{j},\phi_{j}).
\end{equation}
Using definition of  ``$\mathbf{C}$ tensor'', having components~\eqref{eq:ccomp144}, and the definition of {\textit tensor scalar product}, we can rewrite \eqref{eq:addthe} as

\begin{equation}\label{eq:addthe2}
P_{k}(cos\theta_{ij})=\mathbf{C}^{k}(i)\cdot\mathbf{C}^{k}(j),
\end{equation}
and we are able to write the two-electron density functions in a rather useful form as follows
\begin{align}\label {eq:redop3}
\hat{G}_{2}(r_{1},r_{2},\theta)&=\sum_{j<i}\sum_{k=0}^{\infty}\frac{1}{4}(2k+1)^{\frac{3}{2}}(-1)^{k}\{\mathbf{C}^{k}(i)\cdot\mathbf{C}^{k}(j)\}^{0}_{0} \\ \nonumber
&\times \left[\delta(u_{i}-r_{1})\delta(u_{j}-r_{2})+\delta(u_{i}-r_{2})\delta(u_{j}-r_{1})\right]P_{k}(cos\theta),
\end{align}
where we have also used the equation~\eqref{eq:tensororank}, in particular for the ${\textbf C}$ tensor in the form

\[\{\mathbf{C}^{k}(i)\cdot\mathbf{C}^{k}(j)\}^{0}_{0}=(-1)^{k}(2k+1)^{-\frac{1}{2}}\mathbf{C}^{k}(i)\cdot\mathbf{C}^{k}(j),\]

\noindent where $\{\mathbf{C}^{k}(i)\cdot\mathbf{C}^{k}(j)\}^{0}_{0}$ is a scalar operator (or a tensor of rank zero).
Finally, for a two-electron atom (helium isoelectronic series)  equation~\eqref{eq:redop3} reduces to
\begin{align}\label {eq:redop4}
\hat{G}(r_{1},r_{2},\theta)&=\sum_{k}\frac{1}{4}(2k+1)^{\frac{3}{2}}(-1)^{k}\{\mathbf{C}^{k}(1)\cdot\mathbf{C}^{k}(2)\}^{0}_{0} \\ \nonumber
&\times \left[\delta(u_{1}-r_{1})\delta(u_{2}-r_{2})+\delta(u_{1}-r_{2})\delta(u_{2}-r_{1})\right]P_{k}(cos\theta).
\end{align}

\subsection{\label{sec:densityopelem}Two-particle density operator matrix elements}

The two-electron density function can be computed at different levels of approximation. In our case, in terms of the \ac{CI} method and using its variational \ac{WF}, it can be written as

\begin{equation}\label{eq:twodens2}
\rho(r_{1},r_{2},\theta)=\braket{\Psi^{CI}|\hat{G}(r_{1},r_{2},\theta)|\Psi^{CI}}=\sum_{ij}C_iC_j^*\braket{\psi_i|\hat{G}(r_{1},r_{2},\theta)|\psi_j},
\end{equation}
where the integrations involved in the expectation value must be performed over the coordinates $\{u_1,u_2,\theta\}$ in equation~\eqref{eq:redop4}.

Now we use the same procedure followed to compute the inter-electronic Coulomb operator~\eqref{eq:intermulti}, taken as a reference the non-antisymmetrized matrix elements of the operator $G(r_1,r_2,\theta)$

\begin{equation}\label{eq:Gmatrixnonanty}
(\hat{G}(r_1,r_2,\theta))_{ab,cd}^{\aleph}=\bra{(n_al_a)_1(n_bl_b)_2LS}\hat{G}(r_1,r_2,\theta\ket{(n_cl_c)_1(n_dl_d)_2LS}.
\end{equation}
Inserting the equation~\eqref{eq:redop4} in the equation~\eqref{eq:Gmatrixnonanty} we obtain

\begin{align}\label{eq:nonantGmatrix}
(\hat{G}(r_1,r_2,\theta))_{ab,cd}^{\aleph}&=\sum_k\frac{1}{4}(-1)^k(2k+1)^{\frac{3}{2}}R(ab,cd)P_k(cos\theta)\\ \nonumber
&\times\braket{(LS)_{ab}|\{\mathbf{C}^{k}(1)\mathbf{C}^{k}(2)\}^{0}_{0}|(LS)_{cd}}.
\end{align}
where the two-electron radial integral (which now incidentally does not depend on the sum index $k$ at variance with the Coulomb case) is

\begin{equation}\label{eq:funRG}
R(ab,cd)=\int_{0}^{\infty}\int_{0}^{\infty} \mathcal{U}_{a}(u_{1})\mathcal{U}_{b}(u_{2})\gamma(u_{1},u_{2})\mathcal{U}_{c}(u_{1})\mathcal{U}_{d}(u_{2})du_{1}du_{2},
\end{equation}
with
\begin{equation}\label{eq:gamma1}
\gamma(u_1,u_2)=\delta(u_{1}-r_{1})\delta(u_{2}-r_{2})+\delta(u_{1}-r_{2})\delta(u_{2}-r_{1}).
\end{equation}
This integral is straightforwardly calculated yielding 

\begin{equation}\label{eq:funRG1}
R(ab,cd)= \mathcal{U}_{a}(r_{1})\mathcal{U}_{b}(r_{2})\mathcal{U}_{c}(r_{1})\mathcal{U}_{d}(r_{2})+\mathcal{U}_{a}(r_{2})\mathcal{U}_{b}(r_{1})\mathcal{U}_{c}(r_{2})\mathcal{U}_{d}(r_{1}).
\end{equation}
which corresponds to a function of the two radial variables $r_1$ and $r_2$.

Anyway, if we compare the equation~\eqref{eq:nonantGmatrix} with the corresponding for inter-electronic Coulomb matrix elements, equation~\eqref{eq:nonantmatrix}, we realize that both orbital and spin angular momenta integrals are formally equivalent between these equations. Consequently, we may write 

 \begin{align}\label{eq:maelemten87}
(\hat{G}(r_1,r_2,\theta))_{ab,cd}^{\aleph}&= \sum_{k}\frac{1}{4}(-1)^{L-k}(2k+1)\\ \nonumber
&\times R(ab,cd)\left[(2l_a+1)(2l_b+1)(2l_c+1)(2l_d+1)\right]^{\frac{1}{2}}\\ \nonumber
&\times\tj{l_a}{l_c}{k}{0}{0}{0}\tj{l_b}{l_d}{k}{0}{0}{0}\Gj{l_a}{k}{l_c}{l_d}{L}{l_b}P_k(cos\theta).
\end{align}

\subsubsection{Density operator matrix elements with antisymmetric configurations}

Previously we have calculated the Coulomb $1/r_{12}$ matrix elements with both non-antisymmetric, equation~\eqref{eq:maelemten75}, and antisymmetric configurations, see equations~\eqref{eq:eqeqcoul}--\eqref{eq:eqeqcoul3}. In the same way, the non-antisymmetric density operator matrix elements are calculated in the expressions~\eqref{eq:funRG1} and~\eqref{eq:maelemten87}. Therefore, using the antisymmetrized \ac{WF}, equations~\eqref{eq:ansymm33} and~\eqref{eq:ansymm34}, we can write the matrix element of the operator $\hat{G}(r_1,r_2,\theta)$ between two-electron configurations consisting of equivalent or non-equivalent electrons. Then we proceed as follows:
 
\subsubsection*{Equivalent---Equivalent Electrons}

\begin{align}\label{eq:eqeqdens}
(\hat{G}(r_1,r_2,\theta))_{aa,cc}&=\braket{\{(n_al_a)^2\}LS|\hat{G}(r_1,r_2,\theta)|\{(n_cl_c)^2\}LS},\\ \nonumber
&=\braket{(n_al_a)_1(n_al_a)_2LS|\hat{G}(r_1,r_2,\theta)|(n_cl_c)_1(n_cl_c)_2LS},\\ \nonumber
&=(\hat{G}(r_1,r_2,\theta))_{ab,cd}^{\aleph}\\ \nonumber
&=\sum_{k}\frac{1}{4}(-1)^{L-k}(2k+1)R(ab,cd)(2l_a+1)(2l_c+1)\\ \nonumber
&\times\tj{l_a}{l_c}{k}{0}{0}{0}^2\Gj{l_a}{k}{l_c}{l_c}{L}{l_a}P_k(cos\theta).
\end{align}
In a similar way as performed in the two-electron Coulomb integrals, the general antisymmetrized matrix elements $(\hat{G} (r_1,r_2, \theta) )_{ab,cd}$ for the density are summarized below for the rest of configurational cases of two-electron configurations.

\subsubsection*{Equivalent---Non-equivalent Electrons}

\begin{align}\label{eq:eqeqdens1}
&(\hat{G}(r_1,r_2,\theta))_{aa,cd}\\ \nonumber
&=\braket{\{(n_al_a)^2\}LS|\hat{G}(r_1,r_2,\theta)|\{n_cl_c\hspace{4pt}n_dl_d\}LS},\\ \nonumber
&=\bra{(n_al_a)_1(n_al_a)_2LS}\hat{G}(r_1,r_2,\theta)\frac{1}{\sqrt{2}}\left[\ket{(n_cl_c)_1(n_dl_d)_2LS}\right. \\ \nonumber
&+\left.(-1)^{l_c+l_d+L+S}\ket{(n_dl_d)_1(n_cl_c)_2LS}\right],\\ \nonumber
&=\frac{1}{\sqrt{2}}(\hat{G}(r_1,r_2,\theta))_{aa,cd}^{\aleph}+\frac{1}{\sqrt{2}}(-1)^{l_c+l_d+L+S}(\hat{G}(r_1,r_2,\theta))_{aa,dc}^{\aleph},\\ \nonumber
&=\frac{1}{\sqrt{32}}(2l_a+1)[(2l_c+1)(2l_d+1)]^{\frac{1}{2}}\sum_{k}(2k+1)\left[R(aa,cd)(-1)^{L-k}\right.\\ \nonumber
&+(-1)^{l_c+l_d-k+S}\left.R(ab,dc)\right]\\ \nonumber
&\times\tj{l_a}{l_c}{k}{0}{0}{0}\tj{l_a}{l_d}{k}{0}{0}{0}\Gj{l_a}{k}{l_c}{l_d}{L}{l_a}P_k(cos\theta).\\ \nonumber
\end{align}

\subsubsection*{Non-equivalent---Equivalent Electrons}

\begin{align}\label{eq:eqeqdens2}
&(\hat{G}(r_1,r_2,\theta))_{ab,cc}\\ \nonumber
&=\braket{\{n_al_a\hspace{4pt}n_bl_b\}LS|\hat{G}(r_1,r_2,\theta)|\{(n_cl_c)^2\}LS},\\ \nonumber
&=\frac{1}{\sqrt{2}}\left[\bra{(n_al_a)_1(n_bl_b)_2LS}+(-1)^{l_a+l_b+L+S}\bra{(n_bl_b)_1(n_al_a)_2LS}\right] \\ \nonumber
&\times \hat{G}(r_1,r_2,\theta)\ket{(n_cl_c)_1(n_cl_c)_2LS},\\ \nonumber
&=\frac{1}{\sqrt{2}}(\hat{G}(r_1,r_2,\theta))_{ab,cc}^{\aleph}+\frac{1}{\sqrt{2}}(-1)^{l_a+l_b+L+S}(\hat{G}(r_1,r_2,\theta))_{ba,cc}^{\aleph},\\ \nonumber
&=\frac{1}{\sqrt{32}}(2l_c+1)[(2l_a+1)(2l_b+1)]^{\frac{1}{2}}\sum_{k}(2k+1)\left[R(ab,cc)(-1)^{L-k}\right.\\ \nonumber
&+(-1)^{l_a+l_b-k+S}\left.R(ba,cc)\right]\\ \nonumber
&\times\tj{l_a}{l_c}{k}{0}{0}{0}\tj{l_b}{l_c}{k}{0}{0}{0}\Gj{l_a}{k}{l_c}{l_c}{L}{l_b}P_k(cos\theta).\\ \nonumber
\end{align}

\subsubsection*{Non-equivalent---Non-equivalent Electrons}

\begin{align}\label{eq:eqeqdens3}
&(\hat{G}(r_1,r_2,\theta))_{ab,cd} \\ \nonumber
&=\braket{\{n_al_a\hspace{4pt}n_bl_b\}LS|\hat{G}(r_1,r_2,\theta)|\{n_cl_c\hspace{4pt}n_dl_d\}LS},\\ \nonumber
&=\frac{1}{2}\left[\bra{(n_al_a)_1(n_bl_b)_2LS}+(-1)^{l_a+l_b+L+S}\bra{(n_bl_b)_1(n_al_a)_2LS}\right] \\ \nonumber
&\times\hat{G}(r_1,r_2,\theta)\left[\ket{(n_cl_c)_1(n_dl_d)_2LS}+(-1)^{l_c+l_d+L+S}\ket{(n_dl_d)_1(n_cl_c)_2LS}\right] ,\\ \nonumber
&=\frac{1}{2}\left[(\hat{G}(r_1,r_2,\theta))_{ab,cd}^{\aleph}+(-1)^{l_c+l_d+L+S}(\hat{G}(r_1,r_2,\theta))_{ab,dc}^{\aleph}\right.\\ \nonumber
&+\left.(-1)^{l_a+l_b+L+S}(\hat{G}(r_1,r_2,\theta))_{ba,cd}^{\aleph}+\frac{1}{\sqrt{2}}(-1)^{l_a+l_b+l_c+l_d}(\hat{G}(r_1,r_2,\theta))_{ba,dc}^{\aleph}\right].\\ \nonumber
&=\frac{1}{8}[(2l_a+1)(2l_b+1)(2l_c+1)(2l_d+1)]^{\frac{1}{2}}\\ \nonumber
&\times\sum_{k}(2k+1)\left[\left[R(ab,cd)(-1)^{L-k}+(-1)^{l_a+l_b+l_c+l_d+L-k}R(ba,dc)\right]\right.\\ \nonumber
&\times\tj{l_a}{l_c}{k}{0}{0}{0}\tj{l_b}{l_d}{k}{0}{0}{0}\Gj{l_a}{k}{l_c}{l_d}{L}{l_b}P_k(cos\theta)\\ \nonumber
&+\left[R(ab,dc)(-1)^{l_c+l_d+S-k}+(-1)^{l_a+l_b+S-k}R(ba,cd)\right]\\ \nonumber
&\left.\times\tj{l_a}{l_d}{k}{0}{0}{0}\tj{l_b}{l_c}{k}{0}{0}{0}\Gj{l_a}{k}{l_d}{l_c}{L}{l_b}P_k(cos\theta)\right].
\end{align}

\subsection{Two-particle and one-particle electronic density functions of helium-like atoms}\label{sec:densityfunc12}

The two-particle electronic density function of an helium-like atom may be obtained from the rotational trace of the diagonal two-electron density matrix~\citep{Ezra1983}, equation~\eqref{eq:twodens2}. By integrating over the angular coordinate $\theta$ we obtain the two-electron radial density function, which reads
\begin{align}\label{eq:twoparticle}
\rho(r_{1},r_{2})=\int_{-1}^1d(cos\theta)\rho(r_1,r_2,\theta).
\end{align}

We finally obtain the one-particle probability density by integrating equation~\eqref{eq:twoparticle} over the radial coordinate $r_2$

\begin{align}\label{eq:oneparticle}
\rho({r})=\int_0^\infty  r_2^2 dr_2\int_{-1}^1d(cos\theta)\rho(r_1=r,r_2,\theta).
\end{align}

\subsection{One-particle electronic density function for bound states of Helium atom}

After these computational details, we are now able to explore the one- and two-electron radial densities in helium, which are mathematical distribution functions after all, subject to any topological scrutiny by means of information entropic measures.

\begin{table}[h]
\centering
\begin{tabular}{cccccccc}
 \hline\hline 
& & &\multicolumn{2}{c}{$\rho(0)$}& &\multicolumn{2}{c}{$E(a.u.)$}  \\
\cline{4-5}\cline{7-8}
& & &  \citeauthor{Saavedra1995} & Present&  & \citeauthor{Saavedra1995} & Present\\ \hline
\multirow{3}{*}{ $^{1}S^{e}$}&1 & & $3.620858$ & $3.620790$ & & $-2.903724$ & $ -2.903508$ \\
 &2 & & $2.618920$ & $2.618926$  & & $-2.145974$ & $ -2.145960$  \\ 
  &3& & $2.566253$ & $2.566259$  & &    $-2.061272$ & $-2.061267$\\ \hline
  \multirow{2}{*}{ $^{3}S^{e}$}&1& & $2.640710$ & $2.640708$  & & $-2.175229$ & $ -2.175228$ \\
 &2 & & $2.570120$ & $2.570117$  & & $-2.068689$ & $ -2.068688$  \\  \hline\hline  
\end{tabular}
\caption[Values for the one-electron density at the nucleus $\rho(r=0)$ and energies for the
lowest three $L=0$ bound states (singlets and triplets)]{\label{tab:table3}Values for the one-electron density at the nucleus $\rho(r=0)$ and energies for the
lowest three $L=0$ bound states (singlets and triplets). Our results are compared with previous work by~\citep{Saavedra1995}.}
\end{table}

\begin{figure}
\centering
\hspace{-0.2cm}\includegraphics[width=0.63\textwidth]{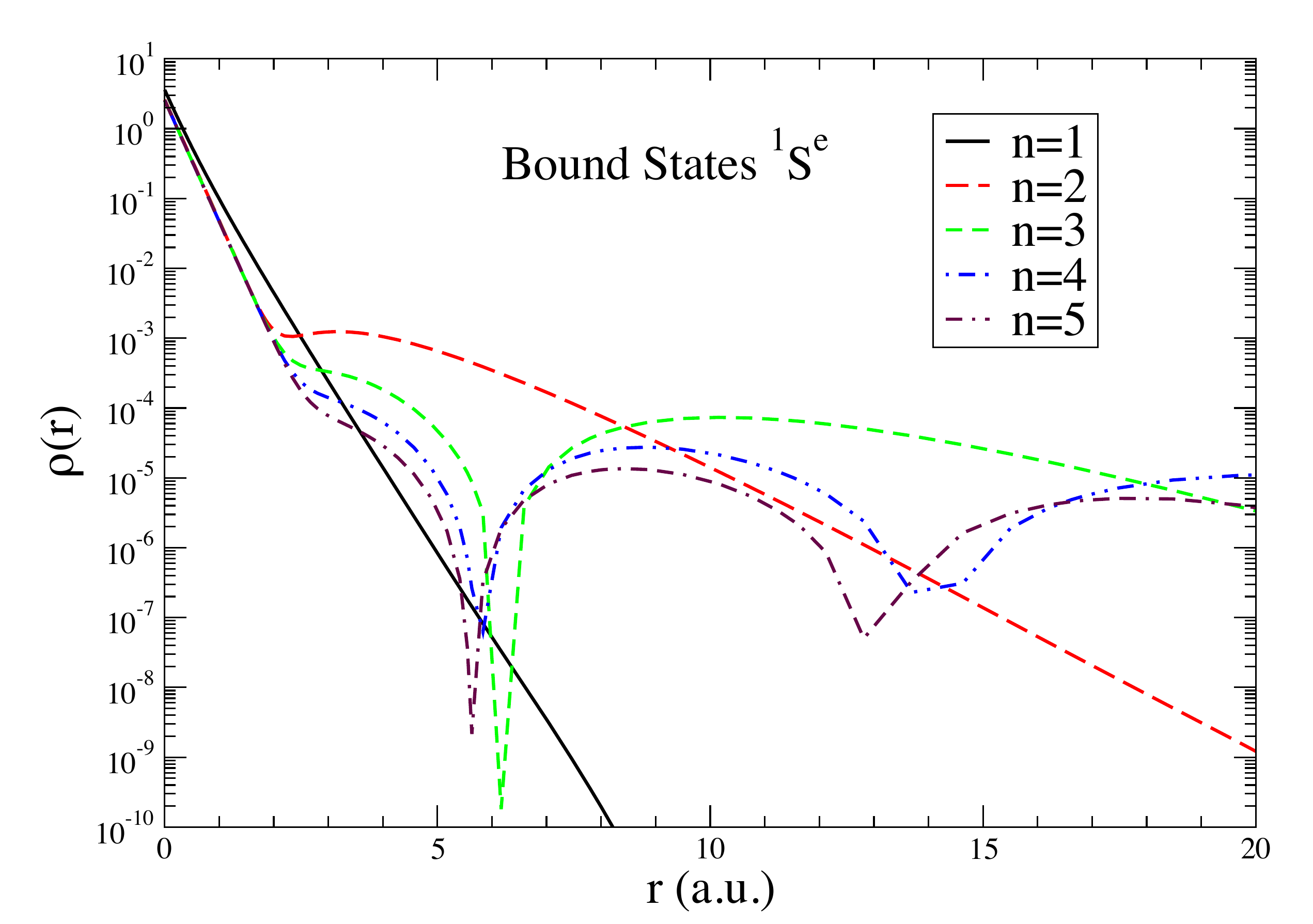}\\
\includegraphics[width=0.62\textwidth]{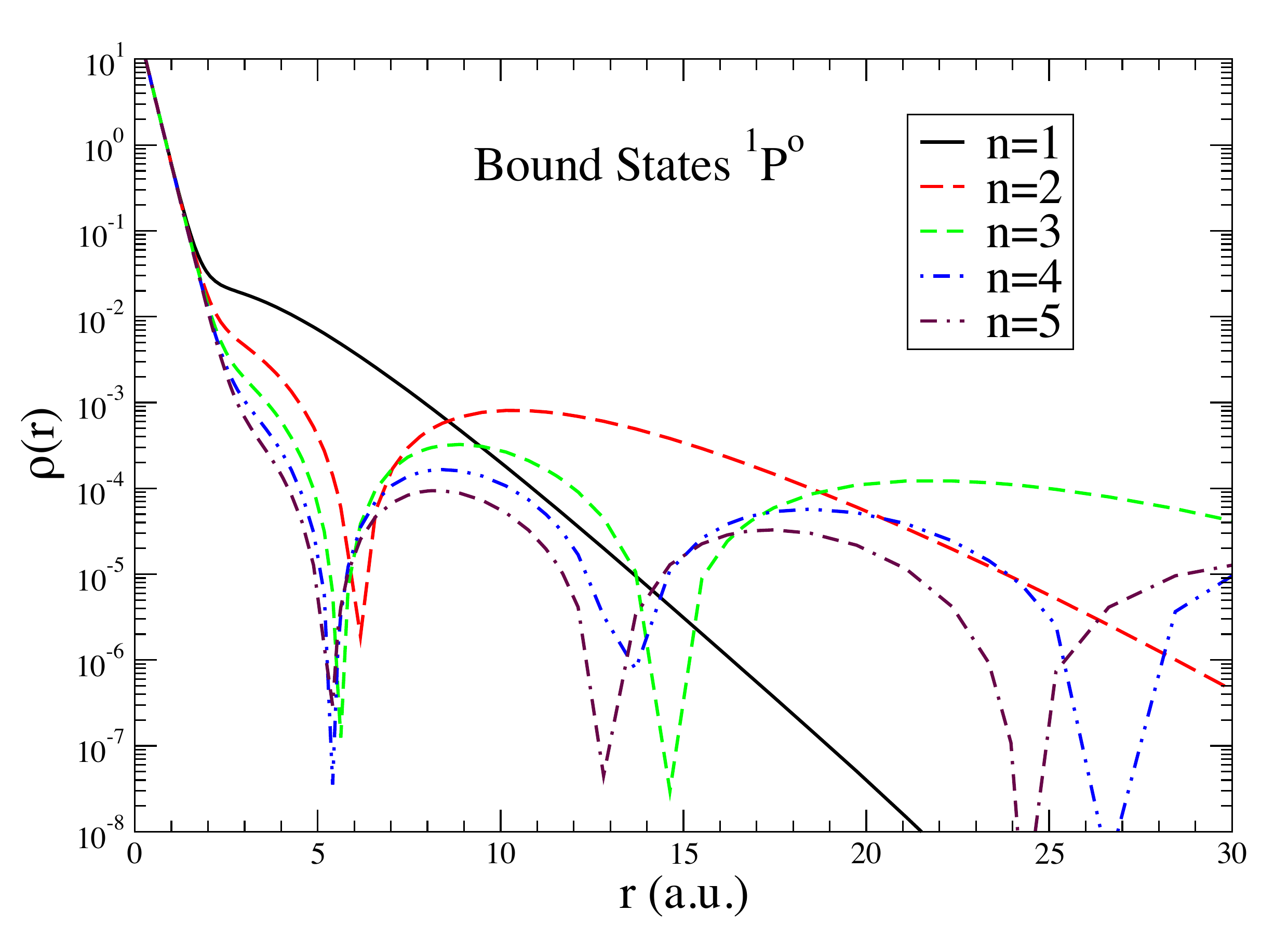}\\
\includegraphics[width=0.61\textwidth]{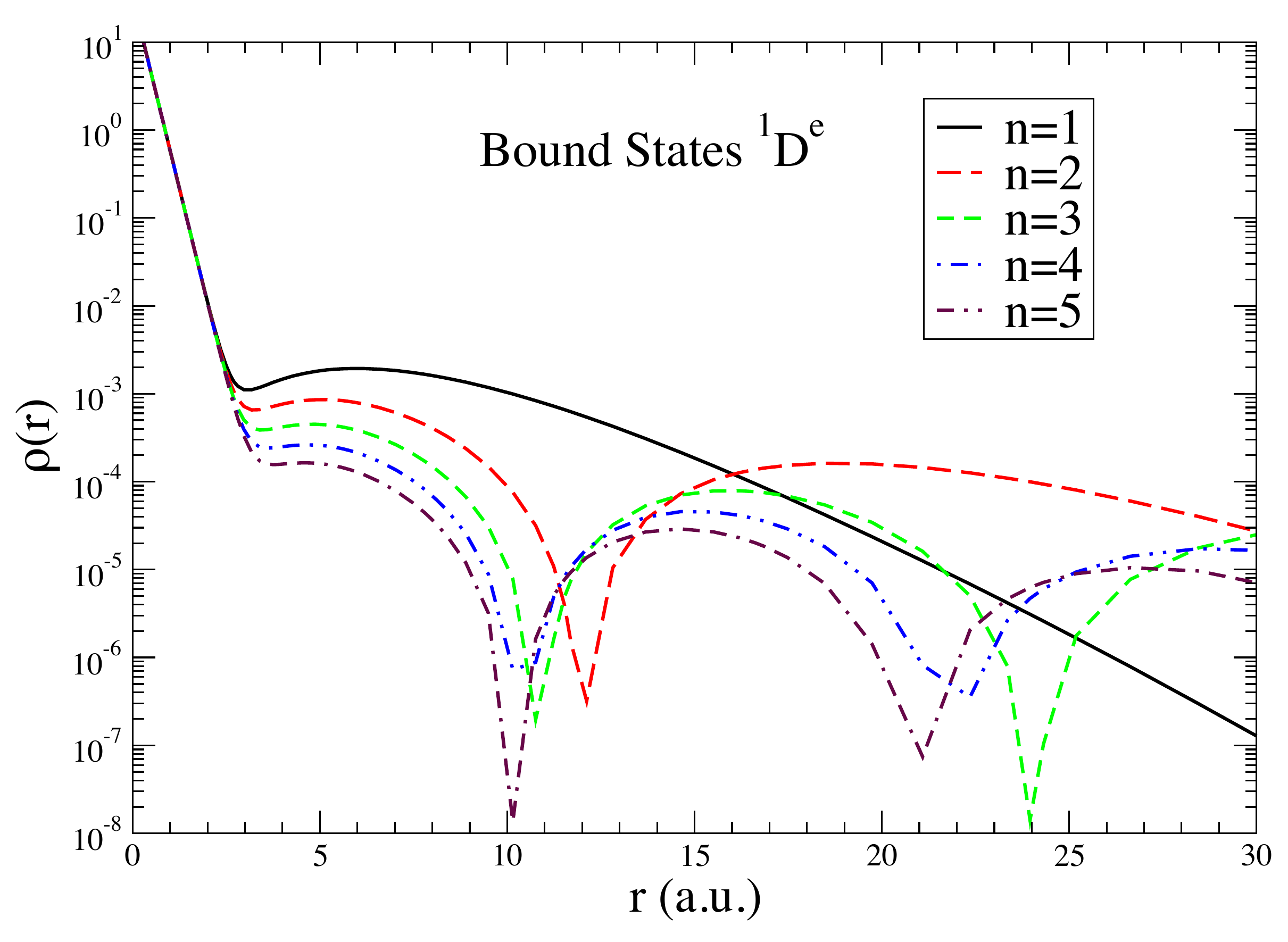}\\
\caption[Electronic one-particle density $\rho({r})$ for the lowest five bound states in the spectroscopic symmetries $^{1}S^{e}$, $^{1}P^{o}$ and $^{1}D^{e}$]{\label{fig:onedens12}Electronic one-particle radial density $\rho({r})$ for the lowest five bound states in the spectroscopic symmetries  $^{1}S^{e}$, $^{1}P^{o}$ and $^{1}D^{e}$. Note that the y-axis is in logarithmic scale.}
\end{figure}

We begin by assessing the quality of our results by comparing them with previous ones in the literature. Our energies and values for the one-electron radial density at $r=0$ for the lowest $^1S^e$ and $^3S^e$ states in helium are included in table~\ref{tab:table3}, and compared with those available from~\citep{Saavedra1995}. The values reported by the latter authors were obtained using explicitly correlated \ac{WF} following the work by~\citep{Pekeris1958,Pekeris1959,Pekeris1962} with perimetric coordinates. Evidently, our \ac{CI}-\ac{WF} has not such degree of sophistication but, increasing our angular correlation, we can reproduce up to six figures in the energies and up to five in the radial densities. This nice comparison endorses our computational procedure, which is not aimed at obtaining precise numerical results for bound states but for highly lying \ac{DES}, for which explicitly correlated methods are less indicated. Anyway, we are mostly interested on the qualitative behavior. The one-particle density calculated at different levels of theory should present differences probing the effects of electron correlation. For instance, the clear effects of electron correlation are visible when comparing the one-electron densities obtained with explicitly correlated configurations and with an uncorrelated Hartree-Fock method (see figure 2 in~\citep{Saavedra1995}). The figure~\ref{fig:onedens12} depicts the  one-particle densities $\rho({r})$  obtained here with our \ac{CI} method. Finally, we want to mention the following interesting qualitative result: only the ground state of Helium $1^1S^e$ and the excited state  $1^1P^o$ have a monotonically decreasing behavior; on the other hand, a non-monotonically decreasing behavior is observed for all of the remaining excited states, this phenomenon was previously observed by~\citep{Rigier1984} and~\citep{Saavedra1995}, and they incidentally report a monotonically decreasing \ac{HF} density function $\rho{}$ for the excited state $2^1S^e$. But this difference with the present \ac{CI}-\ac{WF} density is fundamentally due to the lack of a properly described electron correlation in the \ac{HF} method.

\subsection{Two-particle electronic density function for doubly excited states of Helium atom}

The two-electron radial (two-dimensional dependence) probability is calculated by means of equation~\eqref{eq:twoparticle}. It only involves the spatial coordinates, i.e, we have traced over the angular degrees of freedom. Since we deal with indistinguishable particles we expect that $\rho(r_1,r_2)$ be symmetric about the bisector line of plane $(r_1,r_2)$ (i.e., the line $r_1=r_2$) under the permutation of the particle index. 

The properties of electron correlation on \ac{DES} of two-electron atoms is a problem of considerable theoretical interest~\citep{Cooper1963, Lin1974, Sinanoglu1974, Herrick1975, Ezra1982, Ezra1983}. In order to analyze the electron correlation in the density distribution,~\citeauthor{Ezra1982} have undertaken a detailed study of the two-electron density $\rho(r_1,r_2,\theta_{12})$ via the associated conditional probability  $\rho(r_1,\theta_{12},| r_2=\alpha)$ which is the probability of finding an electron at a distance $r_1$ from the nucleus with interelectronic angle $\theta_{12}$ given that the other electron is at distance $\alpha$ from the nucleus. They conclude that a qualitative examination of the conditional density of the two-electron atoms, calculated via a \ac{CI} approach using Sturmian functions, enables them to find a remarkable degree of collective rotor-vibrator behavior in the $N=2$ shell, showing that the molecular interpretation of the doubly excited spectrum due to~\citep{Kellman1980} is a useful qualitative picture.

\begin{figure}
\centering
\includegraphics[width=0.35\textwidth]{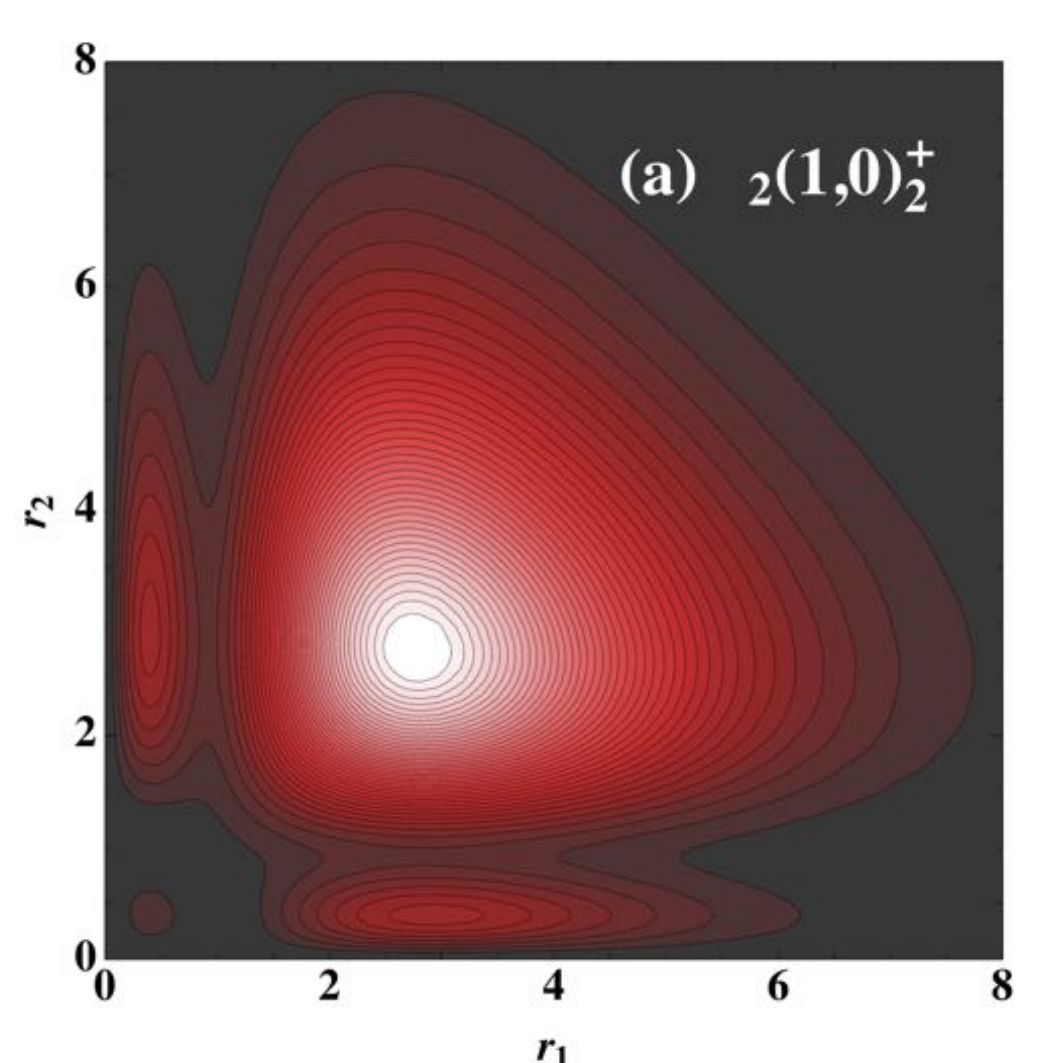}
\includegraphics[width=0.35\textwidth]{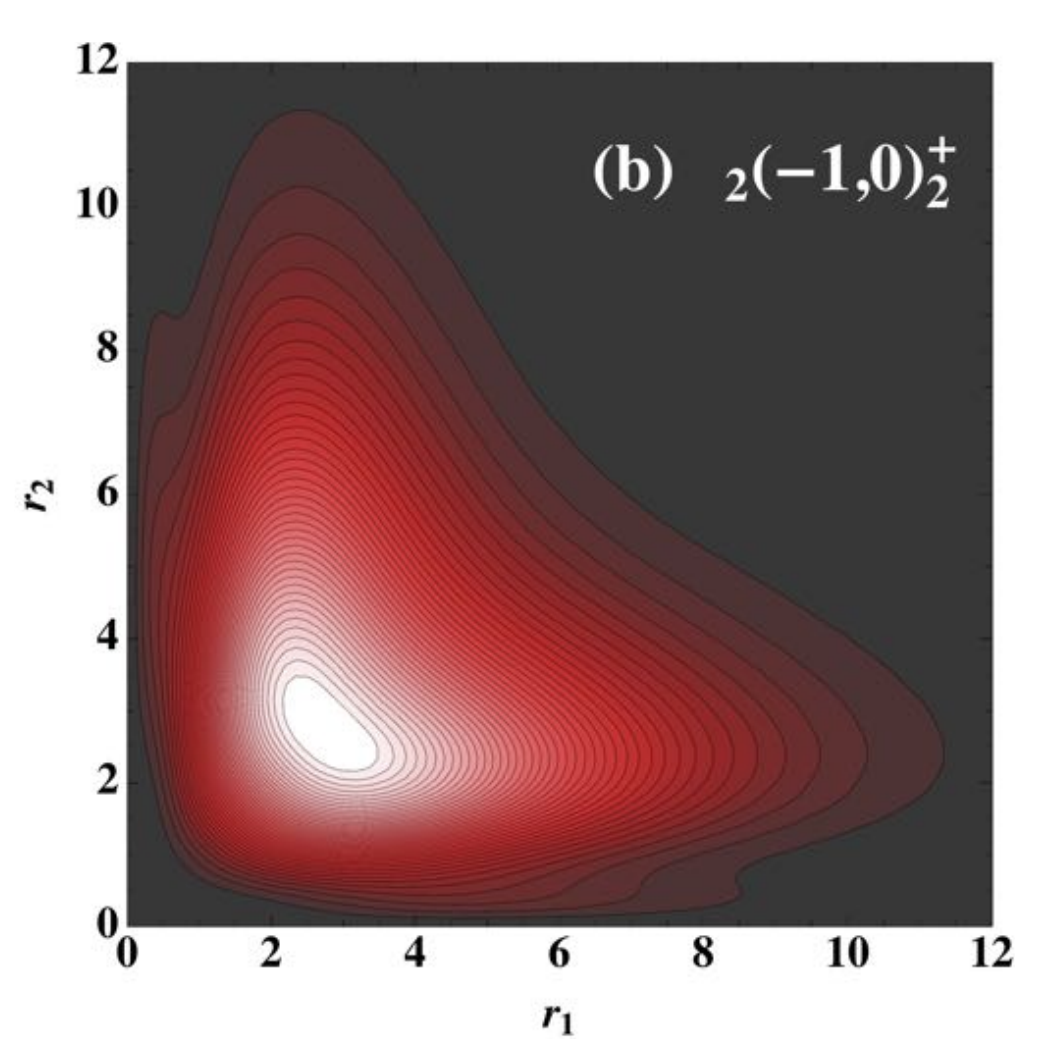}\\
\includegraphics[width=0.35\textwidth]{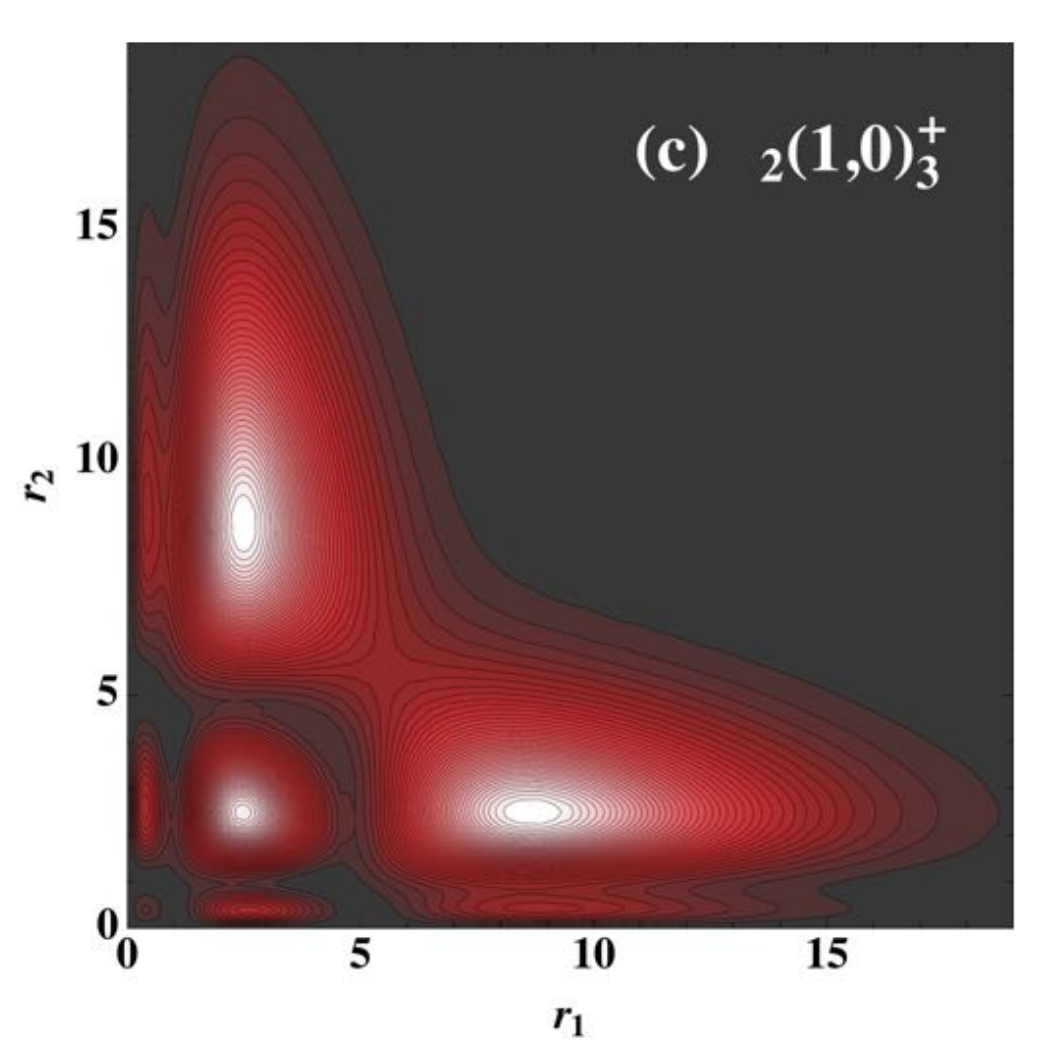}
\includegraphics[width=0.35\textwidth]{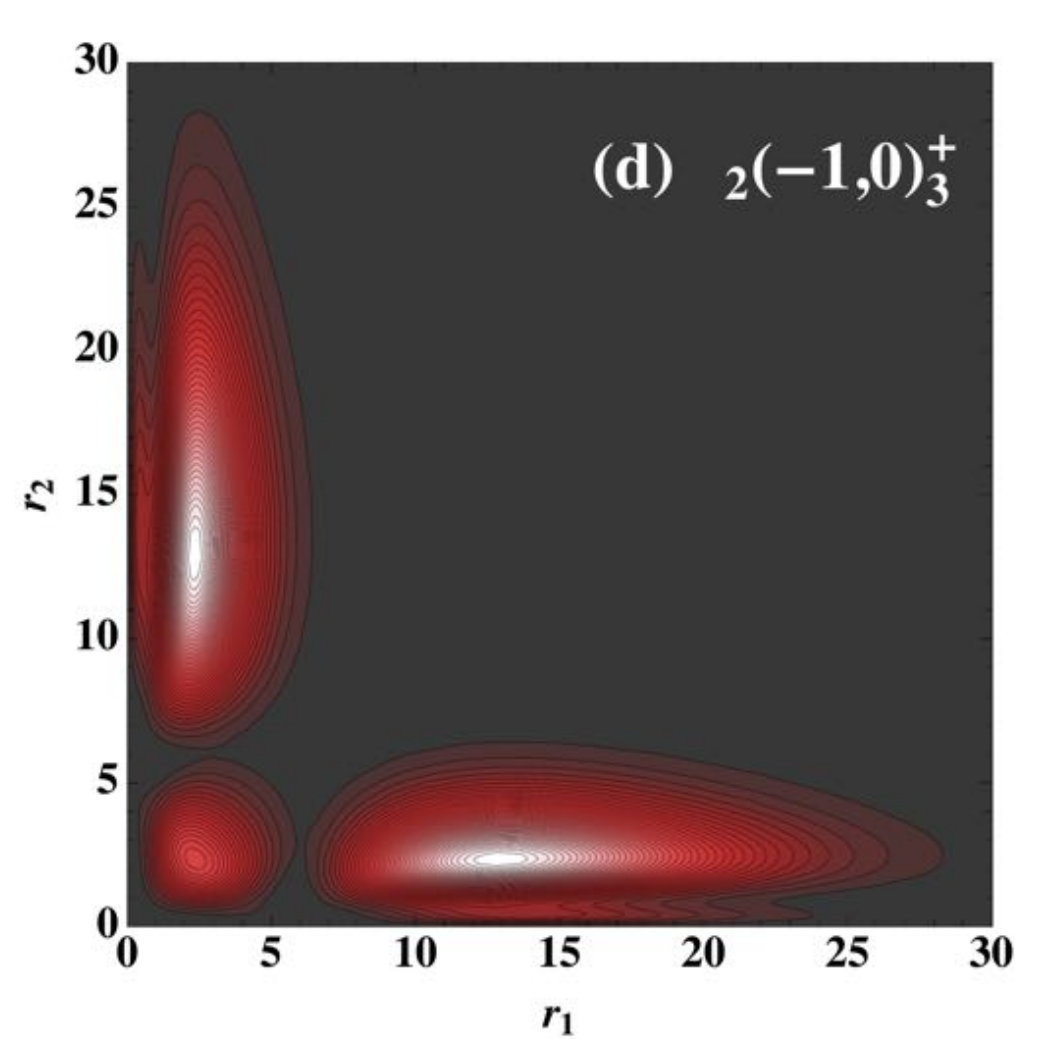}\\
\includegraphics[width=0.35\textwidth]{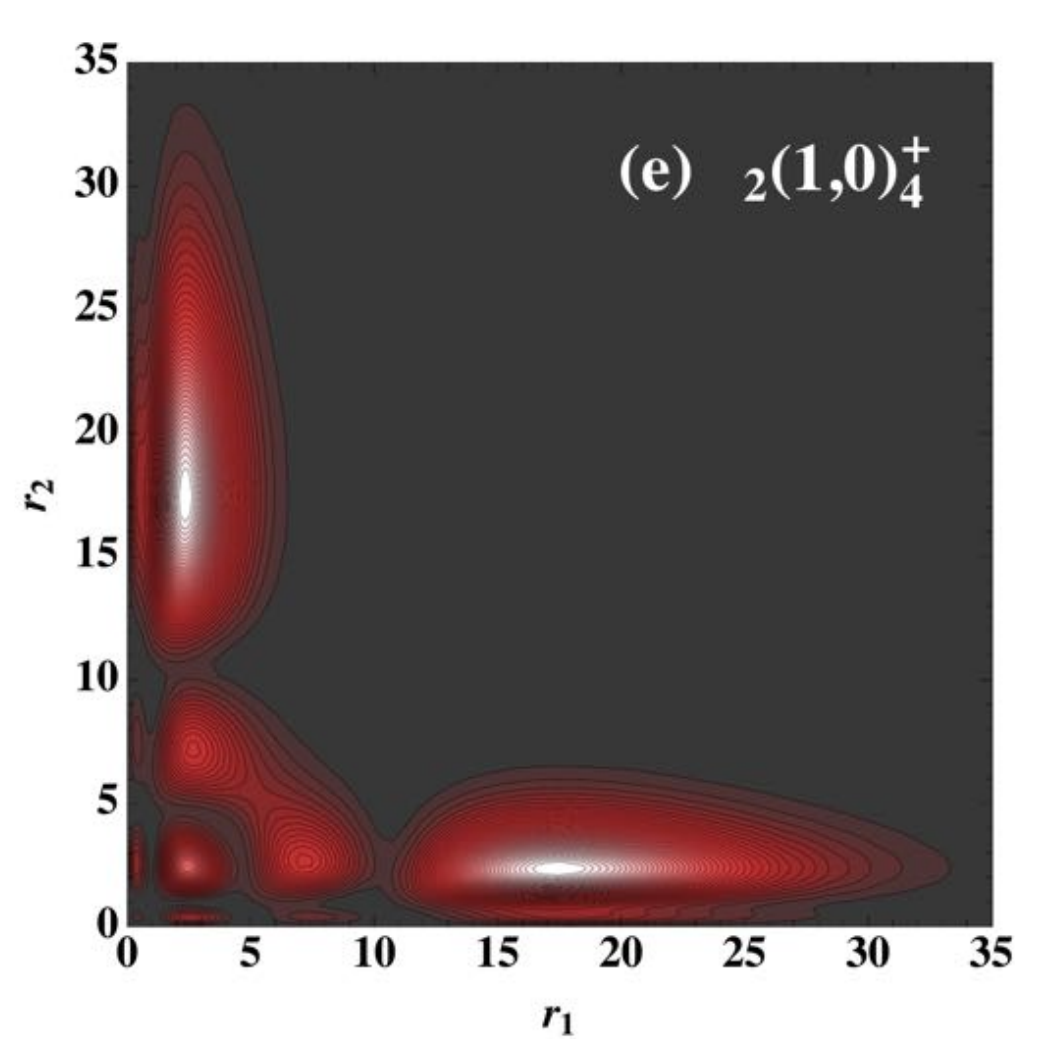}
\includegraphics[width=0.35\textwidth]{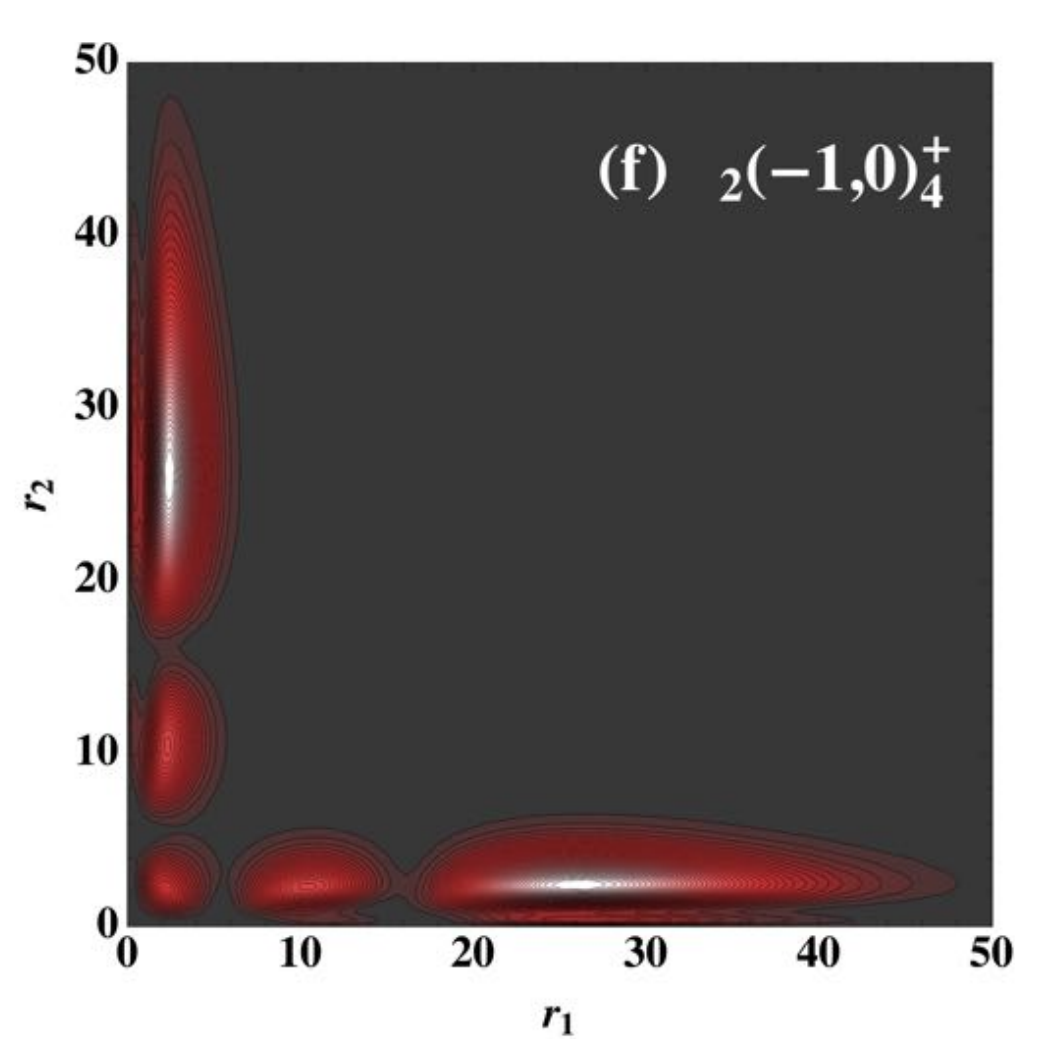}\\
\includegraphics[width=0.35\textwidth]{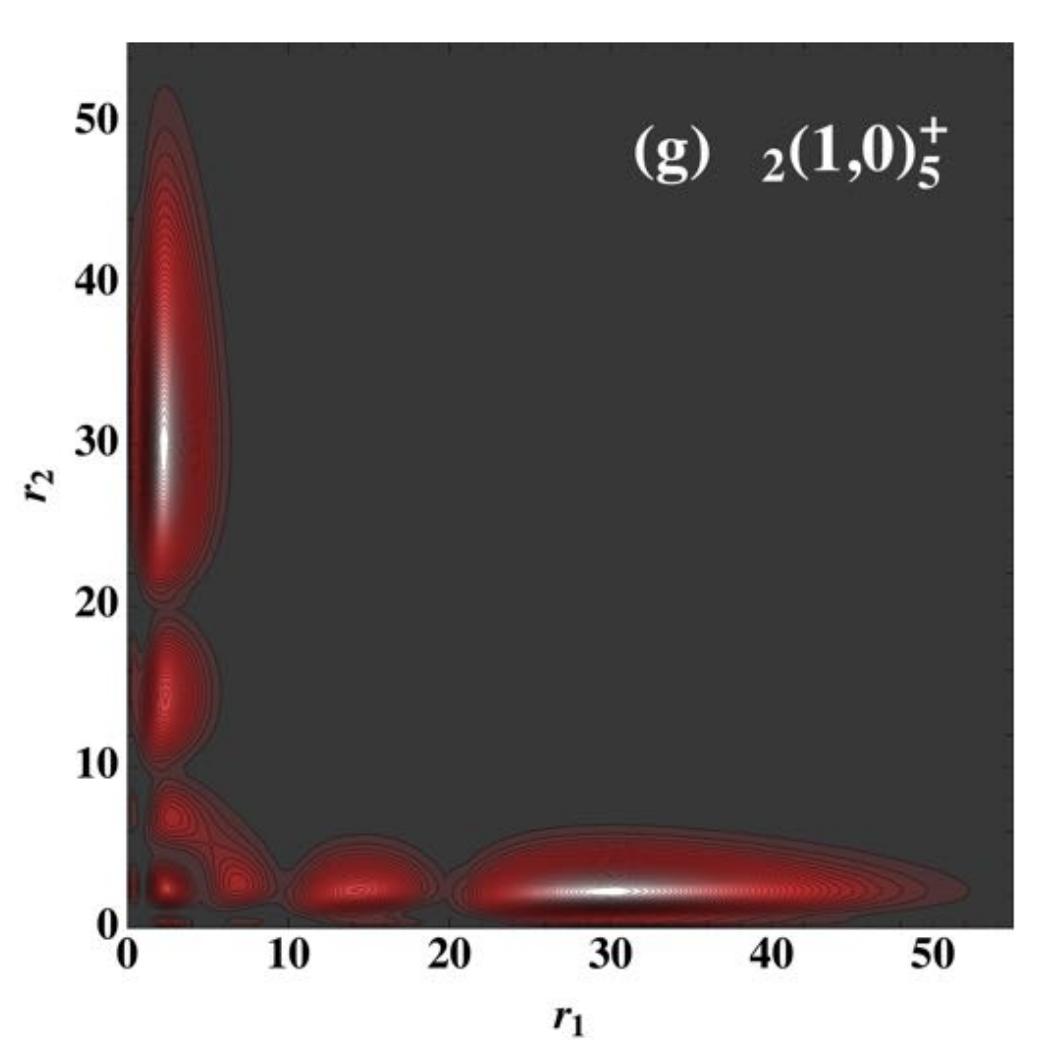}
\includegraphics[width=0.35\textwidth]{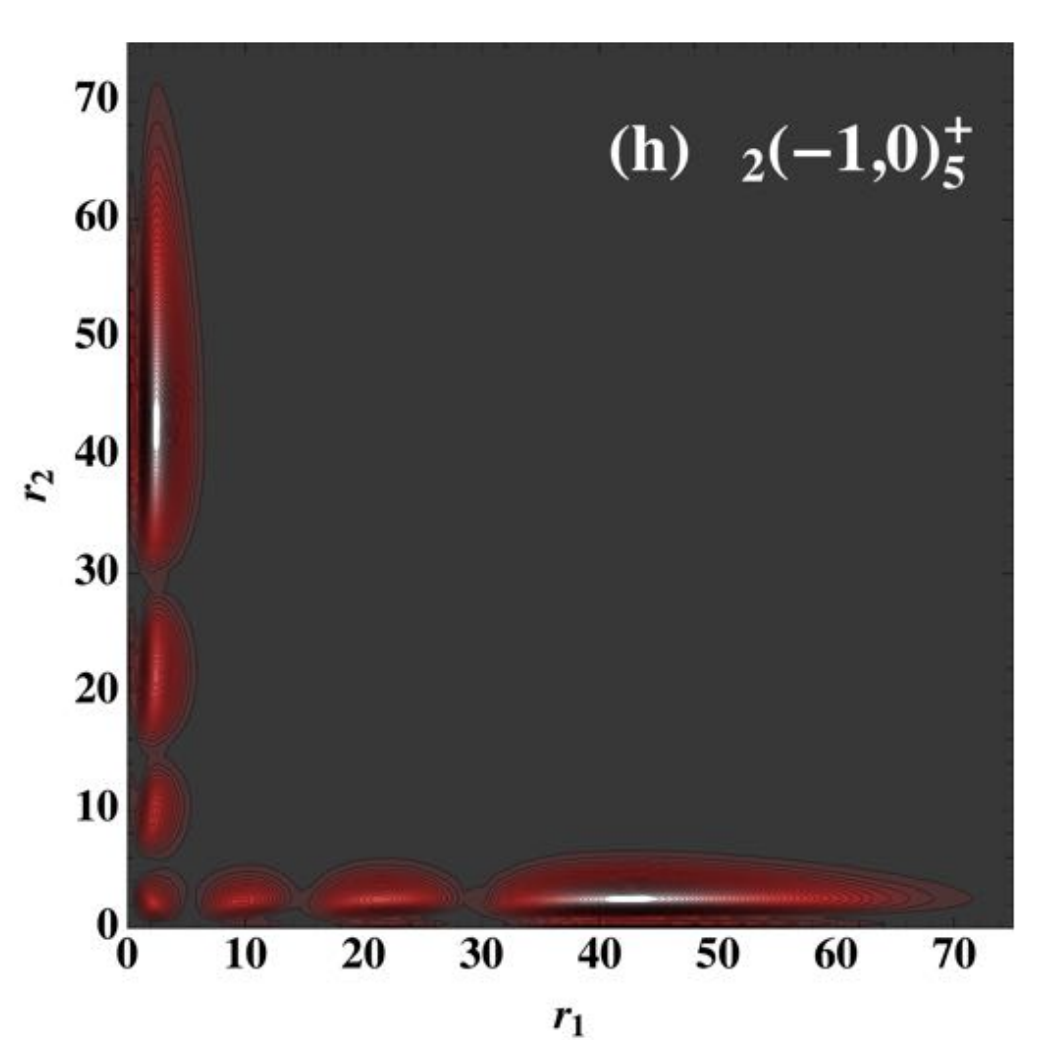}
\caption[Two-electron radial density $\rho(r_{1},r_{2})r_{1}^{2}r_{2}^{2}$ for the lowest eight resonant $^1S^e$ states in He]{\label{fig:wide1}Two-electron radial density $\rho(r_{1},r_{2})r_{1}^{2}r_{2}^{2}$ for the lowest eight resonant $^1S^e$ states in He, located below the second ionization threshold. Resonances are labelled according to the classification proposed by~\citep{Lin1983} using $_{n_1}(K,T)_{n_2}^A$. The energy ordering of the resonances is indicated by the alphabet labels inside the plots.}
\end{figure}

In order to establish a deep qualitative understanding of the structure and classification of \ac{DES} of two-electron atoms  we will focus on studying the two-dimensional electronic density, $\rho(r_{1},r_{2})r_{1}^{2}r_{2}^{2}$, where detailed information of the structure of resonant states can be found. In an earlier work~\citep{Cortes1993}, the authors introduce a multipole expansion of the density of the resonances. They could obtain a description of the electron correlations from the \ac{WF}, using the so called "correlation $Z$ diagrams"~\citep{Macias1989,Macias1991}. Consequently, they concluded that, in general, the electronic density plots of the $^1P^o$ \ac{DES} are roughly scaled pictures of each other and their classification offers no difficulty, e.g., the $(K,T)$ labels may be used throughout the whole $Z$ diagram. Here, we have calculated the two-electron radial density $\rho(r_1,r_2)r_1^2r_2^2$ with the interest of establishing a qualitative and comparative understanding of  the~\citep{Herrick1975,Lin1983} classification  scheme of \ac{DES} in the He atom.

\begin{figure}
\centering
\includegraphics[width=0.35\textwidth]{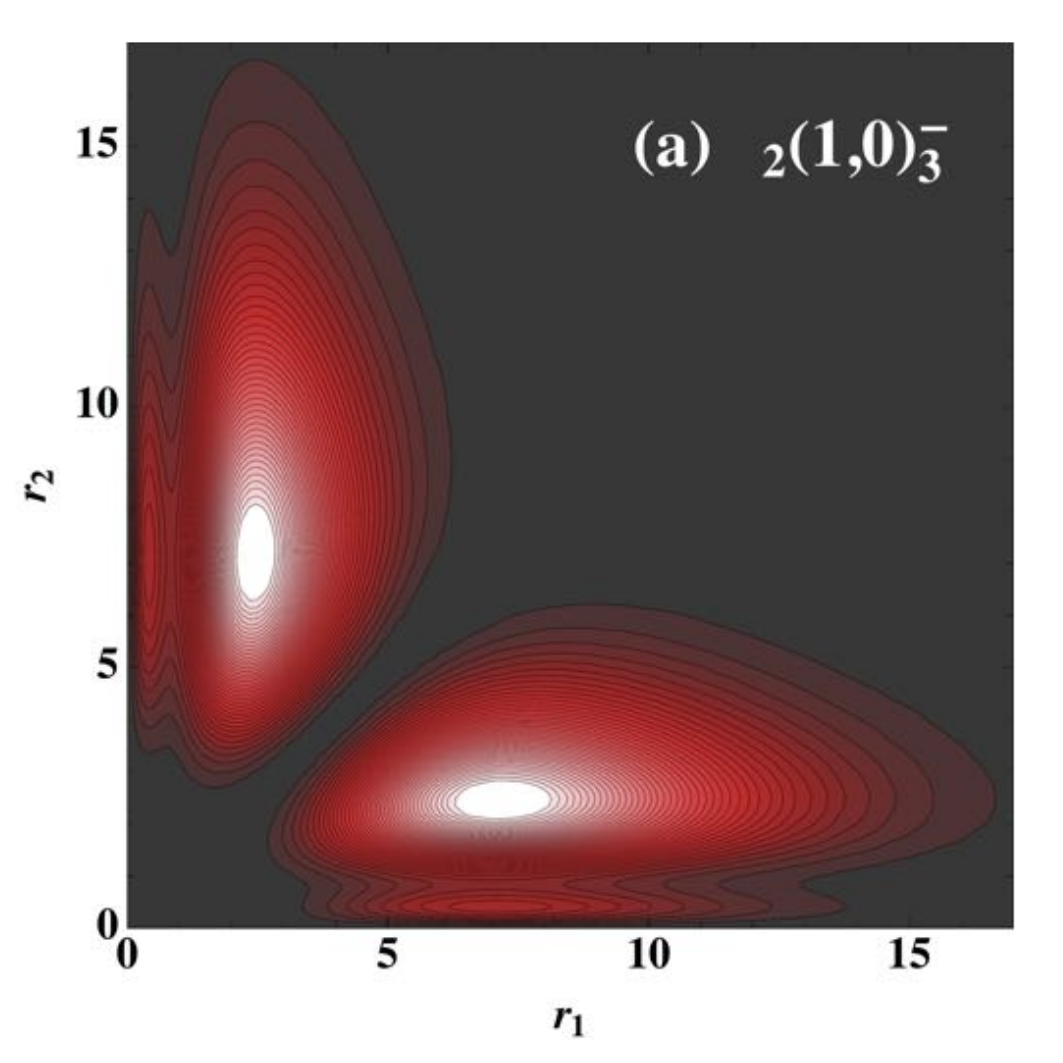}
\includegraphics[width=0.35\textwidth]{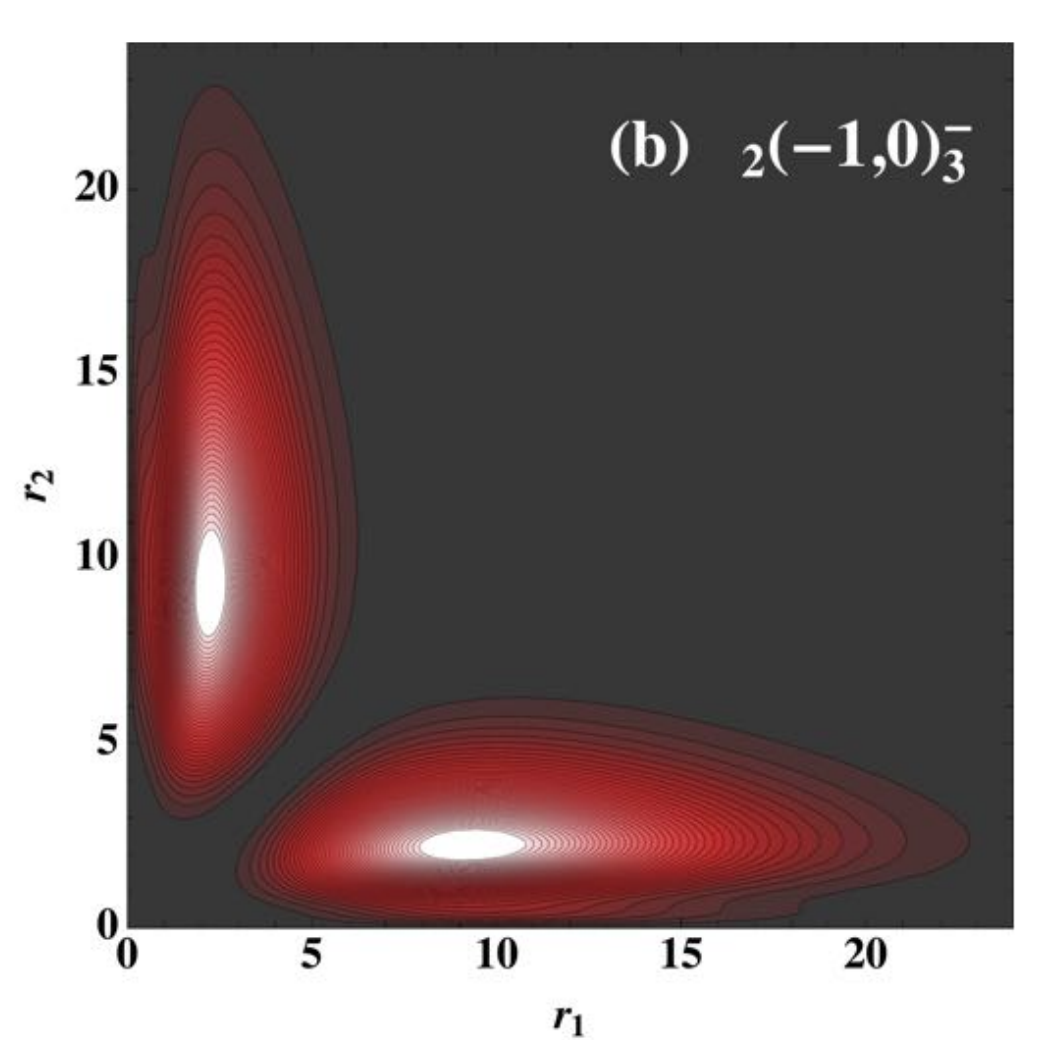}\\
\includegraphics[width=0.35\textwidth]{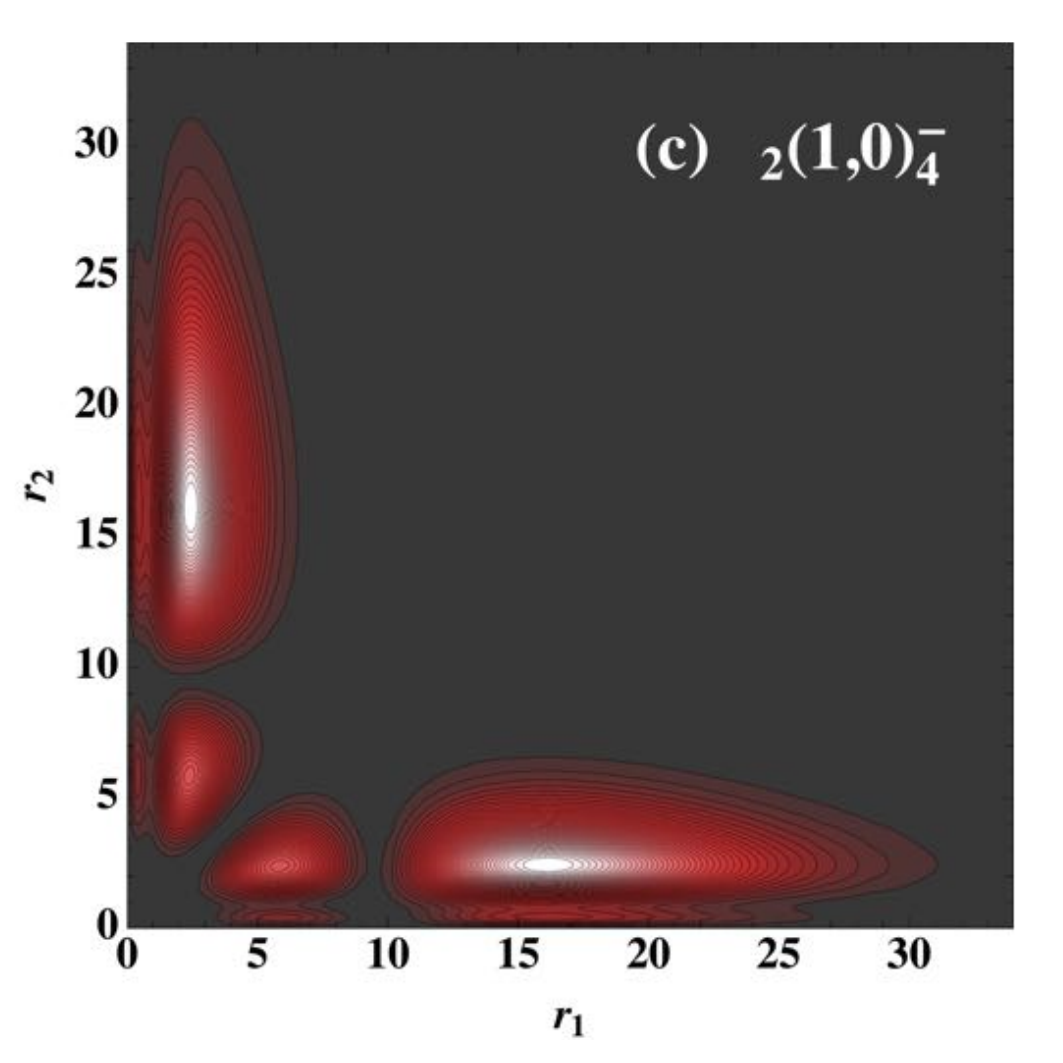}
\includegraphics[width=0.35\textwidth]{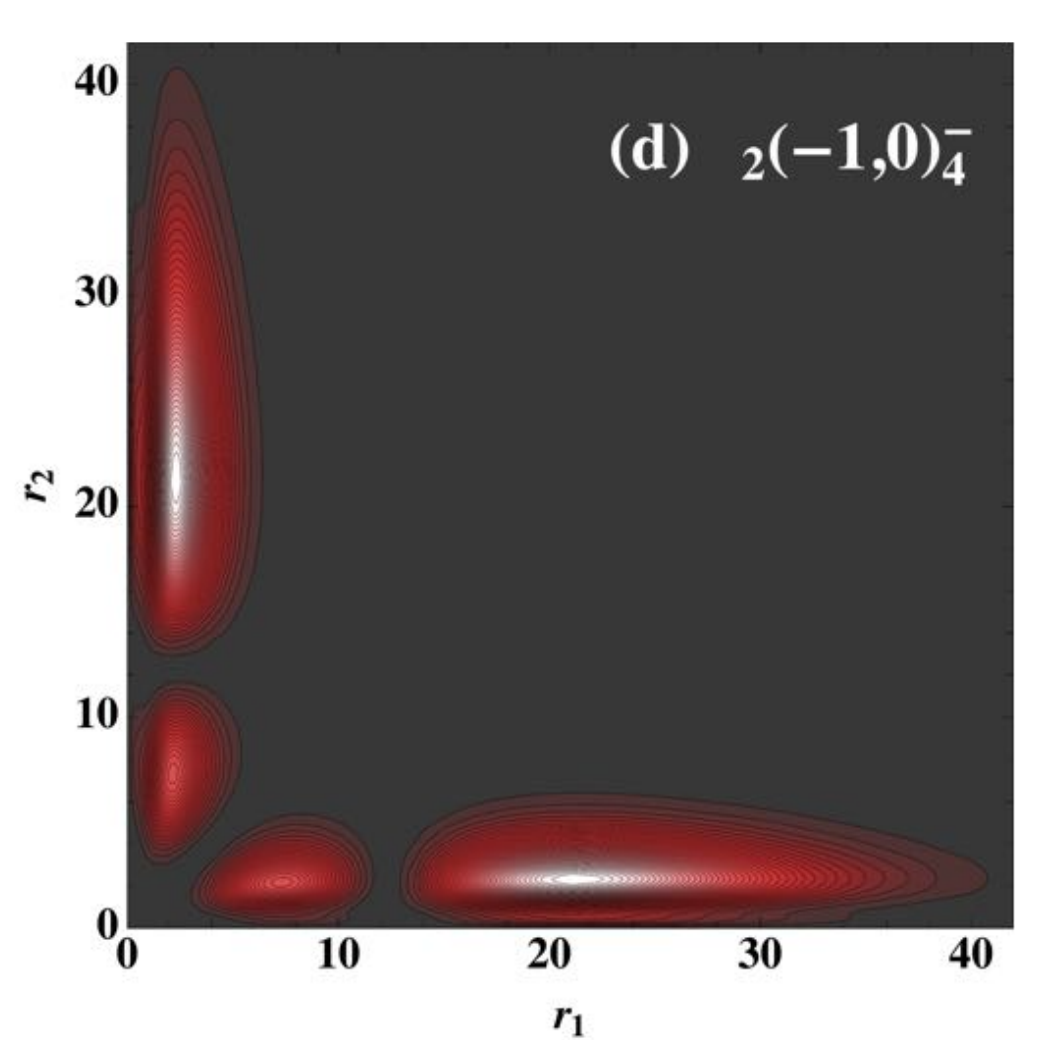}\\
\includegraphics[width=0.35\textwidth]{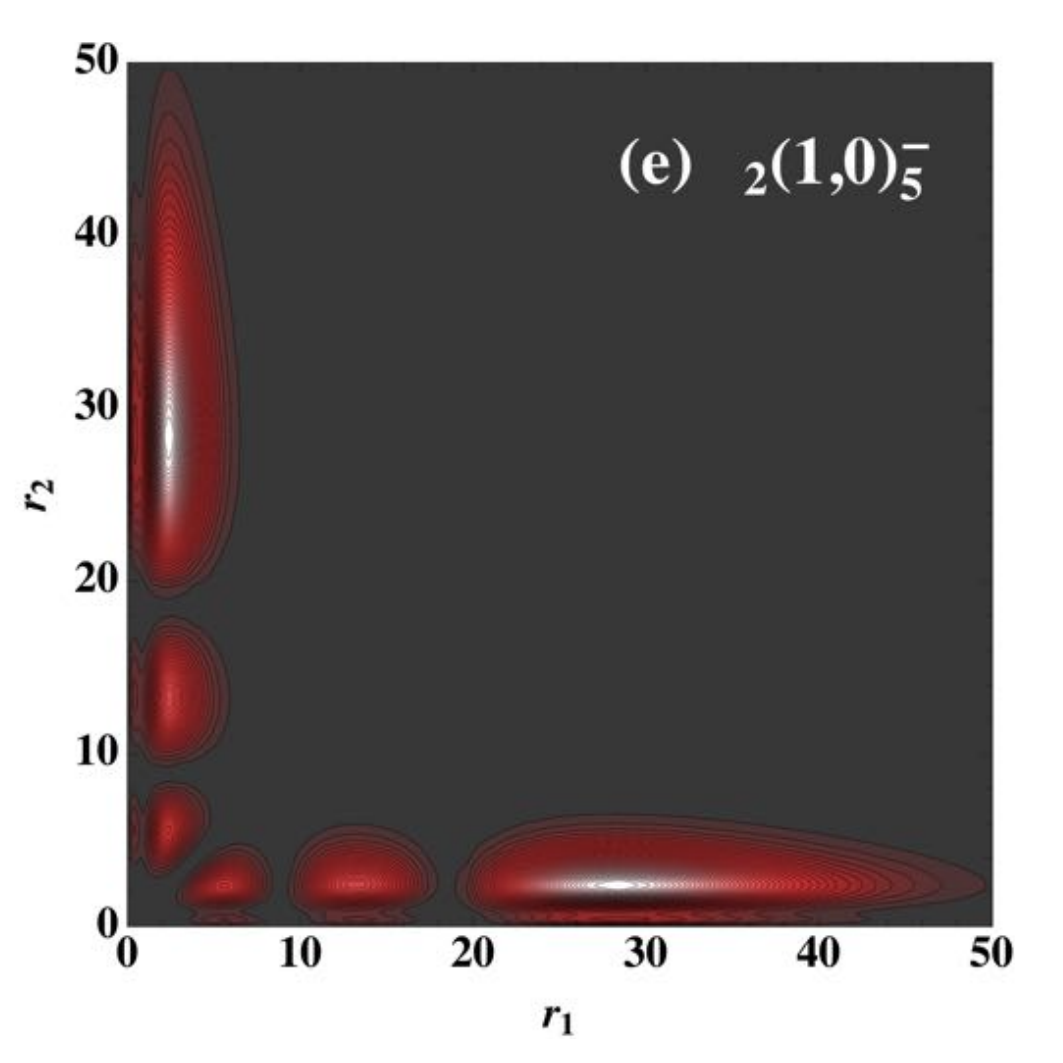}
\includegraphics[width=0.35\textwidth]{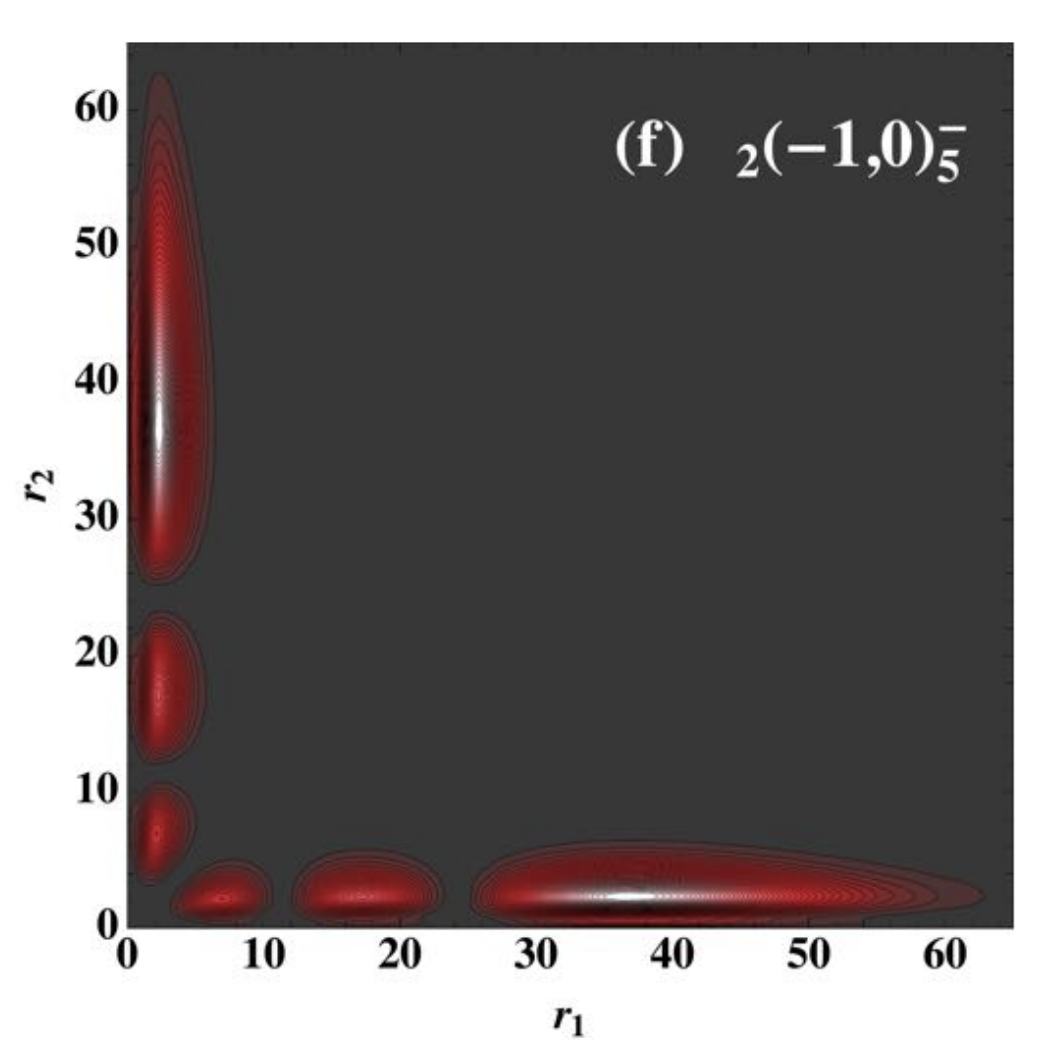}\\
\includegraphics[width=0.35\textwidth]{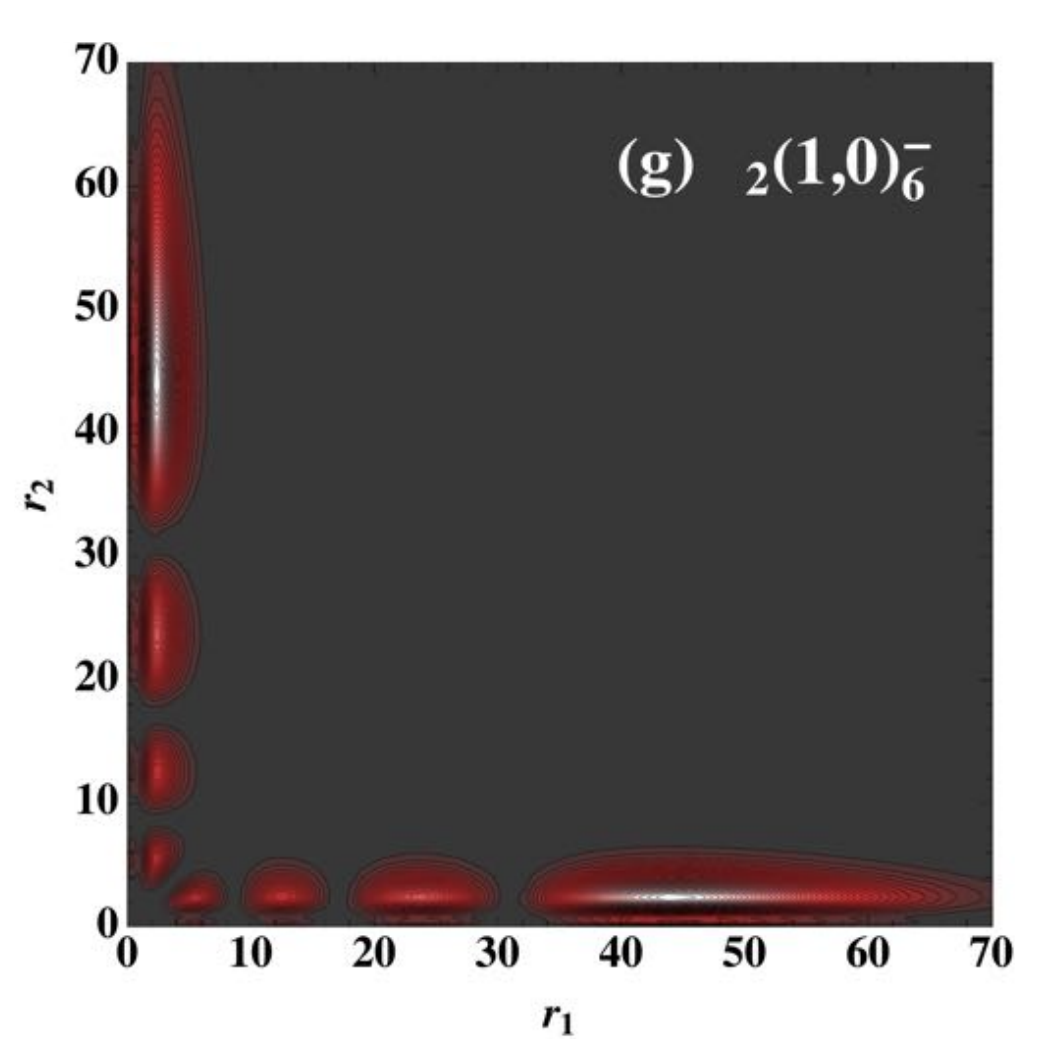}
\includegraphics[width=0.35\textwidth]{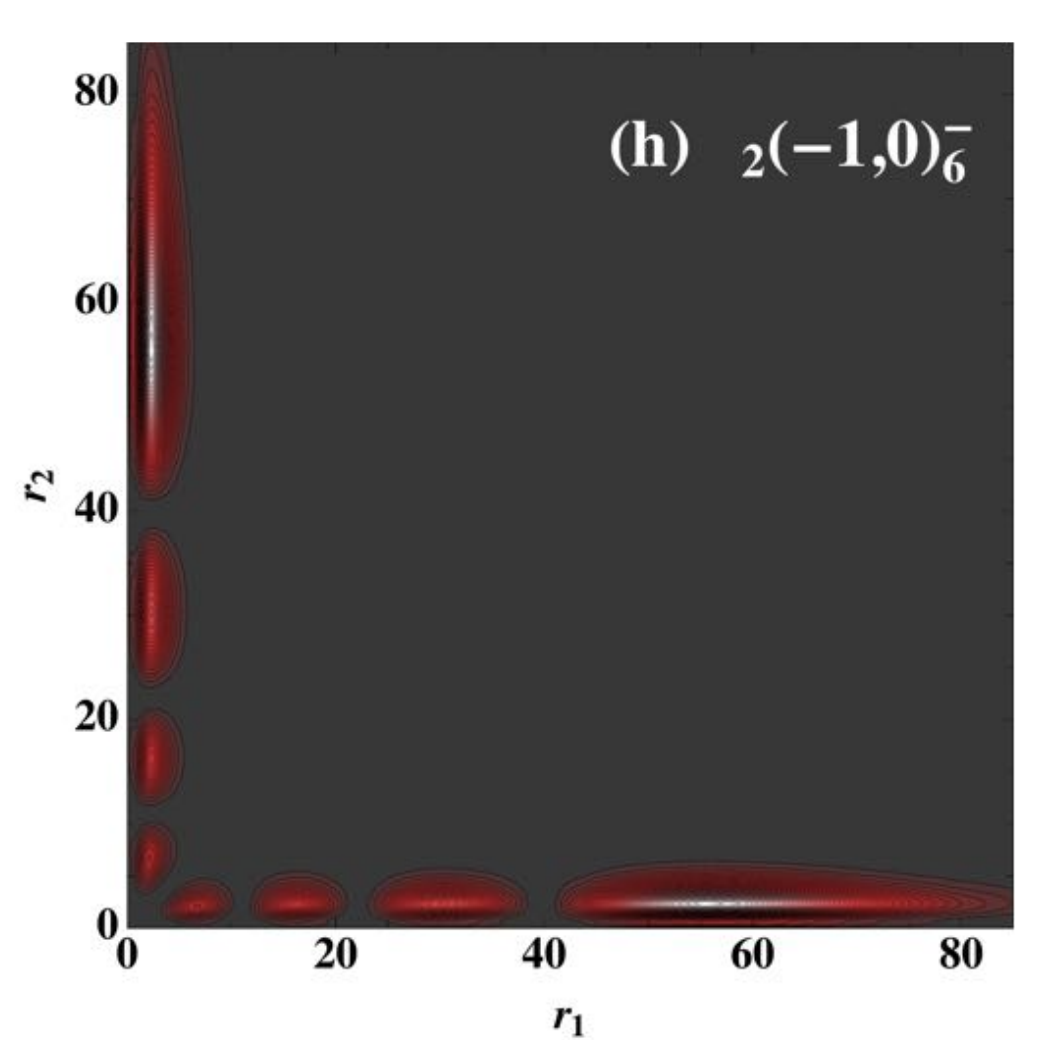}
\caption[Two-electron radial density $\rho(r_{1},r_{2})r_{1}^{2}r_{2}^{2}$ for the lowest eight resonant $^3S^e$ states in He]{\label{fig:wide2}Two-electron radial density $\rho(r_{1},r_{2})r_{1}^{2}r_{2}^{2}$ for the lowest eight resonant $^3S^e$ states in He, located below the second ionization threshold. Resonances are labelled according to the classification proposed by~\citep{Lin1983} using $_{n_1}(K,T)_{n_2}^A$. The energy ordering of the resonances is indicated by the alphabet labels inside the plots.}
\end{figure}

\begin{figure}
\centering
\includegraphics[width=0.28\textwidth]{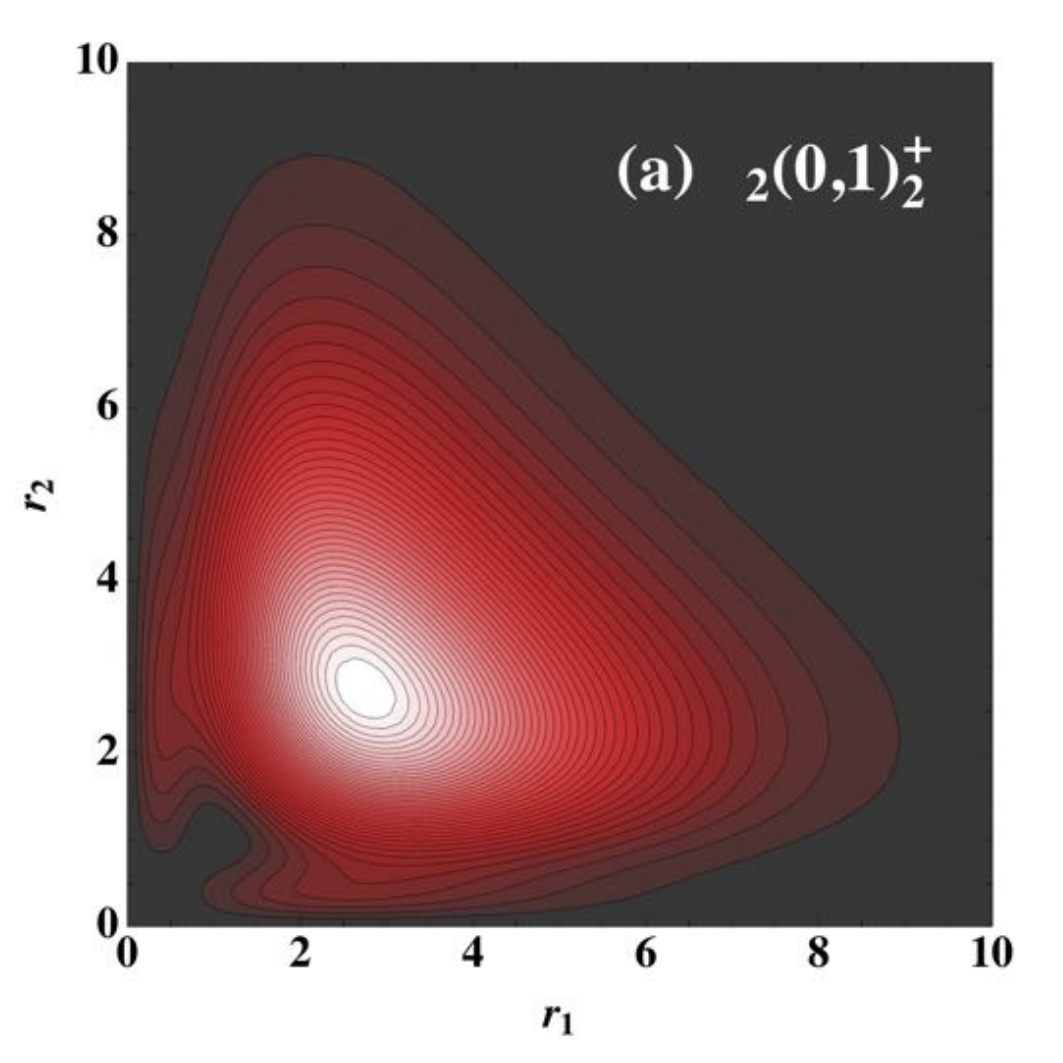}
\includegraphics[width=0.28\textwidth]{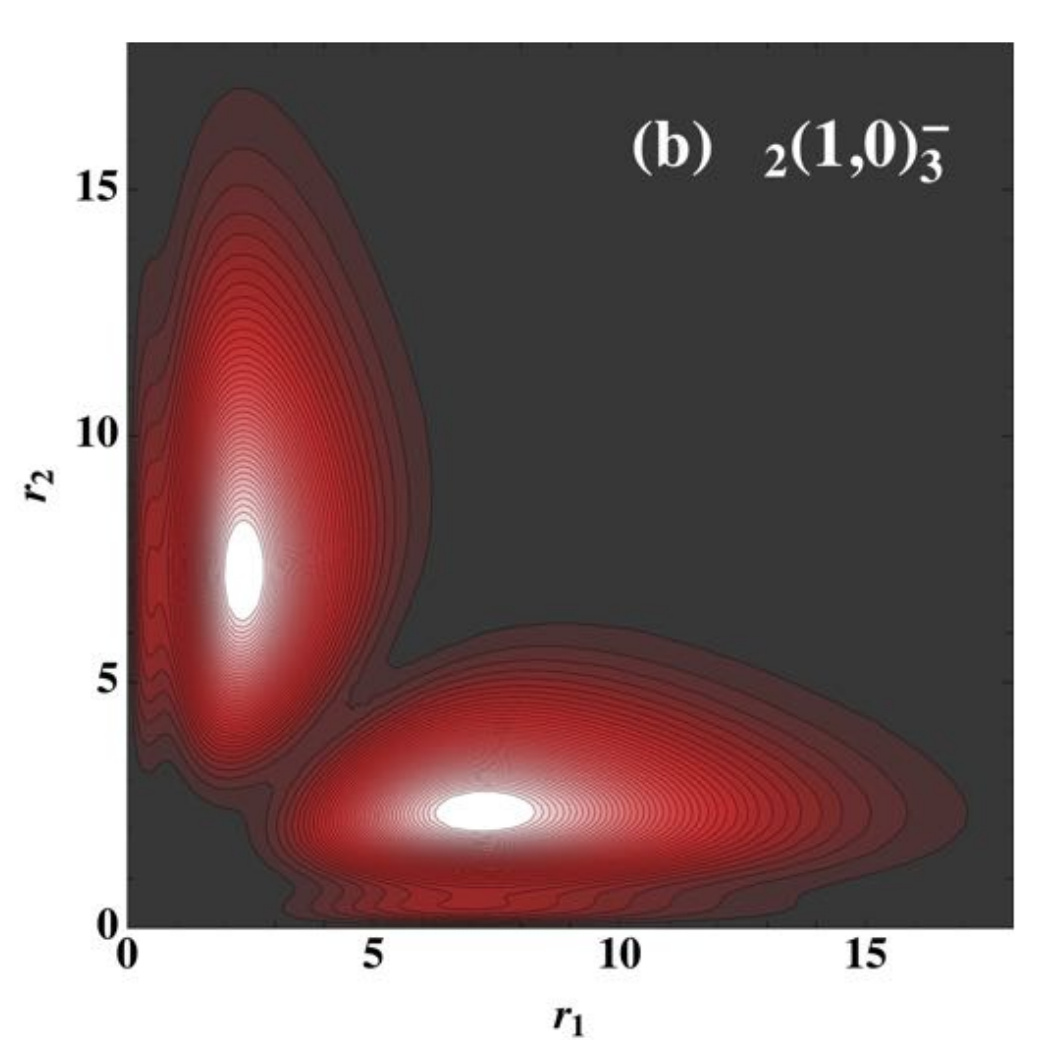}\\
\includegraphics[width=0.28\textwidth]{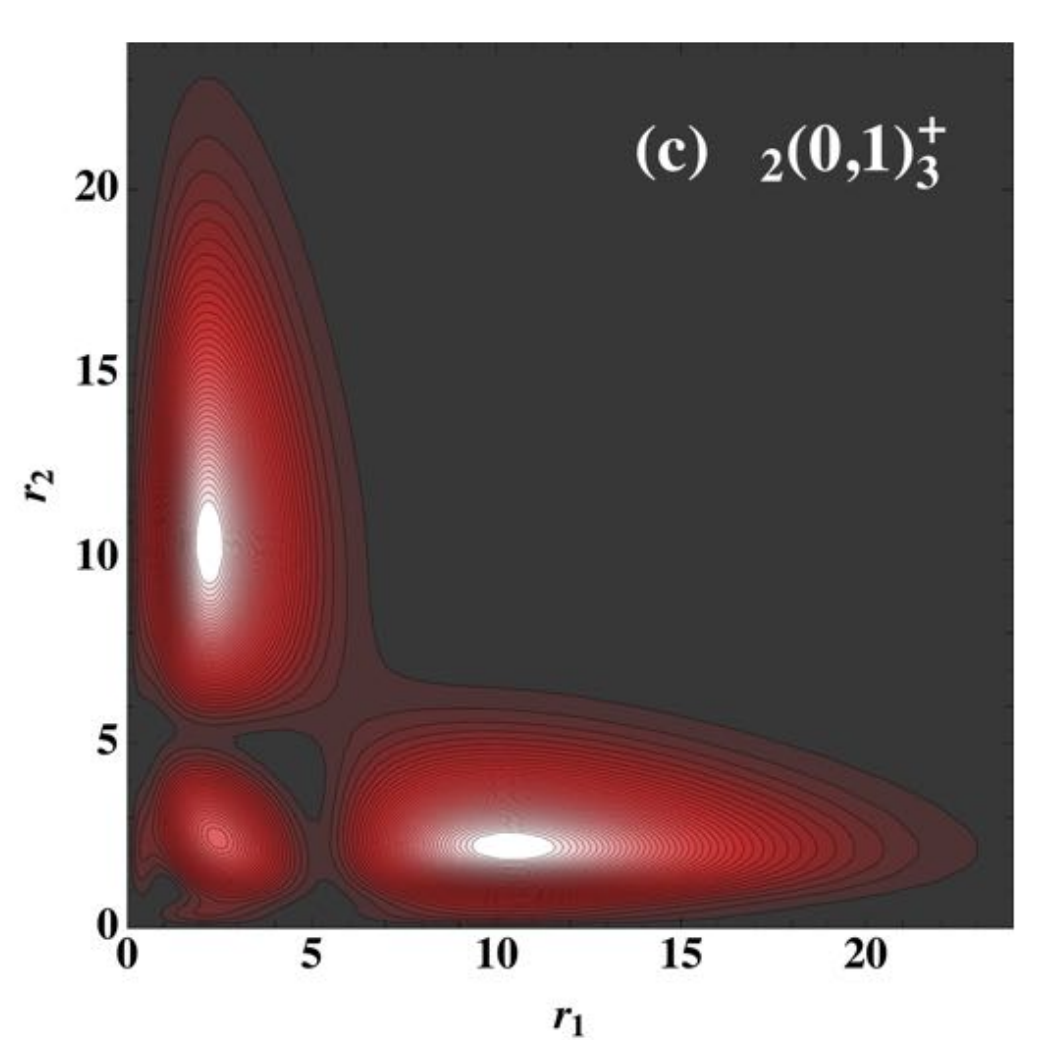}
\includegraphics[width=0.28\textwidth]{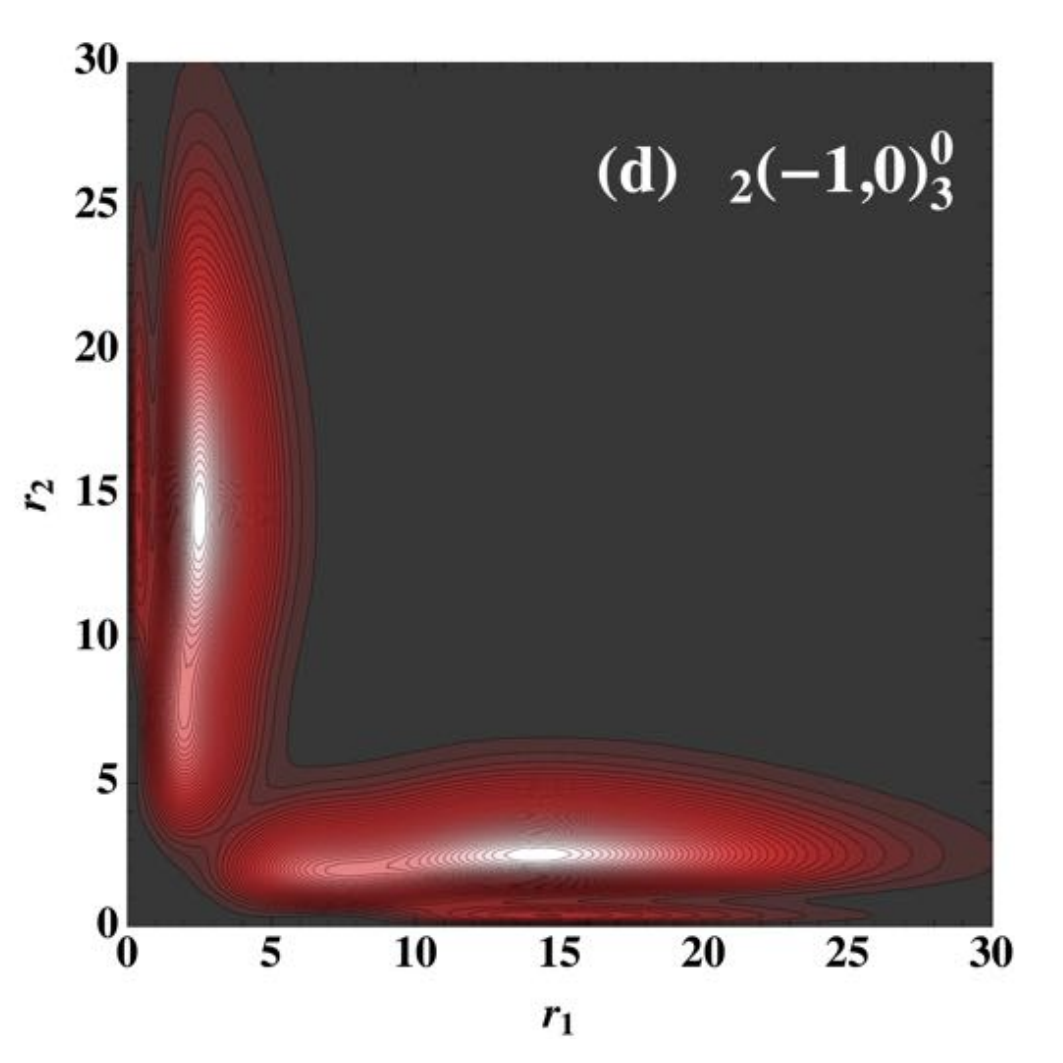}\\
\includegraphics[width=0.28\textwidth]{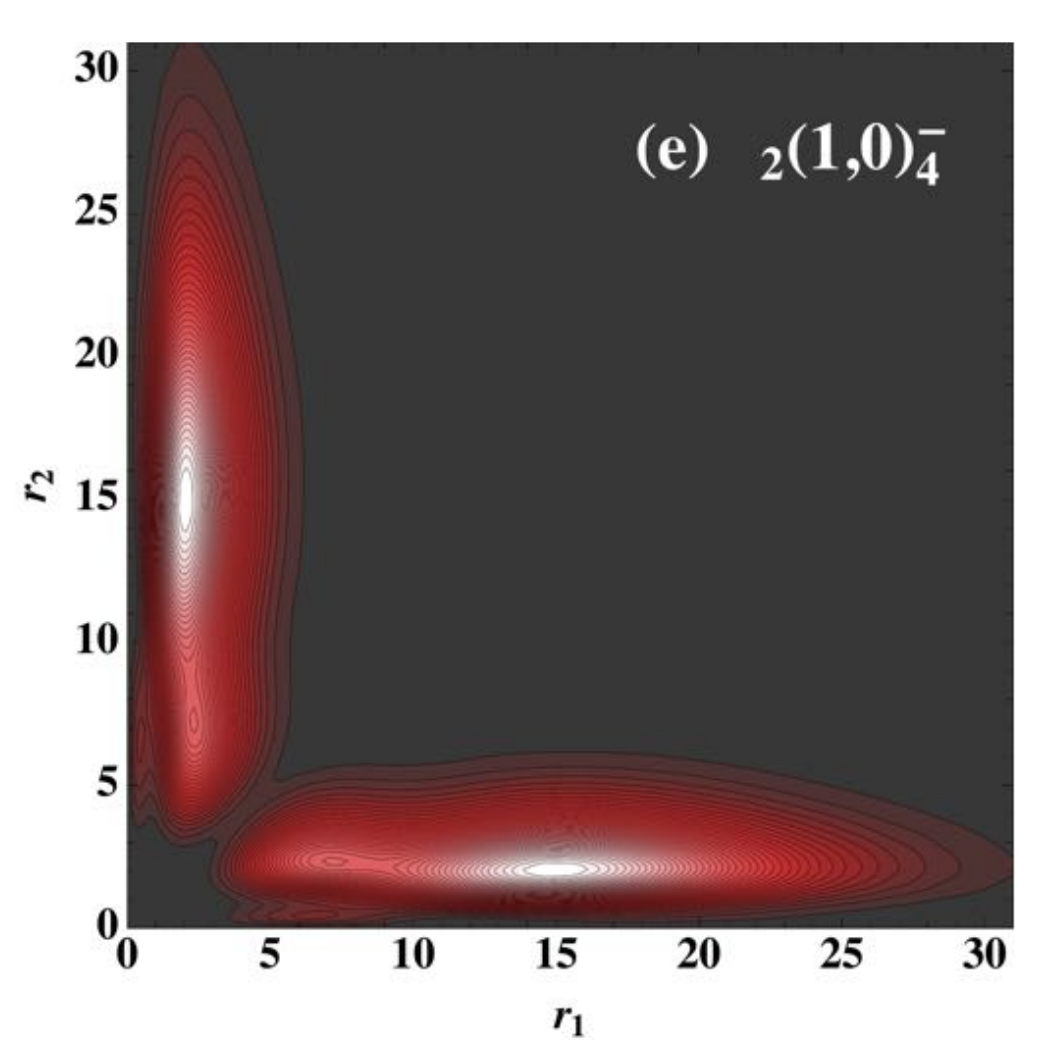}
\includegraphics[width=0.28\textwidth]{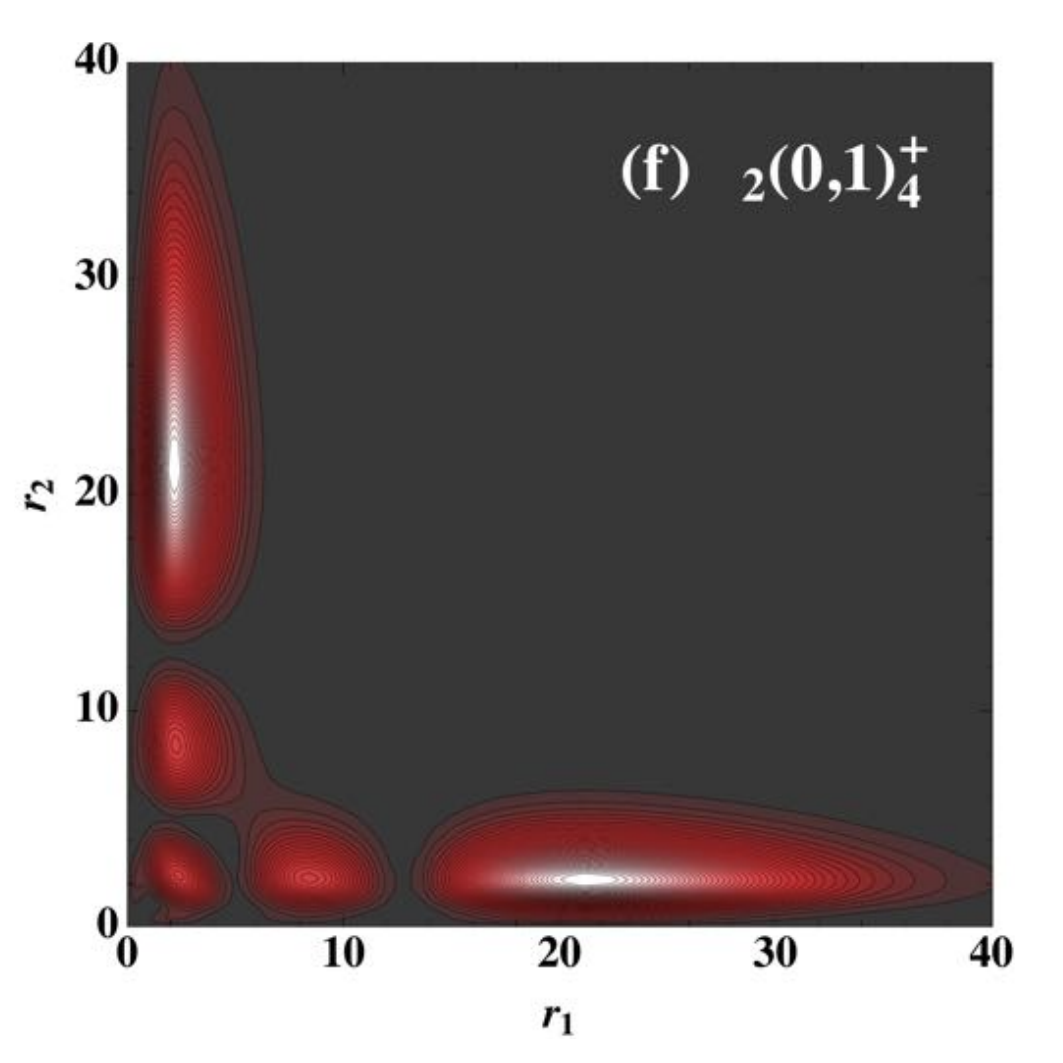}\\
\includegraphics[width=0.28\textwidth]{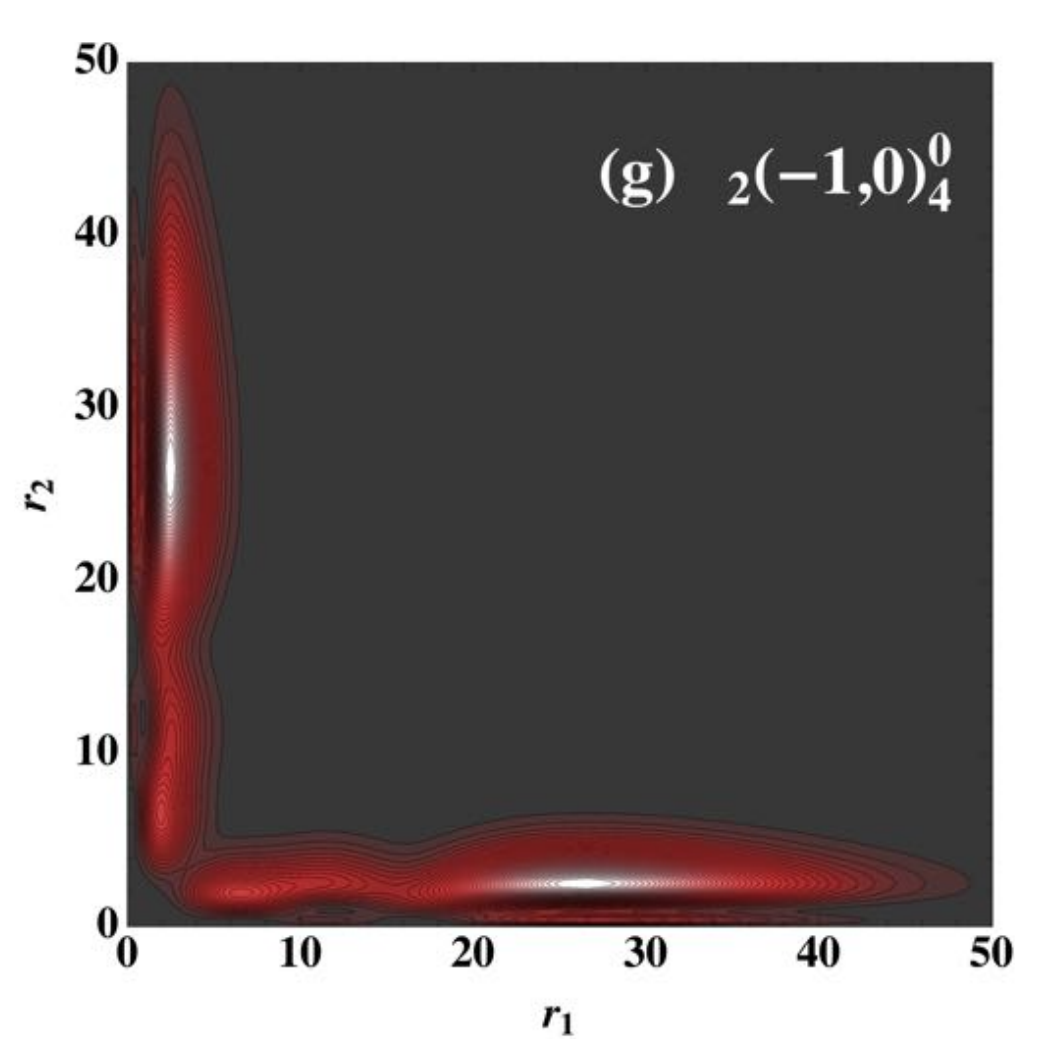}
\includegraphics[width=0.28\textwidth]{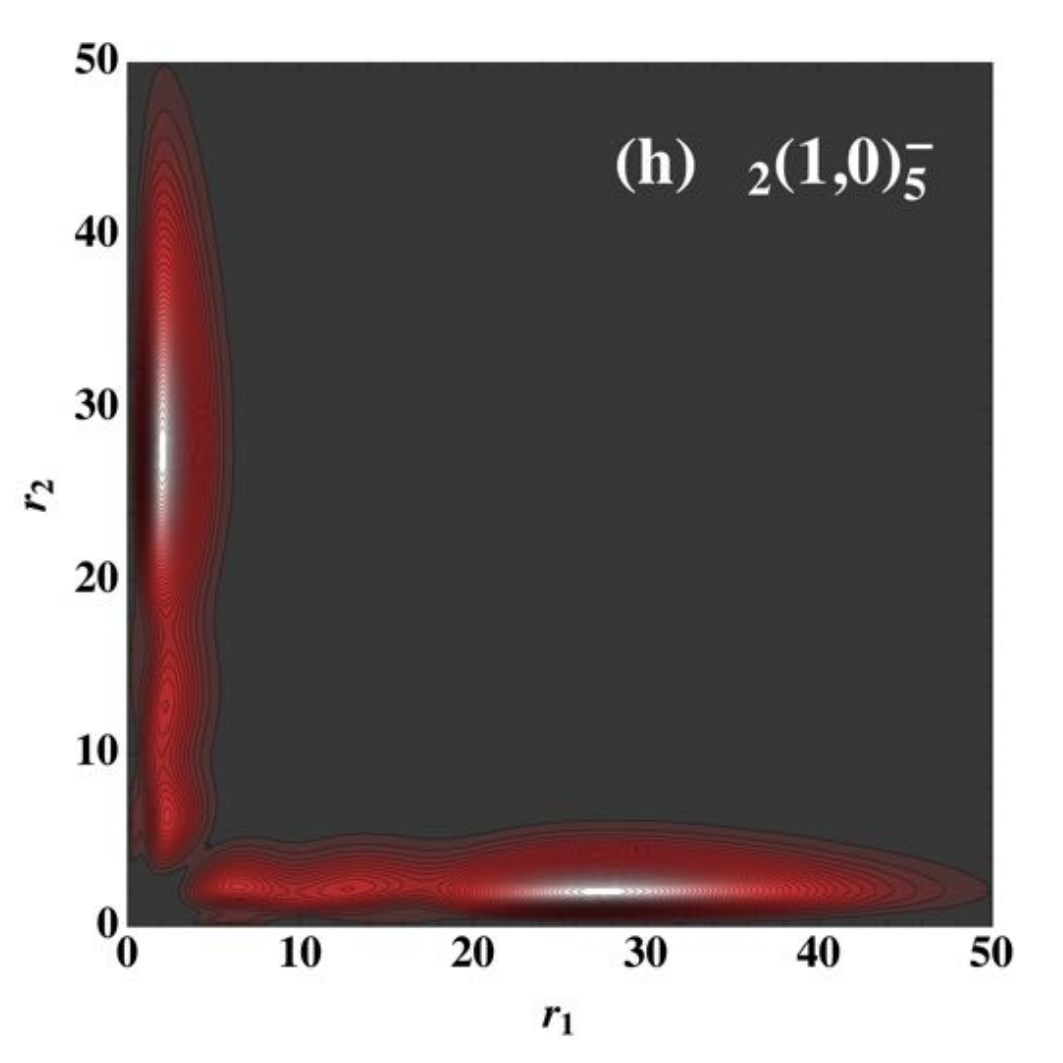}\\
\includegraphics[width=0.28\textwidth]{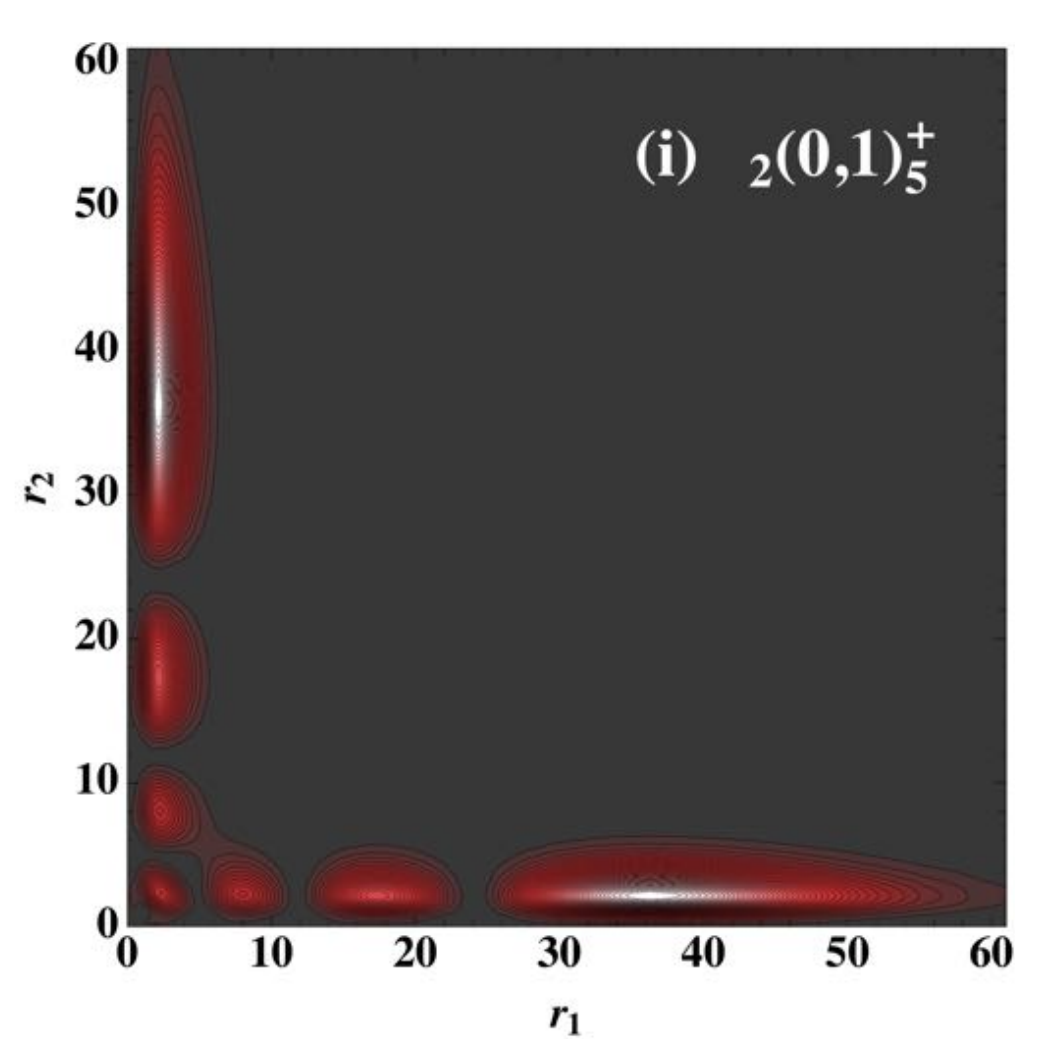}
\includegraphics[width=0.28\textwidth]{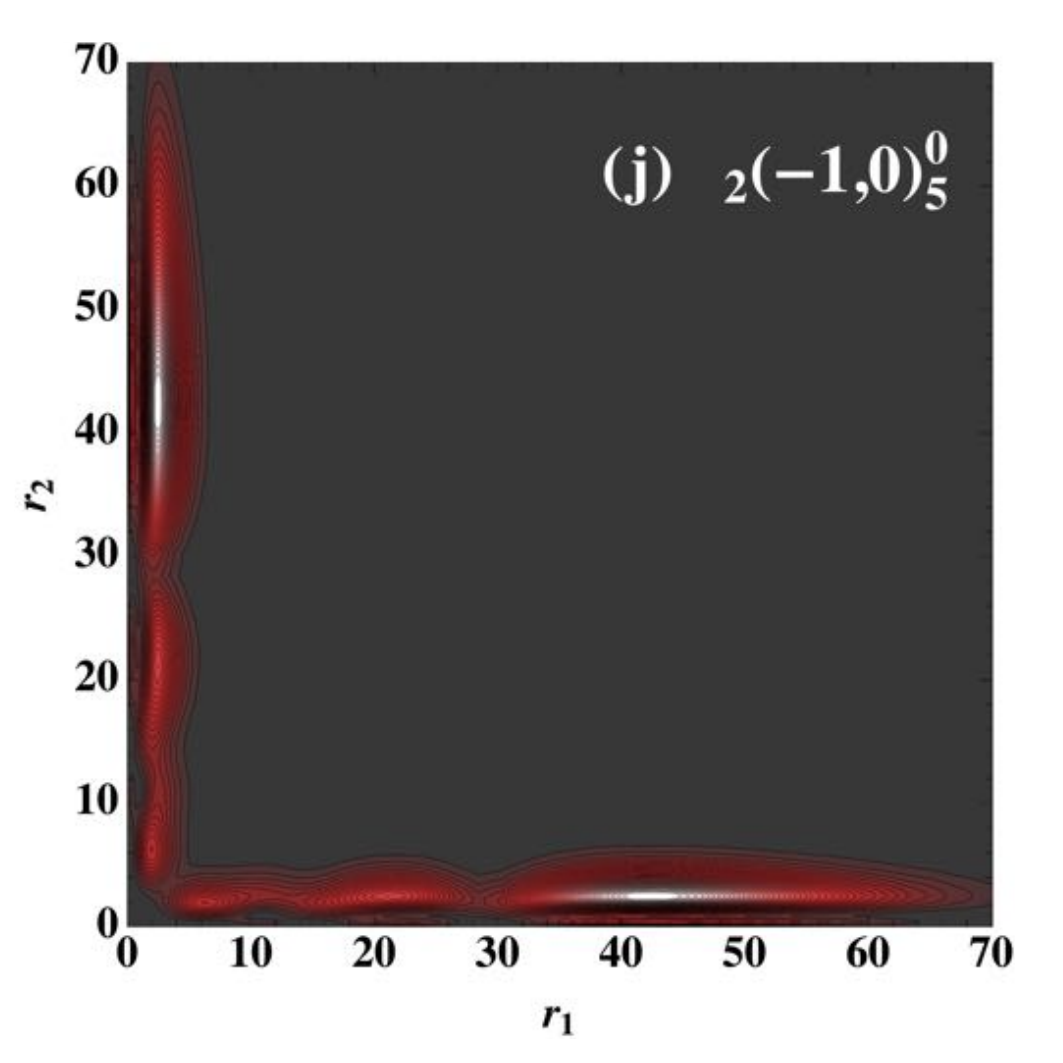}
\caption[Two-electron radial density $\rho(r_{1},r_{2})r_{1}^{2}r_{2}^{2}$ for the lowest ten resonant $^1P^o$ states in He]{\label{fig:wide3}Two-electron radial density $\rho(r_{1},r_{2})r_{1}^{2}r_{2}^{2}$ for the lowest ten resonant $^1P^o$ states in He, located below the second ionization threshold. Resonances are labelled according to the classification proposed by~\citep{Lin1983} using $_{n_1}(K,T)_{n_2}^A$. The energy ordering of the resonances is indicated by the alphabet labels inside the plots.}
\end{figure}

\begin{figure}
\centering
\includegraphics[width=0.28\textwidth]{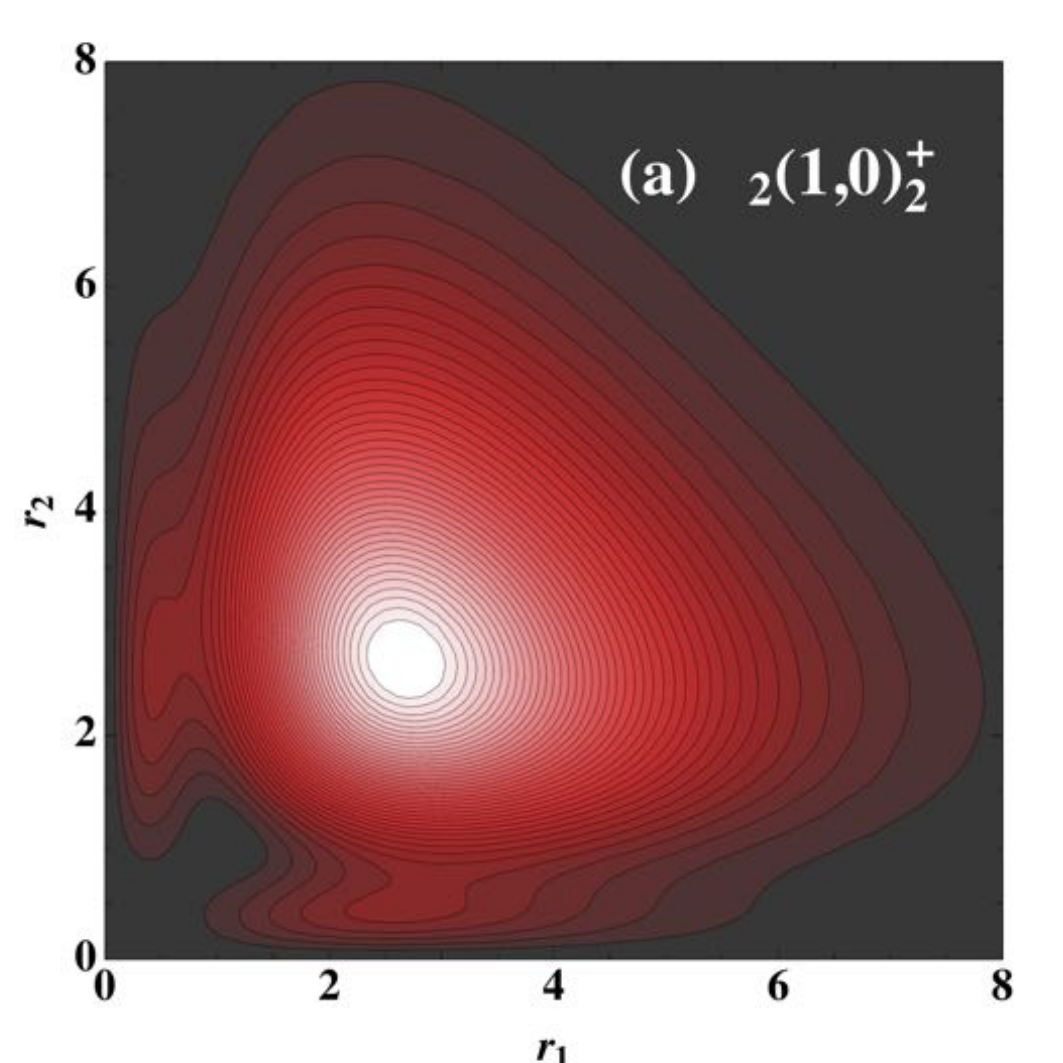}
\includegraphics[width=0.28\textwidth]{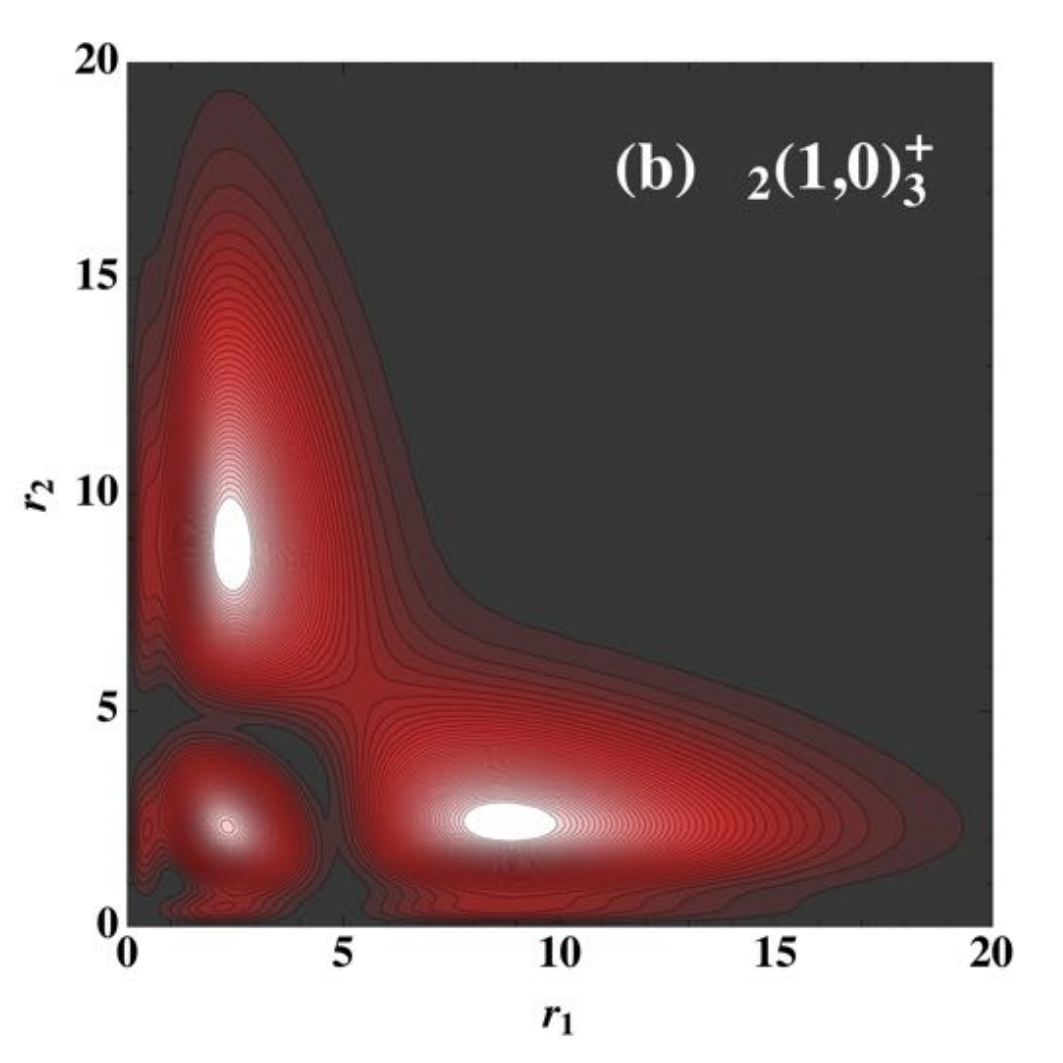}\\
\includegraphics[width=0.28\textwidth]{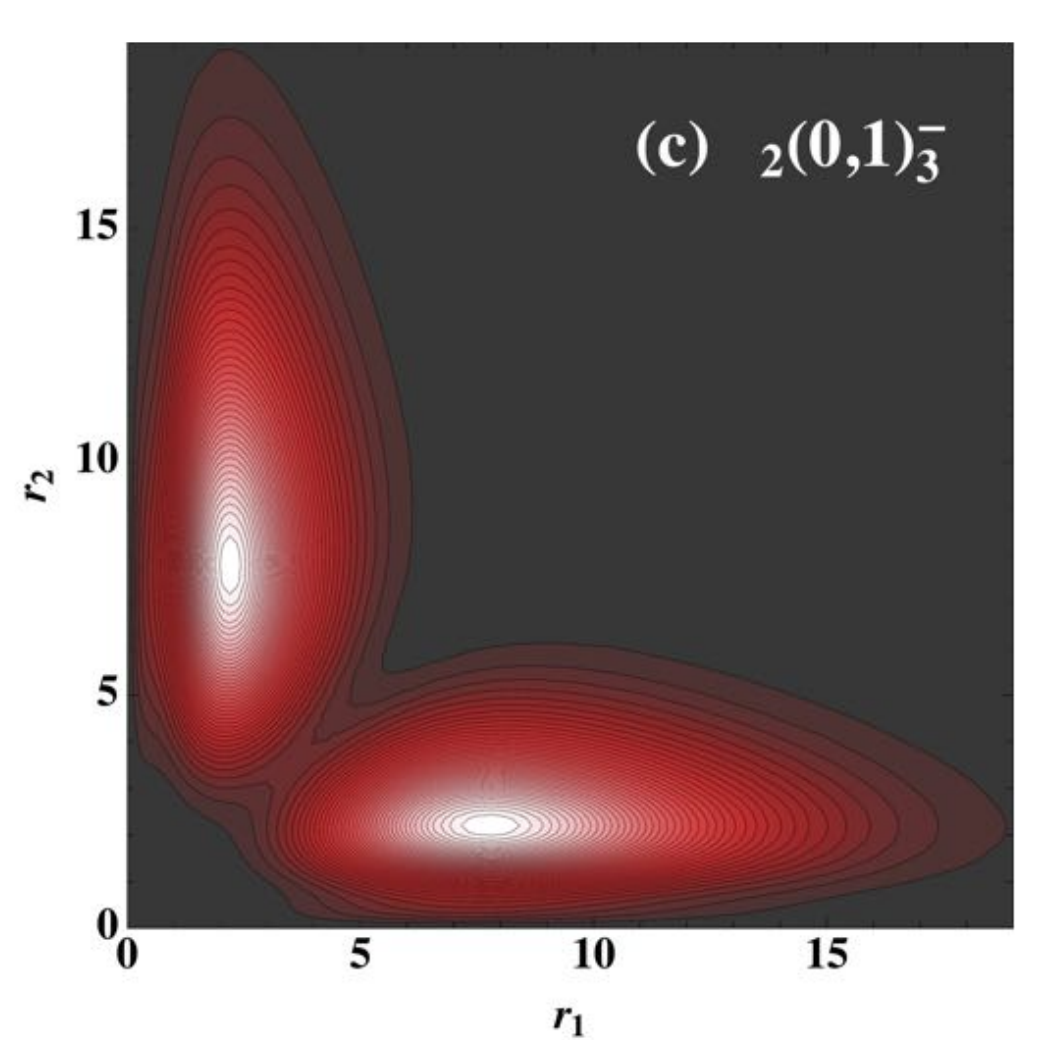}
\includegraphics[width=0.28\textwidth]{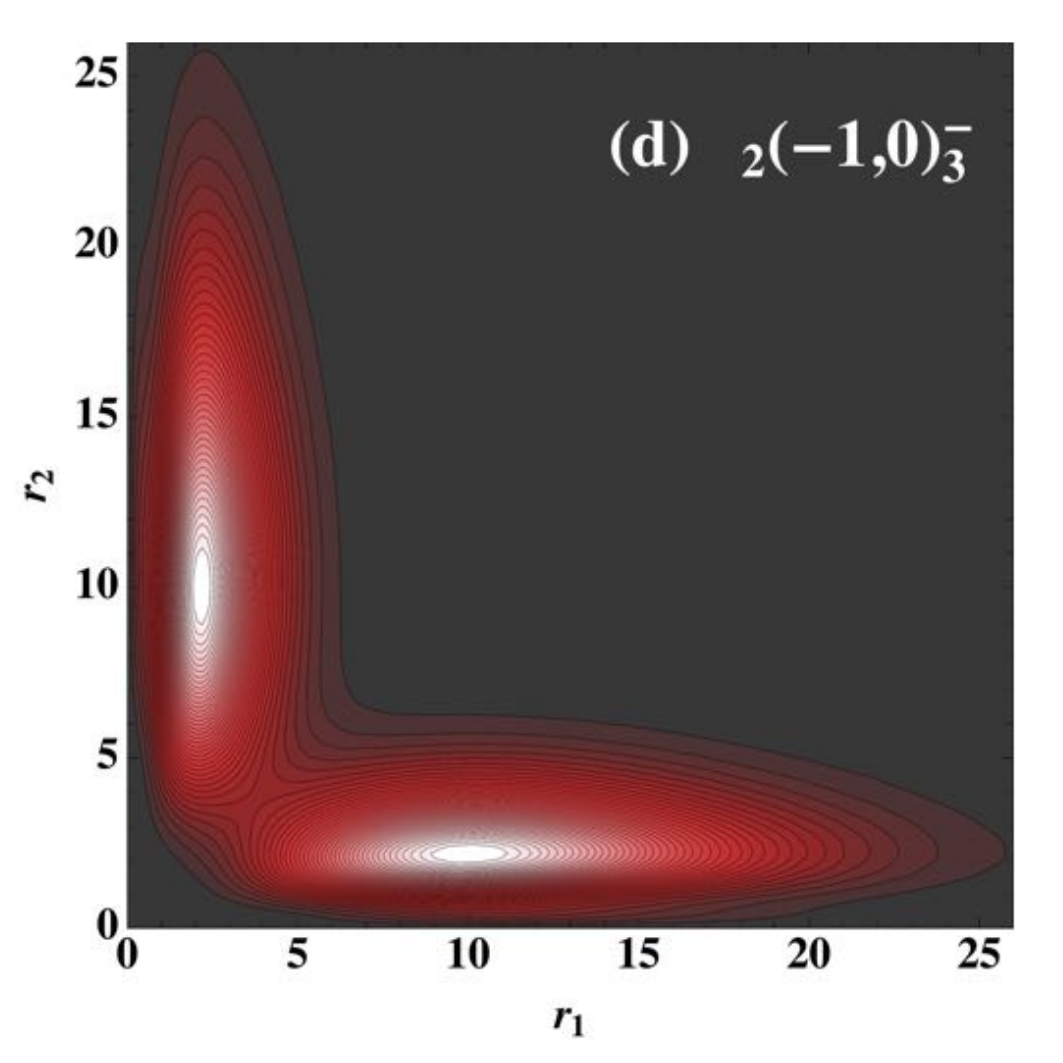}\\
\includegraphics[width=0.28\textwidth]{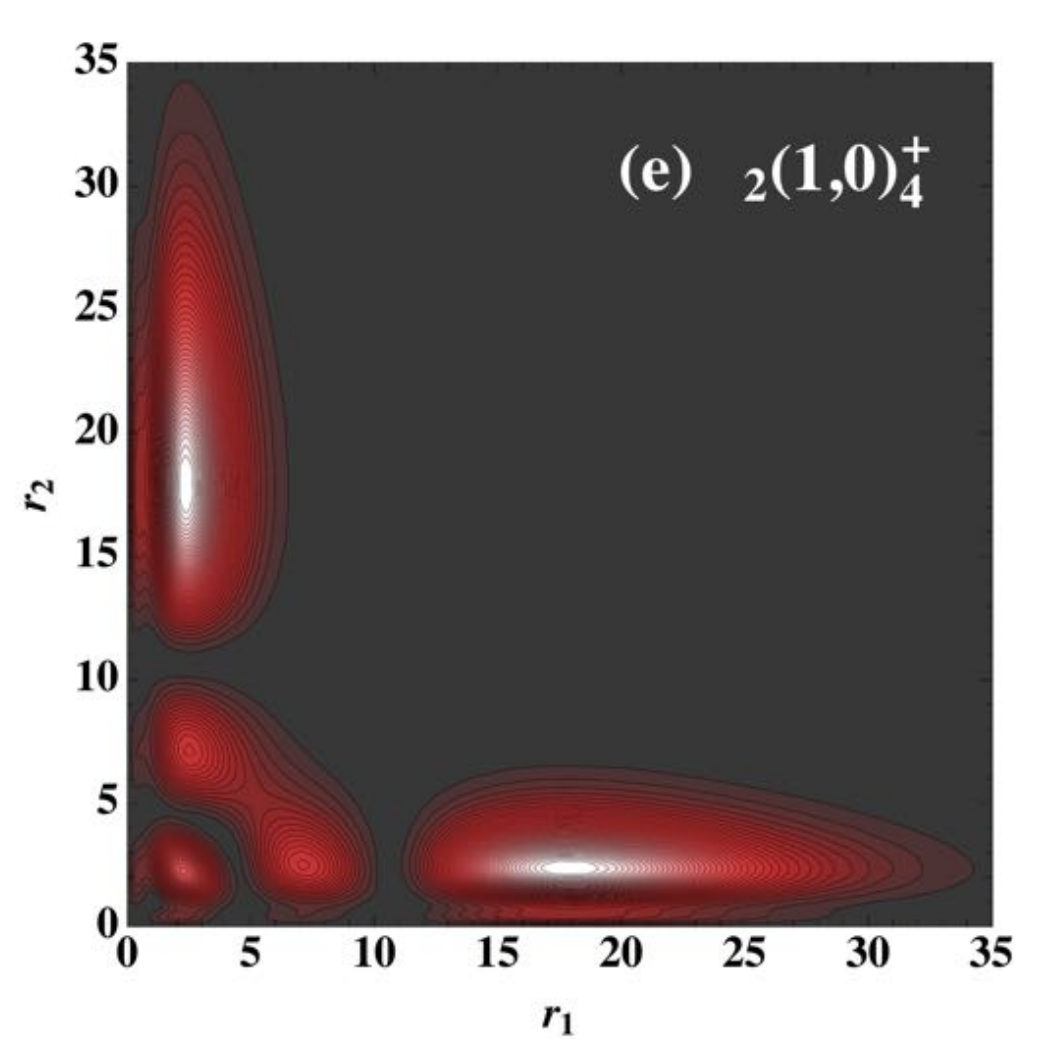}
\includegraphics[width=0.28\textwidth]{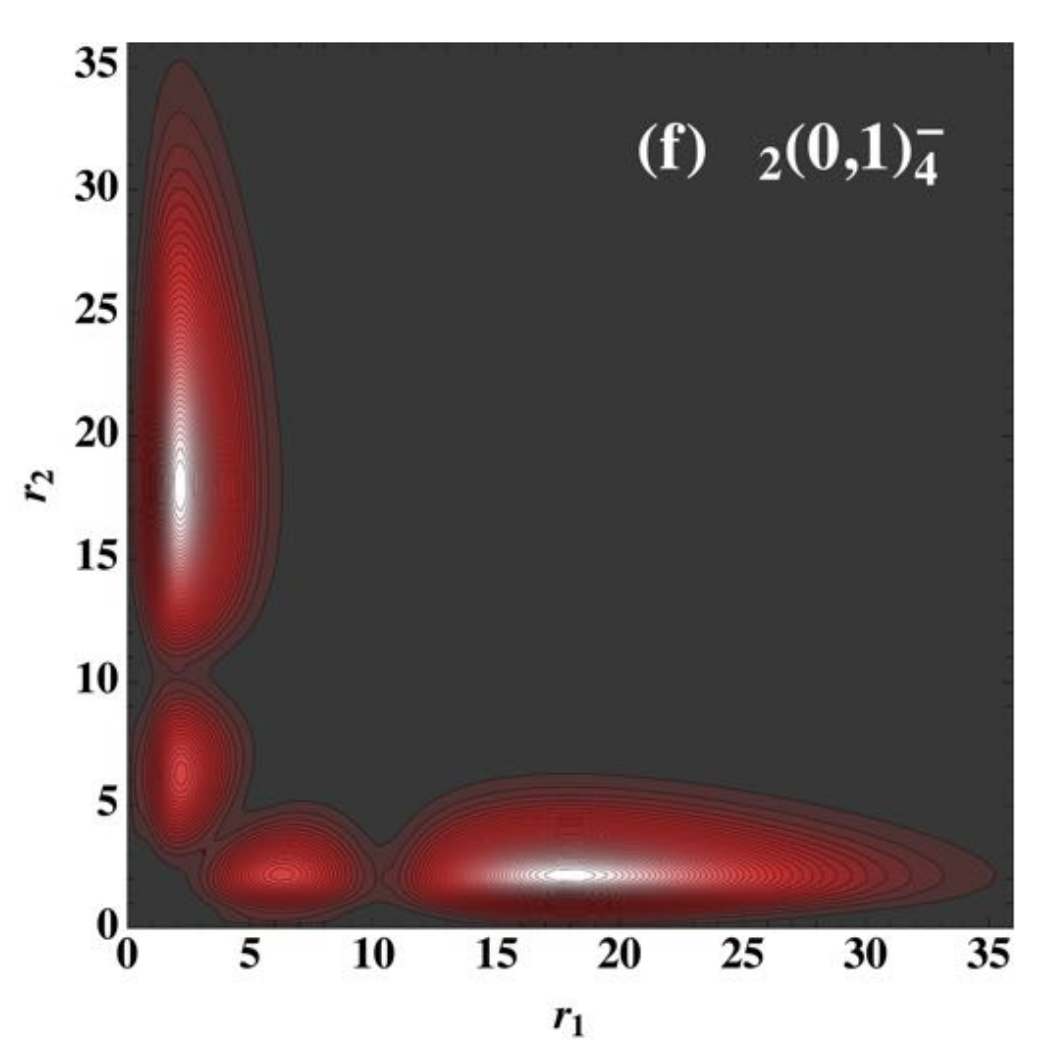}\\
\includegraphics[width=0.28\textwidth]{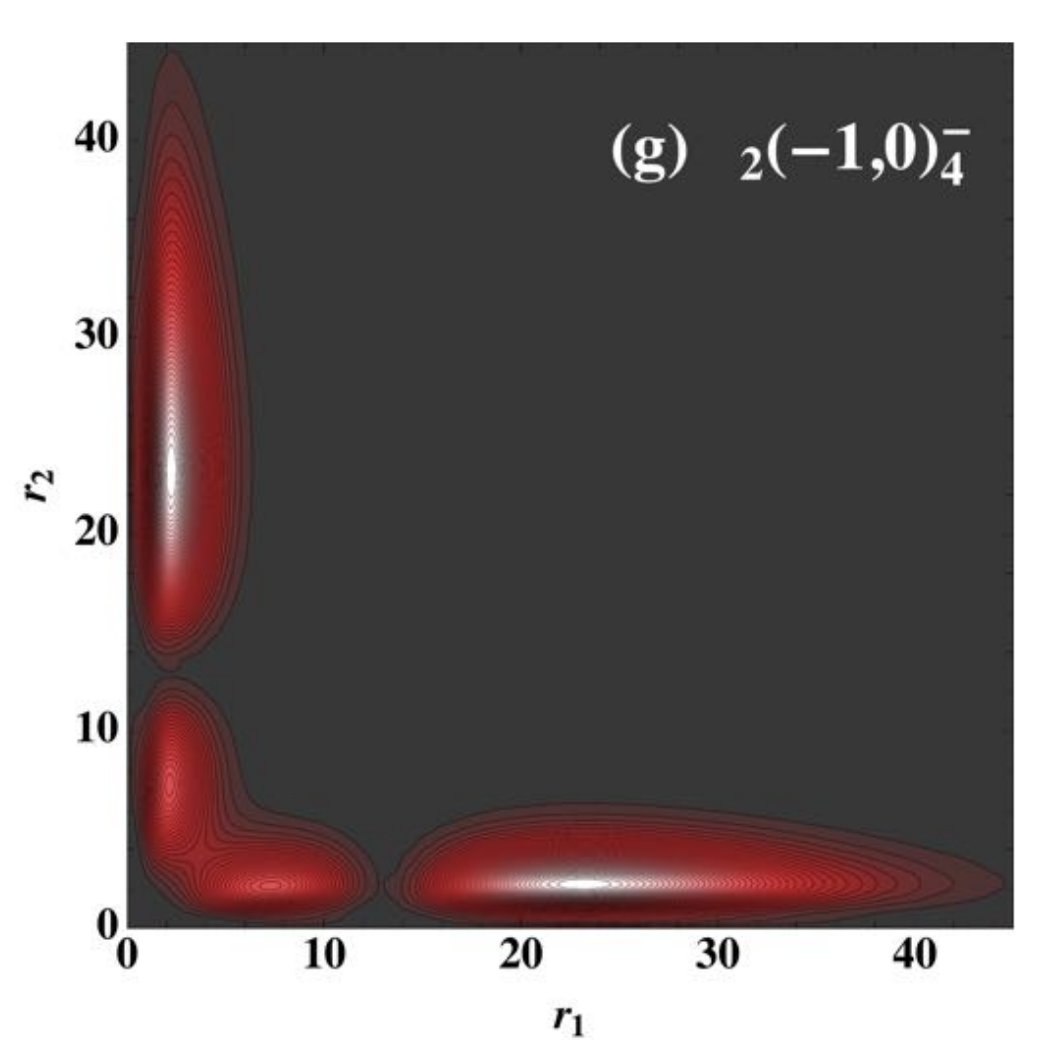}
\includegraphics[width=0.28\textwidth]{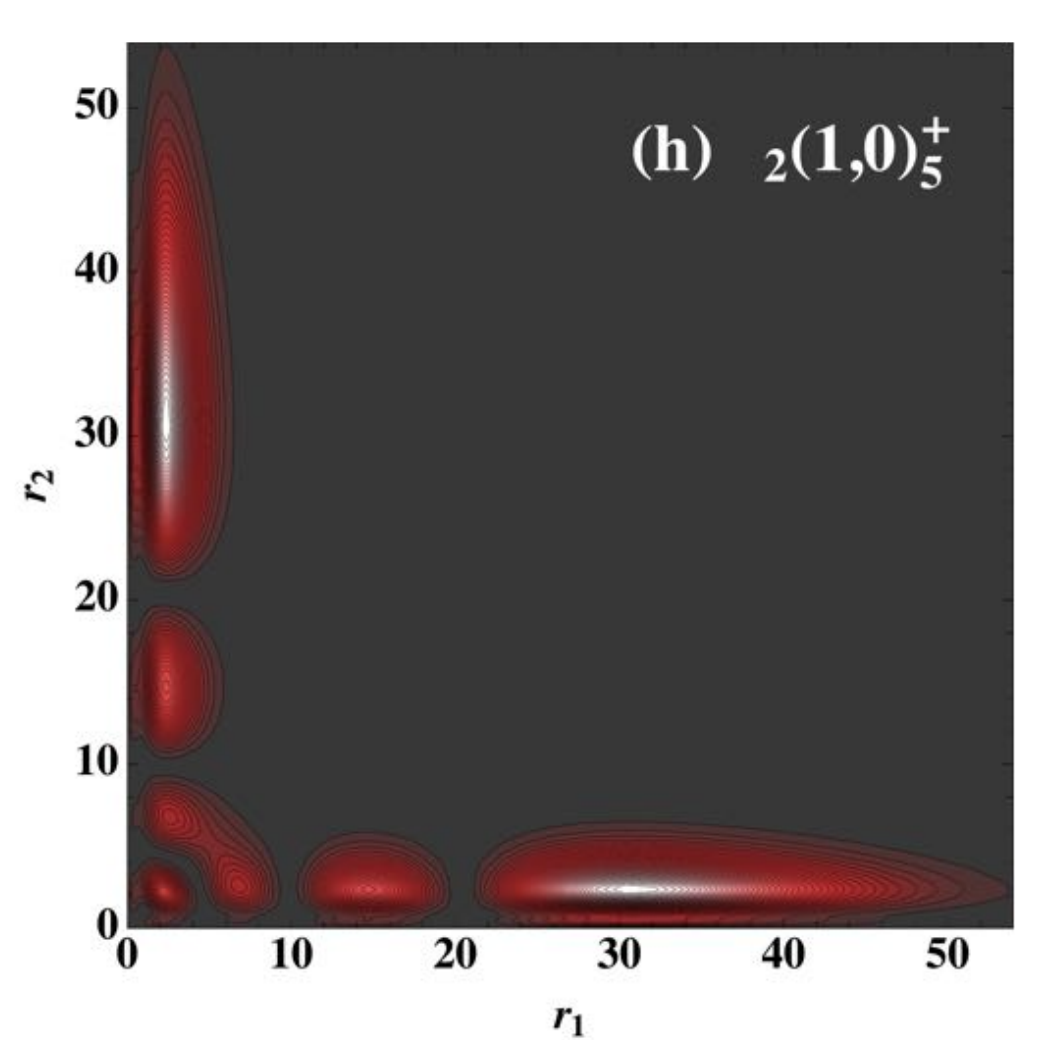}\\
\includegraphics[width=0.28\textwidth]{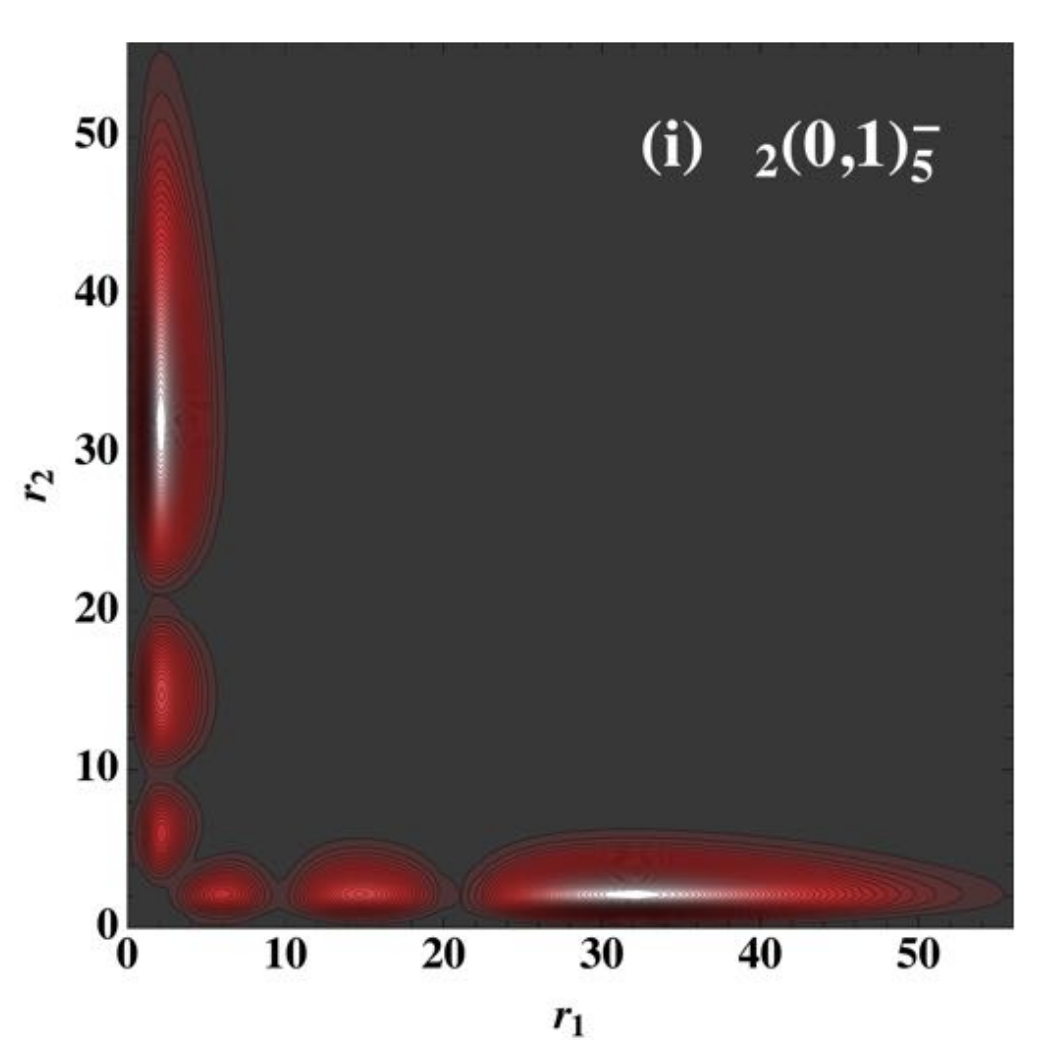}
\includegraphics[width=0.28\textwidth]{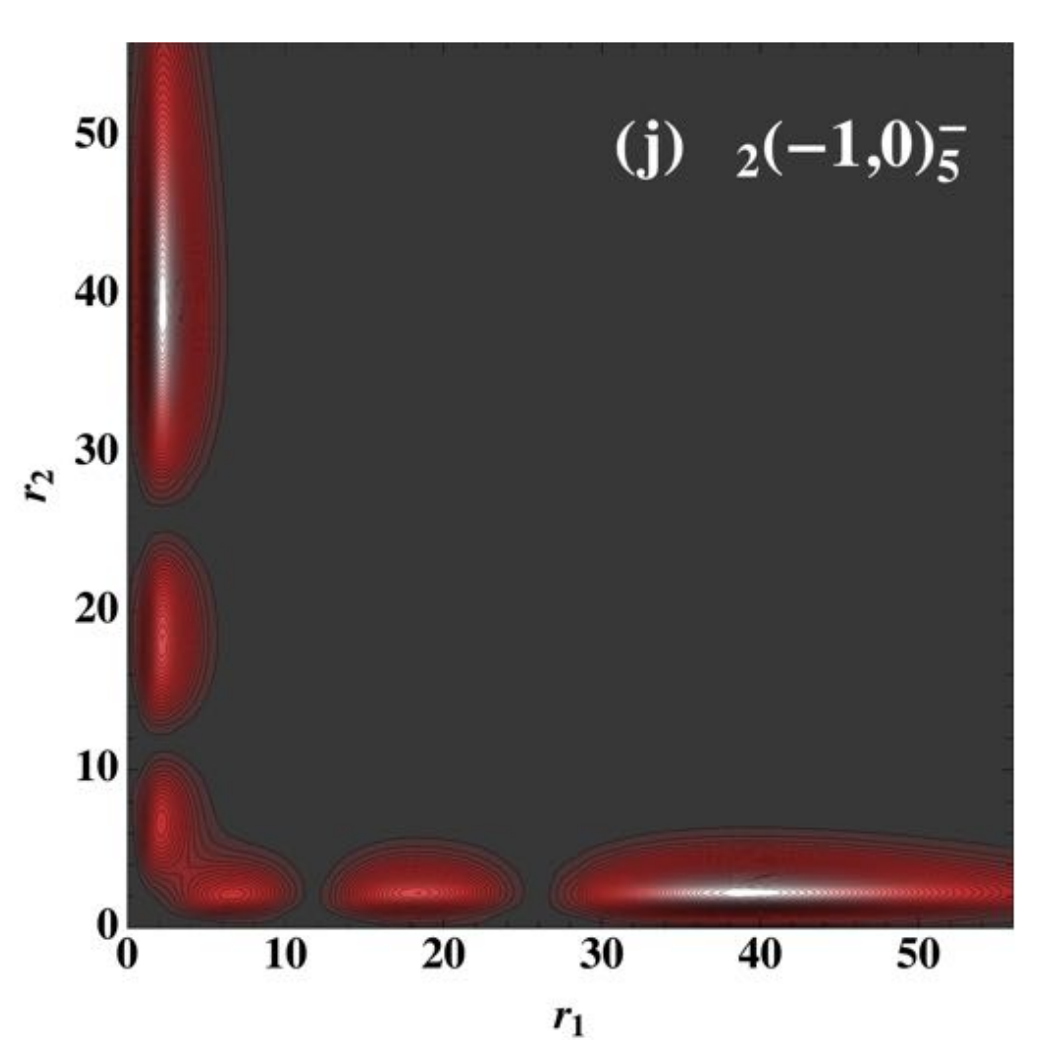}
\caption[Two-electron radial density $\rho(r_{1},r_{2})r_{1}^{2}r_{2}^{2}$ for the lowest ten resonant $^3P^o$ states in He]{\label{fig:wide4}Two-electron radial density $\rho(r_{1},r_{2})r_{1}^{2}r_{2}^{2}$ for the lowest ten resonant $^3P^o$ states in He, located below the second ionization threshold. Resonances are labelled according to the classification proposed by~\citep{Lin1983} using $_{n_1}(K,T)_{n_2}^A$. The energy ordering of the resonances is indicated by the alphabet labels inside the plots.}

\end{figure}

\begin{figure}
\centering
\includegraphics[width=0.28\textwidth]{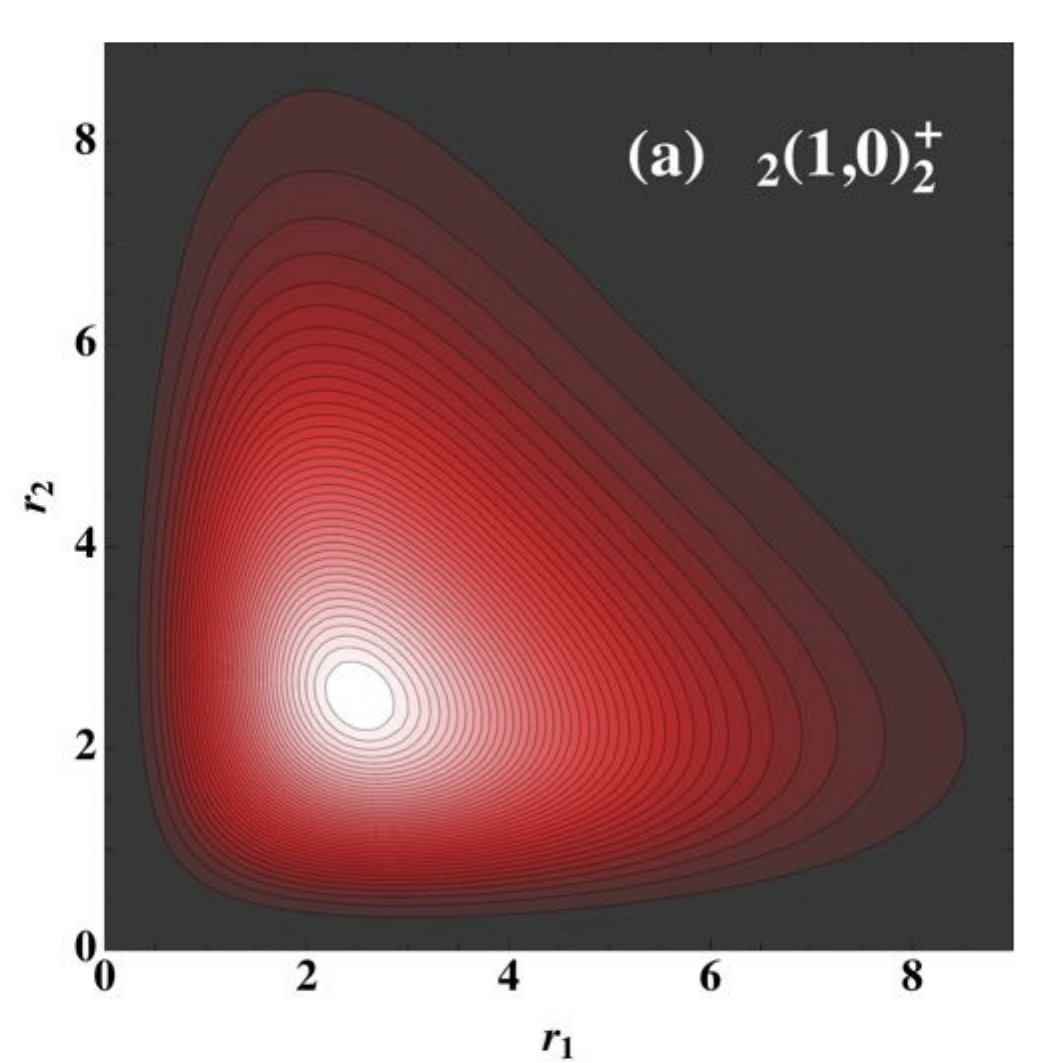}
\includegraphics[width=0.28\textwidth]{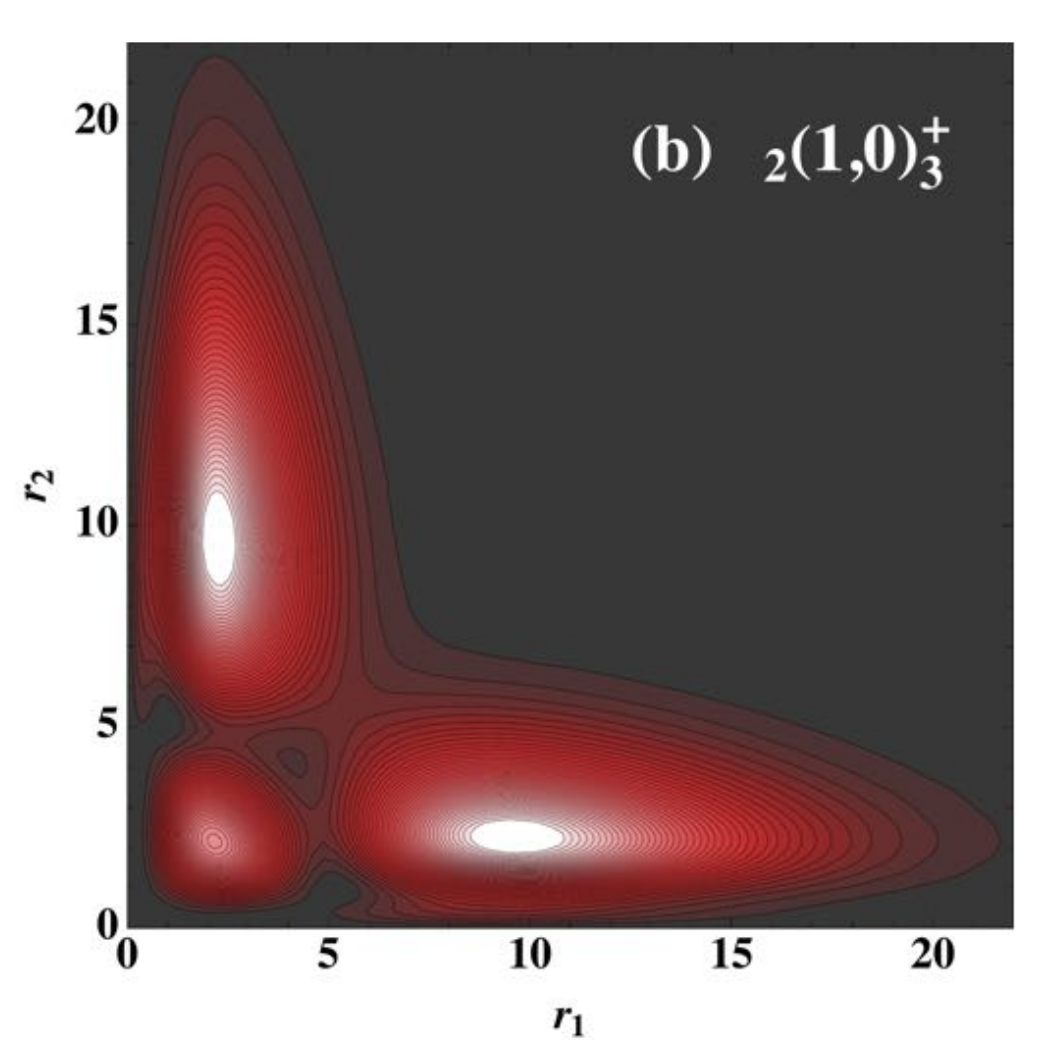}\\
\includegraphics[width=0.28\textwidth]{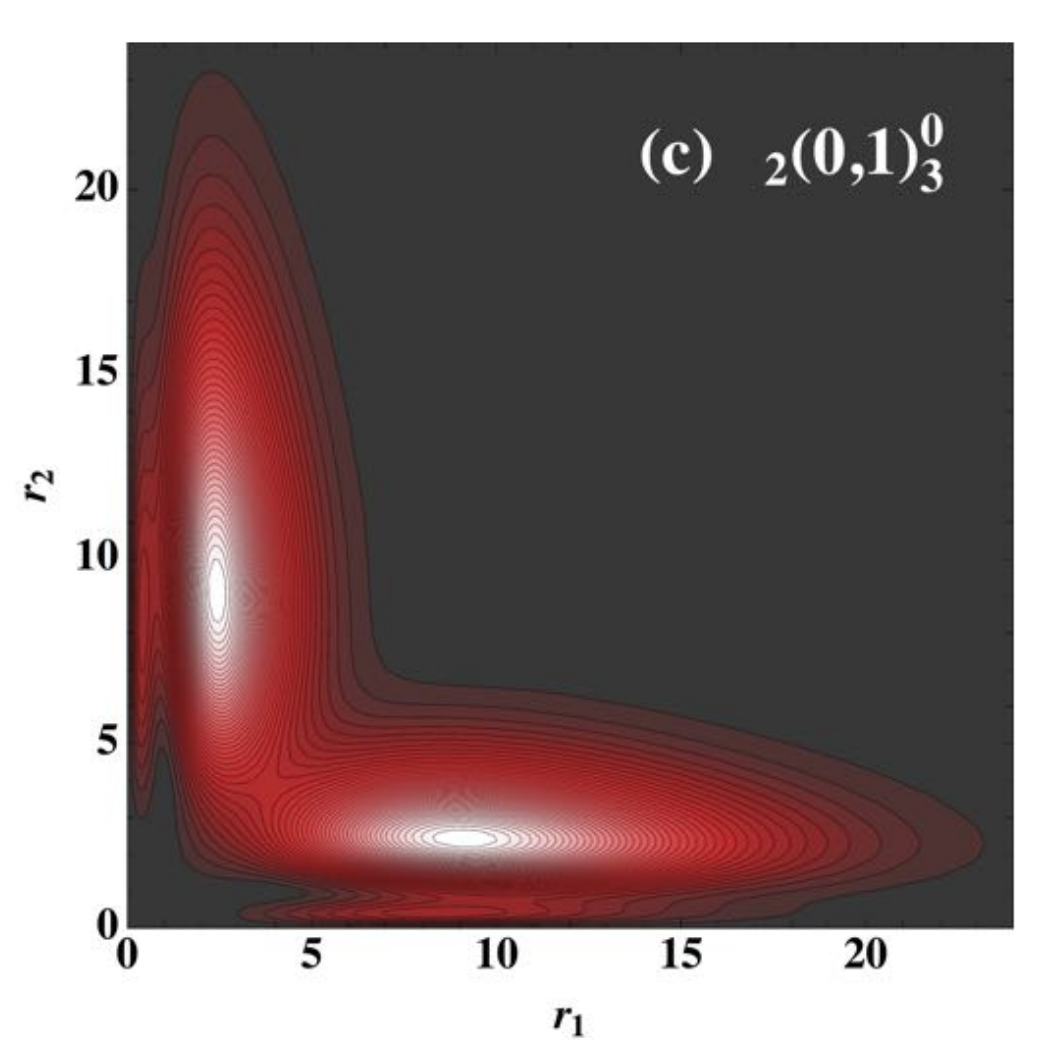}
\includegraphics[width=0.28\textwidth]{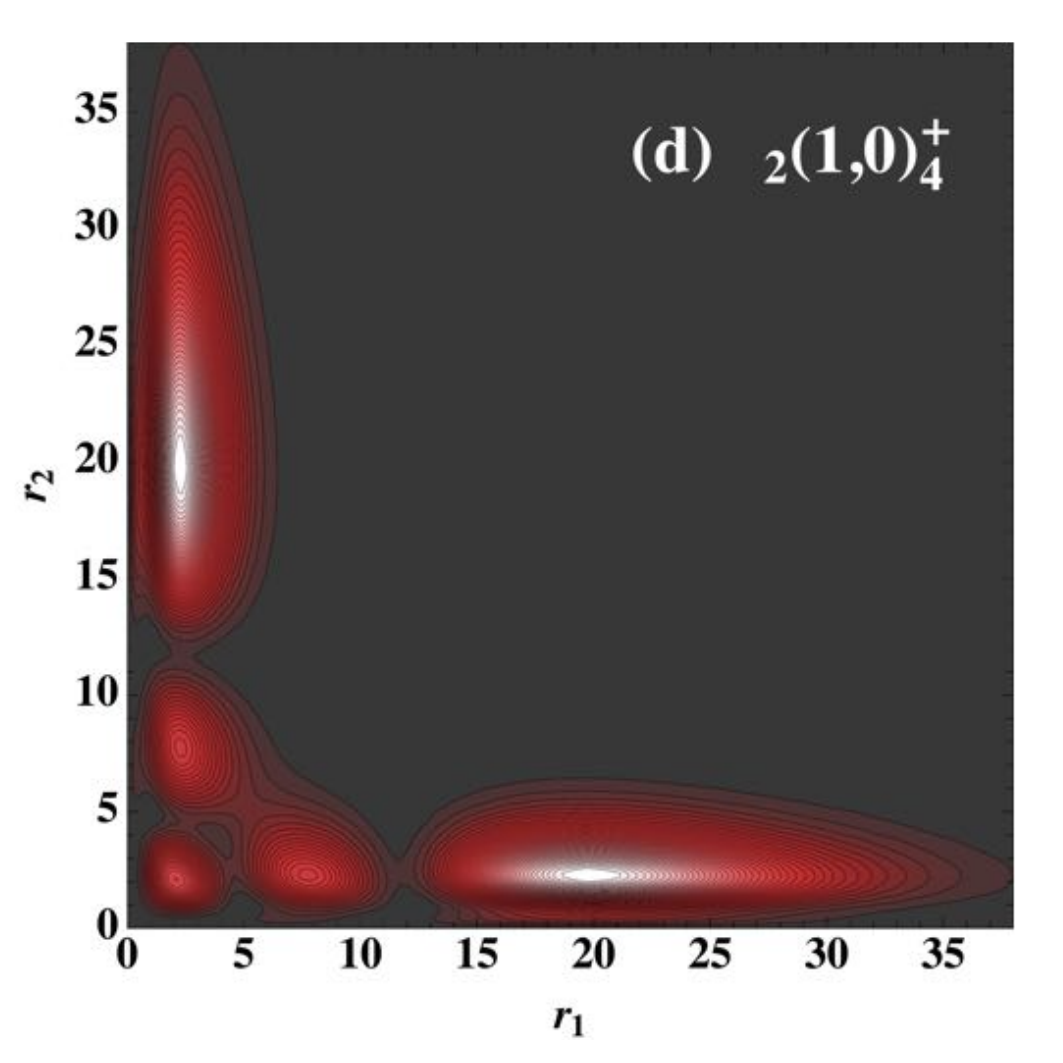}\\
\includegraphics[width=0.28\textwidth]{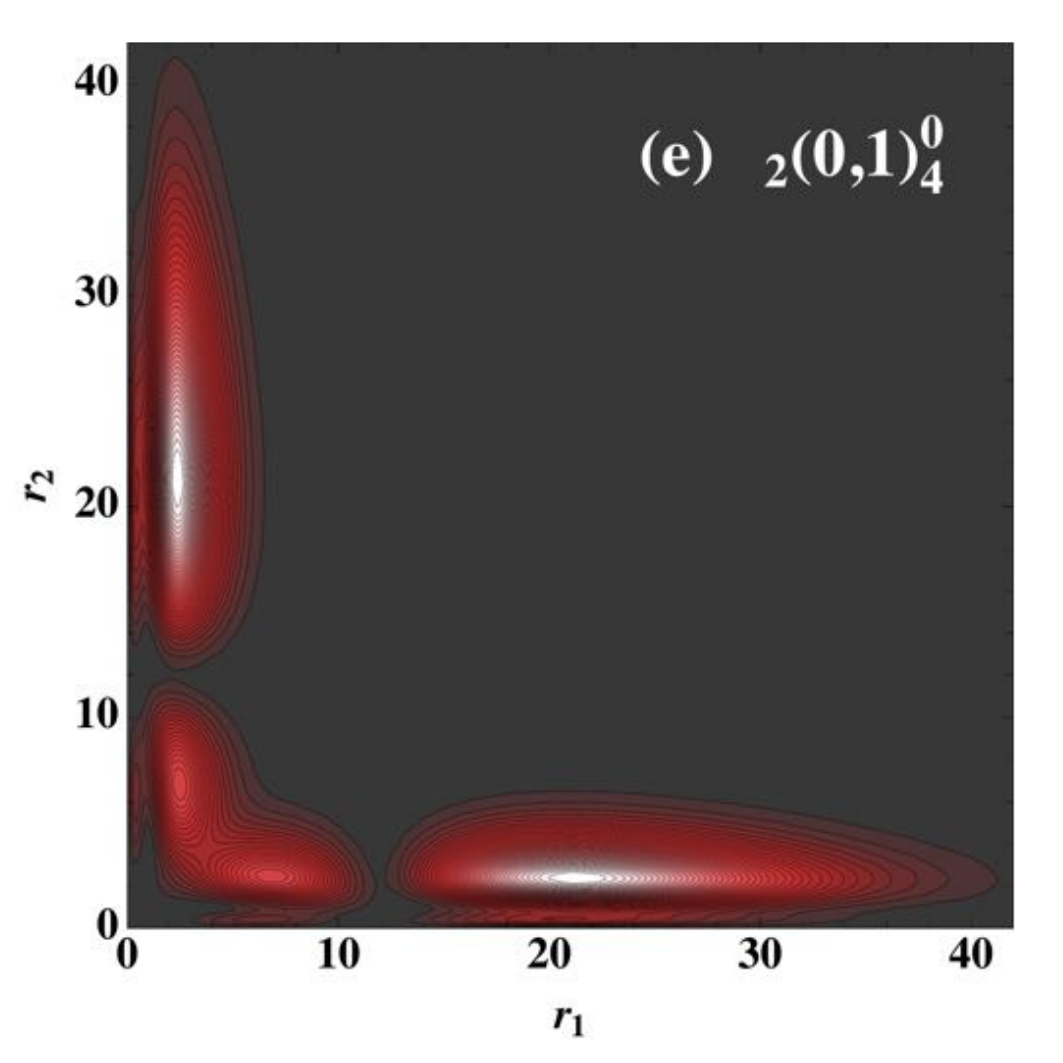}
\includegraphics[width=0.28\textwidth]{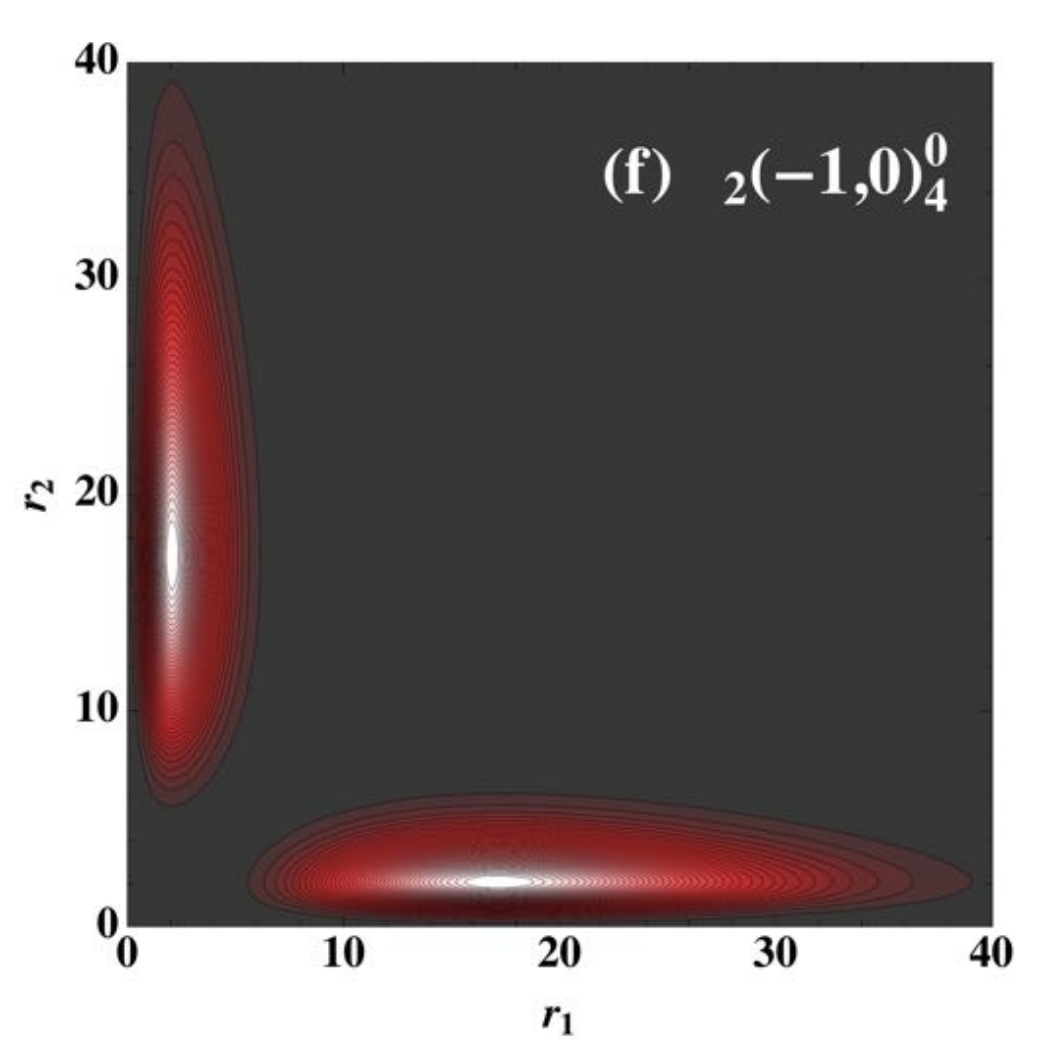}\\
\includegraphics[width=0.28\textwidth]{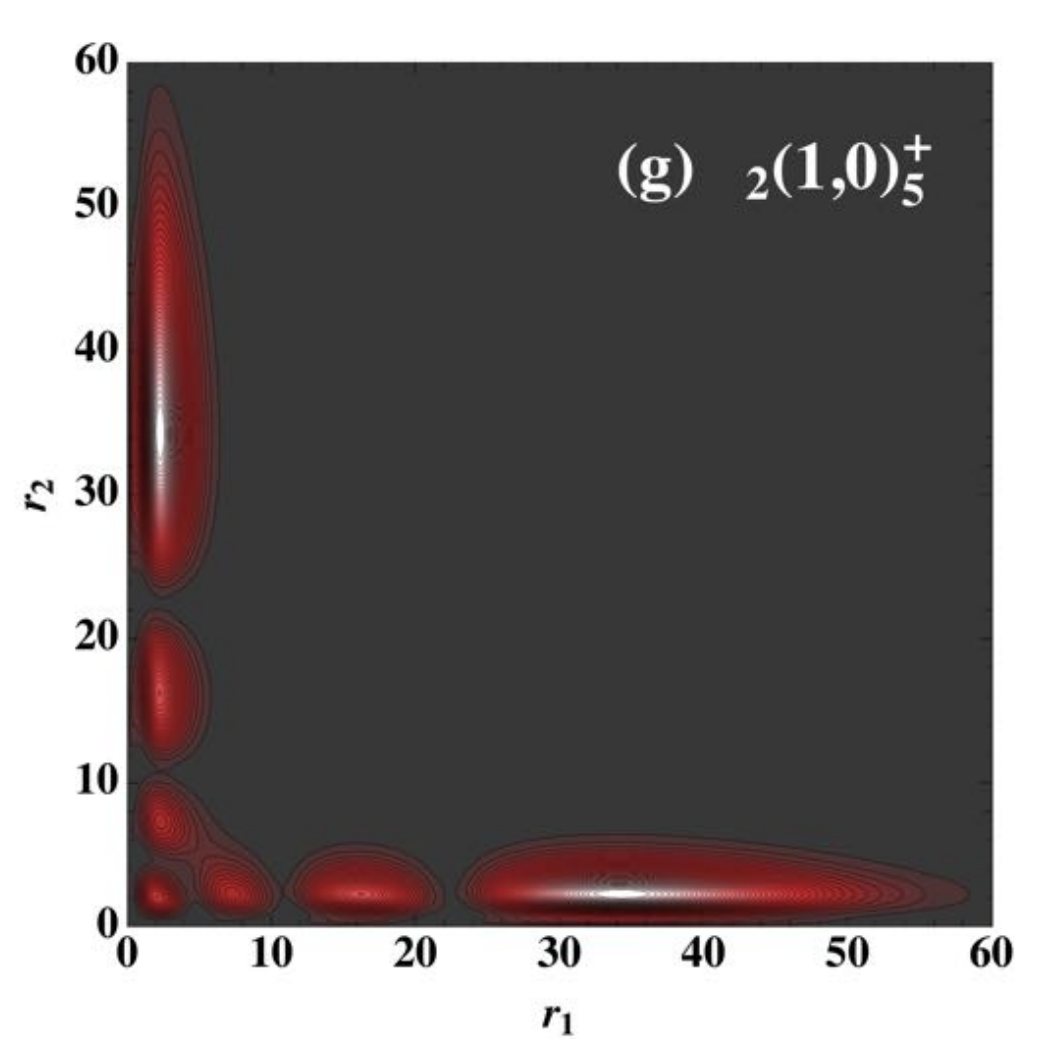}
\includegraphics[width=0.28\textwidth]{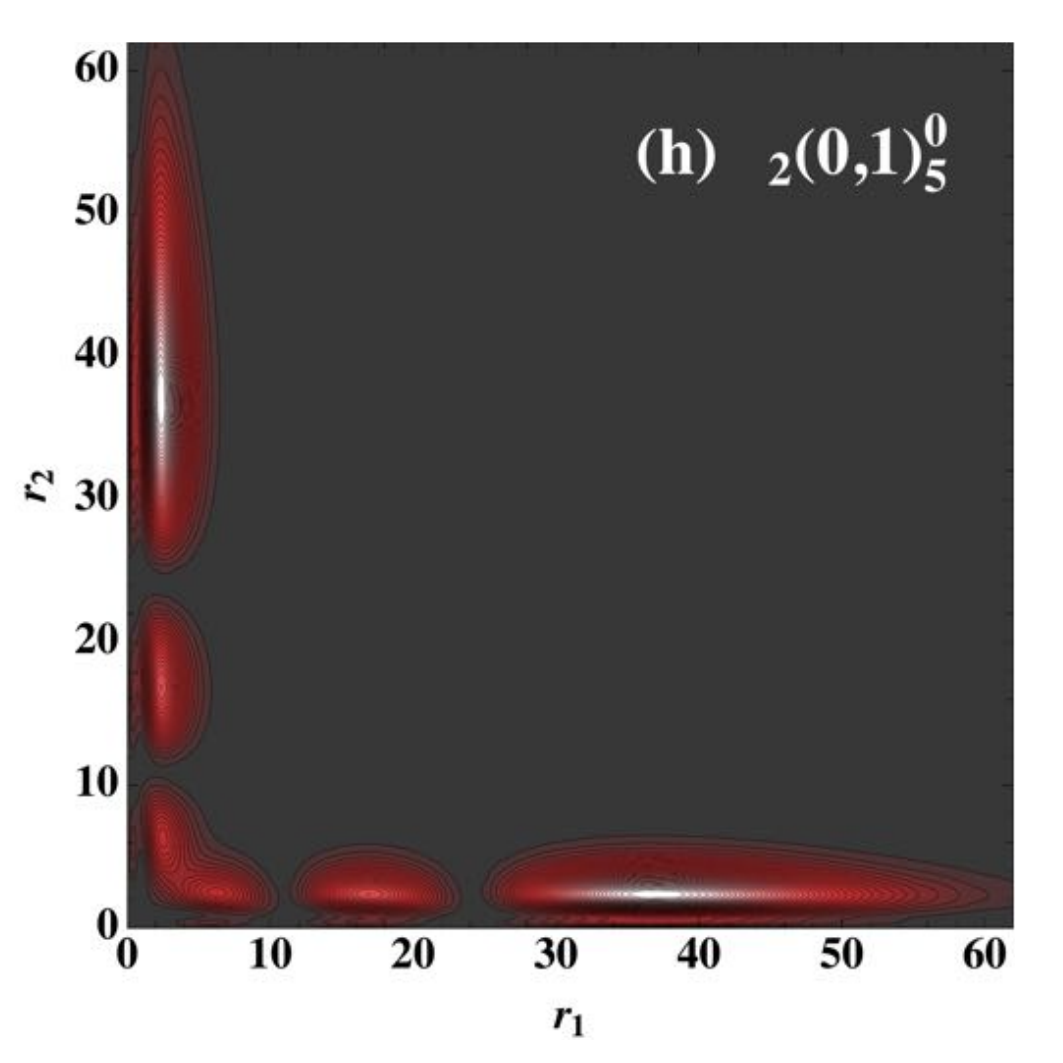}\\
\includegraphics[width=0.28\textwidth]{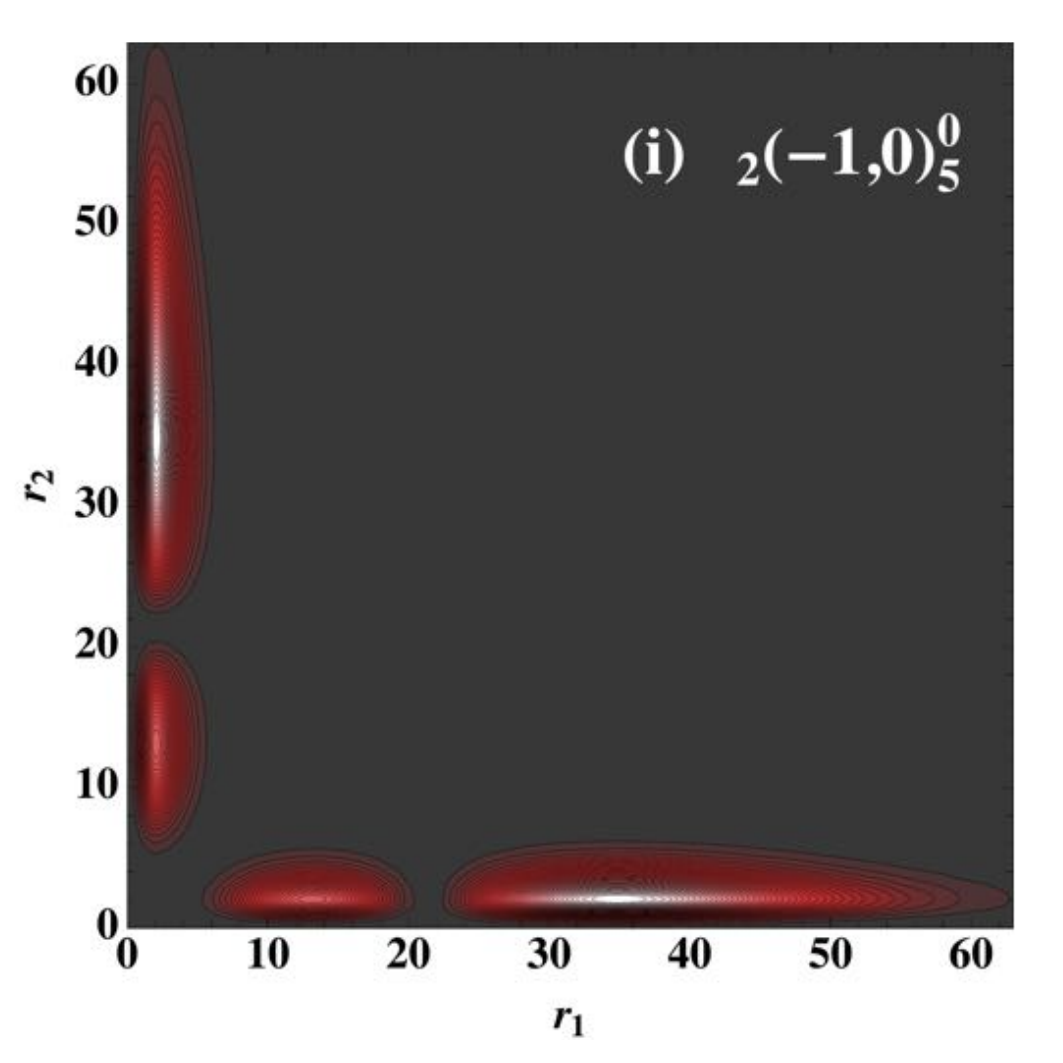}
\includegraphics[width=0.28\textwidth]{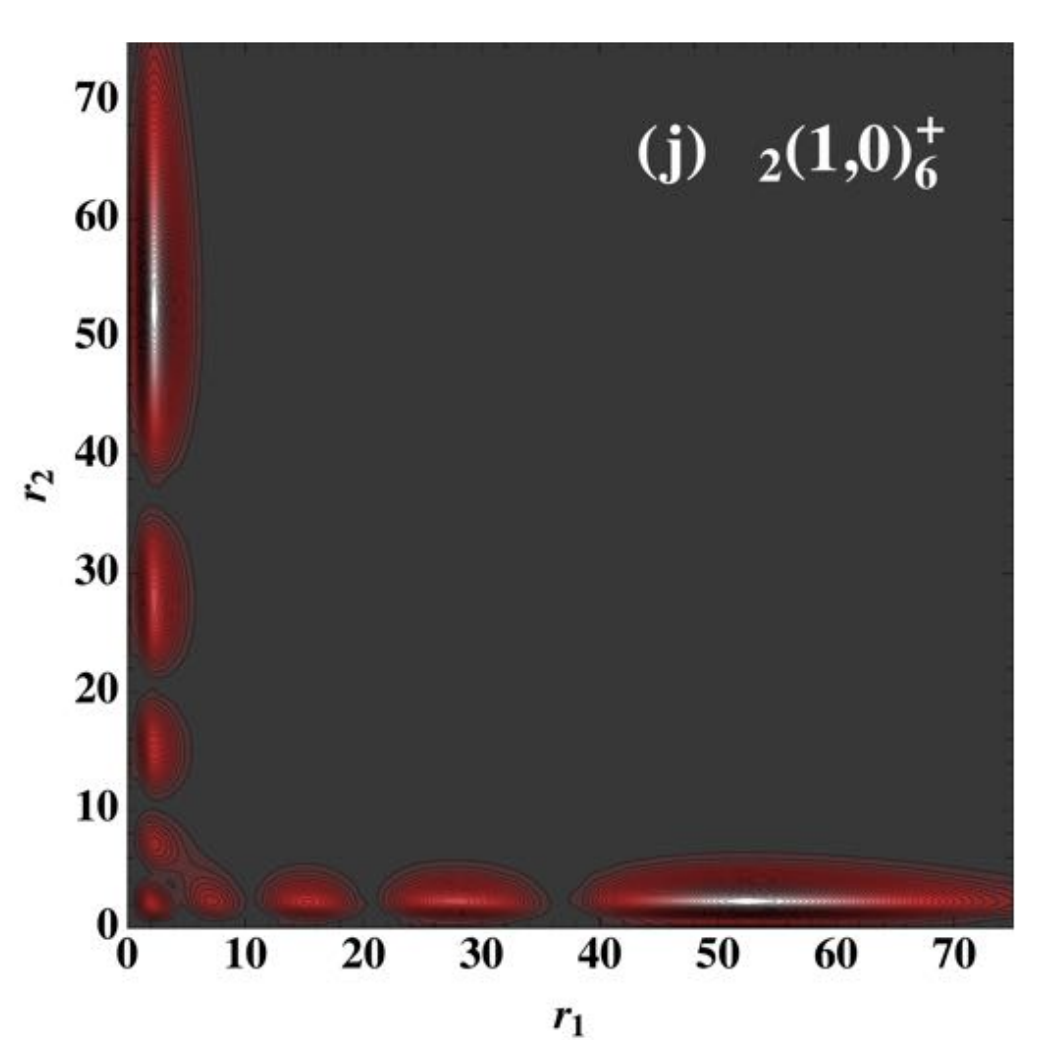}
\caption[Two-electron radial density $\rho(r_{1},r_{2})r_{1}^{2}r_{2}^{2}$ for the lowest ten resonant $^1D^e$ states in He]{\label{fig:wide5}Two-electron radial density $\rho(r_{1},r_{2})r_{1}^{2}r_{2}^{2}$ for the lowest ten resonant $^1D^e$ states in He, located below the second ionization threshold. Resonances are labelled according to the classification proposed by~\citep{Lin1983} using $_{n_1}(K,T)_{n_2}^A$. The energy ordering of the resonances is indicated by the alphabet labels inside the plots.}
\end{figure}

\begin{figure}
\centering
\includegraphics[width=0.28\textwidth]{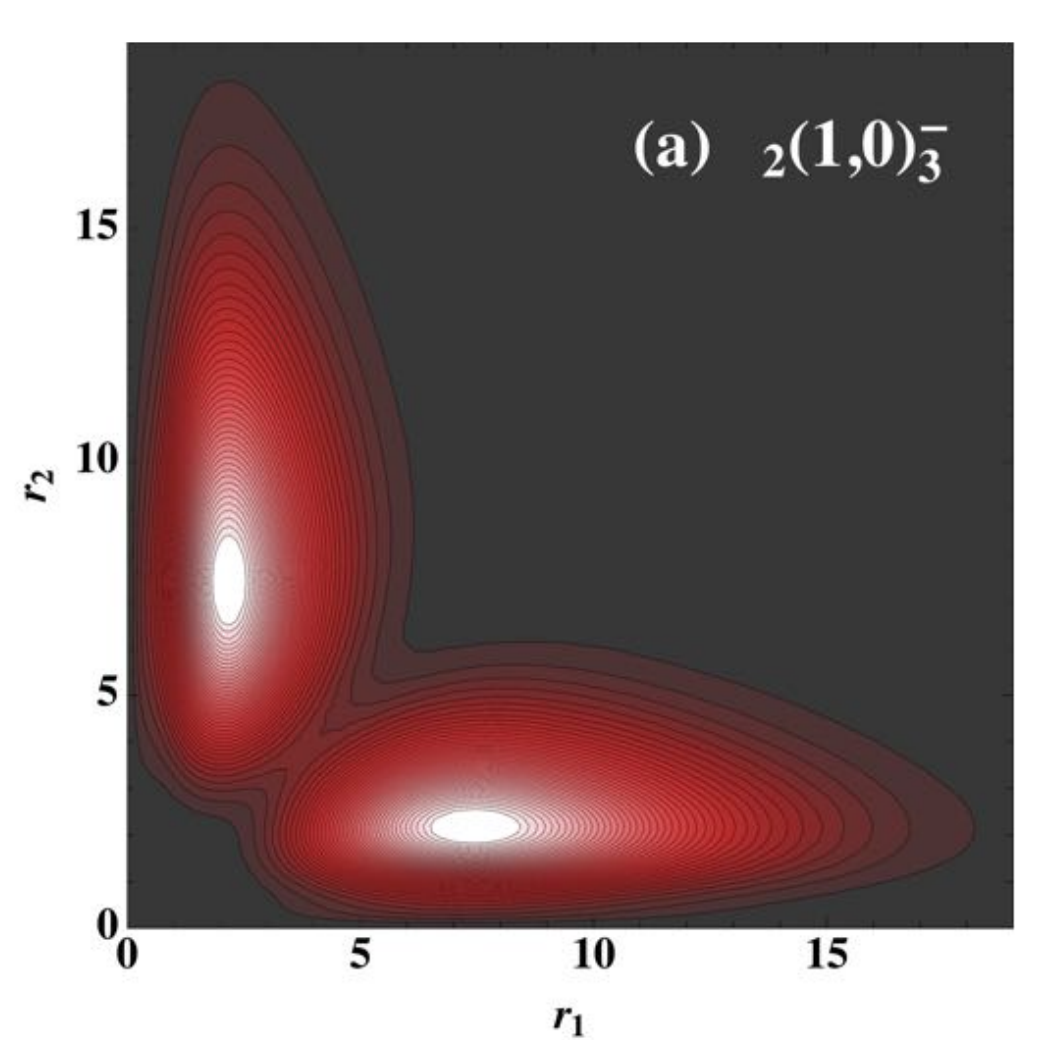}
\includegraphics[width=0.28\textwidth]{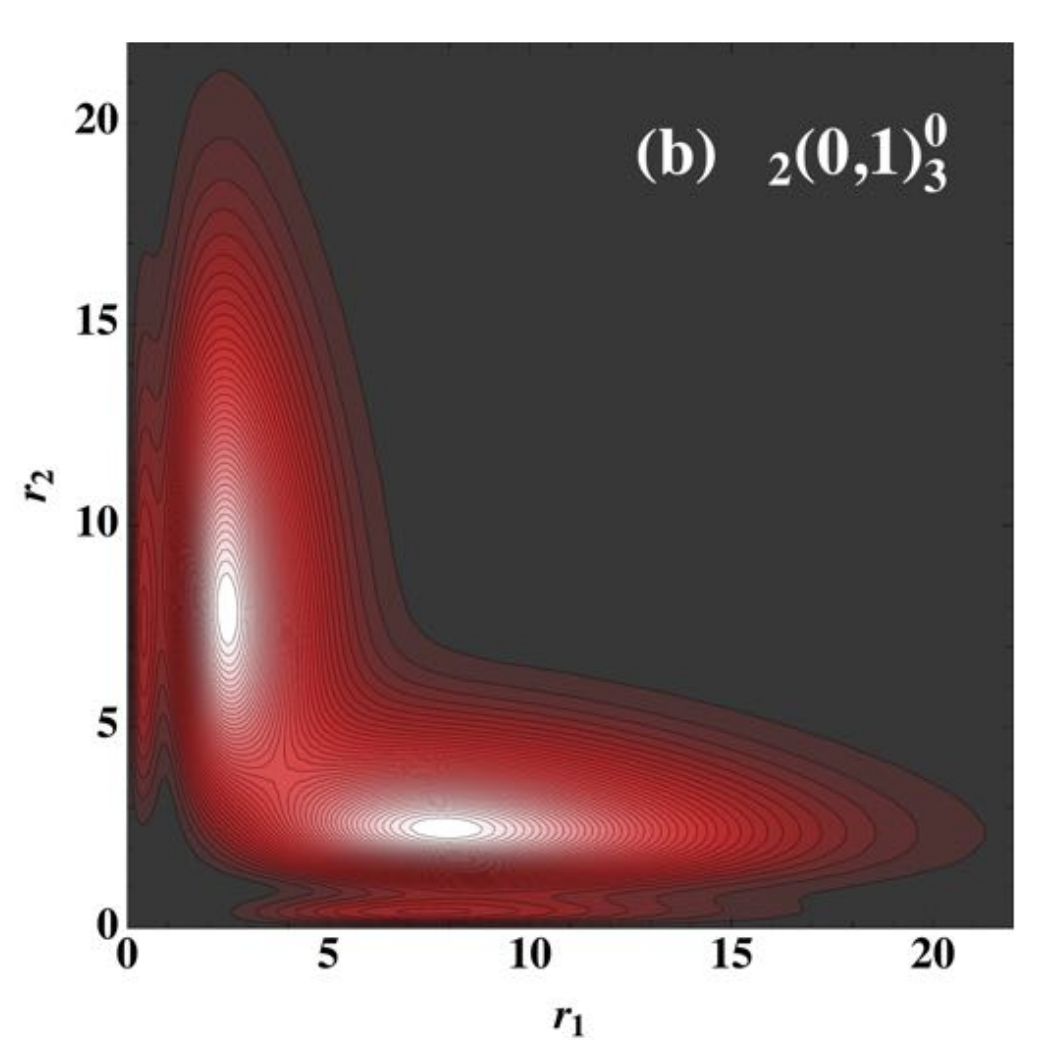}\\
\includegraphics[width=0.28\textwidth]{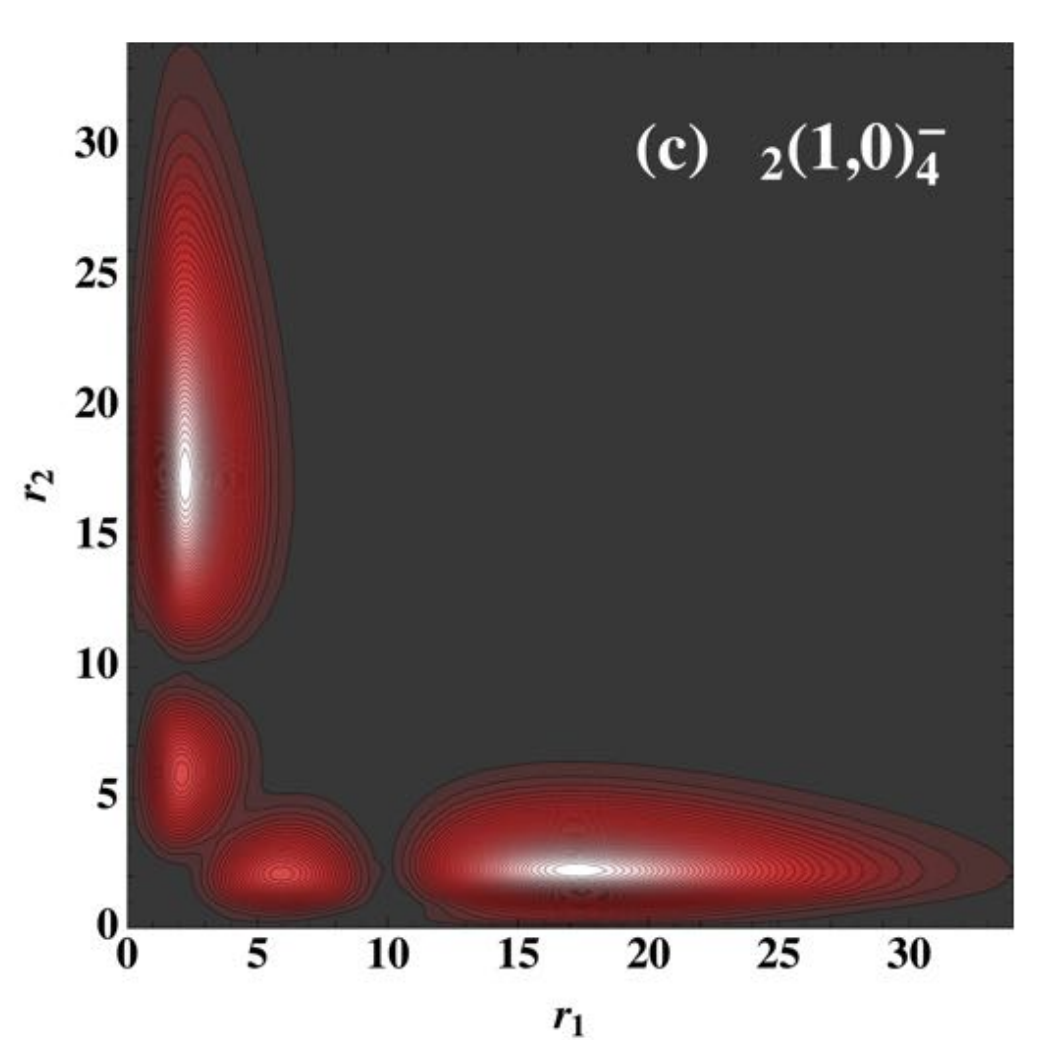}
\includegraphics[width=0.28\textwidth]{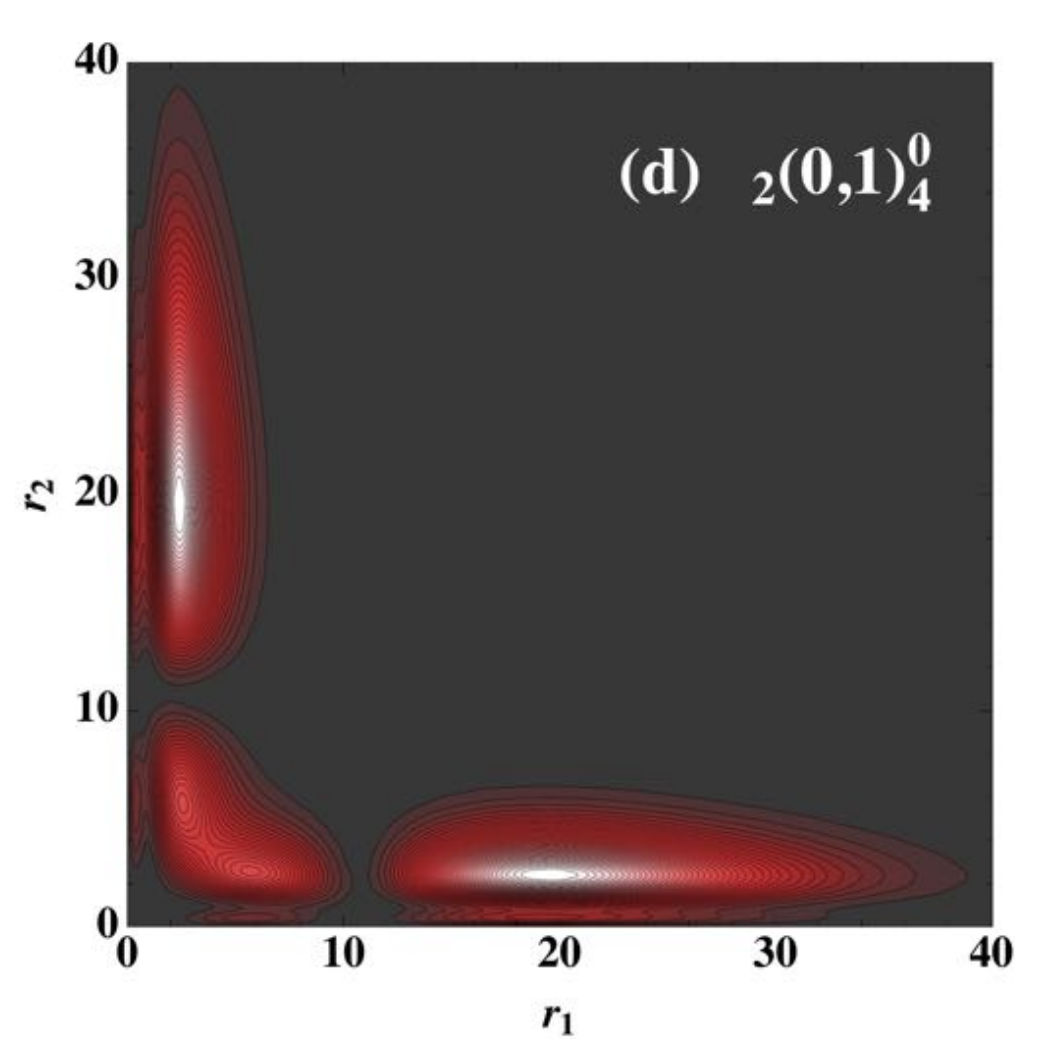}\\
\includegraphics[width=0.28\textwidth]{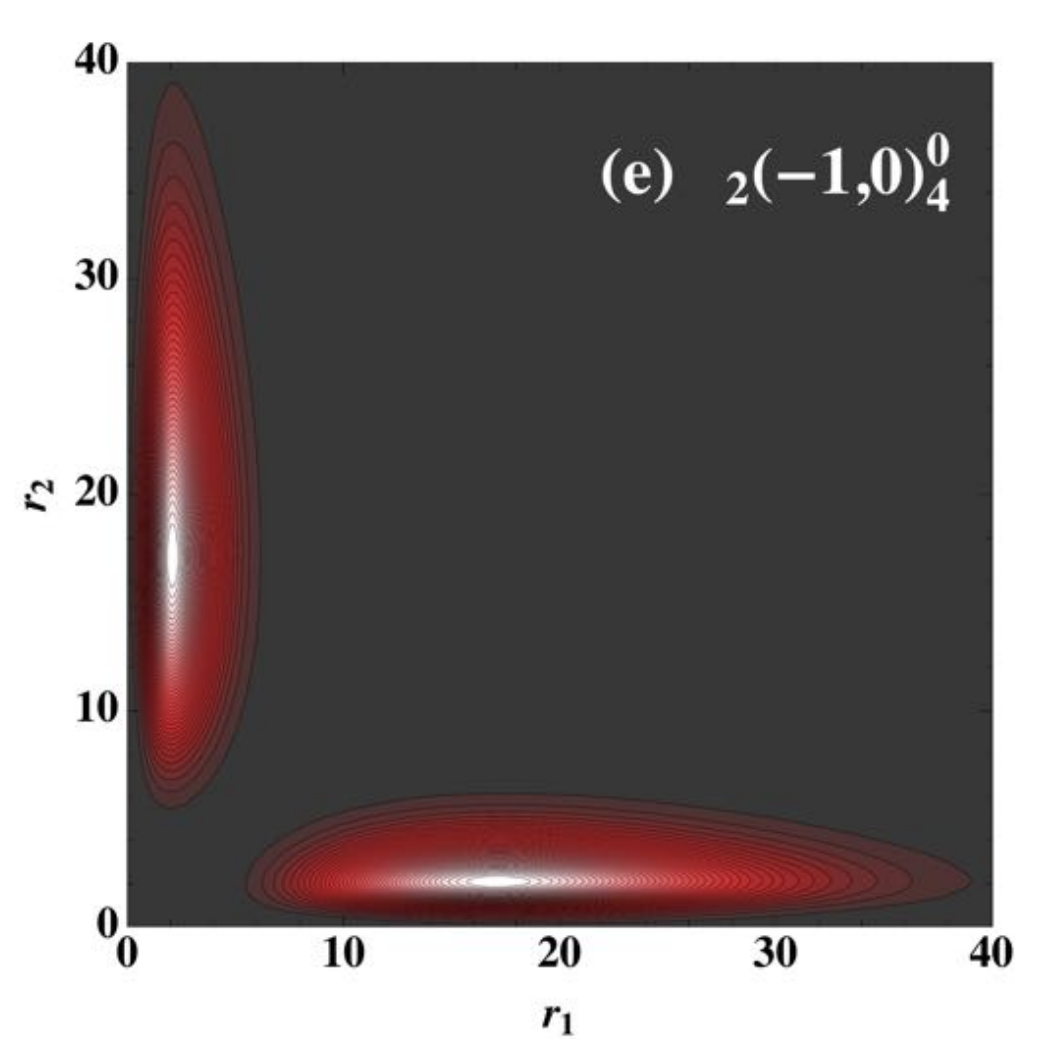}
\includegraphics[width=0.28\textwidth]{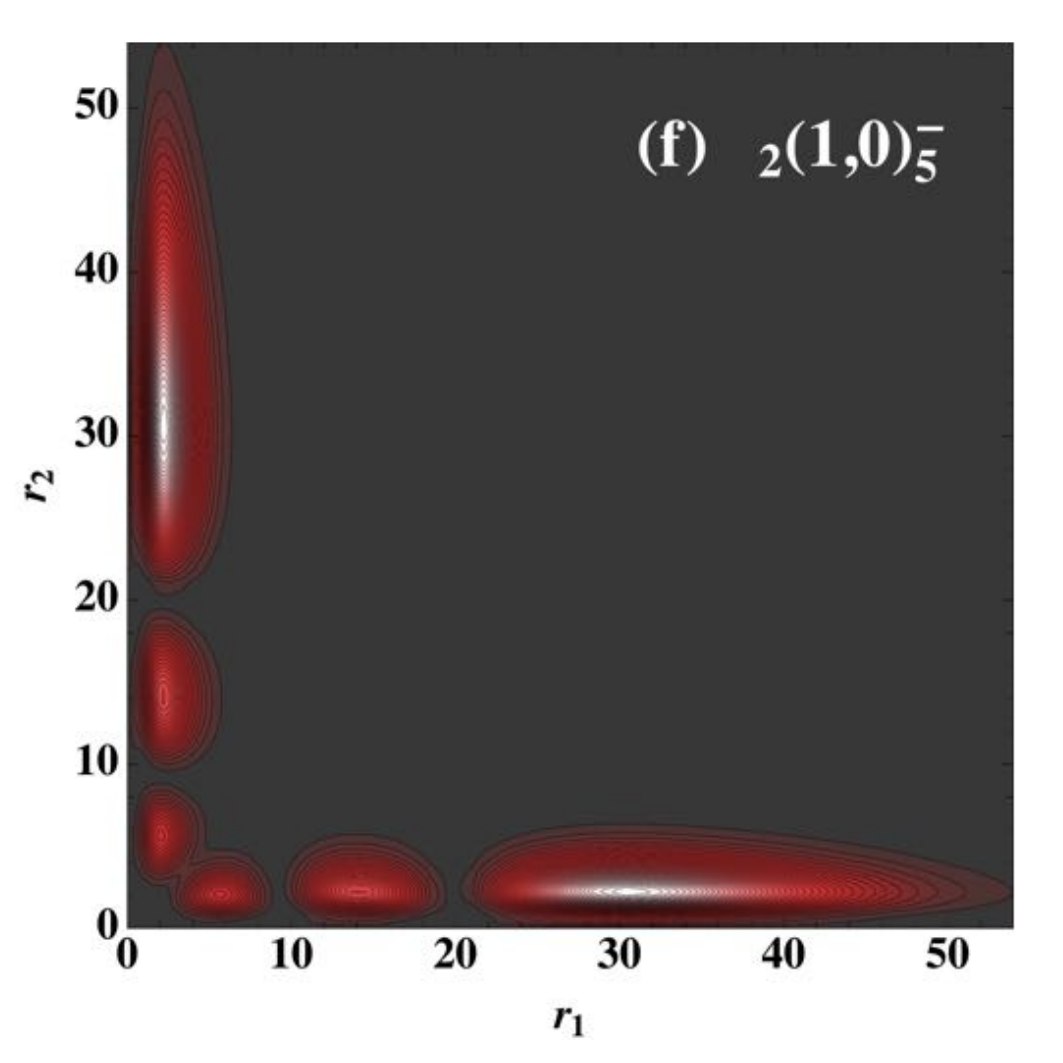}\\
\includegraphics[width=0.28\textwidth]{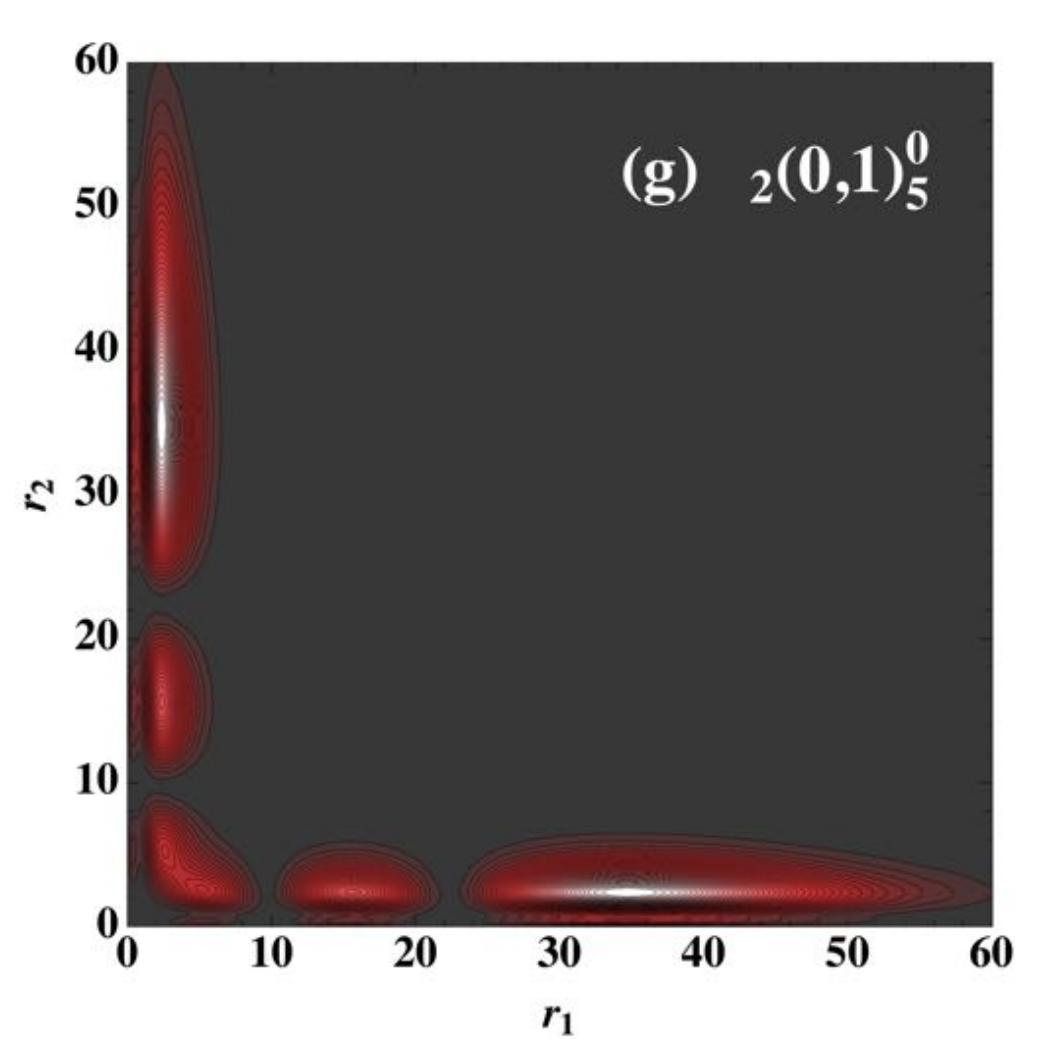}
\includegraphics[width=0.28\textwidth]{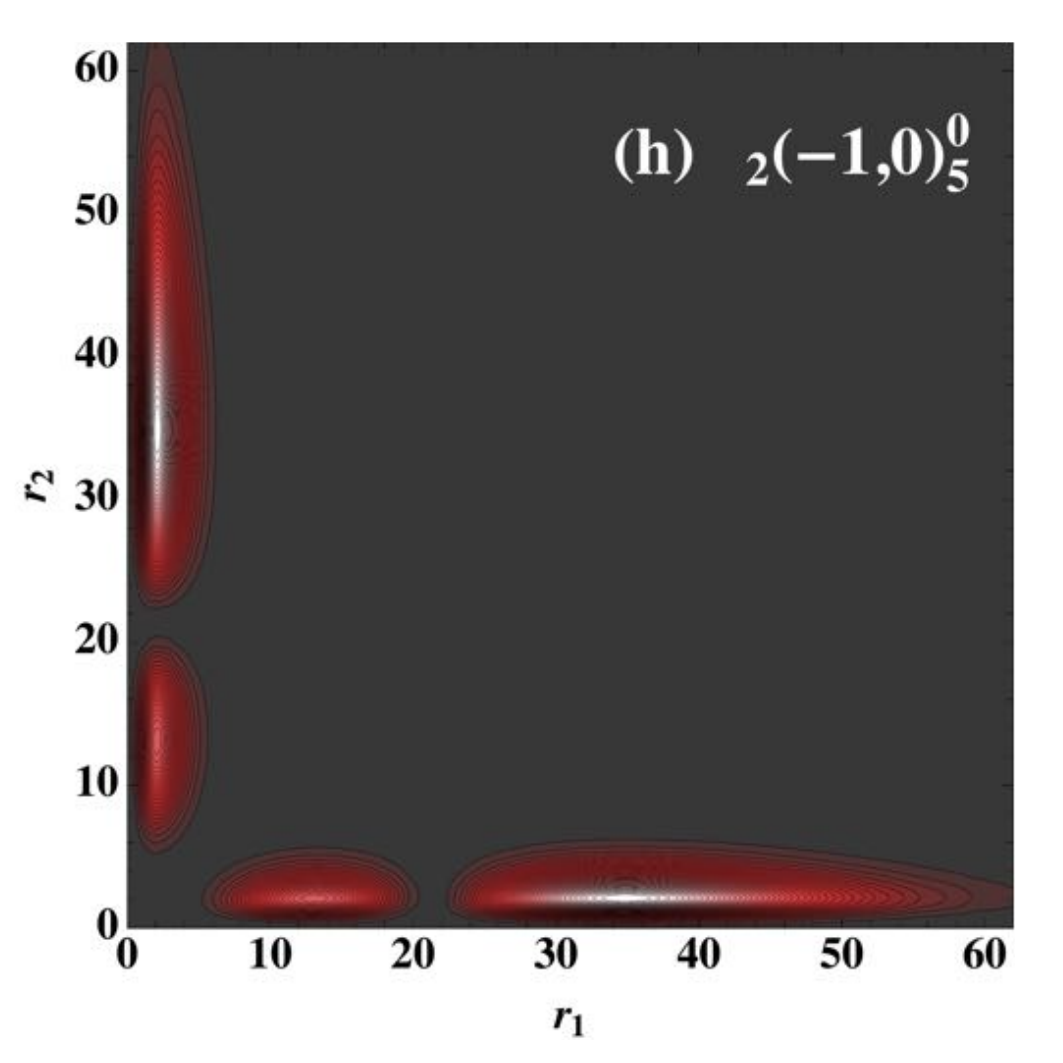}\\
\includegraphics[width=0.28\textwidth]{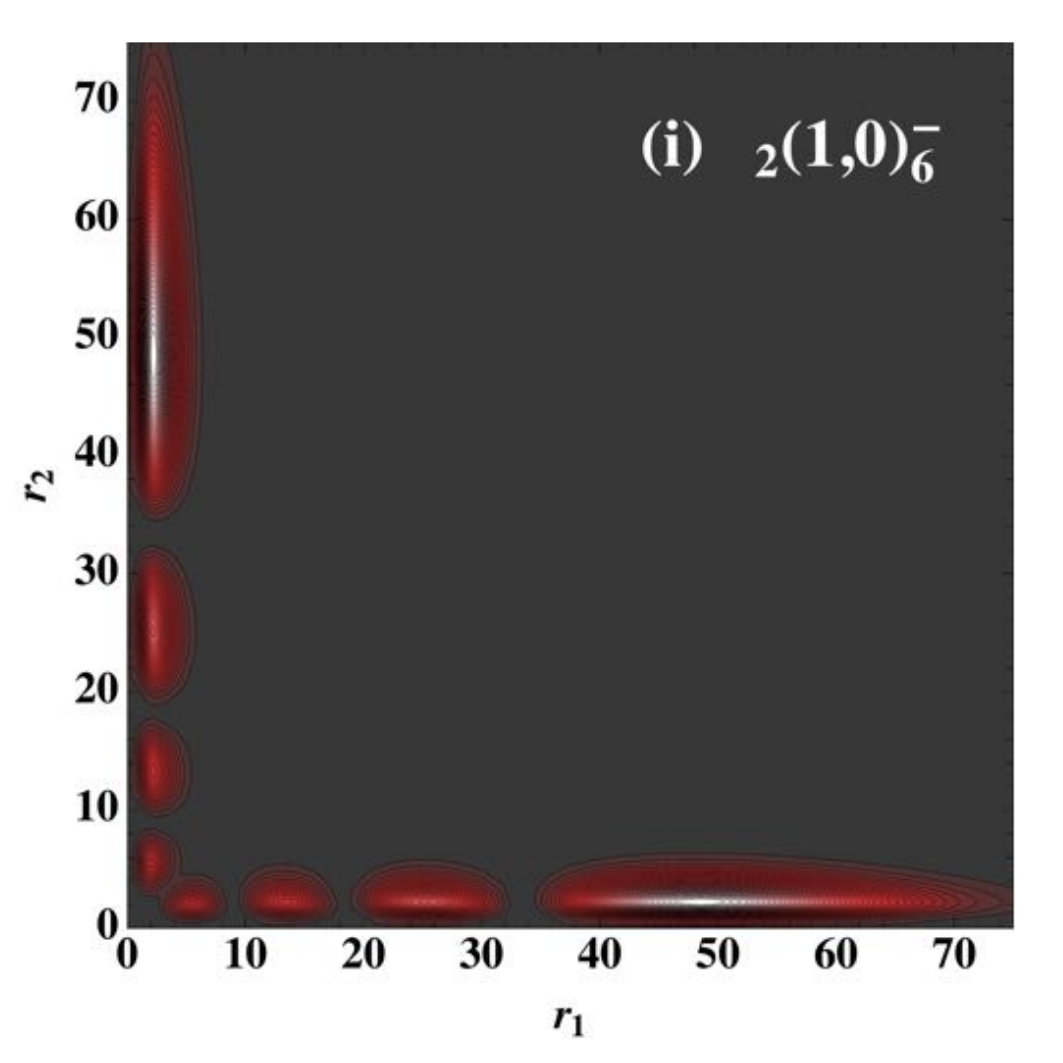}
\includegraphics[width=0.28\textwidth]{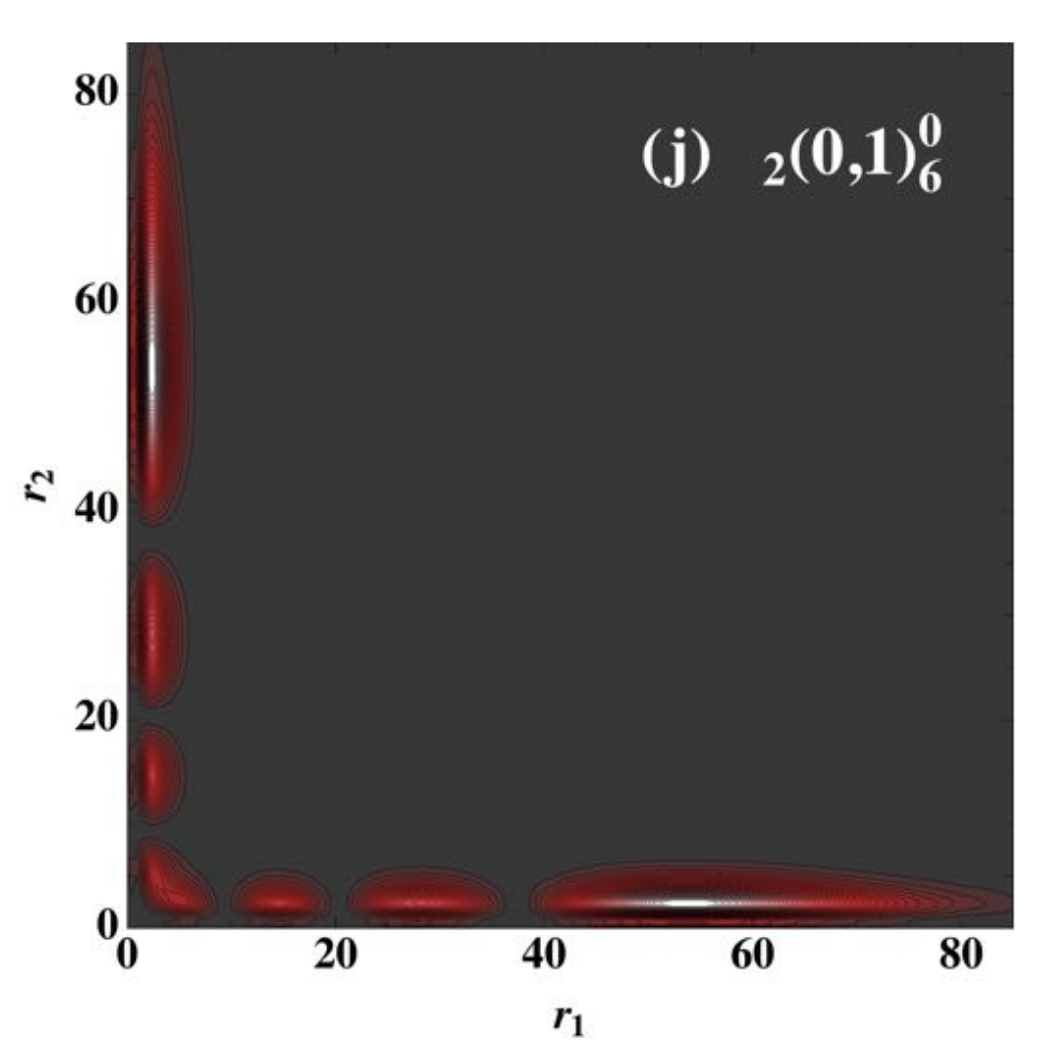}
\caption[Two-electron radial density $\rho(r_{1},r_{2})r_{1}^{2}r_{2}^{2}$ for the lowest ten resonant $^3D^e$ states in He]{\label{fig:wide6}Two-electron radial density $\rho(r_{1},r_{2})r_{1}^{2}r_{2}^{2}$ for the lowest ten resonant $^3D^e$ states in He, located below the second ionization threshold. Resonances are labelled according to the classification proposed by~\citep{Lin1983} using $_{n_1}(K,T)_{n_2}^A$. The energy ordering of the resonances is indicated by the alphabet labels inside the plots.}
\end{figure}

In the figures~\ref{fig:wide1},~\ref{fig:wide2},~\ref{fig:wide3},~\ref{fig:wide4},~\ref{fig:wide5}, and ~\ref{fig:wide6} we show the electronic probability density $\rho(r_{1},r_{2})r_{1}^{2}r_{2}^{2}$ of resonant \ac{DES} of helium located below the second ionization threshold He$^+$ $(n_1=2)$ for the total symmetries $^{1,3}S^e$, $^{1,3}P^o$, $^{1,3}D^e$, respectively. The resonances are organized under a criterion of increasing energy and are labelled according to the classification proposed by~\citep{Lin1983} and ~\citep{Herrick1975}.

By the way, the radial correlation which is described by the $A$ quantum number introduced by~\citep{Lin1984} is evidenced in these two-dimensional electronic densities. From the figures it is clearly seen that the density has an anti-node at the line  $r_1=r_2$ for $A=+1$ and a node for $A=-1$, i.e, the quantum number $A$ describes the even or odd symmetry of the \ac{WF} with respect to the line $r_1 = r_2$ and reflects the Pauli principle~\citep{Brandefelt1996}. In figures~\eqref{fig:wide1} and~\eqref{fig:wide2} for $^{1,3}S^e$ states, where only $T = 0$ is allowed, it is shown that  $A =+1$ corresponds to the spin singlet states which show an anti-node at $r_1=r_2$; on the other hand $A =-1$  labels the spin triplet states, which now have a node at the line $r_1=r_2$ as can be expected.  The symmetries $^{1,3}P^o$ have a more complicated behavior that is pictured in figures~\eqref{fig:wide3} and~\eqref{fig:wide4}. The singlet $^1P^o$ states have an alternating behavior between the values of the quantum number $A=1,-1,0$ evidencing again anti-node ($A=+1$) and node ($A=-1$) behaviors. The $A=0$ value is also predicted by~\citep{Lin1993} in order to generalize the fact that the third series in the figure~\ref{fig:wide1} is not possible to classify with a label $A=\pm 1$. Then, the triplet $^3P^o$ states have  only an alternating behavior between the values of the quantum number $A=\pm1$. This fact shows again the anti-node ($+1$) and the node behavior ($-1$), as expected. Finally, the states of symmetry $^1D^e$  shown in figure~\ref{fig:wide5} only admits the values $A=+1,0$ and the states of symmetry $^3D^e$ in figure~\ref{fig:wide6} only involve  the values $A=-1,0$. In conclusion, there is a strong relationship between the topological behavior of the electronic density distribution and the quantum label $A$, which ultimately describes de radial correlation of the \ac{DES} (symmetric or asymmetric stretching vibration as a correlated motion of the electron pair with respect to the nucleus).

\section{\label{sec:infortheo}Information-theoretic measures}

The physical and chemical properties of atoms and molecules strongly depend on the topological properties of the electronic  density function.  This function characterise the probability structure of  quantum-mechanical states. The structural (topological) properties are for instance the spreading, uncertainty, randomness, disorder, localization, and the small and strong changes of the probability distribution. In order to analyse and quantify the topological properties of a system we can use the widely known measures of the modern information theory described by~\citep{MacKay2003,Cover2006,Lopez-Rosa2010} and the references therein. The pioneering work of~\citep{Bialynicki-Birula1975} pointed out the importance of applying the methods and concepts of the classical information theory~\citep{Shannon1949,Fisher1972} to the wave mechanics. 

In this section we consider two information theoretic measures which complementarily describe the spreading of a probability density function in a box. First a measure of the global character of the distribution, able to quantify the total extent of the probability density distribution using a logarithmic functional known as Shannon entropy. Secondly, we introduce another interesting  quantity called Fisher information which is a functional of the gradient of the probability density. At variance, this measure is very sensitive to the point-wise analytic behavior of the density. This quantity has a local character. 

Our goal is to apply these two objects of analysis or quantifiers to the one-electron radial densities, see equation~\eqref{eq:oneparticle}, of the resonant \ac{DES} in helium in order to obtain information of the their  classification via the topological structure of the density.  

The application of these two quantifiers over the two-electron density functions $\rho(r_1,r_2)$ is possible but cumbersome. The integration over one of the radial coordinates to obtain the one-electron radial density, projects all the rich subtleties of the two-particle distributions included in figures~\ref{fig:wide1}-\ref{fig:wide6} into one axis ($r_1$ or $r_2$). Nevertheless we find that one-particle density may still have enough information content to discriminate qualitatively different states within a Rydberg series.

\subsection{\label{sec:entropies}Global measure: Shannon entropy}

The birth of modern information theory was due to the pioneering paper of~\citep{Shannon1949}. Claude Shannon in 1940s was investigating, in addition to how to make communication procedures safer, how much people could communicate with each other through a physical system (e.g. a telephone network). Shannon was seeking the way to send two or more calls down a single wire. In order to achieve his pursuit he needed to provide a precise mathematical definition of the information concept.  Shannon came up with the definition that the information content of an event is proportional to the $log$ of its inverse probability $p$ of occurrence~\citep{Vedral2010}:

\begin{equation}{\label{eq:information}}
I_S=ln\frac{1}{p}.
\end{equation}

This definition of information expresses two relevant properties: (1) the fact that less likely events, the ones for which the probability of happening is very small are the ones that carry more information; (2)  the total information in two independent events should be the sum of the two individual amounts of information. For this reason, and the fact that the joint probability of the two independent events are the product of the individual probabilities, the information definition involves a logarithmic function. Shannon originally named his measure of information as "entropy" by a direct suggestion of John von Neumann. We often write the Shannon entropy as a function of a probability distribution, $p_1,\dots,p_n$, i.e., as the expectation value of the expression~\eqref{eq:information}  with this probability distribution

\begin{equation}{\label{eq:shannondisc}}
S(p_1,\dots,p_n)=-\sum_xp_xlnp_x.
\end{equation}

Finally, the Shannon entropy is generalized, for an arbitrarily continuous probability distribution function, as \citep{Catalan2002,Cover2006,Lopez-Rosa2009, Lopez-Rosa2010, Angulo2011, Antolin2011}

\begin{equation}{\label{eq:shannonentropy}}
S[\rho]=-\int\rho(\mathbf{r})ln\rho(\mathbf{r})d\mathbf{r}
\end{equation}

The Shannon entropy is a direct measure of the uncertainty for a probability distribution. It talks about ignorance or lack of information concerning an experimental event or outcome. Nevertheless, as we have noted before, the Shannon entropy is a measure of the amount of information that we expect to gain on performing a probabilistic experiment.

\subsection{\label{sec:inffisher}Local measure: Fisher information}

Instead of providing information on the global structure of the probability distribution function, there are other functionals more sensitive to the local topology, for instance, the Fisher entropy. It is a better indicative of the local irregularities or the oscillatory nature of the density as well as a witness of disorder of the system. Therefore, this function is able to detect the local changes of the density in order to provide a better description of the system in terms of the measure of information in the outcome of an experiment. The Fisher information is defined as~\citep{MacKay2003,Cover2006,Lopez-Rosa2010}

\begin{eqnarray}{\label{eq:fisherinformation}}
I[\rho] & = & \int\left|\mathbf{\nabla}ln\rho(\mathbf{r})\right|^{2}\rho(\mathbf{r})d\mathbf{r}\\ \nonumber
            & = & \int\frac{\left|\mathbf{\nabla}\rho(\mathbf{r})\right|^{2}}{\rho(\mathbf{r})}d\mathbf{r}
\end{eqnarray}

The Fisher information measure has been a successful concept to identify, characterize and interpret numerous phenomena and physical processes such as e.g., correlation properties in atoms, the periodicity and shell structure in the periodic table of chemical elements. It has been used for the variational characterization of quantum equation of motion, and also to re-derive the classical thermodynamics without requiring the usual concept of Boltzmann's entropy, as well as other large variety of applications, see~\citep{Lopez-Rosa2010} and the references therein. 

One of the more remarkable applications of the Fisher information is its deep relationship with \ac{DFT} where it plays a central role. The relevance of Fisher information in quantum mechanics and \ac{DFT}  was first emphasized more than thirty  years ago. It states that the quantum mechanical kinetic energy can be considered a measure of the information distribution, see for instance~\citep{Nagy2003} and the references therein. This well established relationship between the quantum mechanical kinetic energy functional and the Fisher information is called  in the literature the Weizs\"acker kinetic energy functional.

\section{Results and discussions}\label{sec:resultanddiscussions}

In this section we show and discuss the results obtained for both measures introduced before: Shannon entropy and the Fisher information for  the electronic density function of \ac{DES} of helium atom. Additionally, we also analyse the behavior of the one-particle electronic density itself and the differential entropies, i.e., the arguments of the both measures, for the resonant states of symmetries $^{1,3}S^e$, $^{1,3}P^o$, and $^{1,3}D^e$ for the same atom. Nevertheless, in order to obtain an intuitive picture of the meaning for each of the entropic measures, we initially present the same analysis, as an interesting illustration for the bound states of the hydrogen atom.

\subsection{Information measures of Hydrogen atom}\label{sec:hydrogenresults}

The hydrogen atom has been considered to have a central role in quantum physics and chemistry. Its analysis is basic not only to gain a full insight into the intimate structure of matter but also for other numerous phenomena like light-matter interaction~\citep{Bransden2003}, the behavior of heterostructures like quantum-dots, and so on. On the whole, since the birth of quantum mechanics, the hydrogen atom had become as a paradigm, mainly because its Schr\"odinger equation can be solved analytically. 

\begin{figure}[h]
\centering
\includegraphics[width=0.85\textwidth]{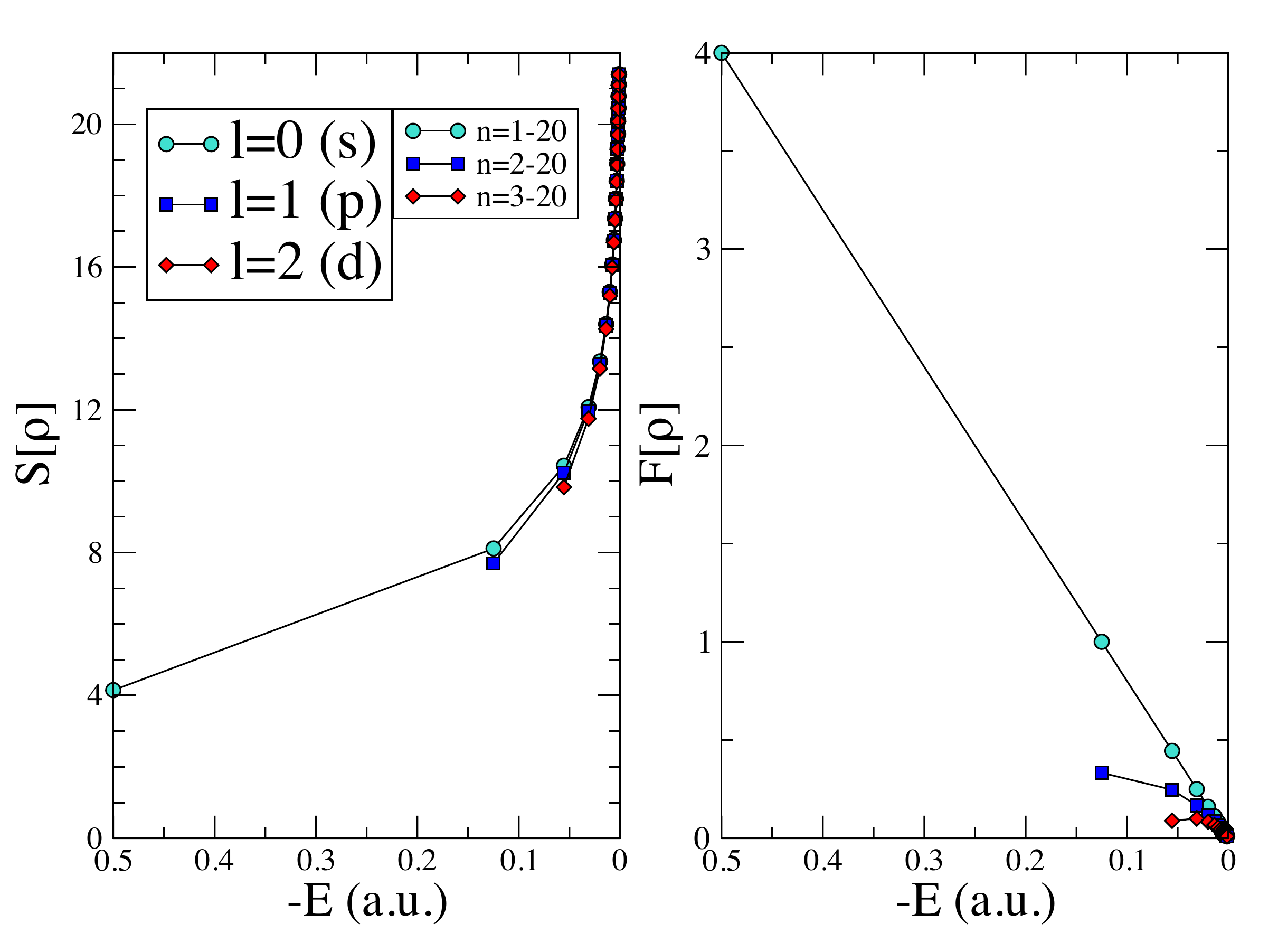}
\caption{\label{fig:shannfisherhyd}Shannon Entropy and Fisher Information or the lowest bound states (for $l=0,1$ and $2$) of the Rydberg series below the ionization threshold in the hydrogen atom.}
\end{figure}

In this section we obtain the information measures, i.e., the Shannon entropy and the Fisher information for the hydrogen atom. Let us now deal with the analytical expression~\eqref{eq:solhyd1} for the \ac{WF} of this atom
\begin{equation}\label{eq:solhyd12} 
\psi_{E,l,m}(r,\theta,\phi)=R_{E,l}({r})\mathcal{Y}^l_{m_l}(\theta,\phi),
\end{equation}
where $R_{E,l}({r})$ is the radial function and $\mathcal{Y}^l_{m_l}(\theta,\phi)$ is the spherical harmonic which describes the angular dependence. We are only interested in calculating the entropic measures on the radial part of the \ac{WF}, therefore we trace over the angular degrees of freedom. Then, the radial one-particle electronic density can be written, in terms of the equation~\eqref{eq:radialf}, as
\begin{align}\label{eq:solhyd13} 
\rho({r})&=|R_{E,l}({r})|^2\\ \nonumber
&=\left|\left\{\left(\frac{2Z}{n}\right)^{3}\frac{(n-l-1)!}{2n[(n+l)!]^3}\right\}^{\frac{1}{2}}e^{-\frac{\rho}{2}}\rho^l L^{2l+1}_{n+1}(\rho)\right|^2,
\end{align}
where $\rho=\frac{2Z}{n}r$ and the $L^i _k(\rho)$ are the associated Laguerre polynomials.
\begin{figure}[h]
\centering
\includegraphics[width=0.85\textwidth]{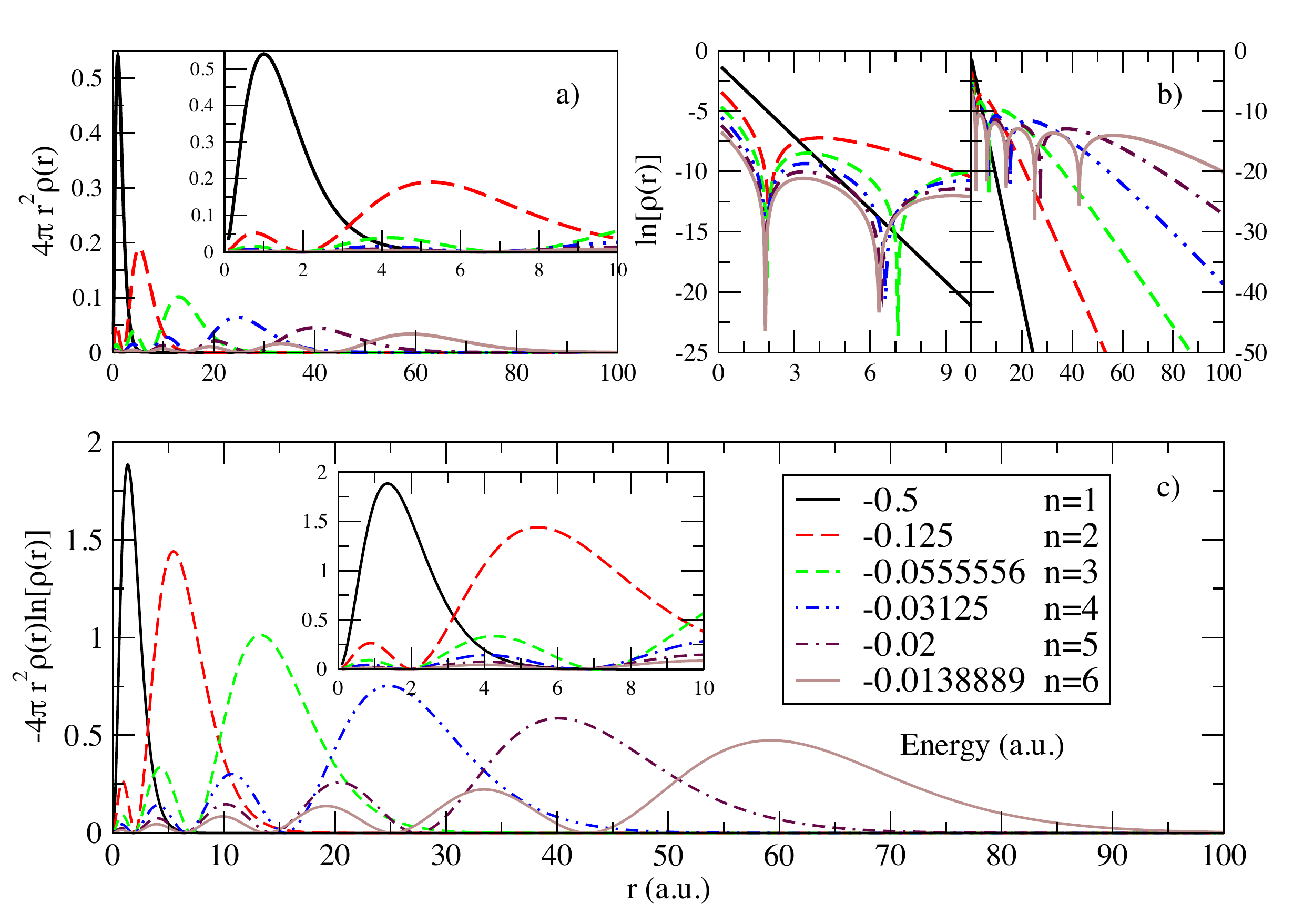}
\caption[Components of the integrand in the Shannon entropy integral formula for the six lowest bound states of hydrogen with $l=0$]{\label{fig:sharg_l0}Components of the integrand in the Shannon entropy integral formula for the six lowest bound states of hydrogen with $l=0$. The panel (a) shows the electronic density times the angular factor $4\pi$ and the volume J-factor $r^2$. Panel b) shows the logarithm of the density at two different scales and panel c) shows the full integrand of the Shannon entropy (or differential Shannon entropy). The inset is a blow-up of the inner radial region.}
\end{figure}

\begin{figure}[h]
\centering
\includegraphics[width=0.85\textwidth]{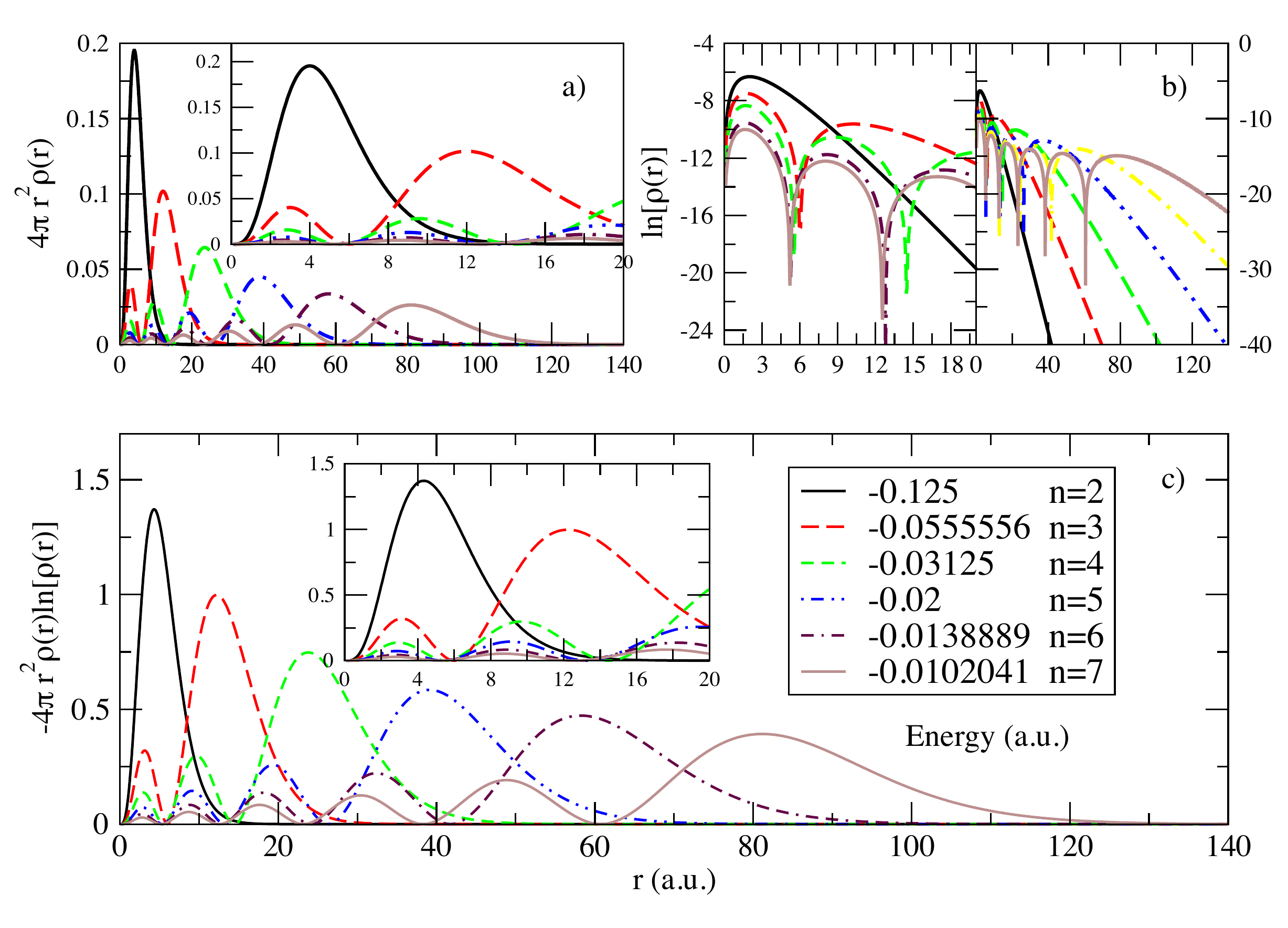}
\caption[Components of the integrand in the Shannon entropy integral formula for the six lowest bound states of hydrogen with $l=1$]{\label{fig:sharg_l1}Components of the integrand in the Shannon entropy integral formula for the six lowest bound states of hydrogen with $l=1$. The panel (a) shows the electronic density times the angular factor $4\pi$ and the volume J-factor $r^2$. Panel b) shows the logarithm of the density at two different scales and panel c) shows the full integrand of the Shannon entropy (or differential Shannon entropy). The inset is a blow-up of the inner radial region.}
\end{figure}

\begin{figure}[h]
\centering
\includegraphics[width=0.85\textwidth]{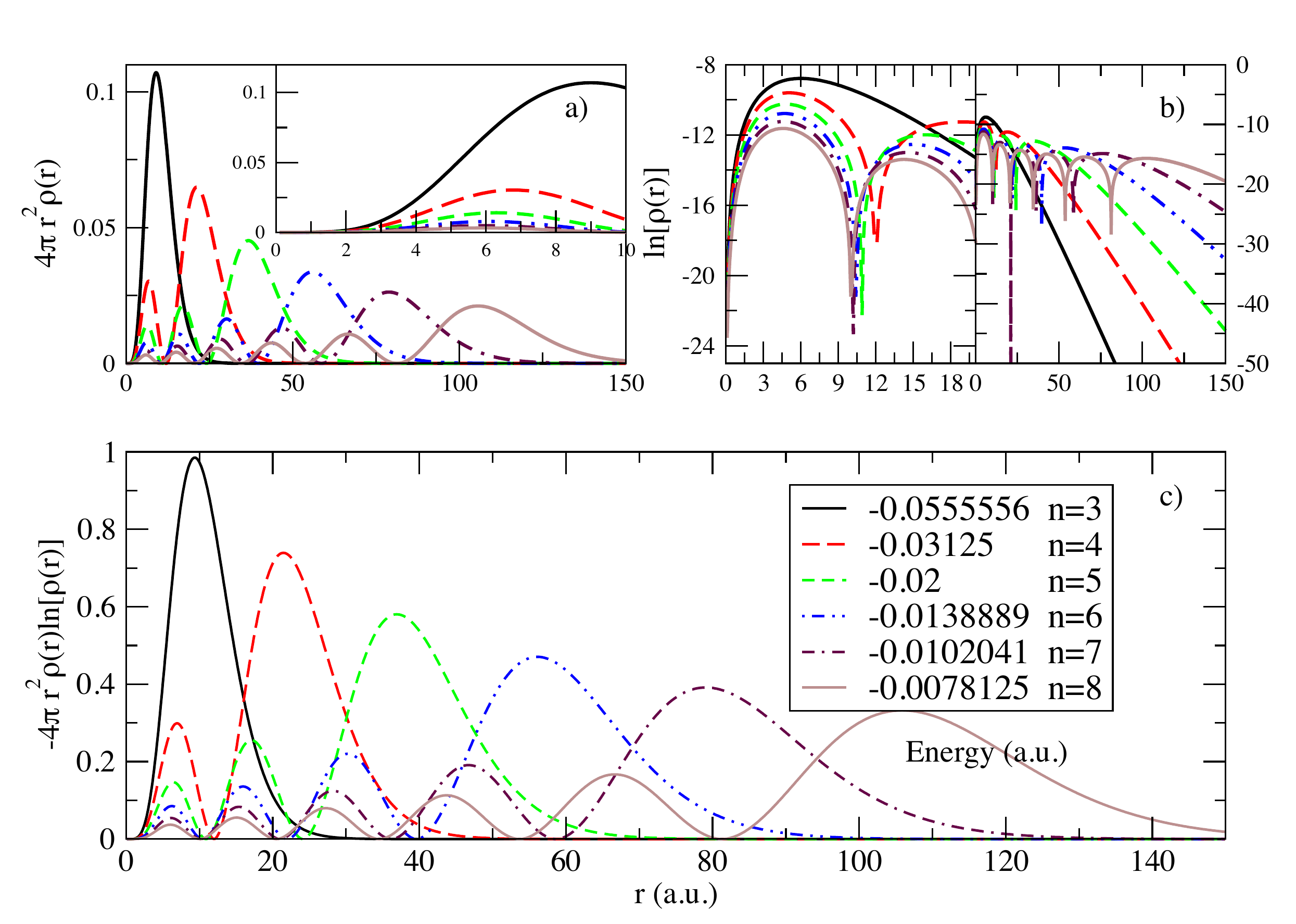}
\caption[Components of the integrand in the Shannon entropy integral formula for the six lowest bound states of hydrogen with $l=2$]{\label{fig:sharg_l2}Components of the integrand in the Shannon entropy integral formula for the six lowest bound states of hydrogen with $l=2$. The panel (a) shows the electronic density times the angular factor $4\pi$ and the volume J-factor $r^2$. Panel b) shows the logarithm of the density at two different scales and panel c) shows the full integrand of the Shannon entropy (or differential Shannon entropy). The inset is a blow-up of the inner radial region.}
\end{figure}

The Shannon entropy for the lowest bound states in the Rydberg series with angular momentum $l=0,1$ and $2$ in H, is shown in figure~\ref{fig:shannfisherhyd}. This quantity is a monotonically increasing function when the energy of the bound states increases, and the curve seems to reach an asymptotic behavior against the location of the ionization threshold at $E=0$ \ac{a.u.}, regardless the value of the angular momentum $l$. This behavior evidences the fact that the density becomes more an more spread with the energy excitation. Even though the ground state and the low energy states has different values of Shannon entropy for different values of the angular momentum $l$, these values tend to converge to the same one for highly excited manifolds ($n \to \infty$) in the Rydberg series, for which all electron densities become highly oscillatory and spread out, regardless of the details at short distances for different angular momentum.  Previous results for the ground state of hydrogen are reported by~\citep{Sen2005} and some analytical expressions are provided by~\citep{Lopez-Rosa2005}. In addition, the figures~\ref{fig:sharg_l0}, \ref{fig:sharg_l1}, and \ref{fig:sharg_l2} show the components of the integrand for the Shannon entropy according to Eq. (2.26) for the lowest eigenstates of hydrogen atom with angular momentum $l=0,1$ and $2$, respectively. In each figure, the panel (a) shows the electronic density, the panel (b) its logarithm, and in panel (c) the complete differential Shannon entropy (the full integrand) is shown. In the figure~\ref{fig:sharg_l0}(a) the electronic density of the ground state of Hydrogen atom is extremely localized close to the nucleus. By inserting the logarithm of the electron density in the definition of the Shannon entropy, many details of the density distributions at large radial distances are incorporated into the entropy. This behavior of the differential Shannon entropy as the energy increases (the peak of the maximum decreases but the distribution spreads out) explains the monotonically increasing character of the Shannon entropy. Incidentally, those states which are degenerated in energy (same $n$ but different $l$, like $2s$ and $2p$ or $3s$, $3p$ and $3d$), in spite of having different spreading of the density, the associated values for the Shannon entropy are similar, a behavior that can be understood from the differential probabilities in figures~\ref{fig:sharg_l0},~\ref{fig:sharg_l1}, and~\ref{fig:sharg_l2}. Finally, it is clear, from direct comparison, that the density of highly excited estates with the same energy and different angular momentum are almost indistinguishable for the Shannon entropy measure.

On the other hand, the Fischer information plot in figure~\ref{fig:shannfisherhyd} for hydrogen  shows that this quantity decreases monotonically to the limit value zero at the ionization threshold ($n\to \infty$), but, at variance with the Shannon entropy, the Fischer information measure shows a distinctive trend for each angular momentum value $l$ (see figure~\ref{fig:shannfisherhyd}). Therefore, it is possible to conclude that the density became almost homogeneous, i.e., it reaches a high oscillatory homogeneous behavior, for highly excited states. Fisher information is higher for the ground state which is more localized and has smaller uncertainty, i.e., the accuracy in estimating the localization of the particle is bigger. This behavior is depicted in figures~\ref{fig:sharg_l0} (a),~\ref{fig:sharg_l1} (a) and~\ref{fig:sharg_l2} (a) evidencing the strong localization of the ground state. Moreover, as is shown in the figures the argument of the Fisher information, or the differential Fisher information, shows that the major contribution to this local measure comes from the regions of the electronic density close to the nucleus for the ground state. However, for the excited states, this contribution becomes more and more unimportant. Some analytical results can be found in~\citep{Lopez-Rosa2005,Lopez-Rosa2010}.

\newpage

\begin{figure}
\centering
\includegraphics[width=0.85\textwidth]{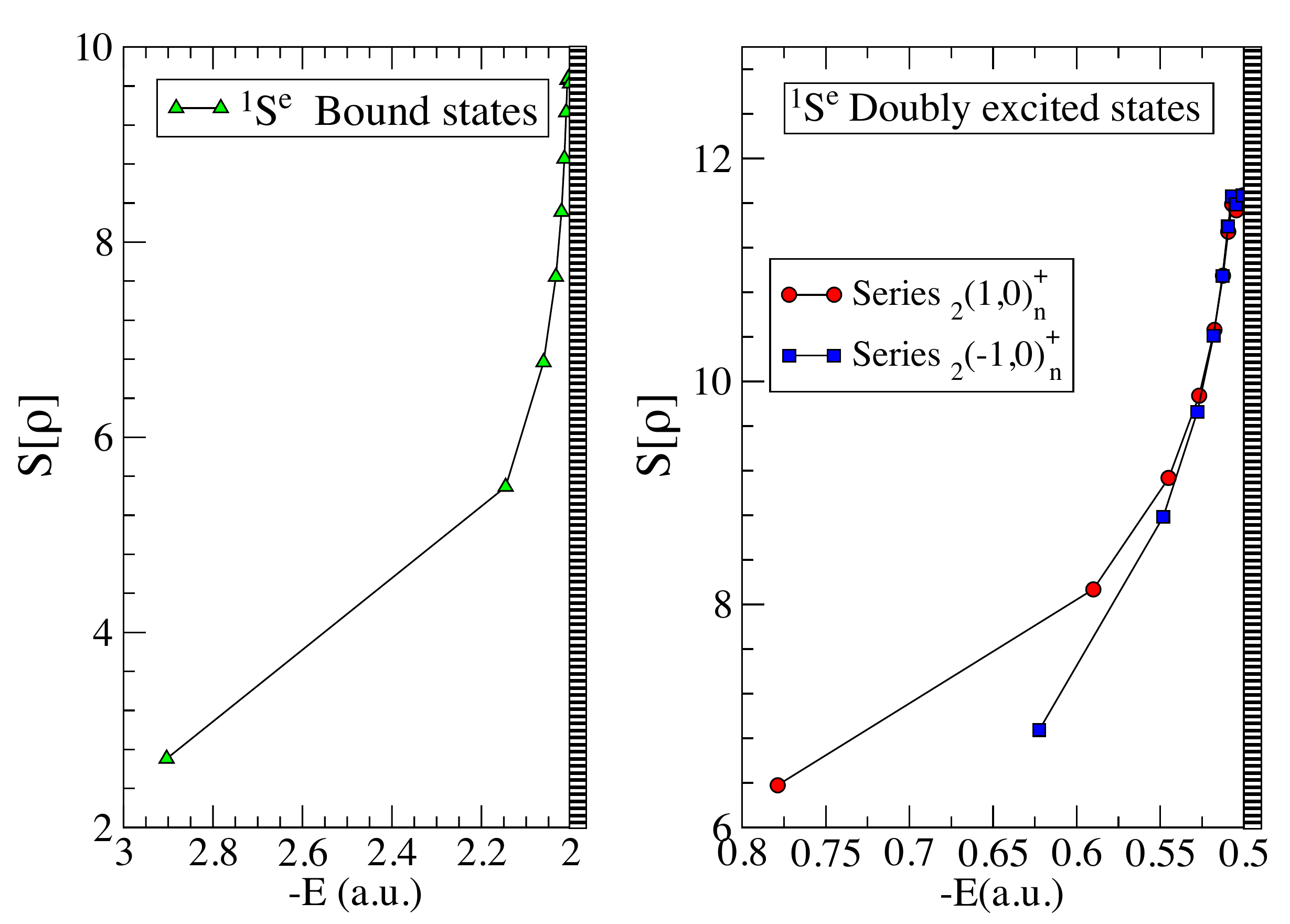}
\caption{\label{fig:shannon1Se}Shannon entropy $S(\rho)$ for the ground state and singly excited $^1S^e$ states in helium (left panel) and doubly excited states belonging to the two series $_2(1,0)^+_n$ and $_2(-1,0)^+_n$ within the $^1S^e$ symmetry of helium (right panel).}
\end{figure}

\subsection{Results of information theory measures for doubly excited states of helium}\label{sec:hydrogen results}

We calculate, using a \ac{CI}-\ac{FM} approach, the eigenenergies and eigenfunctions of the Rydberg series of He \ac{DES} below the second ionization threshold for symmetries $(^{1,3}S^e,^{1,3}P^o,^{1,3}D^e)$. However, as we have said before,  we are not interested in reproducing the highly precise values for the energies already reported in the literature. Instead, we focus our effort to obtain a reasonable good description of the \ac{WF} itself, since our workhorse is related to the radial density and our final results are analyzed more qualitatively than quantitatively.  In addition, we have also calculated  the information-theoretic measures of the  ground state and singly excited states of helium atom, as we have done in the previous section with the bound states of hydrogen atom and we present an analysis. In the following sections we present the results of our numerical studies on the Shannon entropy and the Fisher information integrals for each of the symmetries named before.  In order to obtain all the entropic measures of helium presented through this work, we have used a numerical integration scheme based on the Gauss-Legendre quadrature~\citep{Abramowitz1965,Press2007} which is a very  suitable approximation of the definite integral of an arbitrary function usually stated as a weighted sum of the function values at very specified points within the domain of integration.

\begin{figure}
\centering
\includegraphics[width=0.85\textwidth]{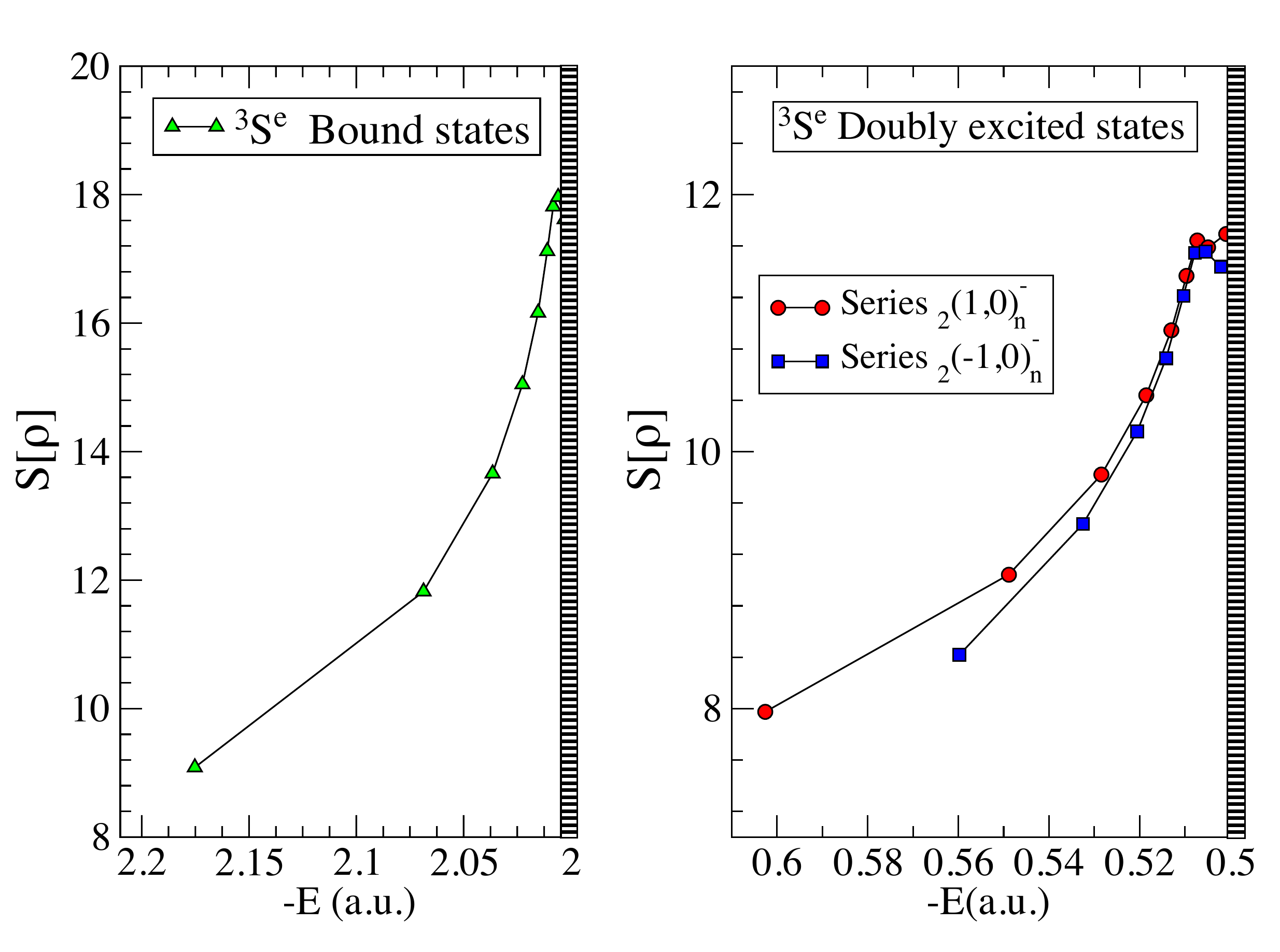}
\caption{\label{fig:shannon3Se}Shannon entropy $S(\rho)$ for singly excited $^3S^e$ states in helium (left panel) and doubly excited states belonging to the two series $_2(1,0)^-_n$ and $_2(-1,0)^-_n$ within the $^3S^e$ symmetry of helium (right panel).}
\end{figure} 

\begin{figure}
\centering
\includegraphics[width=0.85\textwidth]{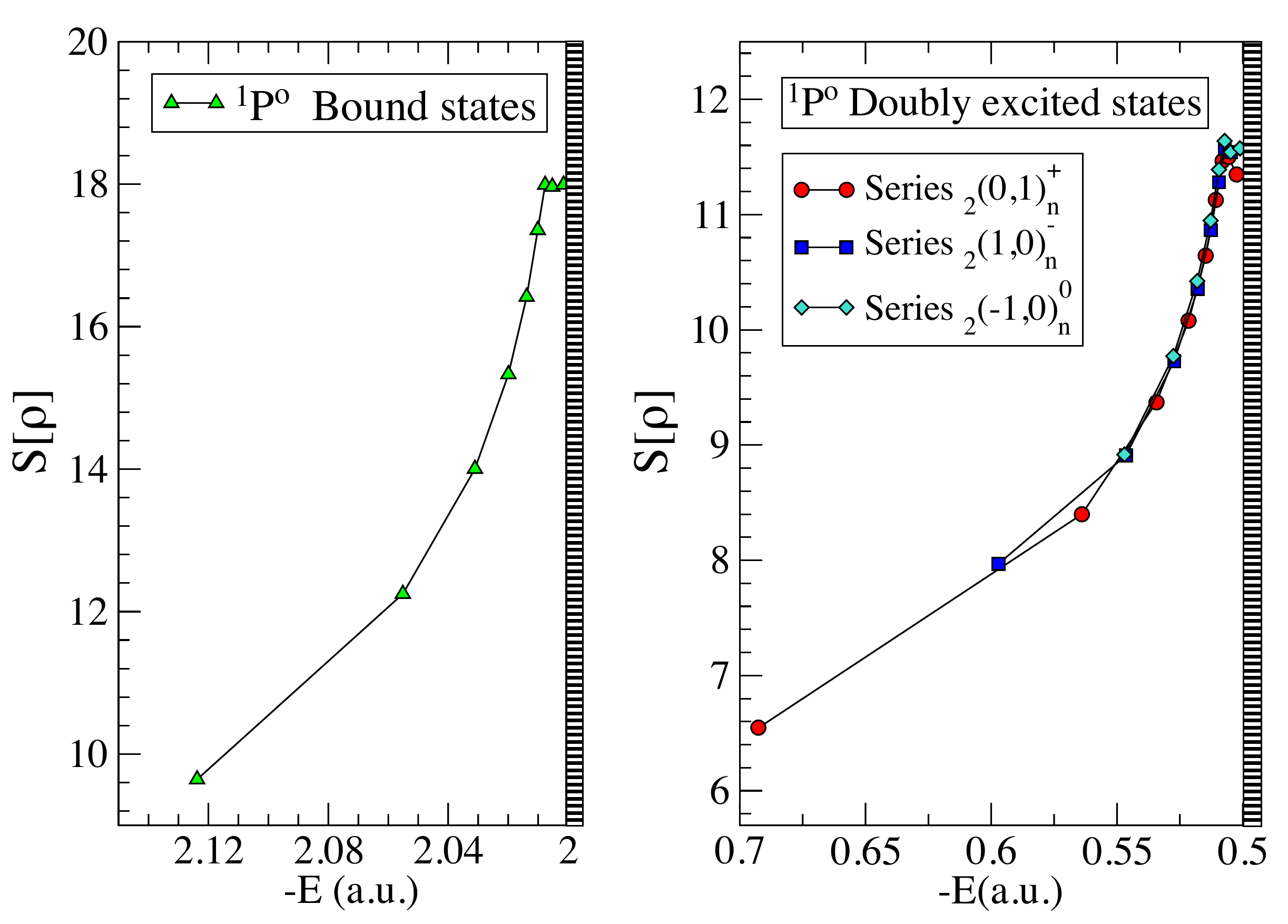}
\caption{\label{fig:shannon1Po}Shannon entropy $S(\rho)$ for the singly excited $^1P^o$ states in helium (left panel) and doubly excited states belonging to the three series $_2(0,1)^+_n$, $_2(1,0)^-_n$ and $_2(-1,0)^0_n$ within the $^1S^e$ symmetry of helium (right panel).}
\end{figure}

\begin{figure}
\centering
\includegraphics[width=0.85\textwidth]{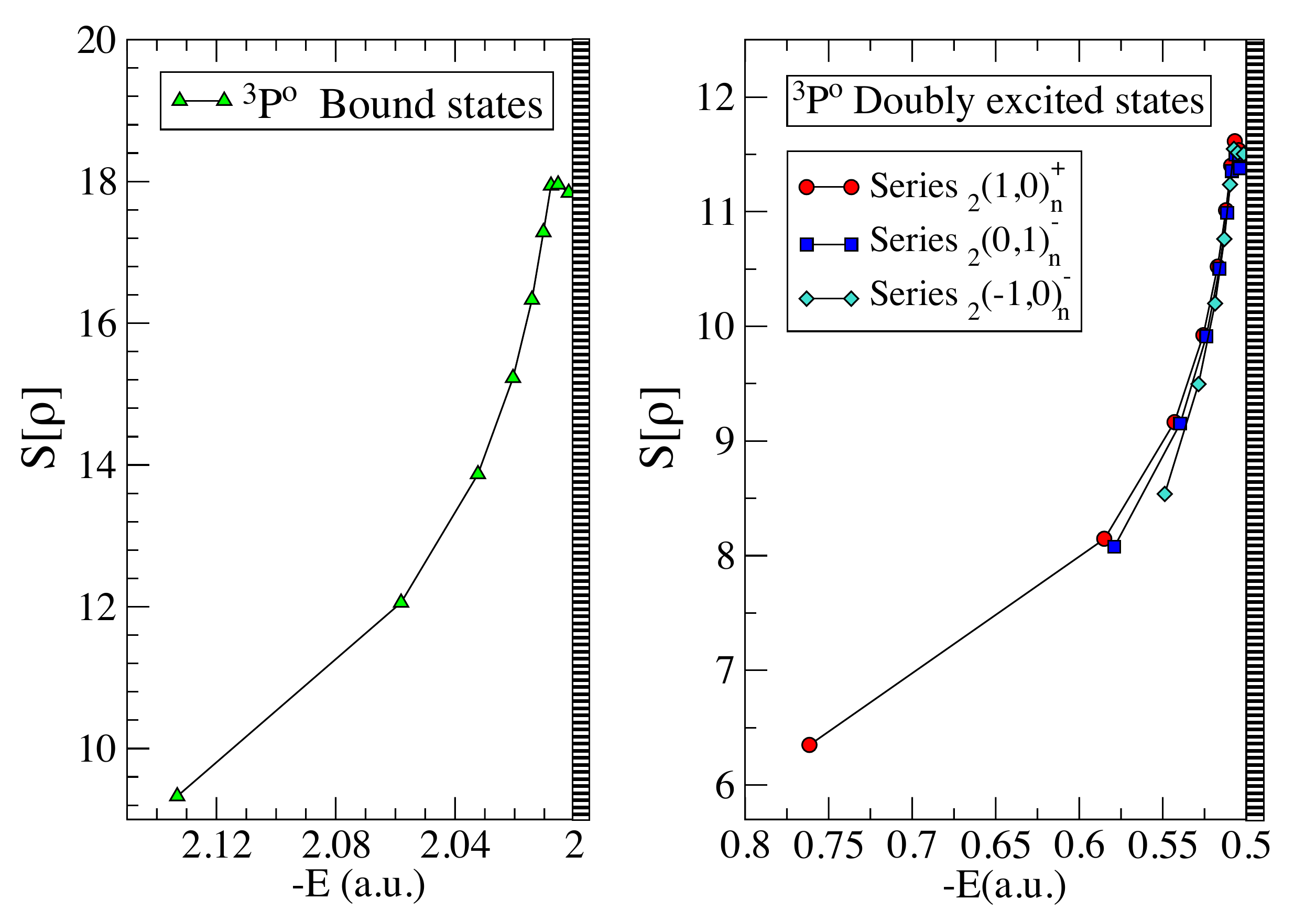}
\caption{\label{fig:shannon3Po}Shannon entropy $S(\rho)$ for the singly excited $^3P^o$ states in helium (left panel) and doubly excited states belonging to the three series $_2(1,0)^+_n$, $_2(0,1)^-_n$ and $_2(-1,0)^-_n$ within the $^3P^o$ symmetry of helium (right panel).}
\end{figure}

\begin{figure}
\centering
\includegraphics[width=0.85\textwidth]{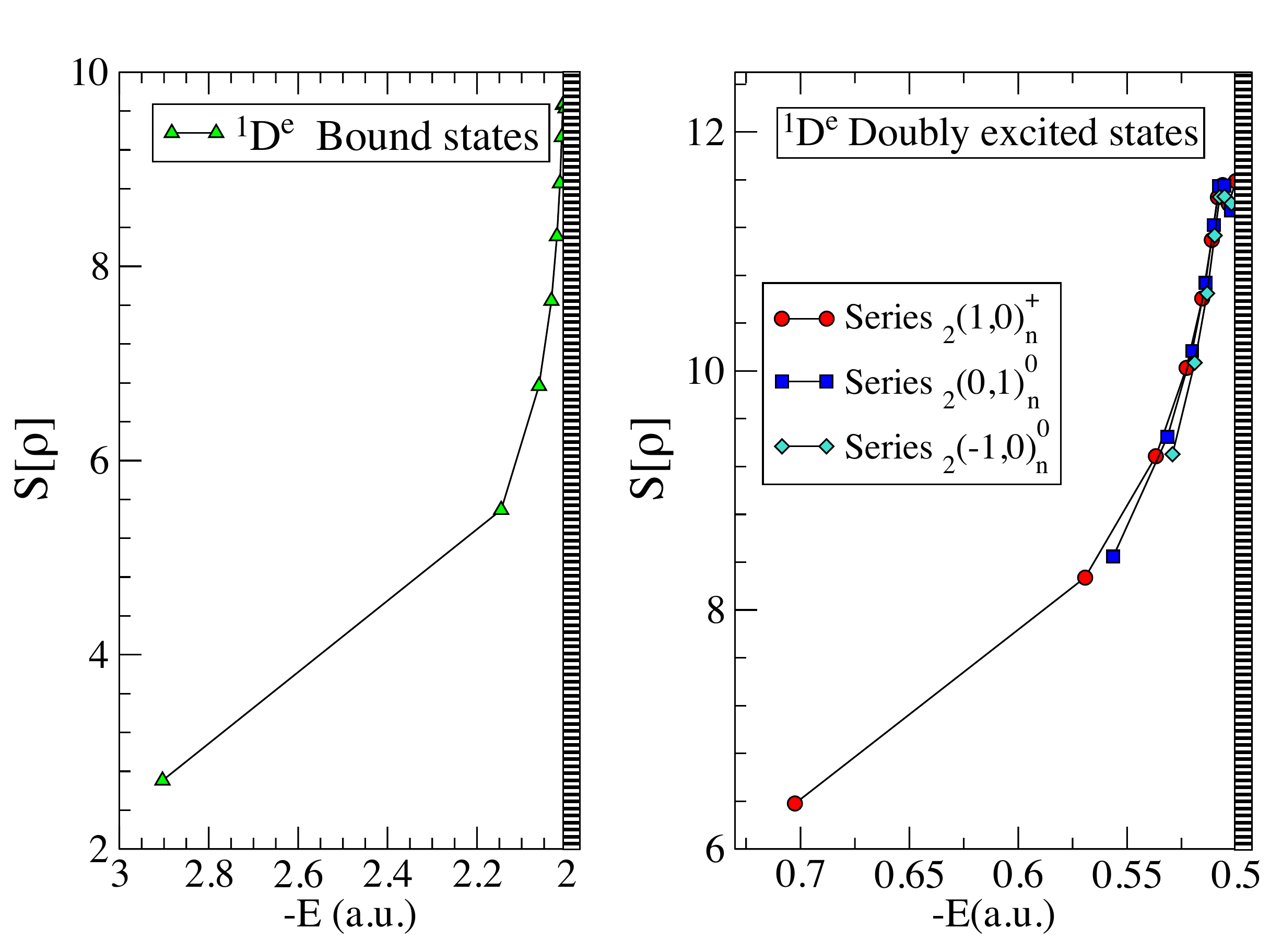}
\caption{\label{fig:shannon1De}Shannon entropy $S(\rho)$ for the singly excited $^1D^e$ states in helium (left panel) and doubly excited states belonging to the three series $_2(1,0)^+_n$, $_2(0,1)^0_n$ and $_2(-1,0)^0_n$ within the $^1D^e$ symmetry of helium (right panel).}
\end{figure}

\subsubsection{Shannon entropy of the $^{1,3}S^e$, $^{1,3}P^o$, and $^{1,3}D^e$ doubly excited states of helium atom}
 
In this section we analyse the Shannon entropy calculated via the equation~\eqref{eq:shannonentropy} using the one-particle radial density 
$\rho({r})$ for the singlet and triplet resonant states of helium atom. In doing so, we show this quantity in figure~\ref{fig:shannon1Se} for the symmetry $^{1}S^e$ where we depict the results for the ground state and the singly excited states on the left panel. The behavior of this quantity for these states is very close to the obtained for the bound states of the hydrogen atom as is evidenced by a comparison with the left panel of figure~\ref{fig:shannfisherhyd}. The Shannon entropy increases monotonically to reach an asymptotic behavior when the energy of the Rydberg series approaches the first ionization threshold. In fact, it seems reasonable to find that the Shannon entropy must diverge to infinity once the ionization threshold is crossed, since the corresponding continuum \ac{WF} becomes fully delocalized in the configurational space. In the same way, as is shown in the  figure, there is  a strong localization of the density of the ground state close to the nucleus. A previous result for the Shannon entropy in the ground state of He is reported by~\citep{Sen2005}. 
 
On the other hand, in the right panel of the figure~\ref{fig:shannfisherhyd} we show the Shannon entropy values for the two $(K,T)$ series of $^1S^e$ \ac{DES}  in helium. Since these states also form Rydberg series in the continuum above the first ionization threshold but below the second one, and we only analyze the contribution of the localized $\mathcal{Q}$ part of the resonance to the Shannon entropy, the behavior of the entropy for \ac{DES} resembles that of the Rydberg series of the bound states. It is important to say that even though we have plotted the two $(K,T)$ series belonging to this symmetry using different  colours for each one (i.e., $_2(1,0)^+_n$ in red and $_2(-1,0)^+_n$ in blue), the Shannon entropy is hardly capable to distinguish them, i.e., it is impossible to say, based on the values of Shannon entropy and without a previous knowledge of the classification, which state belongs to any specific series. By the way, although the Shannon entropy seems to converge to a constant value as the energy approaches the second ionization threshold, this is an apparent behavior due to the finite box approximation in our computations, i.e., all our \ac{WF} are set to zero at the box boundary $r$=$L$, and this edge condition also affects the inner part of the density. This fact produces an inaccurate description of the one-particle electronic density for the highly lying resonances within each Rydberg series. Some figures concerning the differential properties of the Shannon entropy are relegated to Appendix~\ref{ch:supgraphics}. For instance, other figures include the electronic radial density, its logarithm and the radially differential Shannon entropy for the two series  $_2(1,0)^+_n$ and $_2(-1,0)^+_n$ in the $^1S^e$ symmetry, respectively. In conclusion, the behavior of the Shannon entropy for the He resonances with increasing energy reflects the spreading of the density to longer radial distances in the configurational space. The loss of the compactness with higher excitation in the electronic density naturally increases the entropy content. However, this Shannon entropy as an integral measure is not able to discriminate (after integration) the differential subtleties associated to the several $(K,T)$ series within the same total $^{2S+1}L^\pi$ spectroscopical symmetry. It is also evident that the density of highly excited resonances are truncated at $r=L$. However, there are many ways to deal with the described trouble like, inter alia, extending the box size and increasing the number of basis in the \ac{CI} approach, but this is not the scope of the present work. We leave this convergence analysis for another future work which must include the analysis of the  isoelectronic series of the helium atom where. There are no reported values of Shannon entropy for \ac{DES} of helium atom in the literature and this work is a first approach to the subject.

\begin{figure}[h!]
\centering
\includegraphics[width=0.85\textwidth]{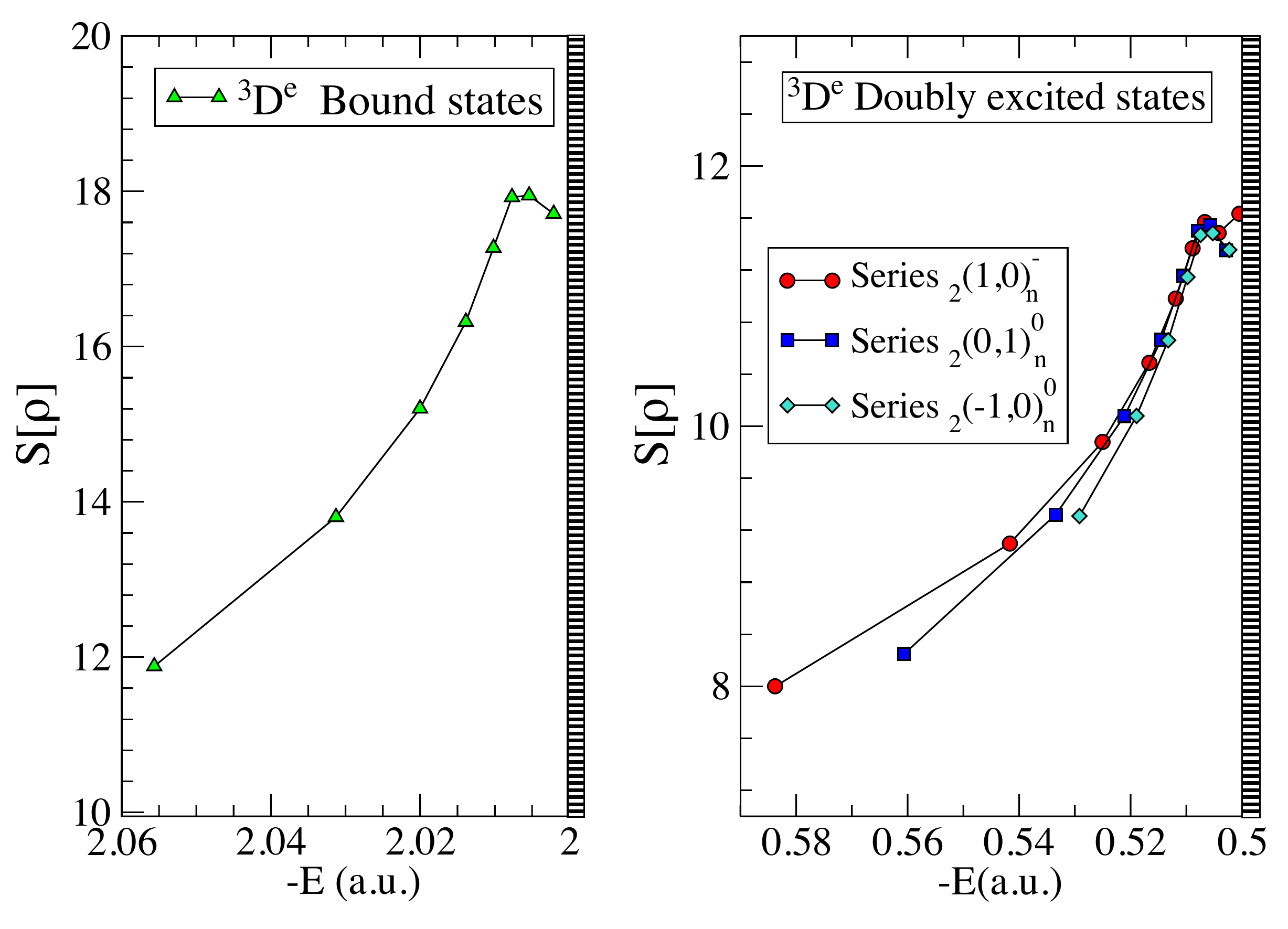}
\caption{\label{fig:shannon3De}Shannon entropy $S(\rho)$ for the singly excited $^3D^e$ states in helium (left panel) and doubly excited states belonging to the three series $_2(1,0)^-_n$, $_2(0,1)^0_n$ and $_2(-1,0)^0_n$ within the $^3D^e$ symmetry of helium (right panel).}
\end{figure}

In a similar way we have calculated the Shannon entropy of the ground state, the singly excited states and the \ac{DES} of the symmetries $^{3}S^e$ plotted in figure~\ref{fig:shannon3Se},  $^{1,3}P^o$ plotted in figures~\ref{fig:shannon1Po} and ~\ref{fig:shannon3Po}, and $^{1,3}D^e$ depicted in figures~\ref{fig:shannon1De} and~\ref{fig:shannon3De}. From these figures it is possible to conclude that the behavior of the Shannon entropy calculated for the states of these symmetries has qualitatively the same characteristics discussed for the symmetry $^{1}S^e$. Both bound states and \ac{DES} increase their Shannon entropy content in a similar trend towards their corresponding upper ionization threshold. Notice that in the resonant case we are dealing only with the bound-like part of the \ac{DES} according to the Feshbach partitioning. Therefore, the Shannon entropy is no more that a witness for the compactness or diffuseness in the inner part of the total resonance \ac{WF}. This physically means that the outer indistinguishable electron becomes more and more delocalized according to its state of excitation. This fact can be evidenced in the figure  where is depicted the one-particle density and the Shannon entropy argument for the bound states of the symmetry $^{1}P^o$, and in the figures where are depicted the same functions of the series  $_2(0,1)^+_n$, $_2(1,0)^-_n$, and $_2(-1,0)^0_n$, respectively, belonging to the symmetry $^{1}P^o$. Finally, it is possible to conclude that Shannon entropy does not provide additional information about \ac{DES} of helium in any particular symmetry. Consequently, this specific measure is unable to extract crucial information about the topological features of the density relevant to the problem of classification of resonances.

\subsubsection{Fisher information of the $^{1,3}S^e$, $^{1,3}P^o$, and $^{1,3}D^e$ doubly excited states of helium atom}\label{sec:fisherinf}

\begin{figure}[h!]
\centering
\includegraphics[width=0.85\textwidth]{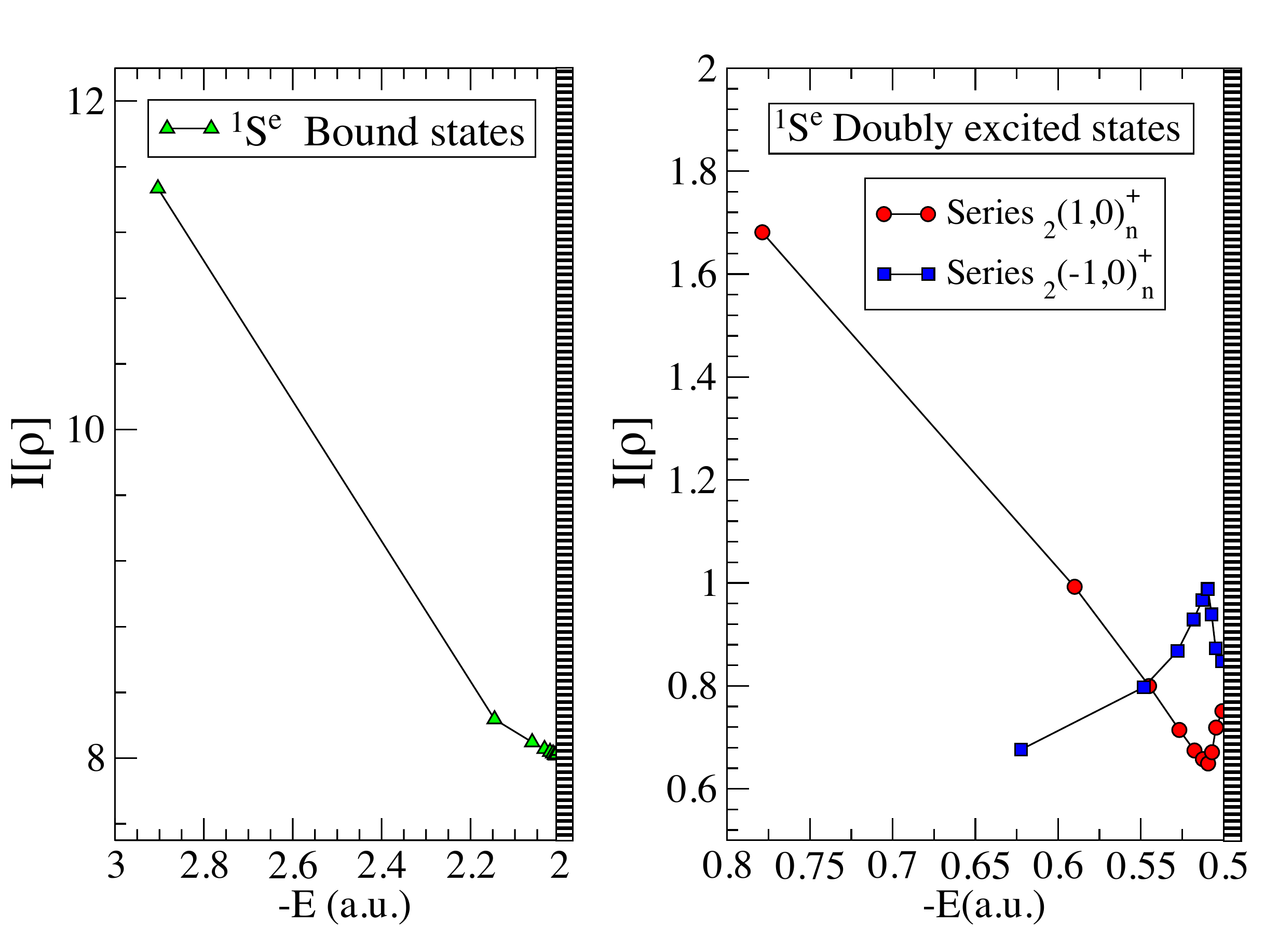}
\caption{\label{fig:fisher1Se}Fisher Information $I(\rho)$ for the ground and singly excited states (left panel) and for the doubly excited states for the two series  $_2(1,0)^+_n$ and $_2(-1,0)^+_n$ belonging to the total symmetry $^1S^e$.}
\end{figure}

\begin{figure}
\centering
\includegraphics[width=0.85\textwidth]{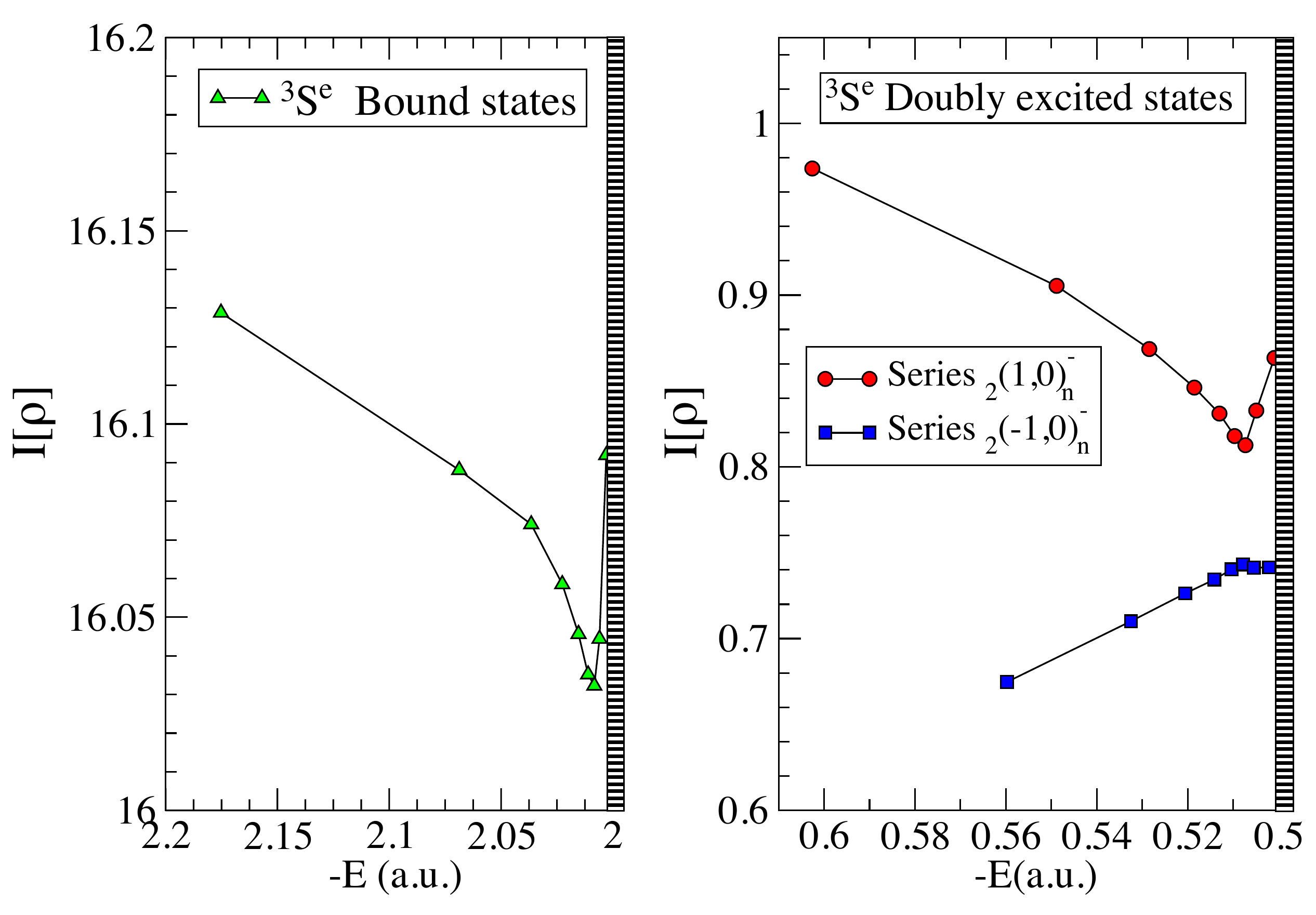}
\caption{\label{fig:fisher3Se}Fisher Information $I(\rho)$ for the singly excited states (left panel) and for the doubly excited states for the two series  $_2(1,0)^-_n$ and $_2(-1,0)^-_n$ belonging to the total symmetry $^3S^e$.}
\end{figure}

The Fisher information, $I[\rho]$, is another important information quantity~\citep{Fisher1972, Cover2006}. It is a measure of the gradient  content of a  distribution function and, for this reason, is a local measure which examines more profoundly  changes in the electronic distribution. For our present purposes, i.e., the analysis and characterization of the topological properties of \ac{DES} of helium we have calculated this quantity using the expression~\eqref{eq:fisherinformation} in terms of their one-particle densities. In addition, we have also calculated the values of this information measure of singly excited states of helium with for the sake of comparison. Let us start with the Fisher information for the bound states with total symmetry $^1S^e$, in the left panel of the figure~\ref{fig:fisher1Se} we show the Fischer information value for the lowest bound $^1S^e$ states against their energy. The Fisher information decreases its value monotonically for increasing excitation energy, similar to the hydrogen bound states in figure~\ref{fig:shannfisherhyd}. The same decreasing behavior is observed for the bound  $^1S^e$ states in the figure~\ref{fig:fisher3Se}. Surprisingly, the limit value for the Fischer information at ionization threshold is zero (see figure~\ref{fig:shannfisherhyd}), but in the case of the He atom the limit value is around $8$. Then, at variance with the Shannon entropy, whose value diverges at threshold, the Fischer information or gradient content seems to provide a discriminating limiting value at threshold. The  show the Fisher information integral argument and from their analysis we can conclude that the Fischer information of the highly excited bound states does not differ significantly, to yield a limiting saturation value.

Similarly, the values of the Fischer information for the $^1S^e$ resonances in helium are included in the right panel of figure~\ref{fig:fisher1Se}. In contrast, this measure as applied to He resonances does have neither a decreasing nor monotonic behavior against the increasing energy. Actually, the Fischer information seems to split into two different paths (one, that corresponds to the series $_2(1,0)^+_n$, decreases their gradient content and the other, associated to the series $_2(-1,0)^+_n$ augments it). The two resonance series seem to tend to a different limiting value at the second ionization threshold, but ultimately they collapse to the same final value around $0.8$.

\begin{figure}
\centering
\includegraphics[width=0.85\textwidth]{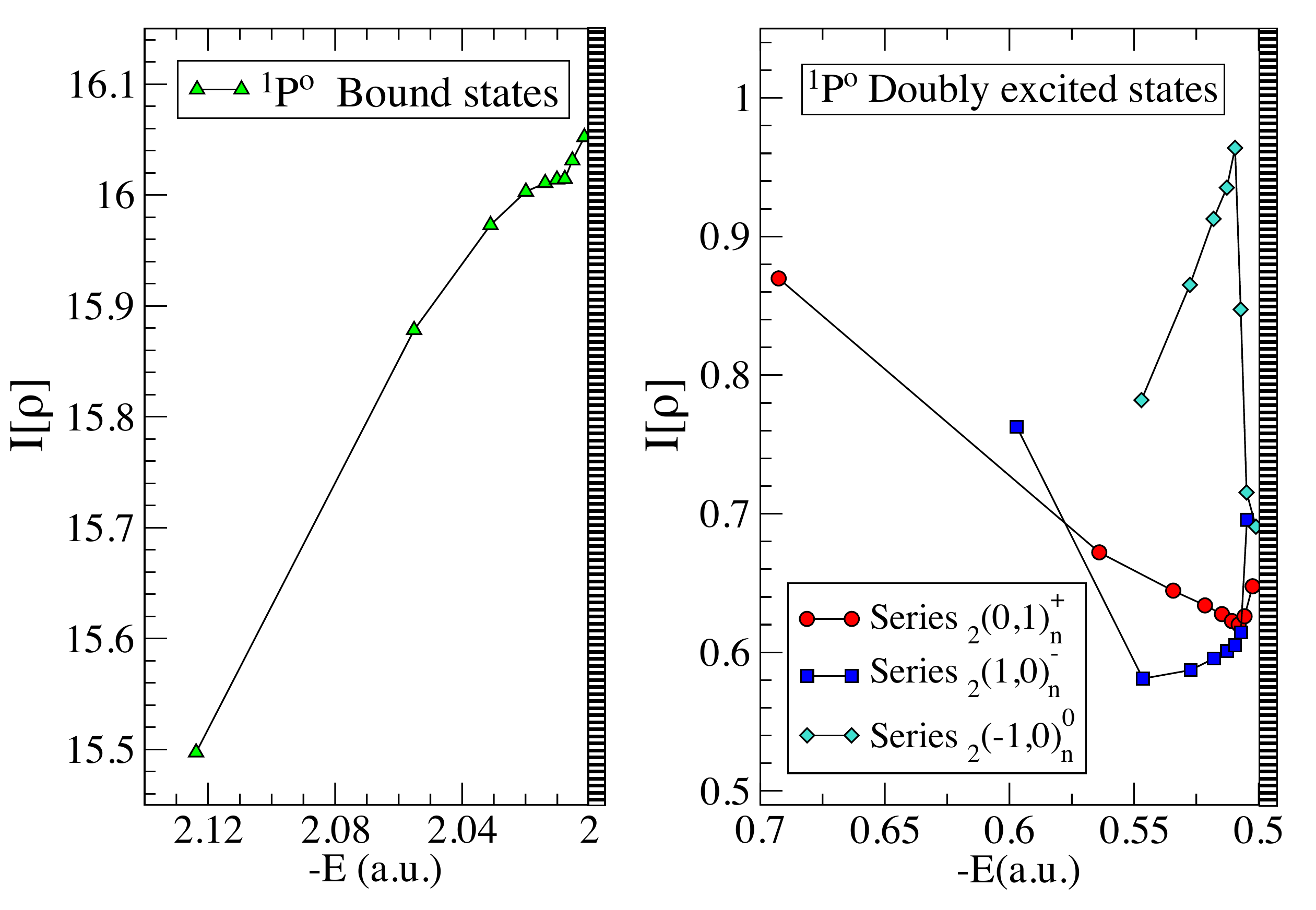}
\caption{\label{fig:fisher1Po}Fisher Information $I(\rho)$ for the singly excited states (left panel) and for the doubly excited states for the two series  $_2(0,1)^+_n$, $_2(1,0)^-_n$ and $_2(-1,0)^0_n$ belonging to the total symmetry $^1P^o$.}
\end{figure}

\begin{figure}
\centering
\includegraphics[width=0.85\textwidth]{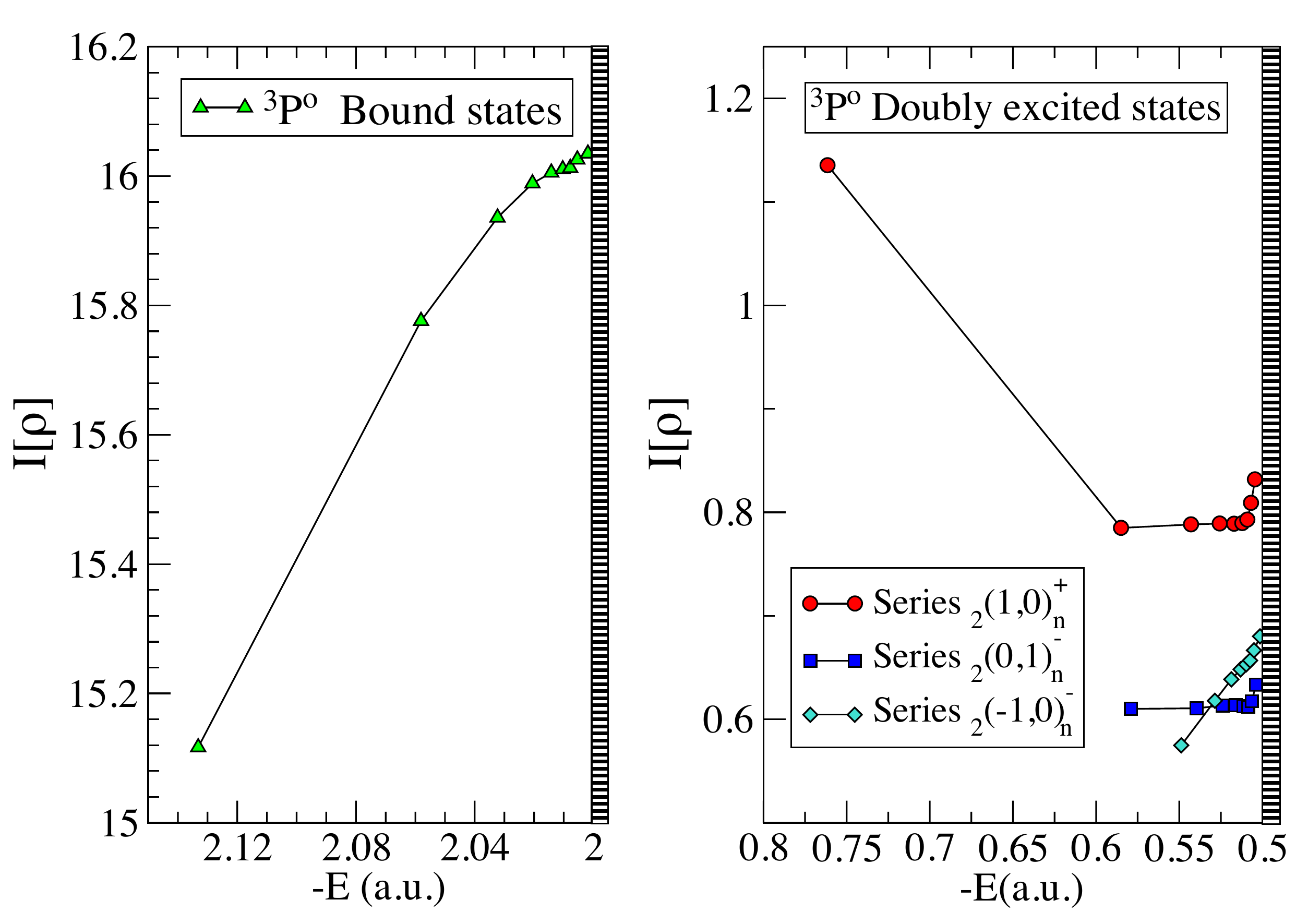}
\caption{\label{fig:fisher3Po}Fisher Information $I(\rho)$ for the singly excited states (left panel) and for the doubly excited states for the two series  $_2(1,0)^+_n$, $_2(0,1)^-_n$ and $_2(-1,0)^-_n$ belonging to the total symmetry $^3P^o$.}
\end{figure}

\begin{figure}

\centering
\includegraphics[width=0.85\textwidth]{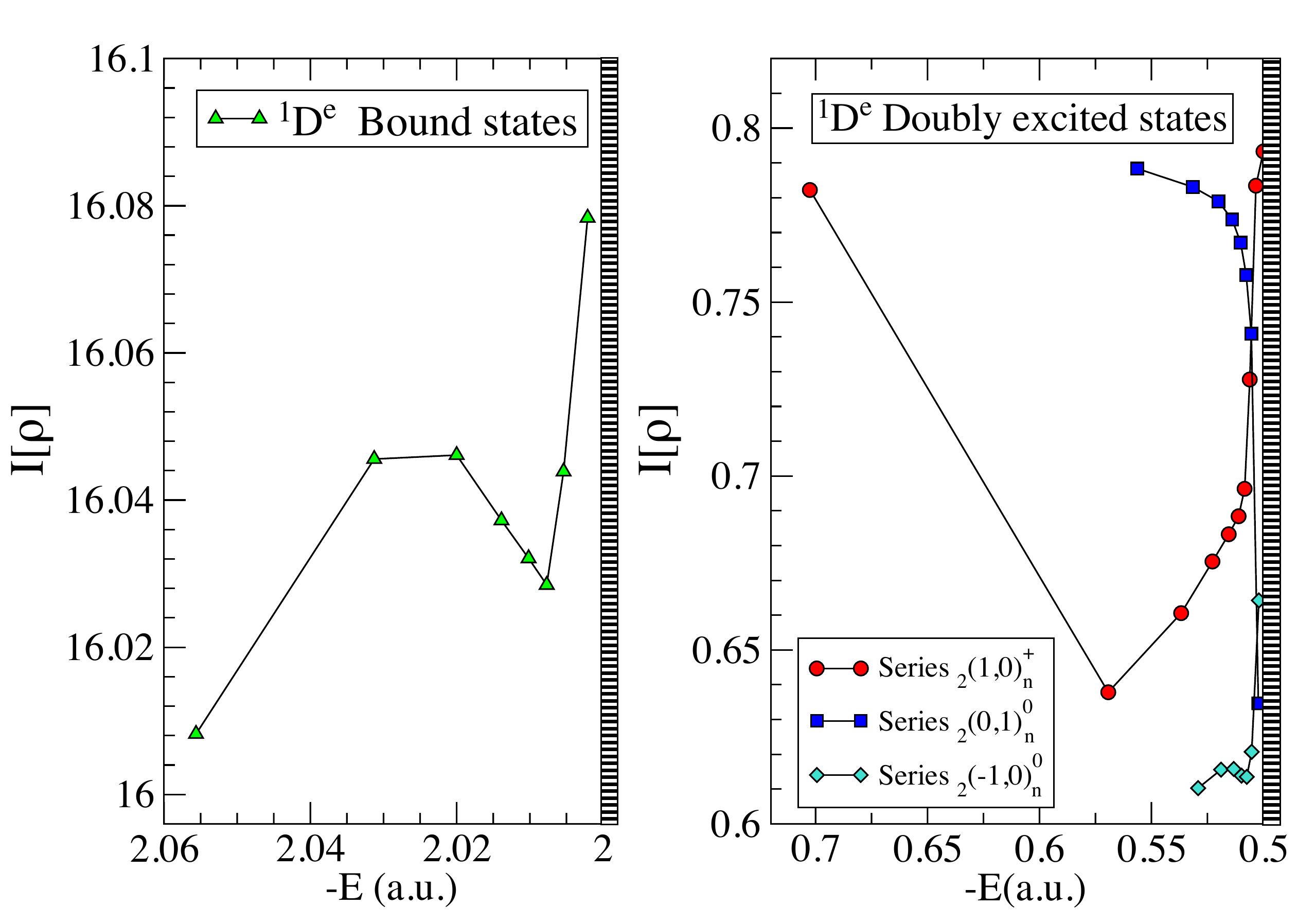}
\caption{\label{fig:fisher1De}Fisher Information $I(\rho)$ for the singly excited states (left panel) and for the doubly excited states for the two series  $_2(1,0)^+_n$, $_2(0,1)^0_n$ and $_2(-1,0)^0_n$ belonging to the total symmetry $^1D^e$.}
\end{figure}

\begin{figure}
\centering
\includegraphics[width=0.85\textwidth]{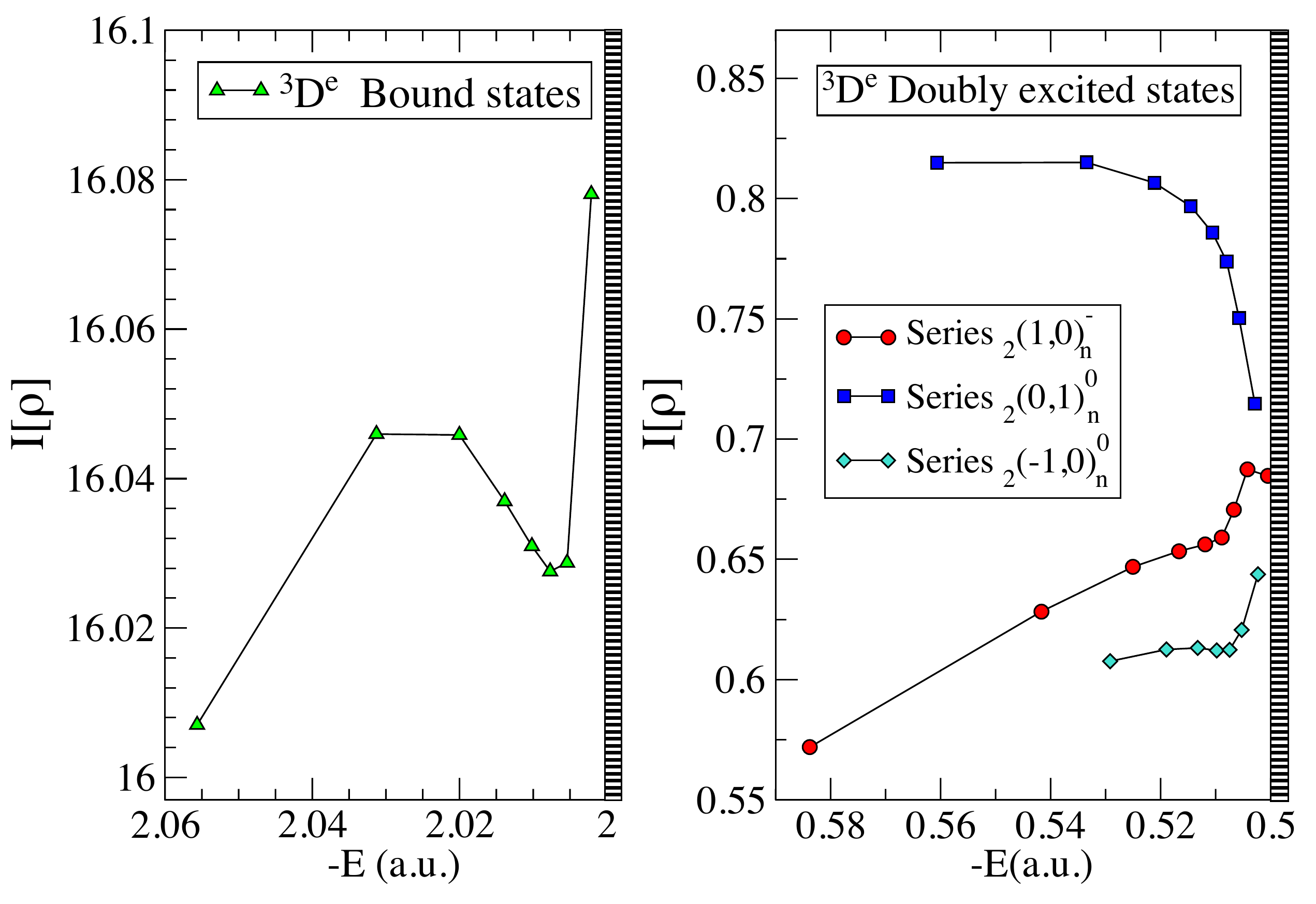}
\caption{\label{fig:fisher3De}Fisher Information $I(\rho)$ for the singly excited states (left panel) and for the doubly excited states for the two series  $_2(1,0)^-_n$, $_2(0,1)^0_n$ and $_2(-1,0)^0_n$ belonging to the total symmetry $^3D^e$.}
\end{figure}

The  figure~\ref{fig:fisher1Po}  corresponds to the Fisher information calculated for the bound states and resonances of helium belonging to the symmetry $^1P^o$. At variance with the $^1S^e$ symmetry in figure~\ref{fig:fisher1Se}, the Fischer information increases with the excitation energy to reach a limiting value $\sim 16$ at the ionization threshold. This behavior will be common to the bound states of the other symmetries of helium $^3S^e$, $^3P^o$ and $^{1,3}D^e$ as can be observed in the left part of figures~\ref{fig:fisher3Se},~\ref{fig:fisher3Po},~\ref{fig:fisher1De}, and~\ref{fig:fisher3De}, respectively. The panel on the right in figure~\ref{fig:fisher1Po} shows the Fisher information values for the three $(K,T)$ series within the symmetry $^1P^o$. It is clearly noticeable that the three $(K,T)$ series follow paths with completely different behavior, to finally collapse to the same point at the second ionization threshold. Although the assignment of individual resonances was associated to a given path from our previous knowledge of the existence of $(K,T)$  series, clearly the Fisher information is able to distinguish local topological properties in the one-particle radial density among the different series. This result is important because it puts in evidence the existence of different resonant series by simply analyzing the electronic density with a selected tool like Fisher information. 

Furthermore, if we analyse other symmetries we find the same behavior. In the right panel of the figures~\ref{fig:fisher3Se},~\ref{fig:fisher3Po},~\ref{fig:fisher1De},~\ref{fig:fisher3De} we have plotted the Fisher information of \ac{DES} for the symmetries $^3S^e$, $^3P^o$, $^1D^e$, and $^3D^e$, respectively. The splitting is particularly evident and strong in the symmetry  $^3D^e$ where the resonances belonging to each of the three series $_2(1,0)^-_n$, $_2(0,1)^0_n$, and $_2(-1,0)^0_n$ can be identified without ambiguity within a given set of points (see figure~\ref{fig:fisher3De}). We may conclude that Fisher information provides a deeper insight into the quantum correlations (interelectronic correlations) that characterizes the behavior of the helium atom prepared in an autoionizing state. This measure can discriminate in the one-particle density the two-electron correlations which are the roots of the quantum properties of \ac{DES} of helium.  On the other hand, in the figures 
we have depicted: a) The gradient part of the Fisher integral argument $r^2|\partial_r\rho({r})|^2$ in two different intervals, b)the one-particle electronic density $4\pi\rho({r})$, c)the derivative of the density $4\pi\partial_r\rho({r})$, and d) the complete Fisher information argument or the differential Fisher information. From the parts a) and d) of these figures we may conclude that the contribution to the total Fisher information divides itself into two regions just at $r=1$ \ac{a.u.}. The main contribution to this quantity seems to come from the first of these regions, i.e., $0\le r\le1$. The parts b) and c) of all these figures show a very peculiar feature of the one-electron density of the \ac{DES}. This density function has a critical point, i.e. a local minimum or maximum, or an inflexion point, just at $r=1$ \ac{a.u.}. Consequently, the derivative of the density, which is shown in c), vanishes at this critical point. This behavior is common for all one-particle electronic density of \ac{DES} regardless of the symmetry. Actually, we do not know yet the reason of this bizarre property  in the density.. Moreover, it seems that there is an underlying universality in the topological structure of the resonances which emerges with the electronic density,  yet to be uncovered.


\chapter{Entanglement on doubly excited states of Helium atom}\label{ch:entanglement} 
Entanglement is understood as a fundamental physical characteristic of the quantum compound systems that inexorably implies its non-separability into the constituents parts. A physical system $S$, composed of two subsystems $S_1$ and $S_2$ and described by a state operator $\rho_S$ ($S$ is called bipartite quantum system), is defined as entangled with respect to $S_1$ and $S_2$ if we can not write the state operator as a convex sum

\begin{equation}{\label{eq:densityopsep}}
\rho_S=\sum_lw_l\hspace{3pt}\rho_{S_1}^l\otimes\rho_{S_2}^l,
\end{equation}
where the weights  $w_l$ satisfy the conditions $w_l\ge0$ and $\sum_lw_l=1$. The singlet state of a pair of two-level system (e.g., a system of two particles with spin $s=1/2$) is  the one of the simplest examples of an entangled state

\begin{equation}{\label{eq:densityopsep1}}
\ket{\Psi}=\frac{1}{\sqrt{2}}\left(\ket{\alpha}_1\otimes\ket{\beta}_2-\ket{\beta}_1\otimes\ket{\alpha}_2\right).
\end{equation}

If the compound (bipartite) system was prepared in an entangled state it cannot be expressed as a factorized state in terms of the individual states of each subsystem, i.e., $\ket{\Psi}\neq\ket{\psi_1}\otimes\ket{\psi_2}$. Consequently, it is not possible to assign a vector state to each subsystem individually. 

The concept of  {\textit entanglement} ({\textit entrelazamiento} in spanish or {\textit verschr\"ankung} in german) was coined by E. Schr\"odinger a sequel of the  \ac{EPR} paper. Schr\"odinger pointed out the striking implications of the entanglement concept~\citep{Schrodinger1935,Schrodinger1936,Wheeler1984}:\\
\\
{\textit When two systems, of which we know the states by their respective representatives, enter into temporary physical interaction due to known forces between them, and when after a time of mutual influence the systems separate again, then they can no longer be described in the same way as before, viz. by endowing each of them with a representative of its own. I would not call that one but rather the characteristic trait of quantum mechanics, the one that enforces its entire departure from classical lines of thought.}

\section{\label{sec:entanglement}Quantum entanglement of indistinguishable particles}

The emergence of new ideas from the classical and quantum information theories provides an alternative perspective about the study of the electronic structure in atoms  and, particularly, the measurement of the degree of quantum entanglement for both fermionic and bosonic indistinguishable particles. Entanglement is a fundamental feature of the compound quantum systems~\citep{Horodecki2009}. Moreover, it has been an important and useful resource for quantum information processing~\citep{Bengtsson2006,Nielsen2000}. It is yet to clarify what is the role of the entanglement in some phenomena like quantum phase transitions~\citep{Zanardi2006,Zanardi2007} and its relationship with the ionization processes  in atoms and molecules, understood as phase transitions. By the way, the fundamental question is how to establish whether a general multipartite quantum state is entangled. For the case of pure states, the Schmidt decomposition is a successful and widely accepted measurement of entanglement~\citep{Peres1995}. Unfortunately, most proposed measures of entanglement for general (mixed) states involves high demanding extremizations which are difficult to handle analytically. Only for a few cases  of physical systems we have found analytical measures of entanglement, e.g., for the general case of a pair of binary quantum objects (qubits) there is a formula for the entanglement of formation, called {\textit concurrence}, as a function of the density matrix~\citep{Wootters1998, Horodecki2009}. There are some additional criteria or witnesses for entanglement based, particularly, on the separability of quantum states~\citep{Horodecki2009,Peres1996,Horodecki1996,Huber2010}. In our case of fermionic systems in atoms and molecules we are interested to characterize the quantum correlations, at short distances from the nucleus, of pure states of a two-particle fermionic system taking into account the indistinguishability of the two electrons. A pure state of indistinguishable particles must be written in terms of Slater determinants which introduces the indistinguishability by means of the {\textit symmetrization postulate}, see section~\ref{sec:symmetrizationpos}. However, this procedure does not introduce the necessarily correlation in order to define a fermionic system as entangled. This is common known as {\textit statistical entanglement} which is a non-distillable kind of entanglement, i.e., it is not useful as a resource in quantum information processing and quantum computation. The symmetrization postulate is introduced over factorized configurations of distinguishable electrons leading to a Slater determinant. This is the case of the \ac{HF} method, which is based in the better description of the \ac{WF} with a single Slater determinant. Furthermore, the additional quantum correlations arise 
when the description of the quantum correlations is done with a \ac{WF} that possesses more than one Slater determinant. In general, this scheme was designed \ac{CI}, see section~\ref{sec:cimethod}. Moreover, the correlation energy is defined as the difference between the limit \ac{HF} energy and the limit \ac{CI} energy (or the exact energy if this exist).  Consequently, in the case of fermions an analogous concept to the Schmidt decomposition is introduced, which is called the Slater decomposition and the Slater rank~\citep{Schliemann2001}.

Following ~\citep{Ghirardi2004}, the general criterion to measure the entanglement of a system of two indistinguishable particles can be jointly established by: i) via the possibility of the decomposition of Slater-Schmidt and the determination of the Slater rank, ii) with the analysis of the von-Neumann entropy or the linear entropy of the reduced density operator of one-particle. This procedures enable us to elucidate whether the fermionic correlations of the states are elementary effects of the indistinguishability of the particles or they are direct evidence of the entanglement. The entanglement notion of a compound system of two-fermions is discussed widely  by~\citep{Schliemann2001}.

\subsection{\label{sec:entanglementhe}Entanglement measure of the helium atom}

The quantum formalism describes the total state of a compound system in a Hilbert space as a {\textit tensor} product of the subsystems spaces, i.e.,  $\otimes^{n}_{l=1}H_{l}$ \citep{Osenda2007,Osenda2008}. The {\textit superposition principle} allows us to write the total state of the system as a sum of antisymmetrized products of spin-orbitals, which conforms the \ac{CI} method~\citep{Yanez2009, Dehesa2012a, Dehesa2012b}. With a basis of spin-orbitals of size $N$ it is possible to write the total \ac{WF} in the particular form as required to determine the Slater rank, as follows~\citep{Schliemann2001}.

\begin{equation} \label{expansionslater}
\ket{\Psi (1,2)} = \sum_i a_i \frac{1}{\sqrt{2}} [ |2i-1\rangle_1 \otimes |2i\rangle_2 - |2i\rangle_1 \otimes |2i-1\rangle_2  ]
\end{equation}
where the index $i$ runs over all the spin-orbitals of the one-electronic basis and the coefficients $a_i$ must satisfy the normalization condition $\sum_ia_i=1$. This is the condition of the Theorem $3.2$ of the reference~\citep{Ghirardi2004}. The number of coefficients $a_i\neq0$ which appear in the expansion~\eqref{expansionslater} is named the Slater number or rank of the state  $\ket{\Psi (1,2)}$.  The relationship between the Slater rank and the concept of entanglement can be defined as: an state is entangled if the Slater number follows the condition $N_S>1$. Therefore, the state whose description is based on a single Slater determinant is not entangled, i.e., the unique correlation present in the state is due to the symmetrization postulate. For this reason, a state described in terms of the \ac{CI} method is, by default, a fermionic entangled state. Anyway, the entanglement information for a bipartite system of two electrons can be found in the reduced density operator $\hat{\rho}_1=Tr_2\hat{\rho}$. This means that we must average over all relevant coordinates of subsystem $2$ by taking the partial trace. The reduced density matrix is calculated by using a partial trace over the second electron in the full density matrix

\begin{align}\label{eq:trace}
\hat{\rho}(\mathbf{r}_1,\mathbf{r}_1^\prime)&=Tr_2\hat{\rho}(\mathbf{r}_1,\mathbf{r}_2;\mathbf{r}_1^\prime\mathbf{r}_2^\prime) \\ \nonumber
&=\int d\mathbf{r}_2 \Psi^{CI}(\mathbf{r}_1,\mathbf{r}_2)\Psi^{*CI}(\mathbf{r}_1^\prime,\mathbf{r}_2).
\end{align}
Now, we can use the following two quantities to measure the amount of entanglement between the particles of a two-electron system: the linear entropy which is also a measure of the purity of the reduced system

\begin{align}\label{eq:linearentropy}
S_L&=1-Tr[\hat{\rho}(\mathbf{r}_1,\mathbf{r}_1^\prime)^2].\\ \nonumber
&=1-\int d\mathbf{r}\hat{\rho}(\mathbf{r},\mathbf{r})^2
\end{align}
and the von-Neumann entropy 

\begin{align}\label{eq:vnentropy}
S_{VN}&=-Tr[\hat{\rho}(\mathbf{r}_1,\mathbf{r}_1^\prime)Log_2\hat{\rho}(\mathbf{r}_1,\mathbf{r}_1^\prime)].\\ \nonumber
&=-\int d\mathbf{r}[\hat{\rho}(\mathbf{r},\mathbf{r})Log_2\hat{\rho}(\mathbf{r},\mathbf{r})].
\end{align}

Since the~\ac{CI} method  is based on the use of (spin and angular momentum) symmetry-adapted two-electron configurations (see equation~\eqref{eq:ciwfunc}), the reduced density matrix $\rho(\mathbf{r},\mathbf{r}^\prime)$ can be calculated in a very simple algebraic form in terms of the \ac{CI} expansion coefficients in equation~\eqref{eq:ciwfunc}; then avoiding the very demanding numerical integration of multidimensional integrals of the density matrix.~\citep{Coe2008, Abdullah2009,Dehesa2012a,Dehesa2012b}. With these considerations and within the \ac{CI} method, the partial trace, the linear entropy and the von-Neumann entropy take the following straightforward form 
\begin{equation}
\hat{\rho}_{ n_{1}l_{1};n_{1}^{\prime}l_{1}^{\prime}}=\sum_{n l}C_{n_{1}l_{1};n l}C^{*}_{n_{1}^{\prime}l_{1}^{\prime};n l}, \label{eq:reduceddden2}
\end{equation}
and
\begin{subequations}
\begin{align}
S_L&= 1-\sum_{nl,n^\prime l^\prime} \hat{\rho}_{ nl;n^{\prime}l^{\prime}} \hat{\rho}_{ n^\prime l^\prime;nl},  \label{eq:linearen2}\\
S_{VN}&= -\sum_i \lambda_i Log_2 \lambda_i, \label{eq:vonnewman2}
\end{align}
\end{subequations}
where  the  $\lambda$s are the eigenvalues of the one-electron reduced density matrix $\hat{\rho}_{ n_{1}l_{1};n_{1}^{\prime}l_{1}^{\prime}}$.

Our goal is to extend these measures of entanglement to the general analysis of the resonant states of the helium atom in order to obtain a deeper insight of their electronic correlation structure, as well as to uncover the classification of the resonant Rydberg series below the second ionization threshold under the scrutiny of entanglement witnesses. Some emerging studies have been recently performed on the analysis of quantum entanglement in two-electron systems~\citep{Yanez2010, Manzano2010}, particularly, in toy models of two-electrons which can be solved analytically  (see also~\citep{Amovilli2003,Amovilli2004}), e.g., Moshisky's atom~\citep{Moshinsky1968},  Hooke's atom ~\citep{Taut1993}, or Crandall's atom~\citep{Crandall1984}. In the Moshinky's atom all the interactions between particles are harmonic. In the Hook's atom, the interelectronic interaction is replaced by a Coulomb interaction and in the Crandall's atom the interaction between electrons are changed by a polarization like interaction $1/r^2_{12}$. None of these toy models allows ionization of electrons, i.e, there are no presence of resonant states which appear when all the interactions are Coulomb like. However, the quantum systems subject to these interaction potentials can be solved exactly, and therefore exact \ac{WF} and density operators are readily available. Consequently, the entanglement measures can be calculated exactly. Nevertheless, there are only entanglement values for the ground state and for a few of excited states in these systems. In the reference~\citep{Manzano2010}, additionally to the analysis of  the entanglement for the Hooke's atom and for the Crandall's atom from the linear entropy, there is a preliminarily attempt to obtain an entanglement analysis in two-electron atoms. To this purpose, the authors employ \ac{CI} of Hylleraas coordinates of the (s,t,u) kind which are explicitly correlated. However, their analysis is focused only to the ground state.

Finally, in the following section we present our calculated measures of entanglement of both bound and \ac{DES} of the helium atom via our high-quality  \ac{CI}-\ac{FM} reduced density matrix.

\subsection{Results of the entanglement amount in the eigenspectrum of the helium atom.}

\begin{table}[h!]

\centering
\begin{tabular}{cccccccc}
\hline\hline
& \citeauthor{Dehesa2012a} &&\multicolumn{2}{c}{ \citeauthor{Benenti2013}}&&\multicolumn{2}{c}{Present}\\ 
\cline{2-2}\cline{4-5}\cline{7-8}
& $S_{L}$&&$S_{L}$&$S_{VN}$&&$S_{L}$&$S_{VN}$\\ \hline
 $\ket{1s1s;^1\hspace{-3.5pt}S^e}$ &0.015914 &&  0.01606 & 0.0785 &&  0.011460 & 0.066475 \\
 $\ket{1s2s;^1\hspace{-3.5pt}S^e}$ &0.48866 && 0.48871 & 0.991099 && 0.487222 & 0.988964 \\
 $\ket{1s3s;^1\hspace{-3.5pt}S^e}$&0.49857 && 0.49724 & 0.998513 && 0.497154 & 0.998530  \\
 $\ket{1s4s;^1\hspace{-3.5pt}S^e}$&0.49892 && 0.49892 & 0.999577 && 0.498909 & 0.999631  \\
 $\ket{1s5s;^1\hspace{-3.5pt}S^e}$&0.4993 && 0.499565 & 0.999838 && 0.499468 & 0.999881   \\ \hline
 $\ket{1s2s;^3\hspace{-3.5pt}S^e}$& 0.50038 && 0.500378  & 1.00494 && 0.500375 & 1.004924  \\
 $\ket{1s3s;^3\hspace{-3.5pt}S^e}$& 0.50019 && 0.5000736 & 1.00114 && 0.500073 & 1.001136 \\
 $\ket{1s4s;^3\hspace{-3.5pt}S^e}$& 0.49993 && 0.5000267 &1.000453 && 0.500026 & 1.000450\\
 $\ket{1s5s;^3\hspace{-3.5pt}S^e}$& 0.50012 && 0.5000125 & 1.000091 && 0.500012 & 1.000227 \\
 \hline\hline
\end{tabular}
\caption[Linear Entropy and von-Neumann Entropy for bound states of helium: Symmetries $^{1}S^e$ and $^{3}S^e$. ]{\label{tab:table1079}Linear Entropy and von Neumann Entropy for bound states of helium: Symmetries $^{1}S^e$ and $^{3}S^e$.  \citep{Dehesa2012a,Dehesa2012b,Benenti2013} and the present work.}
\end{table}

\subsubsection{A comparison with previously reported values of the amount of  entanglement for the $^{1,3}S^e$ bound states of helium atom and results of the entanglement in the bound states of $^{1,3}P^o$ and $^{1,3}D^e$ symmetries.  }

\begin{figure}[h]
\centering
\includegraphics[width=0.75\textwidth]{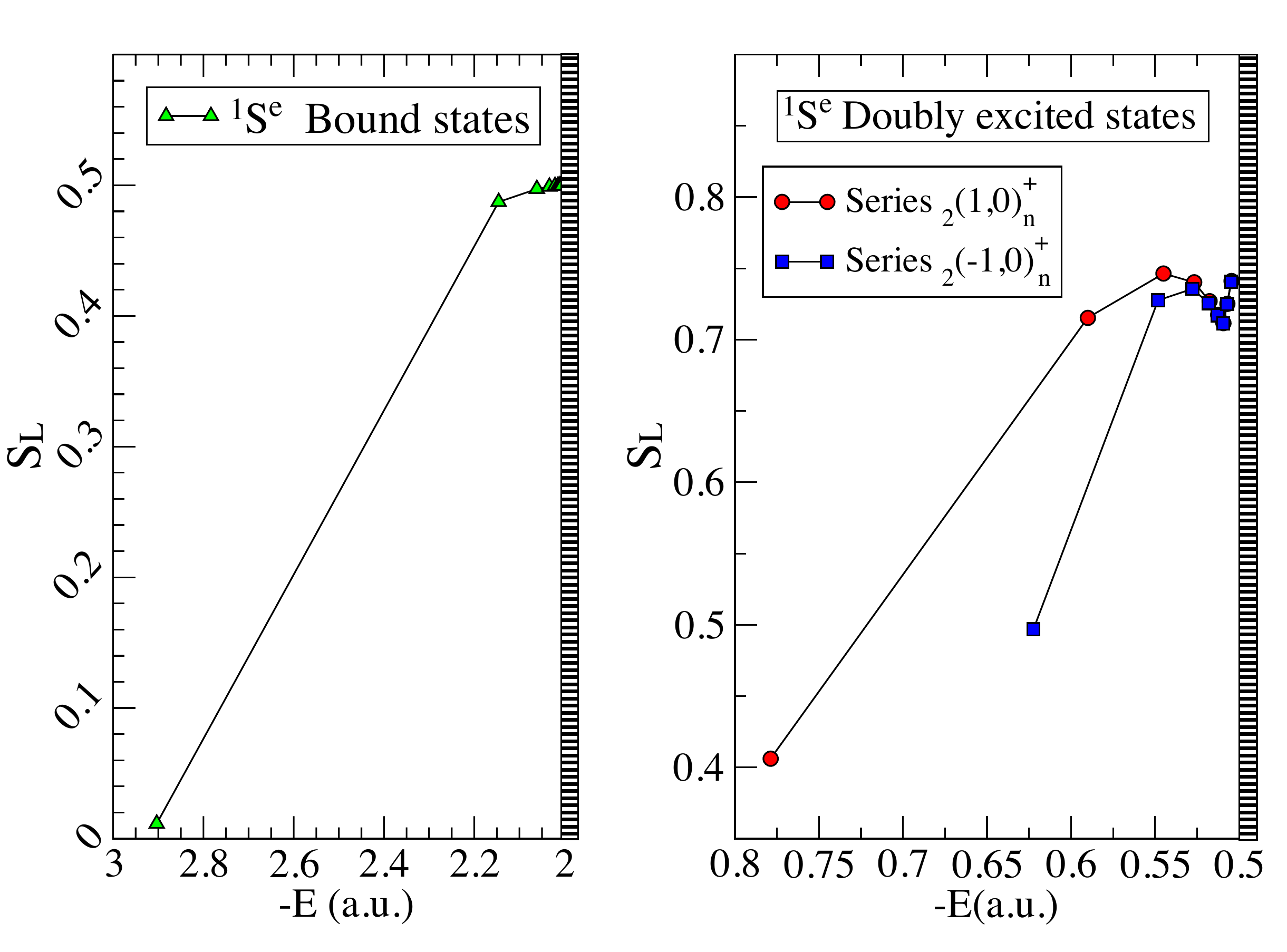}
\caption{\label{fig:entanglement1Se}Linear entropy $S_L(\rho)$ for the ground and the lowest singly excited states for the $^1S^e$ symmetry below the first ionization threshold (left panel) and for the two series  $_2(1,0)^+_n$ and $_2(1,0)^+_n$ of resonances belonging to symmetry $^1S^e$ below the second ionization threshold (right panel).}
\end{figure}

\begin{figure}
\centering
\includegraphics[width=0.70\textwidth]{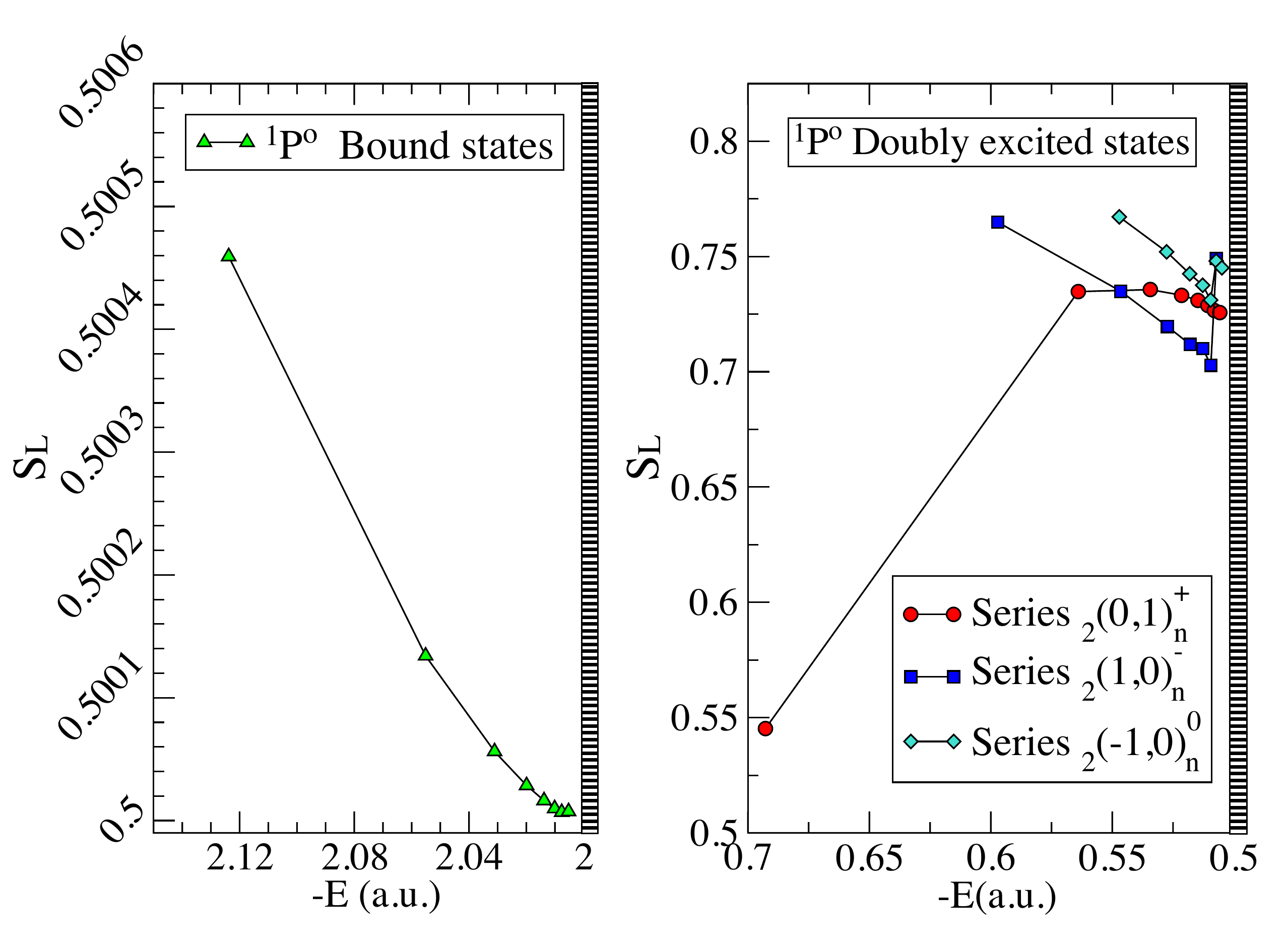}
\caption{\label{fig:entanglement1Po}Linear entropy $S_L (\rho)$ for the lowest singly excited states of $^1P^o$ symmetry below the first ionization threshold (left panel) and for the three series $_2(0,1)_n^+$, $_2(1,0)_n^-$, and $_2(-1,0)_n^0$ of resonant doubly  excited states of symmetry $^1P^o$ below the second ionization threshold (right panel).}
\end{figure}

\begin{figure}
\centering
\includegraphics[width=0.70\textwidth]{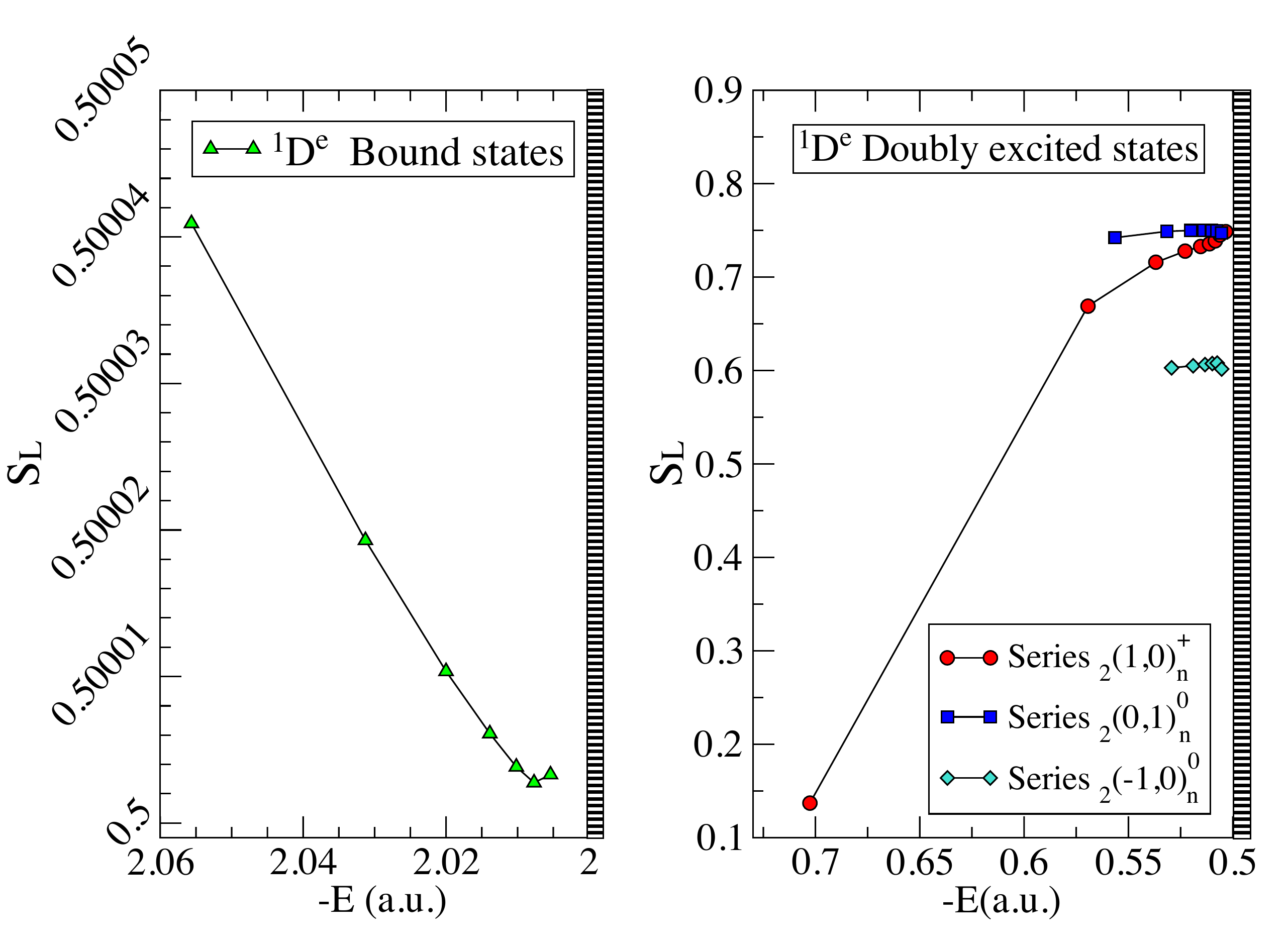}
\caption{\label{fig:entanglement1De}Linear entropy $S_L (\rho)$ for the lowest singly excited states of $^1D^e$ symmetry below the first ionization threshold (left panel) and for the three series $_2(1,0)_n^+$, $_2(0,1)_n^0$, and $_2(-1,0)_n^0$ of resonant doubly  excited states of symmetry $^1D^e$ below the second ionization threshold (right panel).}
\end{figure}

In the first place, we have calculated the linear entropy $S_L$ (equation~\eqref{eq:linearen2}) and the  von-Neumann entropy $S_{VN}$ (equation~\eqref{eq:vonnewman2}) for several bound eigenstates of the helium atom. Our results, along with previously published results by other authors, are included in the table~\ref{tab:table1079}~\citep{Dehesa2012a,Dehesa2012b,Benenti2013}. We present in this table the amount of entanglement for the ground state of helium atom and additionally for some singly excited $^{1,3}S^e$ states. Our results are in a good agreement with the reported ones regardless of the method used to calculate the integrals.  \citeauthor{Dehesa2012a} calculate the integral~\eqref{eq:linearentropy} using a Monte Carlo multidimensional numerical integration of a $12$-dimensional definite integral. They build the electronic density by means of the explicitly correlated Kinoshita-type~\ac{WF}s~\citep{Koga1995,Koga1996}. This high demanding computational method  provides a good description of the density and then of the entanglement amount for bound states of helium atom, However, due to the intrinsic  complexity of this method regardless of the goodness of the Kinoshita-like expansions, we have chosen the alternative \ac{CI}-\ac{FM} to deal with \ac{DES}. On the other hand,~\citeauthor{Benenti2013} also employ a \ac{CI} method, where the radial \ac{WF} is obtained by means of a variational procedure using expansions in terms uncorrelated orthogonal Slater-type orbitals. In this sense, our present method is similar to that of \citeauthor{Benenti2013}, although in our case we use orthogonal hydrogenic orbitals in terms
of B-splines.  The important issue is that by using a \ac{CI} method with orthonormal basis, the expansion variational coefficients straightforwardly allow us to build the density matrix according to the equation~\eqref{eq:reduceddden2}. Consequently, the computation of the linear and von Neumann entropies can be readily performed without explicit complex (Montecarlo) integration procedures. Once again, our results of the entanglement amount are in a very good agreement with the values reported by the other authors. We have found that the entanglement value of bound $^1S^e$ states (see  linear and von Neumann entropies in table~\ref{tab:table1079}) increases with the excitation to reach a saturation value at the first ionization threshold. At variance, for the bound $^3S^e$ states, the entanglement content slightly reduces its value from its maximum corresponding to the lowest triplet state.  Our intuition suggests that the entanglement must be stronger for states of helium atom that keep the electrons localized close to the nucleus. This is in agreement with the result of the $^3S^e$ bound states, but is not with $^1S^e$ singly excited states. The linear entropy in these states seem to increase monotically towards a constant value (the value of maximal mixing of the reduced density matrix). This behavior is shown in the left panel of the figures~\ref{fig:entanglement1Se} and~\ref{fig:entanglement3Se}. Additionally, the increasing behavior of entanglement for the $^1S^e$ bound states seem to be unique; the figures~\ref{fig:entanglement1Po},~\ref{fig:entanglement1De},~\ref{fig:entanglement3Po} and ~\ref{fig:entanglement3De} show that all the remaining bound states of helium have a decreasing behavior of entanglement as a function of the energy of the singly excited states and hence this fact is again in agreement with the intuitive picture of decreasing correlation for high excited states of helium.

\subsubsection{Entanglement amount of the $^{1,3}S^e, ^{1,3}P^o, ^{1,3}D^e$ doubly excited states of helium atom.}

\begin{figure}
\centering
\includegraphics[width=0.75\textwidth]{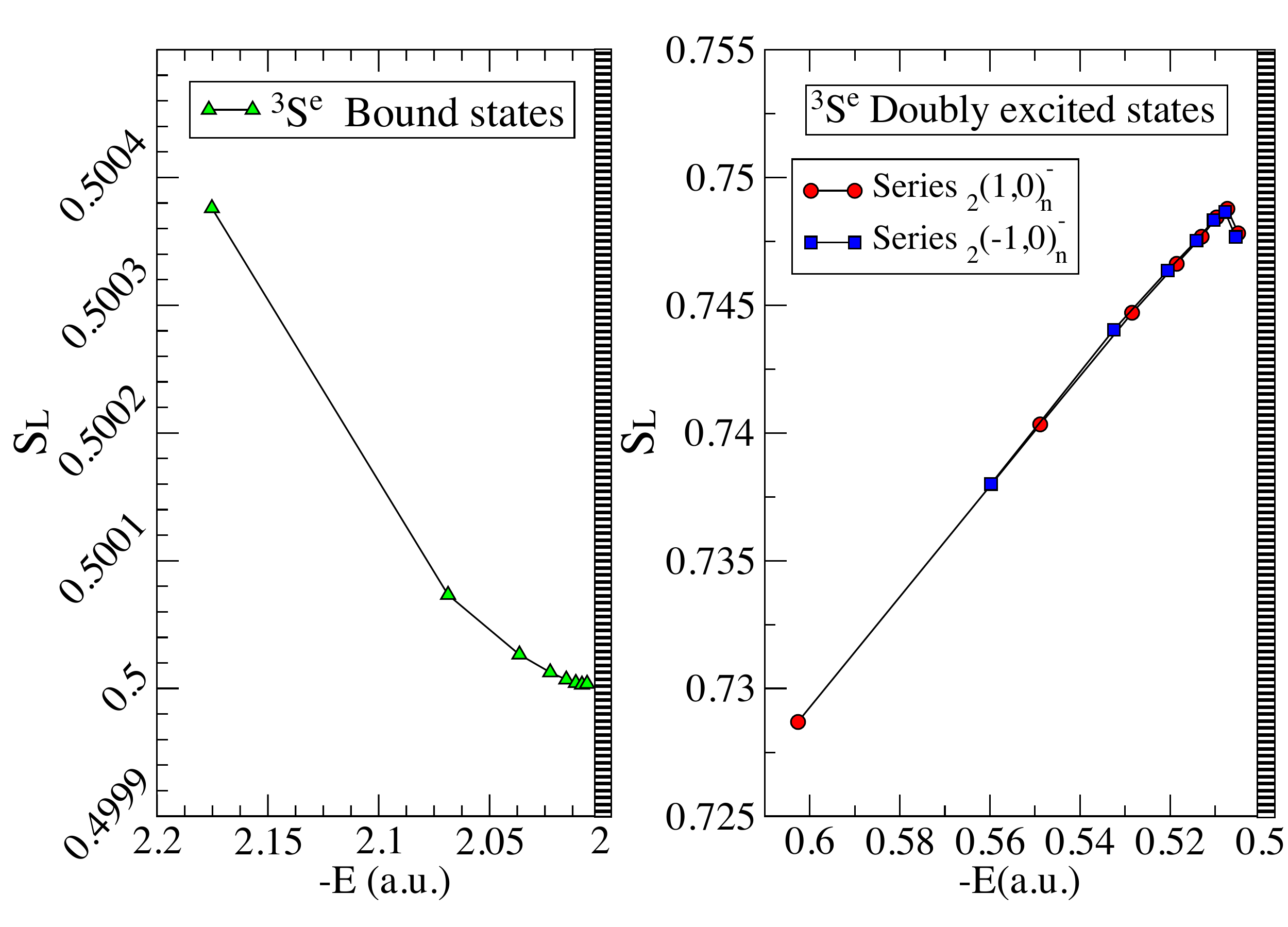}
\caption{\label{fig:entanglement3Se}Linear entropy $S_L (\rho)$ for the lowest singly excited states of $^3S^e$ symmetry below the first ionization threshold (left panel) and for the two series $_2(1,0)^-_n$  and $_2(-1,0)^-_n$ of resonant doubly  excited states of symmetry $^3S^e$ below the second ionization threshold (right panel).}
\end{figure}

\begin{figure}
\centering
\includegraphics[width=0.7\textwidth]{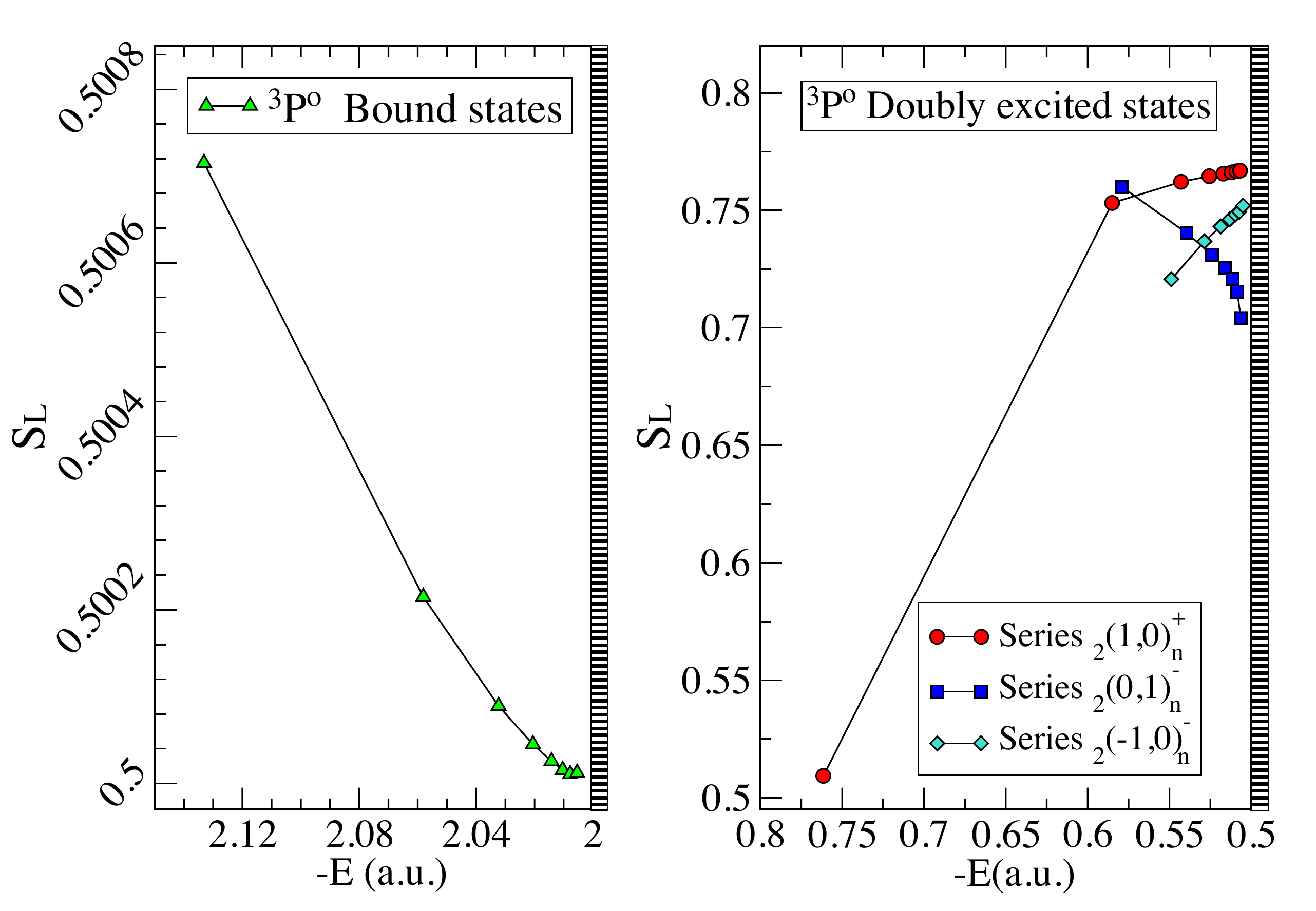}
\caption{\label{fig:entanglement3Po}Linear entropy $S_L (\rho)$ for the lowest singly excited states of $^3P^o$ symmetry below the first ionization threshold (left panel) and for the three series $_2(1,0)^-_n$, $_2(0,1)^0_n$, and $_2(-1,0)^0_n$ of resonant doubly  excited states of symmetry $^3P^o$ below the second ionization threshold (right panel).}
\end{figure}

\begin{figure}
\centering
\includegraphics[width=0.7\textwidth]{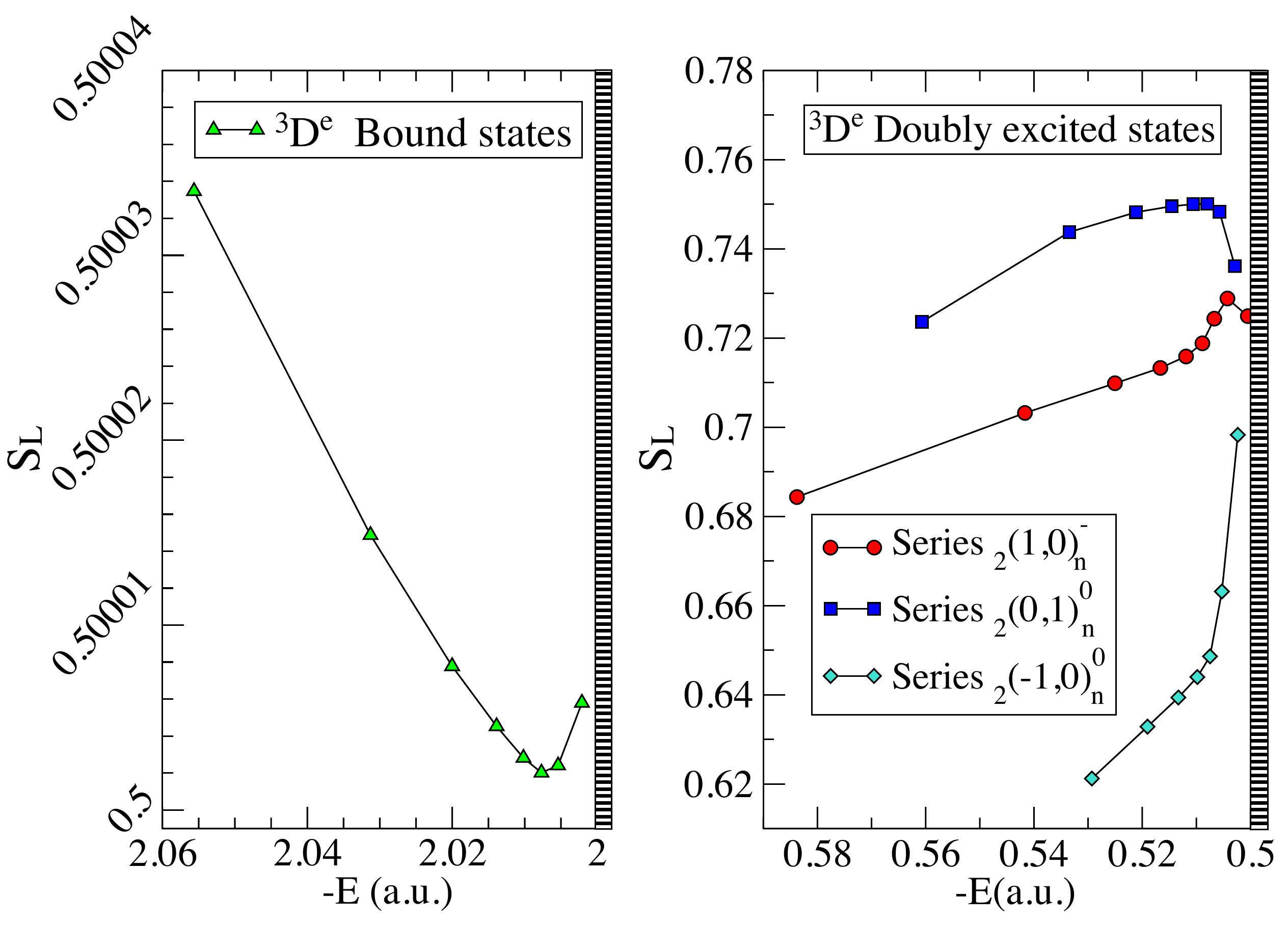}
\caption{\label{fig:entanglement3De}Linear entropy $S_L (\rho)$ for the lowest singly excited states of $^3D^e$ symmetry below the first ionization threshold (left panel) and for the three series $_2(1,0)^+_n$, $_2(0,1)^-_n$, and $_2(-1,0)^-_n$ of resonant doubly  excited states of symmetry $^3D^e$ below the second ionization threshold (right panel).}
\end{figure}

We assume that most of the entanglement content in the resonance \ac{WF} is given by the localized part of the state according to the Feshbach partitioning. This Q-space part is computed with a \ac{CI} approach, that provides a description in terms of a large combination of configurations based upon antisymmetrized products of orthogonal orbitals. This structure of the \ac{WF} corresponds to an entangled state. We provide the numerical values of the amount of entanglement for the He resonances, for all spin singlets and triplets with $L=0,1$ and $2$, below the second ionization threshold. As a entanglement measure we firstly use the linear entropy. Absolute values for the linear entropy in atomic systems are difficult to be interpreted so far, and we are more interested in its relative behavior as a function of the excitation energy in the resonance Rydberg series. In the figures~\ref{fig:entanglement1Se},~\ref{fig:entanglement1Po},~\ref{fig:entanglement1De},~\ref{fig:entanglement3Se},~\ref{fig:entanglement3Po} and~\ref{fig:entanglement3De} we have plotted the linear entropy as a measure of the amount of entanglement of \ac{DES} of helium atom. The behavior of the entanglement of these resonant states differs strongly from the singly excited states. The entanglement neither increases monotonically nor decreases with the energy; instead it has, at first view, a not well defined behavior, almost a random one. If we take no notice of the a priori colour separation in the figures, it may take a long time to pick out any underlying regularity. However, after a more careful scrutiny it is possible to identify two or more independent trends. We have assigned a colour to each of the different paths in the figures to discriminate the independent behavior of each $(K,T)$ series. The fundamental issue is that the linear entropy seems to clearly distinguish the resonance $(K,T)$ series according to the content of entanglement, whose behavior is not monotonic with increasing energy and it is different for each series. This fact becomes as a fundamental tool that enables us to discriminate the resonances into separated set based only in the very strange quantum correlation of non separability of a global quantum state of the helium atom. 

The von Neumann entropies can also be used as a witness of entanglement. we have also calculated the von Neumann entropies for the bound and resonant states in He. We find a behavior for the resonances series very close to the linear entropies and these results are relegated to Appendix~\ref{ch:supgraphics} as a compilation of figures~\ref{fig:entanglementVN1Se},~\ref{fig:entanglementVN1Po},~\ref{fig:entanglementVN1De},~\ref{fig:entanglementVN3Se},~\ref{fig:entanglementVN3Po}, and~\ref{fig:entanglementVN3De}.

As we suggested above in the section~\ref{sec:fisherinf}, the Fisher information also allow us to reach a similar conclusion about the classification of resonances based on a pure topological analysis of the electronic density. The very first question that arises is how a local measure can be compared with essentially a global measure like linear entropy and both provides similar qualitative trends. We can try a very first attempt to respond to this question saying  that all the quantum correlations features of the \ac{DES} were hidden in the off-diagonal elements of the reduced density matrix, i.e, the so called coherences elements of the state operator. We must remember that a measure like Shannon entropy are calculated over the electronic density that represents only the diagonal elements of the state operator.

\part{Summary and Perspectives}

\chapter{Conclusiond and Perspectives}\label{ch:conclusions} 
\begin{enumerate}
\item The Shannon entropy increases monotonically for both bound states and resonances. This quantity is not able to separate the different $(K,T)$ series in the resonant manifold. The global characteristics of the density function cannot be used to classify the doubly excited states.
\item The Fisher information seems to have a trend towards a constant value in each case. This measure as applied to He resonances does have neither a decreasing nor monotonic behavior against the increasing energy. Actually, the Fischer information seems to split into different paths. Because this quantity is sensitive to strong changes on the density function over a small-sized region, these local strong variations allow to identify  each resonance  $(K,T)$ series.
\item Linear entropy measures the amount of entanglement between the two electrons in our system. We can conclude that the linear entropy  clearly may distinguish the resonance $(K,T)$ series according to the content of entanglement, whose behavior is not monotonic with increasing energy and it is different for each series. This fact becomes as a fundamental tool that enables us to discriminate the resonances into separated set based only in the very strange quantum correlation of non separability of a global quantum state of the helium atom.

\item We leave the convergence analysis, due to the finite box approximation, for another future work which must include the analysis of the entanglement in the isoelectronic series of the helium atom.

\item We also leave the analysis of  information theory measures, complexity and quantum entanglement of the resonances belonging to the upper continuum thresholds associated to excited target configurations ( He$^+$ $(n=2,3,4,\dots)+ e^-$) for another future work. 
\end{enumerate}



\appendix
\cleardoublepage
\part{Appendix}
\appendix
\chapter{The basis of B-Splines}

An extended method in quantum mechanics for solving the Schr\"odinger and Dirac equations has been the use of basis sets. Applying the variational method to the  Schr\"odinger equation, Hylleraas obtained a value of $-2.903648$~\ac{a.u.} for the ground state energy of helium using only five trial functions~\citep{Hylleraas1929}. Slater-type orbitals are another kind of basis functions, proposed by~\citep{Slater1930}, used as atomic orbitals in the variational atomic orbital linear combination method. This widely used method enables us to transform the solution of a differential equation into an generalized eigenvalue problem~\citep{Shore1973,Shore1974, Johnson1988, Cheng1994,Bachau2001}. The use of basis sets of many types have been  routinely used in molecular physics and other branches of physics. In atomic physics this method has been systematically used together with another accurate techniques like finite-difference methods.  Nowadays, the great development of high optimized and accurate numerical routines of linear algebra for matrix diagonalization allows for the implementation of several basis solution schemes.

\begin{figure}[h]
\centering
\includegraphics[width=0.85\textwidth]{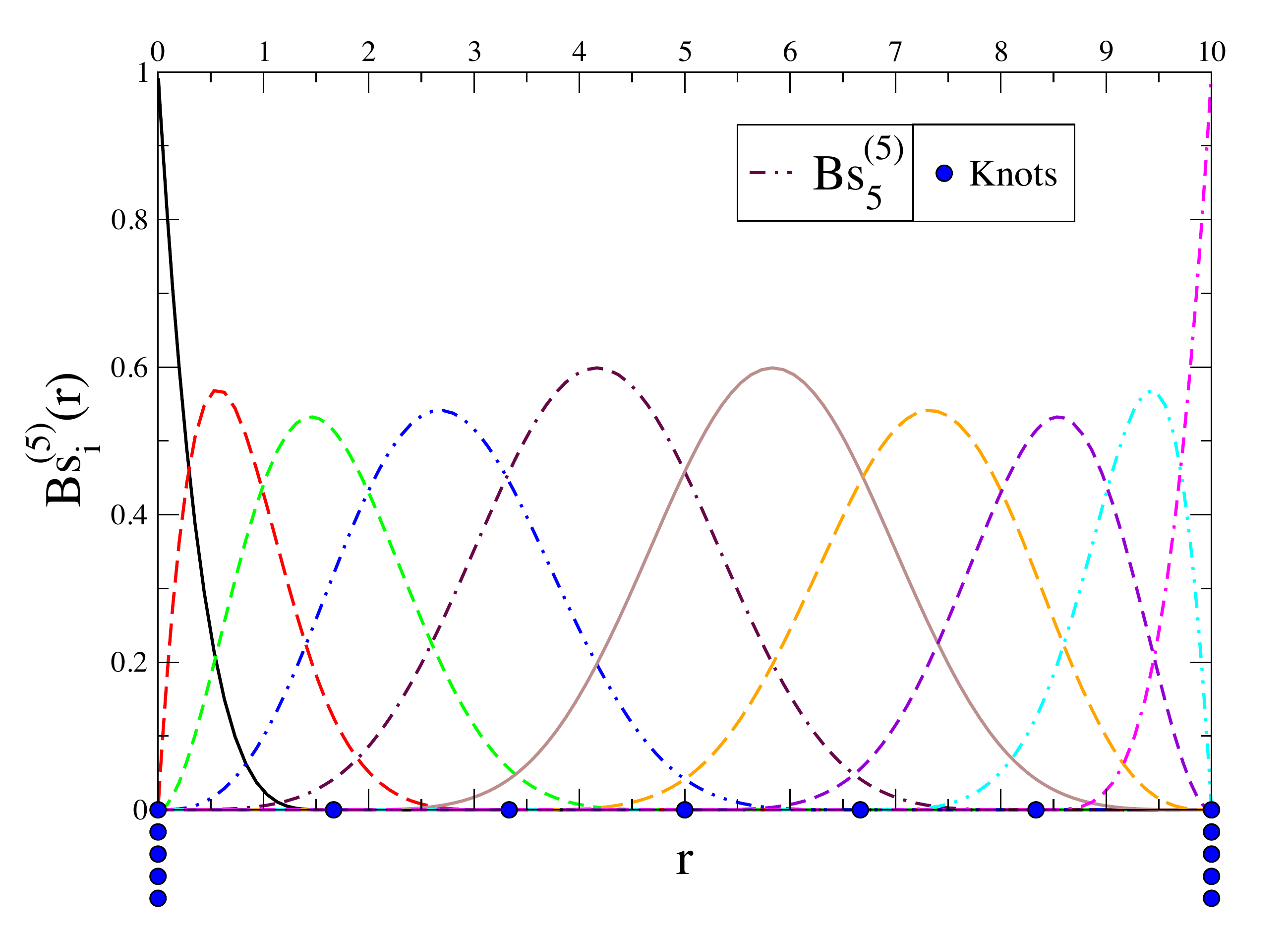}
\caption[Set of ten B-splines of order $k=5$.]{\label{fig:biplaneso}Set of ten B-splines with order $k=5$ originated from the knot sequence $\{0,0,0,0,0,1.6667,3.3333,5,6.6667,8.3333,10,10,10,10,10\}$, within a box of  length $L=10$. Multiple knots are included at the edges of the box to guarantee continuity and smoothness in the derivatives.}
\end{figure}

\section{General features of B-splines}\label{sec:bsplines}

In this appendix we present the B-splines basis set and its most important features. Set of B-splines are designed originally to generalize polynomials for the propose of approximating arbitrary functions~\citep{Schoenberg1946};  an exhaustive description of this particular base set and their properties can be found in the book of~\citep{deBoor1978,Bachau2001}. It is necessary to introduce some mathematical definitions in order to obtain a deep understanding of the B-splines set.

\begin{itemize}
\item The B-splines are piecewise, positive, and compact $L^2$ integrable polynomial functions of order $k$ (degree $k-1$) which are defined in a restricted space known as a {\textit box}. 
\item The polynomials of order $k$ (maximum degree $k-1$) are
\[p(x)=a_0+a_1x+\dots+a_{k-1}x^{k-1}\]
\item Consider an interval $I=[a,b]$ divided into $l$ subintervals $I_j=[\zeta_j,\zeta_{j+1}]$ by a sequence of $l+1$ points $\{\zeta_j\}$ in strict ascending order
\[a=\zeta_1<\zeta_2<\dots<\zeta_{l+1}=b\]
The $\zeta_j$ will be called {\textit Breakpoint} (\ac{bp}).
\item Now, with the aim to define adequate continuity conditions at each interior \ac{bp} $\zeta_j$ let us associate with them a second sequence of non-negative integers, $\nu_j$, $j=2,\dots,l$, $\nu_j\geq0$, which define continuity condition $C^{\nu_j-1}$ (any function which is continuos, on a given interval, together with its derivatives up to order $n$ is said to be of class $C^n$) at the associated  \ac{bp} $\zeta_j$. With the end \ac{bp}s $\zeta_1$ and $\zeta_{l+1}$ we associate $\nu_1=\nu_{l+1}=0$, that is we do not require any continuity.
\item The last  set of points  that we need to introduce is $\{t_i\}$ where each $t_i$ is called a {\textit knot}. This sequence of points is given in ascending order but they are not necessarily distinct $\{t_i\}_{i,\dots,m},$ where $t_1\le t_2\le\dots\le t_m$.  The $\{t_i\}$ sequence are associated with $\zeta_j$ and $\nu_j$ as follows:\\
\begin{eqnarray*}
t_1 &=& t_2 = \dots =t_{\mu_1}=\zeta_1;\hspace{15pt} \mu_1=k\\ \nonumber
t_{\mu_1+1} &=&\dots=t_{\mu_1+\mu_2}=\zeta_2\\ \nonumber
&\dots&\\ \nonumber
t_{p+1} &=&\dots=t_{p+\mu_i}=\zeta_i;\hspace{30pt} p=\sum_{r=1}^{i-1}\mu_r\\ \nonumber
&\dots&\\ \nonumber
t_{n+1} &=&\dots=t_{n+k}=\zeta_{l+1};\hspace{25pt}\mu_{l+1}=k\hspace{20pt}n=\sum_{r=1}^{l}\mu_r \nonumber
\end{eqnarray*}

where $\mu_j$ is the multiplicity of the knots $t_i$ at $\zeta_j$ and is given by $\mu_j=k-\nu_j$. The most common choice for knot multiplicity at inner \ac{bp}s is unity; corresponding to maximum continuity $C^{k-2}$. This is our choice in all calculations throughout the thesis.  With this choice the whole number of B-spline functions is fixed and given by $n=l+k-1$.
\item Any smooth function can be expressed as a linear combination of the $B_i$ and it will be called a \textit{ \ac{pp}-function} over $[a,b]$.
\[f=\sum_{i=1}^nc_iB_i\]
\end{itemize} 

The following qualitative features of B-splines provides us a depth understanding of their behavior:

\begin{enumerate}
\item A single B-spline $B(x)$ is defined by its order $k>0$, and a set of $k+1$ knots, $\{t_i,\dots,t_{i+k} \}$ such that $t_i<t_{i+k}$.
\item $B(x)$ is a \ac{pp}-function of order $k$ over $[t_i,t_{i+k}]$.
\item $B(x)>0$ for $x\in\hspace{2pt}(t_i,t_{i+k})$.
\item $B(x)=0$ for $x\notin\hspace{2pt}[t_i,t_{i+k}]$.
\item The knots need not be equidistant and the shape of B(x) changes smoothly with the change of the knots.
\item The B-splines are normalized as $\sum_iB_i(x)=1$ over the whole $[t_i, t_{i+k}]$.
\item The set of B-splines do not form an orthonormal set of basis functions. 
\end{enumerate}

\begin{figure}
\centering
\includegraphics[width=0.85\textwidth]{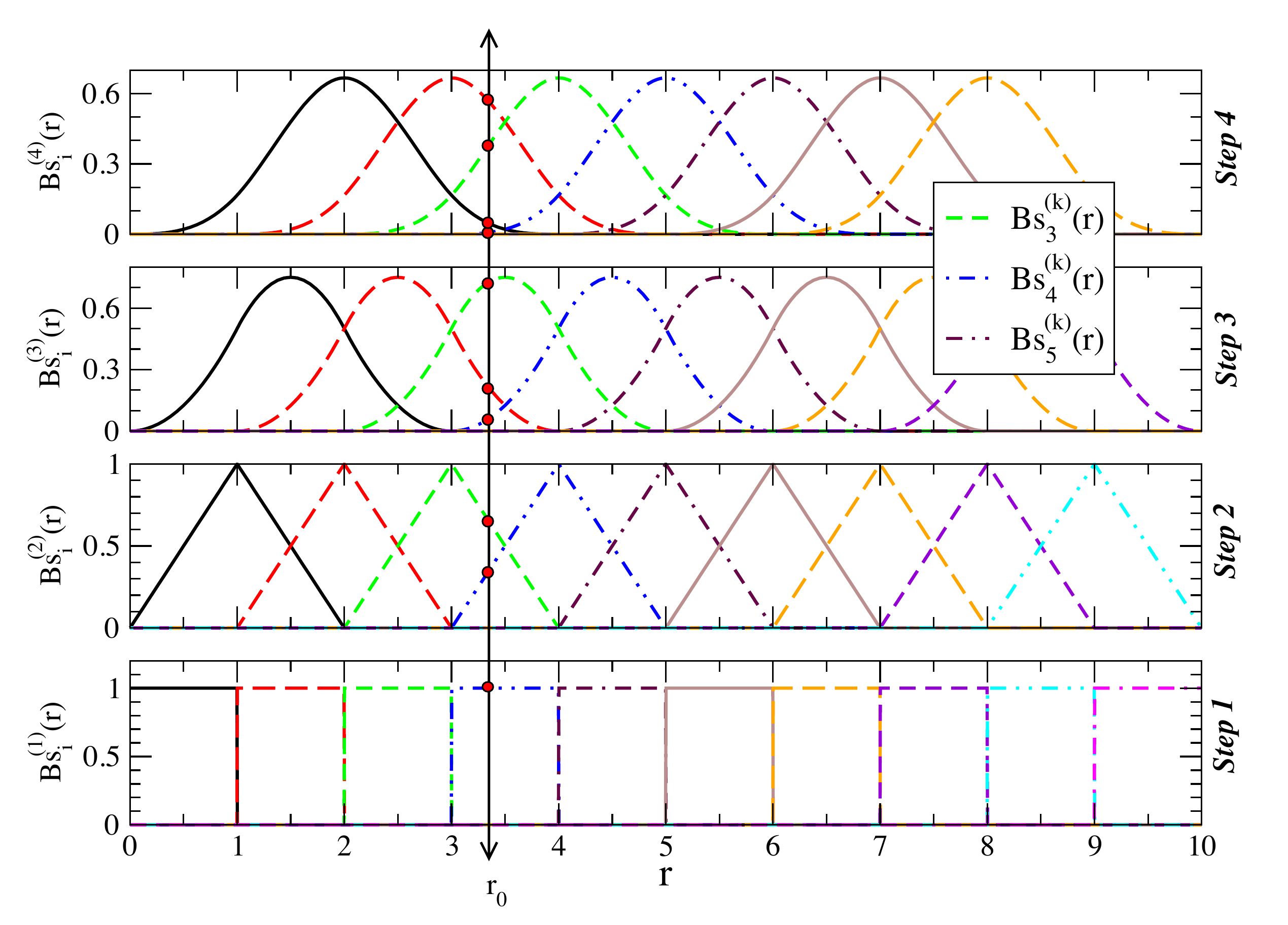}
\caption[Recursive evaluation of B-splines.]{\label{fig:biplanes}The B-splines are generated by a recursive evaluation scheme. This figure sketches the recursive method for evaluating B-splines up to order $k=4$, relative to de knot sequence $\{0,1,2,3,4,5,6,7,8,9,10\}$.   }
\end{figure}

\section{Building up B-splines function set}

Each interval $I_j=[\zeta_j,\zeta_{j+1}]=[t_i,t_{i+1}]$ is characterized by a pair of consecutive knots $t_i<t_{i+1}$. The point $t_i$ is called the left knot of interval $I_j$, and determines de indices of the $B_i$ contributing  over each interval $(t_i, t_{i+k})$. Additionally, over this interval exactly $k$ B-splines are nonzero, $B_j(x)\neq0$ for $j=i-k+1,\dots,i$. The first being $B_{i-k+1}$, which ends at $t_{i+1}$, and the last is $B_i$, which starts at $t_i$. Therefore, we have identically $B_i(x)\cdot B_j(x)=0$ for $|i-j|\geq k$.
In general a family of B-spline functions, $B_i(x)$, $i=1,\dots,n$ is completely defined given $k>0$, $n>0$ and a sequence of knots $\mathbf{t}=\{t_i\}_{i=1,\dots,n+k}$.

The B-spline functions can be generated by a recursive evaluation method. Each function satisfies the following recursion relation

\[B^k_i(x)=\frac{x-t_i}{t_{i+k-1}-t_i}B^{k-1}_{i}(x)+\frac{t_{i+k}-x}{t_{i+k}-t_{i+1}}B^{k-1}_{i+1}(x);\]

\noindent we must define the B-splines of order $k=1$ 
\[B_i^1(x)=1\hspace{15pt}t_i\leq x<t_{i+1}\hspace{15pt}\text{and}\hspace{15pt} B_i^1(x)=0 \hspace{15pt}\text{otherwise,}\]

\noindent this formula becomes the basis for the algorithm employed for the practical evaluation of B-splines: given an arbitrary point $x$,  one may generate, by means of recursion, the values of all the $k$ B-splines which are nonzero at x, as is illustrated in fig~\ref{fig:biplanes}. The derivative of a B-spline or order $k$, being a \ac{pp}-function or order $k-1$, can be also expressed as a linear combination of B-splines of the same order

\[\partial_xB^k_i(x)=\frac{k-1}{t_{i+k-1}-t_i}B^{k-1}_{i}(x)-\frac{k-1}{t_{i+k}-t_{i+1}}B^{k-1}_{i+1}(x).\]

The practical evaluation of the $k$ B-spline function for an arbitrarily value of $x$ is done using the \texttt{FORTRAN} routines originally developed by~\citep{deBoor1978}. The routine \texttt{BSPLVP} provides a numerical implementation of recursion relation both for B-splines and his derivatives; this routine requires as input,  as you can expect, the order $k$ of B-spline, the knot sequence $t_i$, and the desired point $x$ where we want to evaluate the functions.  All this quantities, as we saw previously, defines the B-spline functions. Additionally, we must specify the index of the left-closest knot $t_i$ to x which can be calculated using the routine \texttt{INTERV}. This routine provides the $k$ B-splines functions  that are nonzero at $x$, i.e., $B^{(k)}_{i-k+1}(x),\dots,B^{(k)}_{i}(x)$.

\section{Breakpoint sequences}\label{sec:bps}
 
To successfully achieve the desired behavior of \ac{WF}s that we want to calculate, by means of a variational approach using B-splines, it is necessary to choose the adequate sequence of \ac{bp}s $\zeta_j$.  Most realistic calculations, in atomic and molecular physics,  have made use of different sequences which are selected after a deep consideration of the electron density distribution within a box of length $L$. Actually, there is a collection of different sets  of \ac{bp}s to be chosen. Now, let us review the most commonly used.

\subsection{Linear sequence}

This is the simplest of the all knot sequences that we might consider. It divides, by means of  $l$ \ac{bp}s,  an arbitrary interval $[r_{min}, r_{max}]$ into $(l-1)$ segments of {\textit equal} length. The practical implementation of this sequence is straightforward.

\begin{equation}\label{eq:linearsequence}
\zeta_j=r_{min}+\frac{r_{max}-r_{min}}{l-1}(j-1)\hspace{20pt}j=1,\dots,l.
\end{equation}

\begin{figure}
\centering
\includegraphics[width=0.85\textwidth]{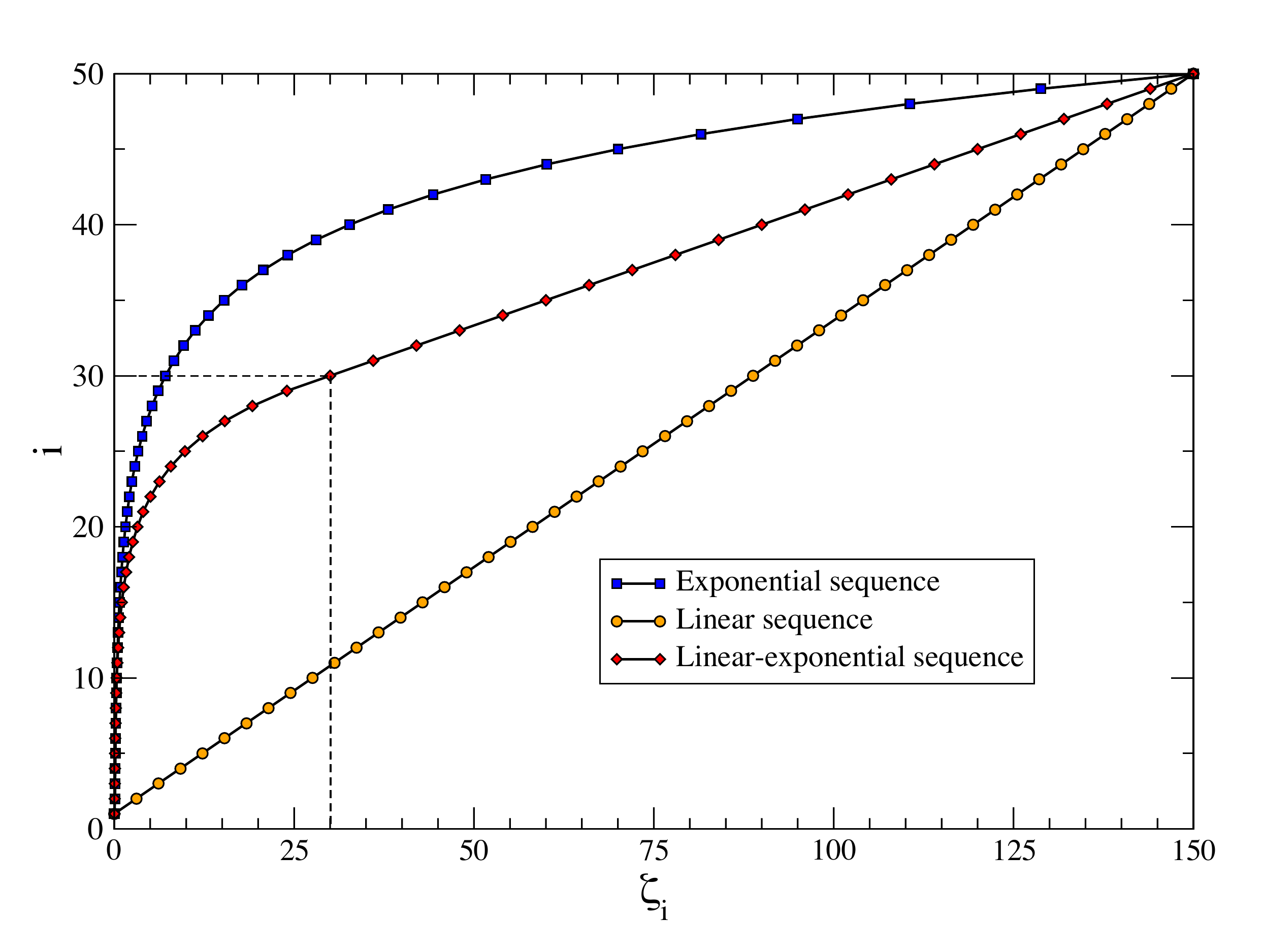}
\caption{\label{fig:bpsequences} Comparison of three sequences of breakpoints: linear sequence (circular points), linear-exponential sequence (diamond points), and the  exponential sequence (square points).}
\end{figure}

This kind of \ac{bp}s is not able to describe appropriately the bound states of  H-like atoms. These particular states are strongly localized in regions near the atomic nucleus~\citep{Castro-Granados2012}. Large radial boxes along with a finite number of \ac{bp}s, equally spaced, makes the radial spacing at short distances inappropriate to describe the fine details of bound states wave functions. On the other hand, linear sequences shows to be superior when describing unbound states, since the equally spaced succession of \ac{bp}s offers a high flexibility for the basis to describe the highly oscillatory behavior of the scattering-like wave functions. In Fig.~\ref{fig:bpsequences} we plot the linear sequence of \ac{bp}s, together with the exponential and linear-exponential sequences.

\subsection{Exponential sequence}\label{sec:expseq}

In many calculations we need to  adequately describe the \ac{WF} or the density of probability for a many-body system in which the particles are strongly localized in some region of space. Bound states of atoms have this localization very close to the atomic nucleus. For a proper description of these states we need to choose a sequence of knots that must be accumulated very close to the atomic nucleus without making strong considerations about the properties of the sequence in some other regions, as is shown in Fig.~\ref{fig:bpsequences}. The exponential sequence can be expressed as

\begin{equation}\label{eq:exponentialsequence}
\zeta_j=\delta e^{\alpha(j-1)}\hspace{20pt}\alpha=\frac{ln\left(\frac{L}{\delta}\right)}{l-1} \hspace{20pt}j=1,\dots,l.,
\end{equation}
where $\delta$ denotes an arbitrarily initial non-zero value for the sequence. It is customarily to choose this point to be $\delta\leq 0.1$ with the appropriate units.  Even though this sequence works fine for bound states, it does not for unbound ones. The exponential spacing at large distances is not suitable for the strong oscillations of the continuum \ac{WF}.  

\subsection{Exponential-linear sequence}

Sometimes it is necessary to deal with problems in which we need to describe several states with radically different behaviors. Consequently, we can define combined sequences to treat these situations. Therefore, if our intention is to describe, in some specific procedure, both bound and unbound states evenly we require a combined sequence with at least two characteristics: accumulation of \ac{bp}s near the nucleus, and a flexible distribution of them in order to get an adequate description of highly oscillatory functions. With the adequate conditions of continuity of the functions that defines the whole sequence and its derivative, we can define a piecewise sequence which changes smoothly between elementary sequences (like linear or exponential) as is shown in Fig.~\ref{fig:bpsequences}.

\chapter{Hydrogen {\textit \`a la} B-Splines}\label{ch:hybsplines}

Before we study the hydrogen atom using a B-spline basis, we briefly review its general full quantum solution. This topic is treated in more detail in almost every introductory textbook of quantum mechanics and atomic physics~\citep{Cohen-tannoudji1977,Bransden2003,Friedrich2005}. 

\section{Full quantum solution for hydrogen atom}\label{sec:fullhy}

The Hamiltonian for a system of two particles, of masses $m_e$ and $m_p$, which are interacting through a time-independent central potential $V({r})$ with $r=|\mathbf{r}_e-\mathbf{r}_p|$ is given by

\begin{equation}\label{eq:hydhamil}
 H=\frac{\hat{\mbf{p}}^{2}_e}{2m_{e}}+\frac{\hat{\mbf{p}}^{2}_p}{2m_{p}}+V({r}),
\end{equation}
where $\mbf{\hat{p}}_j=-i\hbar\nabla_{\mbf{r}_j}$ is the momentum operator for each particle. 

The Schr\"odinger equation for a hydrogen atom can be written as
\begin{equation}\label{eq:schro1}
 i\hbar\frac{\partial}{\partial t}\psi(\mbf{r}_{e},\mbf{r}_{p},t)=
 \left[-\frac{\hbar^2}{2m_{e}}\nabla^2_{\mbf{r}_{e}}-\frac{\hbar^2}{2m_{p}}
 \nabla^2_{\mbf{r}_{p}}+V({r})
 \right]\psi(\mbf{r}_{e},\mbf{r}_{p},t),
\end{equation}
where $m_p$ ($\mbf{r}_{p}$) is the mass (coordinate) of the proton, $m_e$ ($\mbf{r}_{e}$) is the mass (coordinate) of the electron, and $V({r})$ is the Coulomb potential that depends only on  the distance between the proton and the electron. As done in every textbook, we introduce the relative coordinate $\mbf{r}$ and the coordinate of the center of mass $\mbf{R}$, 
\begin{equation}
 \mbf{r}=\mbf{r}_{e}-\mbf{r}_{p},~~\mbf{R}=\frac{m_{e}\mbf{r}_{e}+m_{p}\mbf{r}_{
p}}{M},\label{eq:coor_trans}
\end{equation}
here, $M=m_{e}+m_{p}$ is the total mass of the hydrogen atom.  In these new coordinates, the Schr\"odinger equation~\ref{eq:schro1} becomes
\begin{equation}\label{eq:schro2}
i\hbar\frac{\partial}{\partial
t}\psi(\mbf{R},\mbf{r},t)=\left[-\frac{\hbar^2}{2M}\nabla^2_{\mbf{R}}-\frac{\hbar^2}{2\mu}\nabla^2_{
\mbf{r}}+V({r})\right]\psi(\mbf{R},\mbf{r},t),
\end{equation}
where $\mu=m_{e}m_{p}/(m_{e}+m_{p})$ is the reduced mass. So, two separations of the equation~\ref{eq:schro2} can be made. In the first place, the time dependence shall be separated since the potential is time-independent. After that, the spatial part of the \ac{WF} shall be separated into a product of functions; one of them as a function of  centre of mass coordinate $\mbf{R}$ and the other one as a function of the relative coordinate $\mbf{r}$. Therefore, the whole \ac{WF} can be written as
\begin{equation}
\label{eq:sep}
\psi(\mbf{R},\mbf{r},t)=\psi_{CM}(\mbf{R})\psi_{r}(\mbf{r})e^{-\frac{i}{\hbar}(E_{CM}+E)t}\,,
\end{equation}
where $\psi_{CM}(\mbf{R})$ and $\psi_{r}(\mbf{r})$ satisfy, respectively, 
\begin{align}
-\frac{\hbar^2}{2M}\nabla^2_{\mbf{R}}\psi_{CM}(\mbf{R}) &= E_{CM}\psi_{CM}(\mbf{R}),\label{eq:centermasseqn}\\
-\frac{\hbar^2}{2\mu}\nabla^2_{\mbf{r}}\psi_{r}(\mbf{r})+V({r})\psi_{r}(\mbf{r})  &= E_{r}\psi_r(\mbf{r}).\label{eq:relativeeqn} 
\end{align}

This shows that the motion of a hydrogen atom can be separated into two independent parts: $\psi_{CM}(\mbf{R})$ describes the motion of a free particle with mass $M$ and energy $E_{CM}$, while $\psi_r(\mbf{r})$ describes the motion of a particle of reduced mass $\mu$  in a Coulomb potential $V({r})$. However, this by no means implies that the motions of the electron and the proton can be described by two independent \ac{WF}s. Note that both equations~\ref{eq:centermasseqn} and~\ref{eq:relativeeqn} are time-independent. The total energy of the system is clearly $E_{tot}=E_{CM}+E_r$. 

As suggested above, we are considering an hydrogen-like atom with an atomic nucleus of total charge $Ze$ and an electron of charge $-e$. The Coulomb interaction between   the positive nucleus and the electron is expressed by the Coulomb potential 

\begin{equation}
\label{eq:coulomb}
V({r})=-\frac{Ze^2}{(4\pi\epsilon_0)r},
\end{equation}
where r is the distance between the two particles and $\epsilon_0$ is the vacuum permittivity.  Consequently, using this potential into the equation~\ref{eq:relativeeqn} we obtain, working in the centre of mass system, the following Schr\"odinger equation for the relative motion

\begin{equation}\label{eq:relativeeqn1} 
\left [-\frac{\hbar^2}{2\mu}\nabla^2-\frac{Ze^2}{(4\pi\epsilon_0)r}\right]\psi(\mbf{r})  = E\psi(\mbf{r}).
\end{equation}

Owing to the interaction potential~\ref{eq:coulomb} is central, the Schr\"odinger equation~\ref{eq:relativeeqn1} may be separated in spherical coordinates. In other words we may write the \ac{WF}, as a particular solution of this equation, in the following way

\begin{equation}\label{eq:solhyd1} 
\psi_{E,l,m}(r,\theta,\phi)=R_{E,l}({r})\mathcal{Y}^l_{m_l}(\theta,\phi),
\end{equation}
where $\mathcal{Y}^l_{m_l}(\theta,\phi)$ is a spherical harmonic corresponding to the orbital angular momentum quantum number $l$ and to the magnetic quantum number $m_l$ (with $m=-l,-l+1,\dots,l$). In this way, the functions $\mathcal{U}_{E,l}({r})=rR_{E,l}({r})$ satisfy the radial equation written in \ac{a.u.}

\begin{equation}\label{eq:radschr} 
\left[-\frac{1}{2}\frac{d^2}{dr^2}+V_{eff}({r})\right]\mathcal{U}_{E,l}({r})=E\mathcal{U}_{E,l}({r}),
\end{equation}
where
\begin{equation}
\label{eq:veff}
V_{eff}({r})=-\frac{Z}{r}+\frac{l(l+1)}{2r^2}
\end{equation}
is called the effective potential. Consequently, the problem of solving the Schr\"odinger equation~\ref{eq:schro2} reduces to that of solving the radial one-dimensional equation~\ref{eq:radschr} which corresponds to a particle of mass $\mu$ moving in an effective potential $V_{eff}$. 

The analytic procedure to obtain a particular solution of equation~\ref{eq:schro2} is shown in almost every textbook of quantum mechanics and atomic physics. So, the energies and the normalised radial functions for the bound states of hydrogenic atoms may be written, in \ac{a.u.}, as

\begin{equation}
\label{eq:energy}
E_n=-\frac{Z^2}{2n^2}
\end{equation}

\begin{equation}
\label{eq:radialf}
R_{nl}({r})=-\left\{\left(\frac{2Z}{n}\right)^{3}\frac{(n-l-1)!}{2n[(n+l)!]^3}\right\}^{\frac{1}{2}}e^{-\frac{\rho}{2}}\rho^l L^{2l+1}_{n+1}(\rho),
\end{equation}
with
\begin{equation}
\label{eq:rho}
\rho=\frac{2Z}{n}r,
\end{equation}
and where the $L^i _k(\rho)$ are the associated Laguerre polynomials.

\newpage

\section{Numerical approach to hydrogen atom using the B-spline basis}\label{sec:hybsplines}

In this section we introduce the fundamental scheme for the description of the atomic structure of one-electron systems using B-splines functions~\citep{deBoor1978, Bachau2001}. Even though this scheme has a wide useful range, for instance one-valence electron atoms and closed-shell systems such as rare gases, we only take the case of hydrogen atom, the most elemental atom with only one proton and one electron, and any hydrogen-like ions characterised by the Coulomb potential. One fundamental feature of B-splines basis is its flexibility; the basis allows us to calculate both the energies and the \ac{WF}s in a given central potential. This calculation may be may be performed very efficiently with
arbitrary accuracy up to the machine precision, with a present-day laptop computer.
The radial equation~\ref{eq:radschr} may be solved numerically using the B-spline basis in a subspace, defined by the basis itself, making the assumption that the radial function $\mathcal{U}_{E,l}({r})$, with the initial  condition $\mathcal{U}_{E,l}(0)=0$, can be approximated, in a given box, by B-spline functions. Consequently, we may naturally expand the solution of radial equation belonging to that subspace in terms of the B-spline basis set as 

\begin{equation}
\label{eq:radialbs}
\mathcal{U}_{E,l}({r})=\sum_{i=1}^{N_{max}}c_i^{E,l}B_i({r}),
\end{equation}
where $B_i({r})$ is the $i^{\text{th}}$ B-spline of order $k$ as defined above. The B-spline method requires the definition of a set of knot sequence which depend, as we have said before, on the following parameters: a set of mesh points called the \ac{bp}s defined in $[0,L]$, the order $k$ of the \ac{pp}, and the continuity conditions at each \ac{bp}. We commonly choose these parameters according to the physical problem at hand. In fact,  for the hydrogen-like atoms, we must divide the box in which we solve the problem into segments whose endpoints form the \ac{bp} sequence. As suggested above, we may freely select any  \ac{bp} sequence; the optimal choice will depend on the type of result we are particularly interested in. Take the case of bound states in which the \ac{WF} has a highly drift towards nucleus and it vanishes at some finite distance; an optimal choice for this particular problem has a suitable accumulation of \ac{bp}s in this specific region of the box. The exponential sequence, as can be seen in section~\ref{sec:bps}, may describe with accuracy the behavior of this kind of \ac{WF} due to its strong accumulation of \ac{bp}s when $r\rightarrow0$. On the contrary, the linear sequence, in which the segments between two consecutive \ac{bp}s have the same constant value, can describe properly the \ac{WF} of continuum states of hydrogen-like ions which oscillates indefinitely in their radial part. If, in contrast, our main interest is to describe simultaneously both bound and continuum states, we must choose any mixed \ac{bp}s sequence. Actually, a mixed sequence, made up with two elementary sequences such as an exponential sequence up to some distance from the nucleus and a linear sequence afterwards, might successfully describe both types of \ac{WF}s.  Finally, we may establish the maximum continuity conditions, $\nu_j=k-1$, at each \ac{bp} in the box without any restriction, with this choice the multiplicity of knot is just $\mu_j=1$ corresponding to continuity $C^{k-2}$.  Nevertheless, in order to satisfy the boundary conditions, we require the solution to vanish at the endpoints, tat is $\mathcal{U}_{E,l}(0)=\mathcal{U}_{E,l  }(L)=0$. This conditions are met either by removing the first and the last B-splines from the basis set or by setting the continuity conditions $\nu_j=1$ at the boundaries.

The hydrogen energies and \ac{WF}, corresponding to bound states, which satisfies equation~\ref{eq:radschr} are calculated, for a given fixed angular momentum $l$, by solving the system of $N_{max}$ (size of B-spline basis) linear equations obtained after substituting~\ref{eq:radialbs} into~\ref{eq:radschr} and then projecting on $B_j({r})$. This procedure is formally equivalent, when it is written in matrix form, to solving the following generalized eigensystem problem:

\begin{equation}
\label{eq:radialeigen}
\mathbf{H}_l\cdot\mathbf{c}=E\mathbf{S}\cdot\mathbf{c},
\end{equation}
for $E_{n,l}$, $\mbf{c}^{nl}=\{c^{nl}_i\}$ with $i=1,\dots,N_{max}$, and where

\begin{align}
\label{eq:melemham}
(\mathbf{H}_l)_{ij}&=-\frac{1}{2}\int_0^LB_i({r})\frac{d^2}{dr^2}B_j({r})dr - Z\int_0^L\frac{B_i({r})B_j({r})}{r}dr\\ \nonumber
& +\frac{l(l+1)}{2}\int_0^L\frac{B_i({r})B_j({r})}{r^2}dr,
\end{align}
\begin{equation}
\label{eq:radialeigen1}
(\mbf{S})_{ij}=\int_0^LB_i({r})B_j({r})dr.
\end{equation}

The matrix $\mbf{S}$ is called the overlap matrix. It is originated from the fact that B-splines do not form an orthonormal set of basis functions. The hydrogenic orbitals obtained in terms of B-splines are then used in the configurations interaction method to solve the helium electronic structure, as described in the chapter~\ref{sec:qmd} in this thesis.

\chapter{Supplementary Graphics }\label{ch:supgraphics}
\section{Arguments of the integrals of the theoretic information measures }

In this appendix we present some additional figures which emphasise the conclusions established in earlier chapters in relation with the bound and doubly excited states of helium atom.

\section{Figures of the von Neumann entropy}

\begin{figure}[h]
\centering
\includegraphics[width=0.75\textwidth]{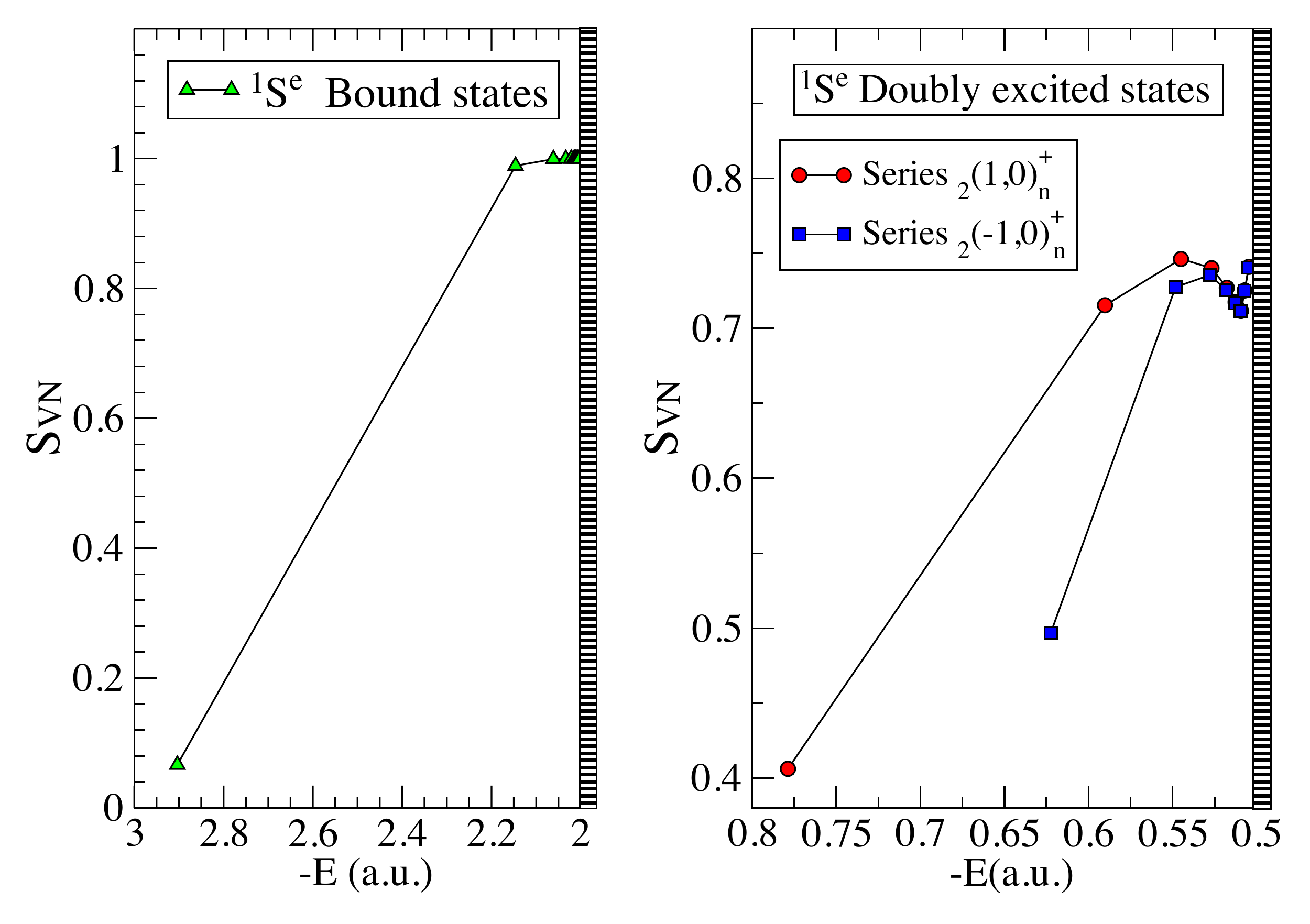}
\caption{\label{fig:entanglementVN1Se}von Neumann entropy $S_{VN}(\rho)$ for the ground and the lowest singly excited states for the $^1S^e$ symmetry below the first ionization threshold (left panel) and for the two series  $_2(1,0)^+_n$ and $_2(1,0)^+_n$ of resonances belonging to symmetry $^1S^e$ below the second ionization threshold (right panel).}
\end{figure}

\begin{figure}[h]
\centering
\includegraphics[width=0.75\textwidth]{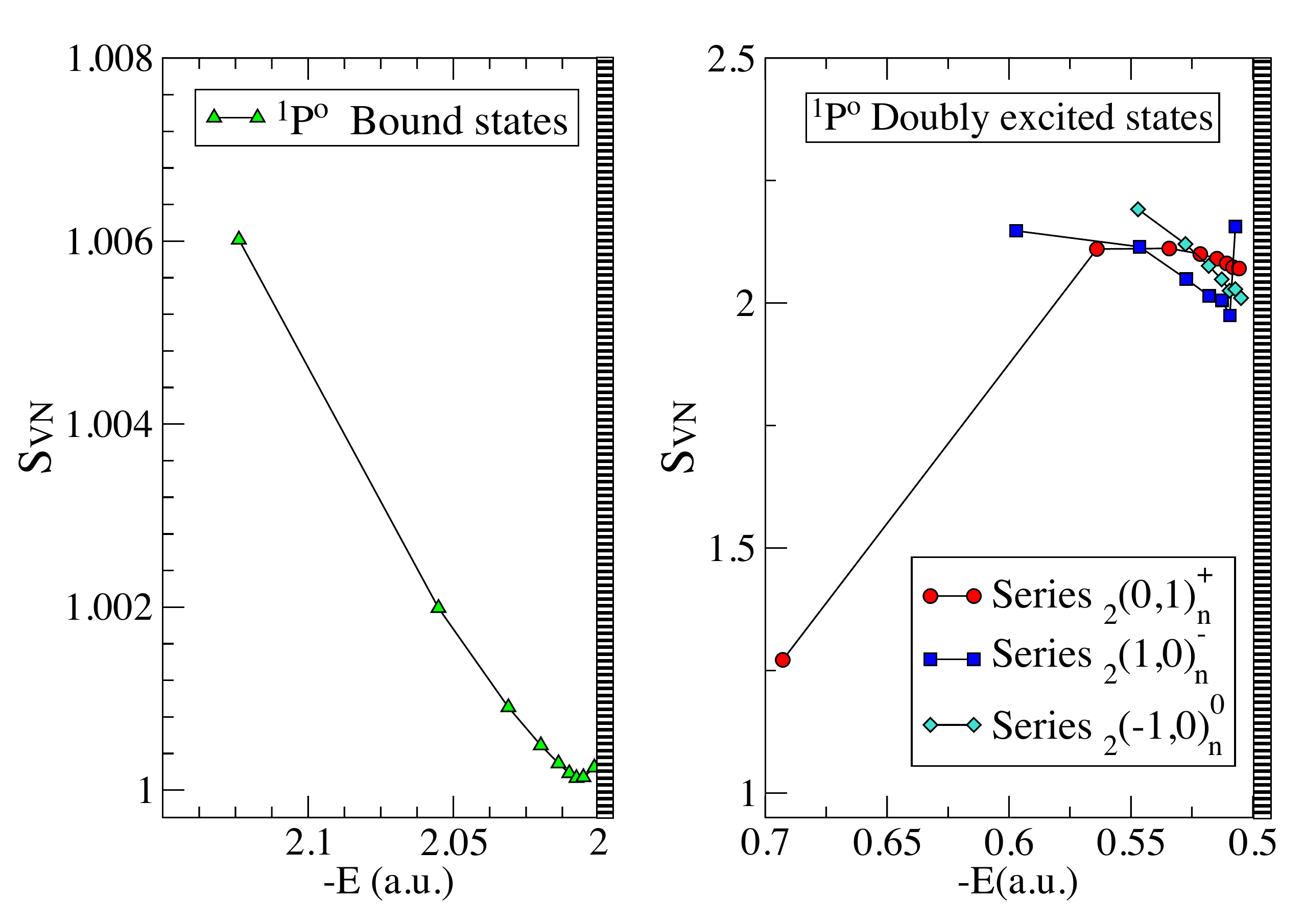}
\caption{\label{fig:entanglementVN1Po}von Neumann entropy $S_{VN}(\rho)$ for the lowest singly excited states of $^1P^o$ symmetry below the first ionization threshold (left panel) and for the three series $_2(0,1)_n^+$, $_2(1,0)_n^-$, and $_2(-1,0)_n^0$ of resonant doubly  excited states of symmetry $^1P^o$ below the second ionization threshold (right panel).}
\end{figure}

\begin{figure}[h]
\centering
\includegraphics[width=0.75\textwidth]{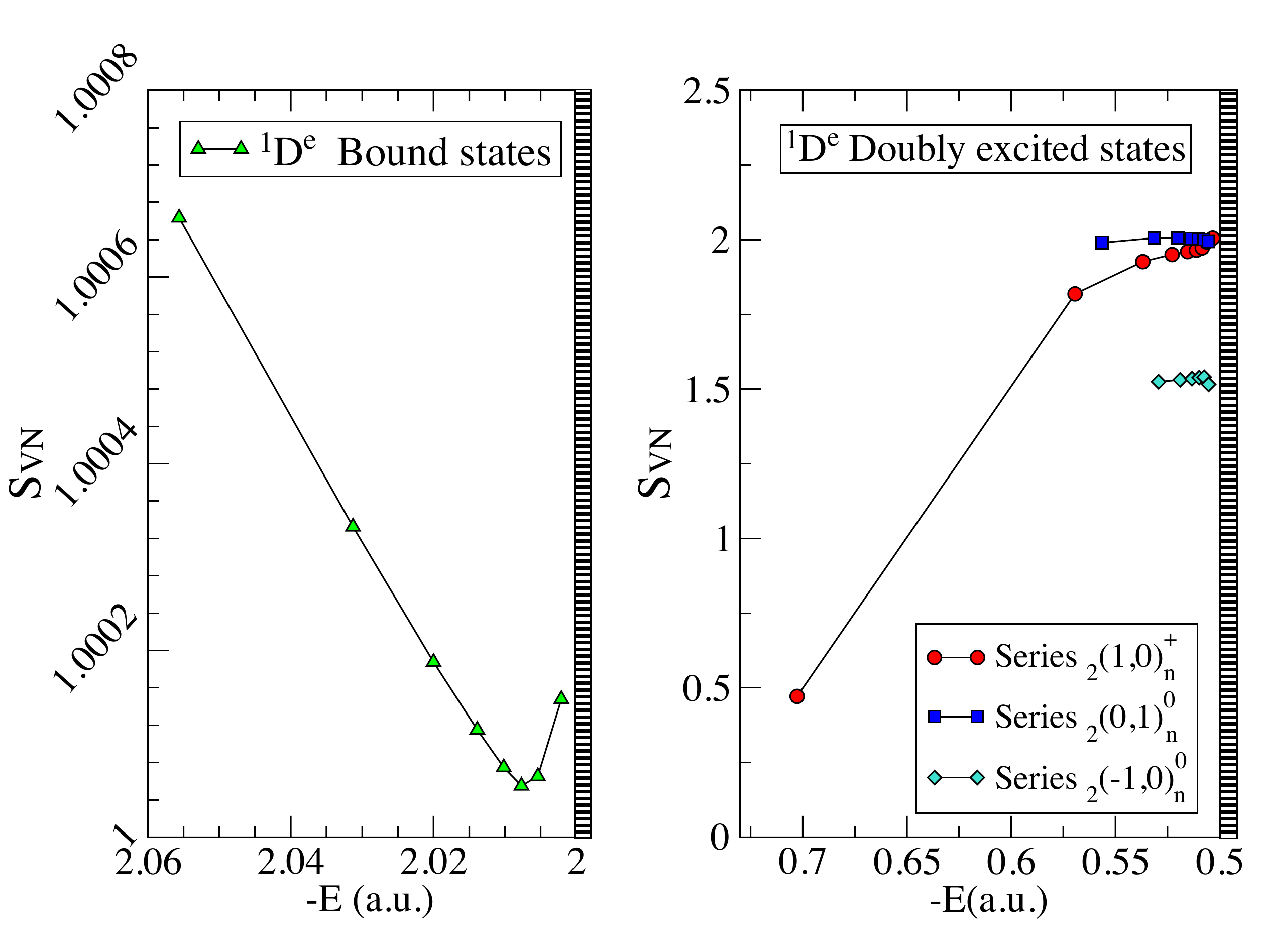}
\caption{\label{fig:entanglementVN1De}von Neumann entropy $S_{VN}(\rho)$ for the lowest singly excited states of $^1D^e$ symmetry below the first ionization threshold (left panel) and for the three series $_2(1,0)_n^+$, $_2(0,1)_n^0$, and $_2(-1,0)_n^0$ of resonant doubly  excited states of symmetry $^1D^e$ below the second ionization threshold (right panel).}
\end{figure}

\begin{figure}[h]
\centering
\includegraphics[width=0.75\textwidth]{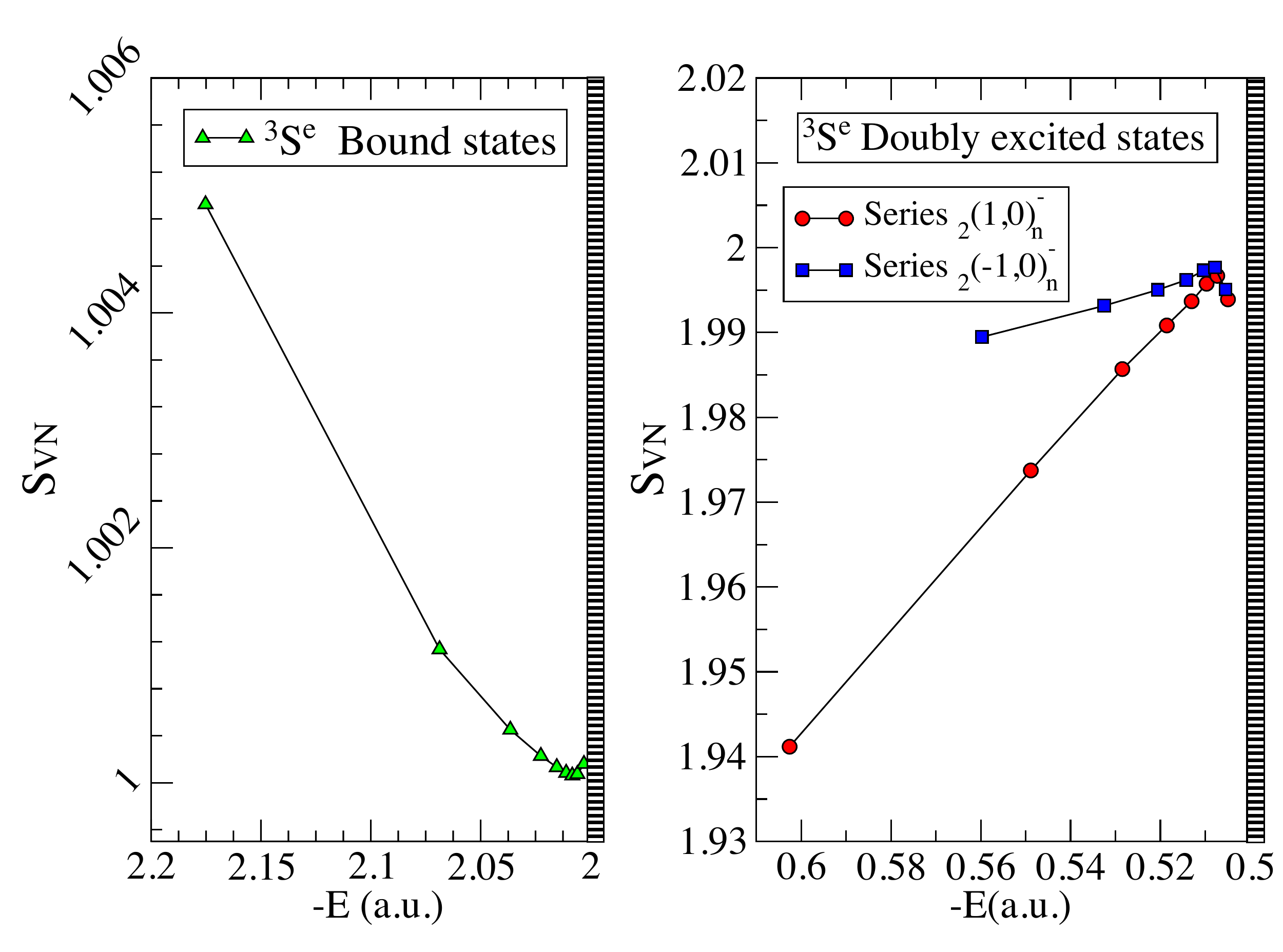}
\caption{\label{fig:entanglementVN3Se}von Neumann entropy $S_{VN}(\rho)$ for the lowest singly excited states of $^3S^e$ symmetry below the first ionization threshold (left panel) and for the two series $_2(1,0)^-_n$  and $_2(-1,0)^-_n$ of resonant doubly  excited states of symmetry $^3S^e$ below the second ionization threshold (right panel).}
\end{figure}

\begin{figure}[h]
\centering
\includegraphics[width=0.75\textwidth]{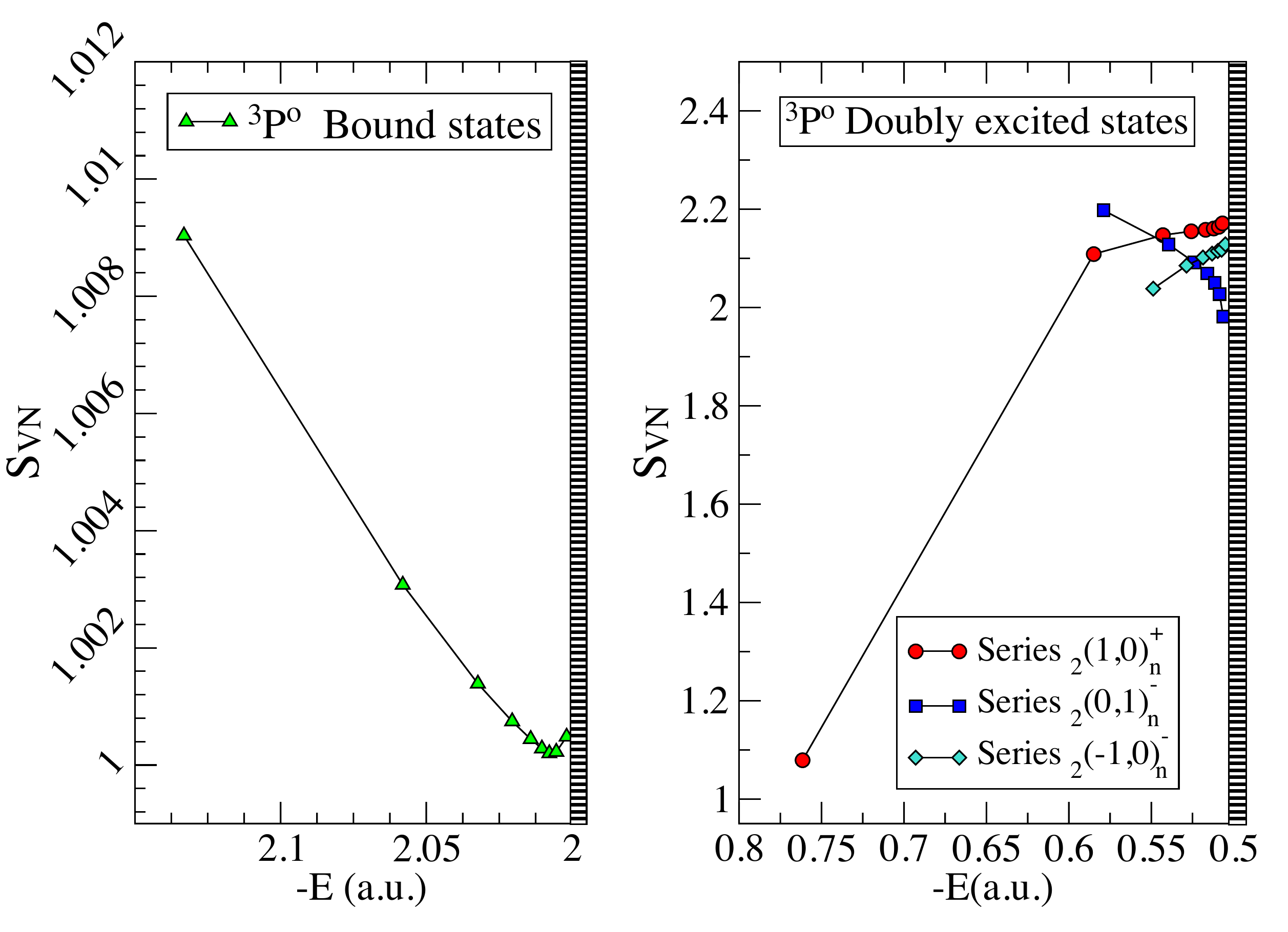}
\caption{\label{fig:entanglementVN3Po}von Neumann entropy $S_{VN}(\rho)$ for the lowest singly excited states of $^3P^o$ symmetry below the first ionization threshold (left panel) and for the three series $_2(1,0)^-_n$, $_2(0,1)^0_n$, and $_2(-1,0)^0_n$ of resonant doubly  excited states of symmetry $^3P^o$ below the second ionization threshold (right panel).}
\end{figure}

\begin{figure}[h]
\centering
\includegraphics[width=0.75\textwidth]{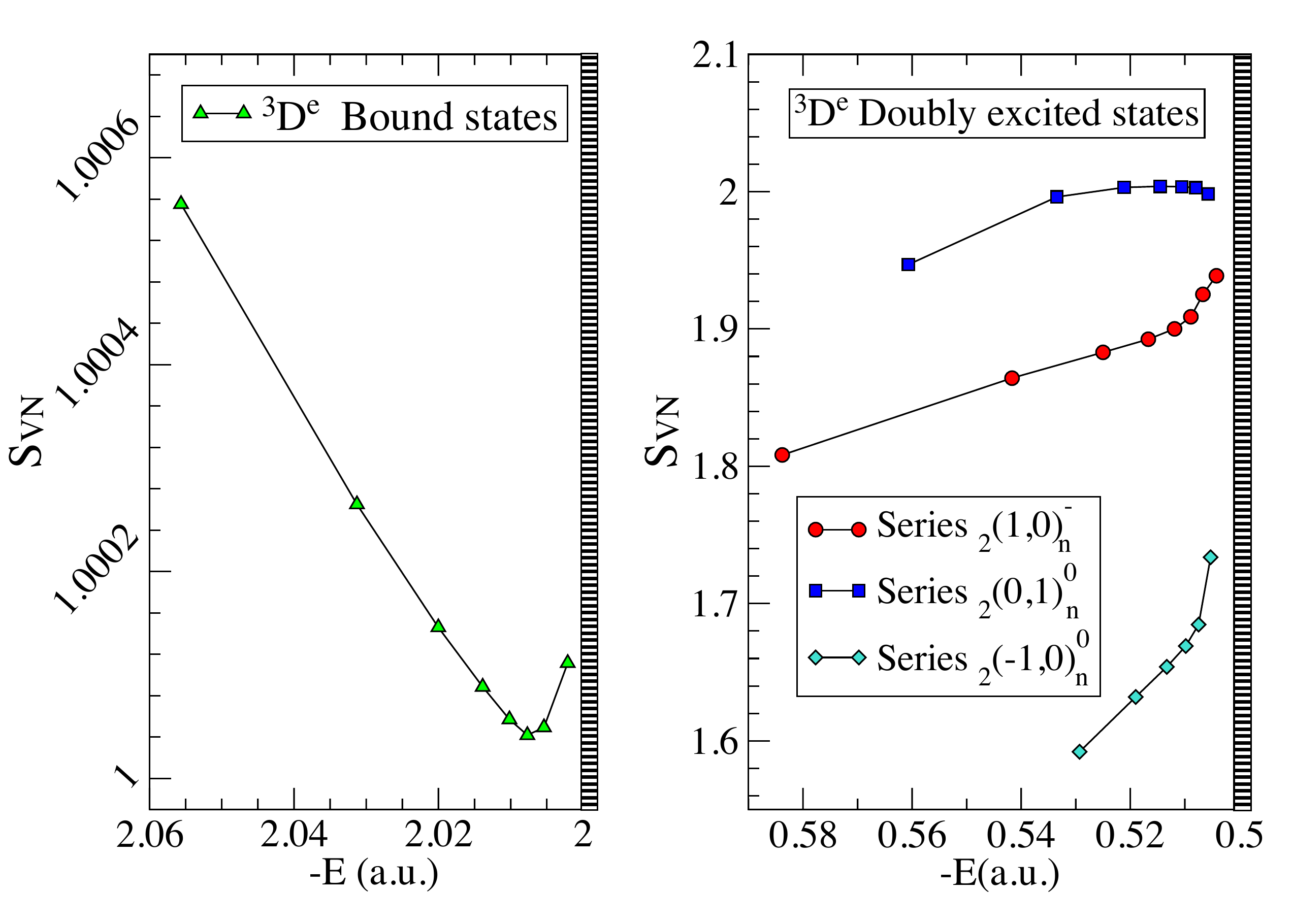}
\caption{\label{fig:entanglementVN3De}von Neumann entropy $S_{VN}(\rho)$ for the lowest singly excited states of $^3D^e$ symmetry below the first ionization threshold (left panel) and for the three series $_2(1,0)^+_n$, $_2(0,1)^-_n$, and $_2(-1,0)^-_n$ of resonant doubly  excited states of symmetry $^3D^e$ below the second ionization threshold (right panel).}
\end{figure}

\cleardoublepage
\manualmark
\markboth{\spacedlowsmallcaps{\bibname}}{\spacedlowsmallcaps{\bibname}} 
\refstepcounter{dummy}
\addtocontents{toc}{\protect\vspace{\beforebibskip}} 
\addcontentsline{toc}{chapter}{\tocEntry{\bibname}}
\bibliographystyle{cell}
\label{app:bibliography} 
\bibliography{MasterThesis}
\end{document}